\documentclass[rmp,amssymb,amsmath,amsfonts,a4paper,
12pt,nofootinbib,eqsecnum]{revtex4}
\usepackage[isolatin]{inputenc}
\usepackage{graphicx}
\usepackage{color}
\bibpunct{[}{]}{,}{a}{,}{,}
\graphicspath{{figures/}}

\newcommand{\JA}{{\mathbb A}}
\newcommand{\JE}{{\mathbb E}}
\newcommand{\JG}{{\mathbb G}}
\newcommand{\JM}{{\mathbb M}}
\newcommand{\JO}{{\mathbb O}}
\newcommand{\JP}{{\mathbb P}}
\newcommand{\JQ}{{\mathbb Q}}

\newcommand{\JU}{{1\hspace{-0.52ex}\mathrm{l}}}

\newcommand{\llongleftarrow}{\leftarrow\!\!\!-\!\!\!-\!\!\!-\!\!\!-\!\!\!-\!\!\!-\!\!\!-\!\!\!-\!\!\!-\!\!\!\!} 
\newcommand{\rmi}{\mathrm{i}}
\newcommand{\rr}{\mathsf{r}}
\newcommand{\pp}{\mathsf{p}}
\newcommand{\xx}{\mathsf{x}}
\newcommand{\yy}{\mathsf{y}}
\newcommand{\zz}{\mathsf{z}}
\newcommand{\oo}{\mathsf{o}}
\newcommand{\kk}{\mathsf{k}}
\newcommand{\vx}{\mathsf{x}}
\newcommand{\vy}{\mathsf{y}}
\newcommand{\tg}{\mathsf{g}}
\newcommand{\CZ}{\mathcal{Z}}
\newcommand{\CD}{\mathcal{D}}
\newcommand{\CE}{\mathcal{E}}
\newcommand{\CM}{\mathcal{M}}

\newcommand{\CR}{\mathcal{R}}
\newcommand{\CO}{\mathcal{O}}
\newcommand{\CV}{\mathcal{V}}

\newcommand{\CS}{\mathcal{S}}
\newcommand{\bren}{b_\mathrm{r}}

\newcommand{\Zfrak}{\mathfrak{Z}}

\newcommand{\Rfrak}{\mathfrak{R}}
\newcommand{\Lfrak}{\mathfrak{L}}

\newcommand{\BBO}{\mathbb{O}} \newcommand{\BBP}{\mathbb{P}}
 \newcommand{\BBI}{\mathbb{I}}
\newcommand{\qq}{\mathsf{q}}
\newcommand{\Sren}{S_{\mathrm{ren}}}
\renewcommand{\bren}{b_{\mathrm{r}}}
\newcommand{\gren}{g_{\mathrm{r}}}
\newcommand{\Lfren}{\mathfrak{L}_{\mathrm{ren}}}
\newcommand{\LfMS}{\mathfrak{L}_{\mathrm{MS}}}

\newcommand{\mvar}{{m_{\mathrm{var}}}}
\newcommand{\E}{\varepsilon}
\newcommand{\A}{\alpha}

\newcommand{\half}{\frac{1}{2}}

\newcommand{\lts}{
    {\raisebox{0ex}{\parbox{0.5ex}{
       \setlength{\unitlength}{0.5ex}
       \begin{picture}(1,12)
          \thinlines
          \put(0.5,-0.25){\line(0,1){12}}
       \end{picture}
    }}}
}
\newcommand{\rmd}{\mathrm{d}}
\newcommand{\rme}{\mathrm{e}}
\newcommand{\ind}[1]{\mathrm{#1}}
\newcommand {\p}{\partial}
\newcommand{\nn}{\nonumber}
\newcommand {\eq}[1]{(\ref{#1})}
\newcommand {\Eq}[1]{Eq.\hspace{0.55ex}(\ref{#1})}
\newcommand {\Eqs}[1]{Eqs.\hspace{0.55ex}(\ref{#1})}
\newcommand {\ds}{\displaystyle}

\newcommand{\tr}{\mbox{tr}}
\newcommand{\fig}[2]{\includegraphics[width=#1]{./figures/#2}}

\newlength{\bilderlength}

\begin{document}

\title{Instanton calculus for the self-avoiding manifold model}
\author{Fran{\c c}ois David\footnote{david@spht.saclay.cea.fr}}
\affiliation{Service de Physique Th\'eorique\footnote{Direction des Sciences de la Mati\`ere, CEA and URA 2306 CNRS/SPM}, CEA-Saclay, 91191 Gif-sur-Yvette, France}
\author{Kay J. Wiese\footnote{wiese@lpt.ens.fr}}
\affiliation{Laboratoire de Physique Th\'eorique\footnote{UMR 8549 CNRS/SPM, ENS}, D\'epartement de Physique,\\ Ecole Normale Sup\'erieure, 24 rue Lhomond, 75005 Paris, France} 
\date{September 28, 2004}
\begin{abstract}
We compute the normalisation factor for the large order asymptotics of perturbation theory for the self-avoiding manifold (SAM) model describing flexible tethered ($D$-dimensional) membranes in $d$-dimensional space,  and the $\epsilon$-expansion for this problem. 
For that purpose, we develop the methods inspired from instanton
calculus, that we introduced in a previous publication (Nucl.\ Phys.\ B 534 (1998) 555), and we compute the functional determinant of the fluctuations around the instanton configuration. 
This determinant has UV divergences and we show that the renormalized action used to
make perturbation theory finite also renders the contribution of the
instanton UV-finite. 
To compute this determinant, we develop a systematic large-$d$ expansion. For the renormalized theory, we point out problems in the interplay between the limits $\epsilon\to 0$ and $d\to\infty$, as well as IR divergences when $\epsilon=0$.
We show that many cancellations between IR divergences occur, and argue that the remaining IR-singular term is associated to amenable non-analytic contributions in the large-$d$ limit when $\epsilon=0$.
The consistency with the standard instanton-calculus results for the self-avoiding walk is checked for $D=1$. 
\end{abstract}
\maketitle
\vfill\hfill Saclay Preprint T04/122
\ \\
\newpage
\tableofcontents

\section{Introduction} 

Flexible polymerized 2-dimensional films (tethered or polymerized
membranes) \cite{NelPel87} have very interesting statistical
properties (for a review see
\cite{Jerusalem87,Jerusalem87-II,Wiesehabil}).  In these objects there
is a competition between entropy which favors crumpled or folded
configurations as for polymers, steric interactions (self-avoidance)
which tend to swell the membranes, and bending rigidity which favors
flat configurations. Internal disorder, inhomogeneities and anisotropy
may also play an important role, that we shall not discuss here (see
the chapters 10-12 in \cite{Jerusalem87-II} and \cite{Wiesehabil} for
a recent review of these effects).

If one does not take into account self-avoidance, theoretical
arguments (mean-field and renormalization-group calculations) and
numerical simulations show two phases: (1) A
high-temperature/low-rigidity crumpled phase, where the membrane is
crumpled with infinite Hausdorff dimension, and where bending rigidity
is irrelevant; (2) a low-temperature (flat) phase with large effective
rigidity and Hausdorff dimension two \cite{KanKarNel86,KanKarNel87}.

In the high-temperature (crumpled) phase, steric interactions
(self-avoidance) are physically relevant, and will swell the membrane,
as for polymers. Two scenarios are possible: Either the membrane is
flat with Hausdorff dimension two (as for high bending rigidity), or
it is crumpled swollen with Hausdorff dimension larger than two.  For
large imbedding dimension $d$ a Gaussian variational approximation can
be argued to become exact, predicting a Haussdorff dimension
$d_{\mathrm{H}}= d/2$. It is non-trivial whether this swollen phase
exists down to $d=3$. Numerical simulations indicate that only a flat
phase exists in $d=3$; for details see the discussion in
\cite{Jerusalem87-II,Wiesehabil}. However, simulations are for rather
small systems. It therefore remains important to have a solid
theoretical understanding. 

The theory in question is a generalization of the Edwards model for
polymers \cite{Edw65,CloJan90}. It was proposed in
\cite{AroLub88,KarNel87,KarNel88}, and is a model of self-avoiding
manifolds (or membranes), hereafter denoted the SAM model.  It is
amenable to a treatment by perturbation theory (in the coupling
constant of steric interactions) and to a perturbative renormalization
group analysis which leads to a Wilson-Fisher like
$\epsilon$-expansion for estimating the scaling exponents and the
critical properties of the swollen phase.

This SAM model is quite interesting at the theoretical level for
several reasons:
\begin{enumerate}
  \item It can only be defined as a non-local field-theory over the
        internal 2-dimensional space of the manifold, with
        infinite-ranged multi-local interactions. Therefore the
        applicability of renormalization theory and of renormalization
        group techniques is a non-trivial issue. A proof of
        perturbative renormalizability to all orders was finally
        given in \cite{DDG3,GBU}.
  \item The model is in fact defined through a double dimensional
        continuation, where both the dimension of space $d$ and the
        internal dimension $D$ of the manifold are analytically
        continued to non-integer values.  The physical case of
        two-dimensional membranes is always in the strong-coupling
        regime where the engineering dimension of the coupling,
        $\epsilon =2D-\frac{2-D}{2}d$ is $\epsilon=4$ for any space
        dimension $d$.
  \item The analytical study of this model at the non-perturbative
        level is still in its infancy, since it is a technically quite
        difficult problem. A first step
        %\footnote{A huge step for us, but a small step for mankind.} 
        was made by the two present authors for the large orders of
        perturbation theory in \cite{DavidWiese1998}.  It is this
        issue of the large-order asymptotics of the SAM model that we
        treat in this paper.
\end{enumerate}
For quantum mechanics \cite{Dyson52,Lam68,BenderWu69} and for local
quantum field theories \cite{Lip76} (such as the
Laudau-Ginzburg-Wilson $\phi^4$ theories) the large-order asymptotics
of perturbation theory are known to be controlled by (in general
complex) finite-action solutions of the classical equation of motion
called ``instantons". More precisely the large-order asymptotics are
described by semi-classical approximations around these instantons. We
refer for instance to \cite{Zin82} for a review of this ``instanton
calculus".

In \cite{DavidWiese1998} we have shown that similarly for the SAM
model there exists an instanton which controls its large-order
asymptotics. This instanton is a scalar field configuration in the
external $d$-dimensional space, which extremizes a highly non-local
effective-action functional, and which cannot be computed exactly.  We
also showed that remarkable simplifications occur in the large-$d$
limit, which suggests that a systematic $1/d$ expansion can be
constructed to study the instanton, but also that already the first
$1/d$ correction to the large-$d$ limit is plagued with infrared (IR)
divergences whose origin was unclear. In \cite{DavidWiese1998} we only
studied the instanton at the classical level, i.e.\ the (non-local)
equation of motion and the properties of its solution, the instanton.

In this article we present the full semi-classical analysis of the
instanton for the SAM model, derive its connection with the
large-order asymptotics, and study the UV divergences and
renormalization necessary for the instanton. For this
purpose, many new calculational techniques had to be developed, hence
the length of the paper and its technical character. More precisely,
the main new results are:
\begin{enumerate}
  \item We first show in much more details than in
        \cite{DavidWiese1998} how the instanton emerges from the
        functional integral which defines the continuum SAM model. In
        particular we treat properly and carefully the zero-modes for
        the instanton, how the contour of functional integration has
        to be deformed in the complex saddle-point method, as well as
        various normalization problems for the functional
        integration. This is done in Sect.~III-A,B.
  \item Using this, we obtain the contribution of the fluctuations
        around the instanton in the semi-classical approximation as the
        determinant of a non-local kernel operator in $d$-dimensional
        space, and derive the normalization factor for the large-order
        asymptotics (Sect.~III-C,D,E).
  \item We analyze completely the UV divergences of this determinant,
        and show that in the renormalized theory these UV divergences
        for the instanton determinant factor are canceled by the
        one-loop perturbative counterterm of the renormalized theory,
        making the final asymptotics UV finite. This is an important
        check of the consistency of the SAM model, since the original
        proof of renormalizability is only valid in perturbation
        theory. (In a field theoretic language it is not a
        background-independent proof). This is done in Sect.~IV. Our
        argument is based on the extension of the perturbative
        renormalizability argument to the general case of ensembles of
        interacting manifolds in an external background potential.
  \item In \cite{DavidWiese1998} the instanton equation was solved
        within a variational approximation. In sect.~V we study how
        this approximation can be applied to the explicit calculation
        of the instanton determinant factor. We first show that a
        direct variational calculation gives a result which is too
        naive, and does not take properly into account the UV
        fluctuations. We then propose a systematic framework to
        construct an expansion around the variational approximation,
        developing ideas that we proposed in \cite{DavidWiese1998}. We
        then show that this framework gives the leading term for the
        instanton determinant factor in the large-$d$ limit.
  \item We are thus able to construct a systematic $1/d$ expansion for
        the instanton calculus, and show that this expansion is well
        defined as long as the SAM is super renormalizable,
        i.e.\ $\epsilon>0$ (no UV divergences in perturbation theory,
        apart from vacuum energy terms). The leading and first
        subleading terms are computed explicitly for the determinant
        factor and the normalization factor of the zero mode of the
        instanton. These calculations involve a new non-trivial
        diagrammatic expansion. This is done in Sect.~\ref{s:1/dcor} .
  \item Finally we study the $1/d$ expansion for the renormalized
        theory at $\epsilon=0$. We show that, at variance with the
        instanton calculus for local field theories, some subtle
        issues arise for the SAM model. Indeed, we show that already
        at leading order in $d$, the limits $d\to\infty$ and
        $\epsilon\to 0$ do not commute, and that some care is needed
        in order to obtain the instanton determinant factor for the
        renormalized theory at large $d$. We then show that the
        subleading terms of the $1/d$ expansion are plagued with IR
        divergences at $\epsilon=0$, generalizing the results of
        \cite{DavidWiese1998}. We analyze completely these IR
        divergences at the first subleading order, and show that many
        compensations occur, leaving a single IR-singular term
        associated with a single eigenmode for the fluctuations around
        the instanton, namely the unstable eigenmode generated by
        global dilation for the instanton. This analysis of the
        renormalized theory is done in Sect.~VII.
\end{enumerate}
To summarize, we have performed a non-trivial check of
the consistency of the model, in particular of its renormalization, in
a non-perturbative regime, and we have developed the tools to compute
the large-order asymptotics of the SAM model.

Appendices contain more technical computations and details about the
normalizations. In particular in appendix C we explicitly check that
in the special case of the self-avoiding polymer ($D=1$) we recover
the large-order asymptotics of the Edwards model obtained by field
theoretical methods (using instanton calculus and the well known
equivalence between the Edwards model and the O($n=0$) $\phi^4$ Landau
Ginzburg model \cite{Gen72,Clo81}). This provides a check of the
consistency of the SAM model.

\section{The model}
\label{s:model}

\subsection{The non-interacting manifold} \label{ss:freemanifold}
First we define the model for the Gaussian non-interacting manifold
(free or phantom manifold). Of course this model reduces to a massless
free field, but we reconsider it closely in order to fix properly the
normalization for the measure and for the definition of the
observables, and for the treatment of the zero modes.

\subsubsection{The model and its action}
\label{sss:freemodact}
We consider a manifold $\mathcal{M}$ with a finite size, as a closed
$D$-dimensional manifold $\mathcal{M}$, with a \emph{fixed} internal
(or intrinsic) Riemannian structure, given by a metric tensor
$\tg=g_{\mu \nu }(\vx)$.  $\vx=(x^\mu ;\ \mu =1,\ldots,D$) describes
(a system of) local coordinates on $\mathcal{M}$.  From now on the
internal volume of $\CM$, $\mathrm{Vol}(\CM)$ and its internal size
$L$ are defined as \begin{equation} \label{2.1} \mathrm{Vol}(\CM)\ =\
\int_\CM \rmd^D\vx\,\sqrt{g} \quad,\quad L\ =\
\mathrm{Vol}(\CM)^{{\frac{1}{D}}}
\end{equation} 
with $g=\det[\tg]$.  The manifold is embedded in external (or bulk)
$d$-dimensional Euclidean space $\mathbb{R}^d$.  This embedding is
described by the field $\rr=\{r^a;\ a=1,\ldots,d\}$
\begin{equation}
\label{M2Rd} \mathcal{M}\to\mathbb{R}^d\ \ ,\ \
\vx\to\rr(\vx) \end{equation} We shall use dimensional regularization
in this paper by considering that the internal dimension $D$ of the
manifold is $0<D<2$ non-integer. See the reference paper \cite{GBU}
for a more precise discussion of how we can define a finite membrane
within dimensional regularization. In practice we can restrict
ourselves to the case of a square $D$-dimensional torus of size $L$,
$\mathbb{T}_D=\left(\mathbb{R}^D/(L\cdot\mathbb{Z})^D\right)$, with
flat metric $g_{\mu \nu }=\delta_{\mu \nu }$.

We first consider the free non-interacting manifold (phantom
membrane).  The manifold may fluctuate freely in external
$d$-dimensional space. Its free energy is given by the Gaussian local
elastic term $S_0$, which is the integral of the square of the
gradient of the field $\rr$
\begin{equation}
\label{freeaction}
S_0[\rr]\ =\ \int_{\mathcal{M}}{1\over 2} (\nabla\rr)^2
\ =\ \int \rmd^D\vx\,\sqrt{g}\,{1\over 2} 
g^{\mu \nu }\partial_\mu \rr\cdot\partial_\nu \rr
   \ .
\end{equation}
  This is nothing but the Euclidean
action for a free massless field (with $d$ components) living on
$\CM$.  The manifold may (and does) freely intersect itself, as does
a random Brownian walk in $d\le 4$  space dimensions.

\subsubsection{The partition function} \label{ss:freepartfun} The
partition function for the free manifold is thus given by the
functional integral
\begin{equation}
\label{Z0free} Z_0\ =\ \int \CD[\rr]\,\rme^{-S_0[\rr]} \end{equation}
where $\CD[\rr]$ is the standard functional measure for the free
massless field $\rr$ (see Appendix \ref{section-measures} for details
and the normalization used in this paper).

There is an infinite factor in $Z_0$ (the volume of bulk space
Vol($\mathbb{R}^d$)) coming from the translational zero mode of the
manifold. This can be isolated by choosing a specific point $\vx_0$ on
the manifold and a specific point $\rr_0$ in bulk space, and by
defining the partition function $\CZ_0$ for a marked manifold
\begin{equation}
\label{CZ0free}
\CZ_0\ =\ \CZ_0(\rr_0)\ =\ \int 
\CD[\rr]\,\delta^d(\rr(\vx_0)-\rr_0)\,\rme^{-S_0[\rr]}
\end{equation}
$\CZ_0(\rr_0)$ is infra-red (IR) finite and does not depend on the
choice of $\rr_0$ or of $\vx_0$.  We have formally
\begin{equation}
\label{Z0toCZ0}
\qquad Z_0\ =\ \int\rmd^d\rr_0\ \CZ_0(\rr_0)\ =\
\mathrm{Vol}(\mathbb{R}^d)\, \CZ_0 
\ .
\end{equation}

The partition function $\CZ_0$ is found  
to be related to the
determinant of the Laplacian operator over $\CM$ through
\begin{equation}
\label{CZ0detLapl}
\CZ_0(\rr_0)\ = \ \Big[{\det}'\left[-\Delta\right] \cdot
2\pi/\mathrm{Vol}(\mathcal{M})\Big]^{-d/2}
\ ,
\end{equation} 
where ${\det}'\left[-\Delta\right]$ is the product of the non-zero
eigenvalues of (minus) the Laplacian operator
$\Delta=g^{-1/2}\partial_\mu g^{\mu \nu }\partial_\nu $ on
$\mathcal{M}$.  $\mathrm{Vol}(\mathcal{M})=\int\rmd^D\vx\,\sqrt{g}$ is
the internal volume of the manifold. This last term comes from the
proper treatment of the translational zero mode of the Laplacian (see
Appendix \ref{section-measures}).

The determinant ${\det}'\left[-\Delta\right]$ is ultra-violet (UV)
divergent, and is defined through a zeta-function regularization (for
a manifold $ \CM$ with non-integer dimension $D\neq 1$ or $2$ this is
equivalent to dimensional regularization)
\begin{equation}
\label{logdet2zeta}
\log({\det}'\left[-\Delta\right])
\ =\ 
-\zeta'(0)
\quad,\quad
\zeta (s)=\mathrm{tr}'(\left(-\Delta\right)^{-s})
\ .
\end{equation} 
The zeta-function $\zeta (s)$
is defined by analytic continuation from $\mathrm{Re}(s)$ large.
$\mathrm{tr}'$ means the trace over the space orthogonal to the kernel
of $\Delta$.
 $\zeta(s)$,
scales with the size of $\CM$ as
\begin{equation}
\label{zetadef}
\zeta(s)\ =\
\left[\mathrm{Vol}(\mathcal{M})\right]^{2s/D}\,\tilde\zeta(s) 
\ ,
\end{equation}
where the ``normalized zeta-function" $\tilde\zeta(s)$ depends on the
shape of the manifold but not on its size (scale invariance).  In the
absence of a conformal anomaly, as this is the case for the generic
case of $D$ non-integer
 we have the exact identity
\begin{equation}
\label{zetade0}
\zeta(0)\ =\ -1\ .
\end{equation}
(This factor comes from the contribution of the subtracted zero mode
in the determinant). Hence the partition function reads
\begin{equation}
\label{CZ0explict}
\CZ_0(\rr_0)\ = \ 
\left[\mathrm{Vol}(\mathcal{M})\right]^{-{d(2-D)\over 2D}}\,
\left[{\rme^{\tilde\zeta'(0)}\over 2\pi}\right]^{d/2}
\ .
\end{equation}
The last term is a ``form factor" depending on the shape of $\CM$.

For 2-dimensional manifolds ($D=2$), the conformal anomaly gives an
additional scale factor of the form ${|L\mu_0|}^{\chi\,d/6}$, where
$L= \mathrm{Vol}(\mathcal{M})^{1/D}$ is the size of $\CM$, $\mu_0$ the
regularization mass scale, required to define properly the measure in
the functional integral (see Appendix \ref{section-measures}), and
$\chi$ the Euler characteristics of the membrane. We shall not discuss
this any further, since this is not relevant for the problem treated
here, where we consider manifolds with $D<2$.

\subsection{The interacting self-avoiding manifold}
\label{ss:interactingmodel}

\subsubsection{The action}
\label{sss:SAMact}
The steric self-avoiding interaction is introduced by adding a 
2-body repulsive contact interaction term of the form 
\begin{equation*}
\label{lf1} \int_\vx\int_\vy\delta^d(\rr(\vx)-\rr(\vy)\ =\ \int_\CM
\rmd^D\vx \sqrt{g(x)}\,\int_\CM \rmd^D\vy
\sqrt{g(x)}\,\delta^d(\rr(\vx)-\rr(\vy))
\end{equation*}
(where $\delta^d(\rr)$ is the Dirac distribution in the external space
$\mathbb{R}^d$ ) to the action, which is now
\begin{equation}
\label{ActionSA} S[\rr,b] =\ \int_{\mathcal{M}}{1\over 2}
(\nabla\rr)^2\ +\ {b\over
2}\int_\vx\int_\vy\delta^d(\rr(\vx)-\rr(\vy))
\end{equation}
$b>0$ is the self-avoidance coupling constant.
This is similar to what is done in the Edwards model for polymers.

\subsubsection{The partition functions}
The partition function for the self-avoiding manifold  is 
\begin{equation}
\label{Z4r}
\CZ(\rr_0,b) =\ \int 
\CD[\rr]\,\delta^d(\rr(\xx_0)-\rr_0)\,\rme^{-S[\rr,b]}
\qquad;\qquad
Z(b)\ =\ \int \rmd^d\rr_0\,\CZ(\rr_0,b)
\ .
\end{equation}
These partition functions are defined in perturbation theory, within a
dimensional regularization scheme, i.e.\ by analytic continuation in
the internal dimension $D$.

If the internal coordinate $\vx$ has engineering dimension $1$, the
external coordinate $\rr$ has engineering dimension $\nu_0$ given by
(i.e.\ $[\rr]\sim[\vx]^{\nu_0}$)
\begin{equation}
\label{nu0expl}
\nu_0\ =\ (2-D)/2
\end{equation}
and the coupling constant $b$ has engineering dimension $-\epsilon$
(i.e.\ $[b]\sim[\vx]^{-\epsilon}$) with
\begin{equation}
\label{epsilonexpl}
\epsilon\ =\ 2D-(2-D)d/2
\end{equation}

As usual in polymer and membrane problems, we shall consider mainly
the normalized partition function $\mathfrak{Z}(b)$, defined by the
ratio of the interacting partition function for the interacting
manifold $\CM$, divided by the partition function for the same
manifold $\CM$, but free. 
\begin{equation}
\label{Zfrakdef}
\mathfrak{Z}(b)\ =\ Z(b)/Z_0\ =\ \CZ(\rr_0,b)/\CZ_0(\rr_0)
\ .
\end{equation}
Let 
\begin{equation}
\label{lf2} L\ =\ (\mathrm{Vol}(\CM))^{1/D}\quad;\qquad
\mathrm{Vol}(\CM)\,=\,\int_\CM \rmd^D\vx\,\sqrt{g(\vx)}
\end{equation}
be the internal size of the manifold.  The normalized partition
function has a perturbative series expansion in powers of $b$, of the
form 
\begin{equation}
\label{Zfrakpertexp}
 \mathfrak{Z}(b)\ =\ 1\,+\,\sum_{k=1}^{\infty}
\mathfrak{Z}_k\,\left(bL^\epsilon\right)^k \end{equation} where the
coefficients $\mathfrak{Z}_k$ depend only on the shape of the
manifold, on its internal dimension $D$, and on the external dimension
$d$.  These coefficients are given by the expectation value in the
massless free theory defined by the free action $S_0$ of the bi-local
operators corresponding to the interaction term
\begin{equation}
\label{lf3} \mathfrak{Z}_k\ =\ {1\over
k!\,2^k}\,\iint_{\vx_1,\vy_1}\cdots\iint_{\vx_k,\vy_k}\ \big\langle
\prod_{i=1,k}\delta^d(\rr(x_i)-\rr(y_i))\big\rangle_0 \ ,
\end{equation}
with $\langle \dots \rangle_{0}$ the expectation value w.r.t.\ ${\cal
S}_{0}[r]$, see (\ref{freeaction}).

\subsubsection{Observables and correlation functions in external
space} \label{sss:corrfunobs} We shall be mainly interested in
correlations functions which correspond to observables which are
global for the manifold (i.e.\ do not depend on the internal position
of specific points on the manifold), but which may be local in
external space (i.e.\ do depend on the position of specific points in
the external space).  These observables are the simplest ones.  In
particular for the case $D=1$  (polymers) these observables have a
direct interpretation in terms of correlation functions of local
operators in the corresponding local field theory in external
$d$-dimensional space.

The observables involve the manifold density $\rho(\rr)$.  We define
the manifold density at the point $\rr_1$, $\rho(\rr_1)$, as the
functional of the field $\rr(x)$
\begin{equation}
\label{defRho}
\rho(\rr_1;\rr)\ =\ \int_\vx \delta^d(\rr(\vx)-\rr_1)
\ .
\end{equation}
The $N$-point density correlator $\CR^{(N)}(\rr_1\cdots\rr_N)$ for
the interacting manifold is defined as
\begin{equation}
\label{defCRN}
\CR^{(N)}(\rr_1\cdots\rr_N;b)\ =\ \int\CD[\rr]\,\prod_{i=1}^N\, 
\rho(\rr_i;\rr)\,\rme^{-S(\rr,b)}
\ .
\end{equation}
Obviously the one-point density correlator is related to the partition
function (for a one-point marked manifold) by
\begin{equation}
\label{CR1Z}
\CR^{(1)}(\rr_0;b)\ =\ \mathrm{Vol}(\CM)\,\CZ(\rr_0;b)
\ .
\end{equation}
Ratios of density correlators define expectation values of
densities. For instance, the e.v. (expectation value) of a product of
$N$ density operators $\rho(\rr_i)$ for a manifold constrained to be
attached to a point $\rr_0$ is the ratio
\begin{equation}
\label{vevRhoCR}
\langle\rho(\rr_1)\cdots\rho(\rr_N)\rangle_{\rr_0;b} \ =\ 
\CR^{(N+1)}(\rr_0,\rr_1\cdots\rr_N;b)/\CR^{(1)}(\rr_0;b)
\ .
\end{equation}

As for the partition functions, it is more convenient to normalize the
density correlators with respect to the partition function for the
{\em free} manifold.  We thus define the normalization
for the
normalized  density correlators, by
\begin{equation}
\label{FrakRtoCR}
\mathfrak{R}^{(N)}(\rr_1\cdots\rr_N;b)\ =\ 
\CR^{(N)}(\rr_1\cdots\rr_N;b)
{\Bigg/}
\mathrm{Vol}(\CM)\,\CZ_0
\ .
\end{equation}
In particular the normalized 1-point correlator coincides with the
normalized partition function
\begin{equation}
\label{FrakR1toCZ}
\mathfrak{R}^{(1)}(\rr_1;b)\ =\ \mathfrak{Z}(b)
\end{equation}
and is independent of $\rr_1$.

These observables have a perturbative series expansion in the coupling
constant $b$.  In particular they scale with the size $L$ of the
manifold as
\begin{equation}
\label{CRscaling}
\CR^{(N)}(\rr_1\cdots\rr_N;b,L)\ =\ 
L^{N(\epsilon-D)}
\cdot\CR^{(N)}(\rr_1\,L^{-\nu_0},\cdots,\rr_N\,L^{-\nu_0};bL^{\epsilon}) 
\end{equation}
and 
\begin{equation}
\label{FrakRscaling}
\Rfrak^{(N)}(\rr_1\cdots\rr_N;b,L)\ =\ 
L^{(N-1)(\epsilon-D)}
\cdot\Rfrak^{(N)}(\rr_1\,L^{-\nu_0},\cdots,\rr_N\,L^{-\nu_0};bL^{\epsilon}) 
\end{equation}

\subsubsection{Global quantities and gyration radius moments}
\label{sss:globgyr}
We define the moments of order $k$ for the gyration radius (in short
the $k$-th gyration moment) 
$\mathsf{R_{gyr}^{(k)}}$ 
of the manifold by
\begin{equation}
\label{defGyrMom}
\mathsf{R_{gyr}^{(k)}}
={\int_{\vx_1} \int_{\vx_2} \left|\rr(\vx_1)-\rr(\vx_2)\right|^k \over
\int_{\vx_1} \int_{\vx_2} 1}
\ .
\end{equation}
The standard gyration radius is $\mathsf{R}_{\mathrm{gyr}}=
\sqrt{\mathsf{R_{gyr}^{(2)}}}$.
The expectation value $\mathcal{R}_{\mathrm{gyr}}^{(k)}$ of the $k$-th
gyration moment $\mathsf{R_{gyr}^{(k)}}$
for the interacting manifold is thus (for $k>0$)
\begin{equation}
\label{GyrMomk} 
\mathcal{R}_{\mathrm{gyr}}^{(k)}\ =\ \langle\mathsf{R_{gyr}^{(k)}}\rangle
\ =\
{1\over\mathrm{Vol}(\CM)^2}\,\int_{\rr_1}\int_{\rr_2}
\left|\rr_1-\rr_2\right|^k \, \CR^{(2)}(\rr_1,\rr_2;b)
\ .
\end{equation}

\subsection{UV divergences and perturbative renormalization}
\label{ss:UVrenormalization} Using dimensional regularization, the
perturbative expansion for the partition function and the observables
is known to be UV finite for
\begin{equation}
\label{lf4}
0<D<2\quad\text{and}\quad D<\epsilon \qquad\text{i.e.}\qquad d<{2D\over 2-D}
\ .
\end{equation}
As long as we deal with finite-size manifolds ($L<\infty$),
perturbation theory is free from infra-red divergences (which occur
for infinite manifolds since perturbation theory is made around the
free-manifold theory, which is a free massless scalar field in $D\le
2$ dimensions).

The perturbative expansion suffers from short-distance (UV)
divergences when
\begin{equation}
\label{lf5}
D\,\le \,\epsilon
\ .
\end{equation}
These UV divergences come from the short-distance behavior of the
expectation values which appear as integrands in the integrals, when
the distance between several points $\vx_i$ and $\vy_j$ goes to zero.
Using dimensional regularization these divergences appear as poles in
$\epsilon$ ($d$ being fixed), or equivalently as lines of singularity
in the $(d,D)$ plane.

As proved in \cite{GBU}, these UV divergences can be studied within a
multi-local operator product expansion (MOPE) which generalizes
Wilson's OPE of local field theory.  As a consequence, these UV
divergences are proportional to the insertion of multi-local
operators, and are amenable to renormalization theory.

The MOPE formalism and dimensional analysis show that for
$0<\epsilon\le D$ there is a finite number of divergences,  with
poles at 
\begin{equation} 
\label{epspoles}
\epsilon\ =\ D/n\ ,\ n\in\mathbb{N_+}
\ .
\end{equation} 

These divergences are proportional to insertions of the identity
operator $\mathbf{1}$ (with dimension $0$).  The model is
super-renormalizable for $\epsilon>0$
and these divergences are subtracted by adding to the action a local
counterterm proportional to the volume of the manifold (i.e.\ to the
integral of the identity operator $\mathbf{1}$).
\begin{equation} 
\label{vacenct}
\Delta S(\rr)\ \propto\ 
\int_\vx \mathbf{1} \ =\ \mathrm{Vol}(\CM)\ \ .
\end{equation} 
These divergences and the corresponding counterterm are constant
terms, independent of the configuration of the manifold, i.e.\ of the
field $\rr$, and they disappear in the observables given by ratios of
correlators such as the e.v.
$\langle\rho(\rr_1)\cdots\rho(\rr_N)\rangle_{\rr_0}$ and the
normalized correlators $\mathfrak{R}^{(N)}$.

For $\epsilon=0$ the model has an infinite number of divergences.
These divergences are proportional to the insertion of the two
operators present in the original action $S$.  This means that the
theory is renormalizable, and that it can be made UV-finite by adding
to the action
counterterms of the same form than those of the original action.  In
other words, one can construct in perturbation theory a renormalized
action
\begin{equation}
\label{renact1}
S_\mathrm{r}(\rr;\bren,\mu)\  =\  {Z(\bren )\over 
2}\int_\vx\left(\nabla\rr\right)^2\,
+\, {\bren \mu^\epsilon Z_b(\bren)\over 
2}\int_\vx\int_\vy\delta^d(\rr(\vx)-\rr(\vy))
\ ,
\end{equation}
where $\bren$ is the dimensionless renormalized coupling constant,
$Z(\bren)$ and $Z_b(\bren)$ the wave-function and the
coupling-constant renormalization factors, and $\mu$ is the
renormalization momentum scale, while $\rr(\vx)$ is now the
renormalized field.  This renormalized action is such that the
renormalized correlation functions
\begin{equation}
\label{rencorrf}
\CR_r^{(N)}(\rr_1\cdots\rr_N;\bren,\mu)\ =\ \int \CD[\rr]\ 
\prod_{i=1}^N\rho(\rr_i;\rr)\,\rme^{-S_r(\rr;\bren,\mu)}
\end{equation}
have a perturbative series expansion in $\bren$ which is UV finite for
$\epsilon\ge 0$ and stays finite for $\epsilon=0$.  For a finite
manifold with size $L$ the renormalized perturbation theory is still
IR finite.

From the standard arguments of renormalization group (RG) theory the
renormalized theory describes the universal large-distance scaling
behavior of self-avoiding manifolds.  One can write RG equations which
tell how the observables scale with the size of the manifold for
$\epsilon>0$.  When expressed in terms of the renormalized observables
and the renormalized coupling, these RG equations have a regular limit
(at least in perturbation theory) as $\epsilon\to 0_+$.  As a
consequence one can construct an $\epsilon$-expansion \`a la
Wilson-Fisher for the scaling exponents.

\subsection{Effective non-local model in external space}
\label{ss:effectiveaction}

As shown in \cite{DavidWiese1998}, to study the large-order behavior
of the SAM model as well as its large-$d$ behavior, it is necessary to
reformulate the model as an effective non-local model for an auxiliary
composite field $V(\rr)$ in the external $d$-dimensional space.We
 recall this reformulation.

\subsubsection{Auxiliary fields and  effective action}
\label{auxfeffa}
First we recall the auxiliary field $\rho(\rr)$ (local manifold
density) defined in (\ref{defRho}), 
\begin{equation}
\label{lf6}
\rho ( {\rr})=\int_{\vx \in \CM} \delta^d(\rr(\vx)-\rr)
\end{equation}
and its conjugate field $V (\rr)$, which is the Lagrange multiplier for
the above constraint, such that
\begin{equation}
\label{unityresolve} 1\ =\ \int \CD[V]\,\CD[\rho]\,\exp\left( {\int
\rmd^d\rr\,V(\rr)\left[\rho(\rr)-\int_\CM
\delta^d(\rr(\vx)-\rr)\right]} \right)\ .
\end{equation}
$\rho$ is a real field, while $V$ is imaginary, and has to be
integrated from $-\rmi\infty$ to $+\rmi\infty$ in the functional
integral. Equivalently the functional measures for $\rho$ and $V$ are
formally
\begin{equation}
\label{lf7} \int \CD[\rho]\ =\ \int_0^\infty \prod_\rr \rmd\rho(\rr)\
\equiv\ \int_{-\infty}^\infty \prod_\rr \rmd\rho(\rr) \qquad;\qquad \int
\CD[V]\ =\ \int_{-\rmi\infty}^{\rmi\infty} \prod_\rr {\rmd V(r)\over
2\rmi\pi} \ .
\end{equation}
We now insert \eqref{unityresolve} in the functional integral.  
Since the interaction term can be written as
\begin{equation}
\label{lf8} \int_\vx \int_\vy \delta^d(\rr(\vx)-\rr(\vy)) \ =\
\int_\rr \rho(\rr)^2\ ,
\end{equation}
the integral over the field $\rho$ is Gaussian and can be performed
explicitly. We obtain for the partition function
\begin{equation}
\label{normDV}
Z(b) = \int \CD[\rr]\,\CD[V]\, \exp
\left[
-\int_\vx \left(\half (\nabla_{\!\vx} \rr)^2 +V(\rr)\right)
\,+\, {1\over 2b} \int_\rr V(\rr)^2 \right]
\ .
\end{equation}
Note that 
the functional measure $\CD[V]$ over $V[\rr]$ is now normalized so that
\begin{equation}
\label{lf9}
\int \CD[V]\,\exp \left( {1\over 2b} \int_\rr V(\rr)^2 \right)
\ =\ 1
\ 
\end{equation}
and depends explicitly on the coupling constant $b$.

This functional integral describes a free (not self-interacting)
manifold fluctuating in a random annealed potential $V(\rr)$. This is
a simple generalization of the reformulation of the SAW problem into a
random walk in a random annealed potential.

Now we integrate over the field $\rr(\vx)$ and define the effective
free energy $F_\CM[V]$ for the non-interacting (phantom) manifold
$\CM$ in the external potential $V(\rr)$ by
\begin{equation}
\label{lf10}
\exp\left(-\,F_\CM[V] \right)\ =\ \int \CD[\rr]\,\exp
\left[
-\int_\vx \left(\half (\nabla_{\!\vx} \rr)^2 +V(\rr)\right)
\right]
\ .
\end{equation} 
We are left with 
the effective action for the field $V$, $S_\CM[V]$, which is given by 
\begin{equation}
\label{lf11}
S_\CM[V]\ =\ F_\CM[V]\,-\, {1\over 2b} \int_\rr V(\rr)^2
\end{equation}
and is a non-local functional of the potential $V$.  The partition
function is now given by a functional integral over the potential $V$
alone
\begin{equation}
\label{Z4V}
Z(b)\  =\  \int \CD[V]\, \exp
\left[-\,F_\CM[V]\,+\, {1\over 2b} \int_\rr V(\rr)^2 \right]
\ =\  \int \CD[V]\, \exp\left[-\,S_\CM[V]\right]
\ .
\end{equation}

\subsubsection{Correlation functions for global observables}
\label{effcorrfun} The same transformation can be used to compute the
density correlators $\CR^{(N)}$ and the corresponding correlation
functions as expectation values of observables with the effective
action $\CS[V]$.
Indeed inserting a density operator $\rho(\rr)$ in the original
functional integral (\ref{Z4r}) over $\rr(\vx)$ amounts to insert a
functional derivative with respect to the conjugate field $V(\rr)$ in
the functional integral (\ref{Z4V}) over $V(\rr)$.
\begin{equation}
\label{lf12}
\rho(\rr)\ \to\ {\delta\over\delta V(\rr)}
\end{equation}
so that
\begin{equation}
\label{CorrV}
\CR^{(N)}(\rr_1,\cdots,\rr_N;b)\ =\ \int \CD[V]\,\exp\bigl[-F_\CM[V]\bigr]\,{\delta\over\delta V(\rr_1)}\cdots
 {\delta\over\delta V(\rr_N)}\,
 \exp\left[{1\over 2b}\,\int_\rr
V(\rr)^2\right]
\ .
\end{equation}
For instance the 2-point correlator is
\begin{eqnarray}\label{lf13}
\CR^{(2)}(\rr_1,\rr_2;b) &=& \int \CD[V]\, \left[{1 \over
b^2}\,V(\rr_1)V(\rr_2)\,+\,{1\over b}\,\delta^d(\rr_1-\rr_2)\right] \,
\rme^{-F_\CM[V]+{1\over 2b}\,\int_\rr V(\rr)^2}
\nonumber 
\\
 &=& {1\over b^2}\,Z(b)\,\langle V(\rr_1)V(\rr_2)\rangle\,+\,{1\over
 b}\,\delta^d(\rr_1-\rr_2)
\end{eqnarray}
Similarly, for the moments of order $k$ of the gyration radius
(defined by (ref{defGyrMom})) we get
(for $k>0$)
\begin{equation}
\label{lf17} \mathcal{R}^{(k)}_{\mathrm{gyr}} \ =\ {1\over b^2}\,
{1\over\mathrm{Vol}(\CM)^2}\,\int_{\rr_1}\int_{\rr_2}
\left|\rr_1-\rr_2\right|^k
\,\big\langle V(\rr_1)\,V(\rr_2)\big\rangle
\ ,
\end{equation}
where $\langle\ \ \rangle$ denotes the average over $V$ with the
effective action $S_\CM[V]$ given by \Eq{Z4V}.

\section{Large orders of perturbation theory and instanton calculus}
\label{s:large orders} \subsection{Instanton and large orders in
Quantum (Field) Theory} \label{s:large orders.a}
\subsubsection{Instanton semi-classics} To fix our normalizations let
us first recall the basics of instanton calculus in quantum mechanics
and quantum field theory.  We consider a model defined by the
functional integral over a field $\phi(\rr)$ with a classical action
$S[\phi]$ and a (dimensionless) coupling constant $g$.  The partition
function is
\begin{equation}
\label{Z4Phi}
    Z\ =\ \int\CD[\phi]\,\rme^{-{1\over g}S[\phi]}
\ .
\end{equation}
The functional measure $\CD[\phi]$ over $\phi$ is defined from the
so-called DeWitt metric $G$ on classical field configuration (super)space.
We choose it to be local and translationally invariant, so it must be
of the form 
\begin{equation}
\label{Metric4Phi}
G(\delta\phi,\delta \phi)
  \ =\ 
  {\mu_{0}^{2}\over 2\pi g}\,\int \rmd^d{\rr}\,|\delta \phi(\rr) |^2
   \ =\ 
 {\mu_{0}^{2}\over 2\pi g}\, \|\delta \phi\|^{2}_{\scriptscriptstyle 2}
\end{equation}
$\|\cdot\|_{\scriptscriptstyle 2}$ is the $L_{2}$ norm.  This metric
depends explicitly on an (arbitrary) normalization mass scale
$\tilde\mu_{0}=\mu_{0}/\sqrt{g}$. 
The corresponding measure over field (super)space is (formally)
$\CD[\phi]=\prod_{\rr}\rmd\phi(\rr)\sqrt{\det G}$.  It is such that
\begin{equation}
\label{measure4Phi}
    \int \CD[\phi]\,\rme^{-{\mu_{0}^{2}\over 2g}\int_{\rr}\,\phi^{2}}\,=\,1
\ .
\end{equation}
(Note that the factor of $g$ has been
introduced for convenience, to have the same functional dependence on
$g$ for the measure in (\ref{measure4Phi}) and the Boltzmann factor in 
(\ref{Z4Phi}).)

We assume that there is a classical vacuum (field configuration)
$\phi_{0}$ which minimizes the action $S$, which is constant
($\phi_0(\rr)=\phi_0$) and which is unique (no zero modes around the
classical vacuum).  In the semi-classical approximation the
contribution of $\phi_{0}$ to the partition function is simply
\begin{equation}
\label{ZfromVaccuum} 
Z\quad \stackrel{\text{~~~classical vacuum}}
{\llongleftarrow}\quad \rme^{-{1\over g}S[\phi_{0}]}\,
\left(\mathrm{Det}\left[S''[\phi_{0}]/\mu_{0}^{2}\right]\right)^{-1/2}
\ ,
\end{equation}
where $S''$ is the Hessian operator, with kernel
\begin{equation}
\label{Hessdef}
    S''[\phi](\rr_{1},\rr_{2})\ =\ 
    {\delta^{2}S[\phi]\over\delta\phi(\rr_{1})\delta\phi(\rr_{2})}
\ .
\end{equation}
Now we assume that there are also instanton configurations which
contribute to the functional integral.  These instantons are
non-constant field configurations $\phi_{\mathrm{inst}}(\rr;z^{a})$
which are classical solutions of the field equations, and thus local
extrema of the action $S$, i.e.\ $S'[\phi_{\mathrm{inst}}]=0$, with a
finite action $S_{\mathrm{inst}}=S[\phi_{\mathrm{inst}}]$.  In
general, the set of instantons with action $S_{\mathrm{inst}}$ is a
finite-dimensional subspace, called the instanton moduli space.  We
denote $z=\{z^a, a=1,m\}$ a (local system of) collective coordinates
on the $m$-dimensional moduli space of the instantons with action
$S_{\mathrm{inst}}$.  The collective coordinate $z$ must include the
position of the instanton $\rr_{\mathrm{inst}}$ ($d$ moduli), its size
if the action $S$ is scale invariant, and in addition the internal
degrees of freedom of the instanton if needed.

The contribution of the instanton to the functional integral is also
given by a semi-classical formula. We must separate the integration
over the instanton moduli space $\CM_{z}$ from the integration over
the field fluctuations transverse to the moduli space $\CM_{z}$, since
the Hessian $S''[\phi_{\mathrm{inst}}]$ has now
$\delta_{\mathrm{inst}}$ zero-modes
$\partial_{a}\phi_{\mathrm{inst}}={\partial\phi_{\mathrm{inst}}[z]\over\partial
z^{a}}$.  The moduli space integration is then done explicitly. For
that purpose, we must consider the restriction of the metric $G$ to
the instanton moduli space $\CM_{z}$. The corresponding metric tensor
$h_{ab }$ in the coordinate system $z$ is defined by
\begin{equation}
\label{ds4Moduli}
    \|\rmd\phi_{\mathrm{inst}}\|^{2}\ =\ \ \rmd z^a\,\rmd z^b\,h_{ab}(z)
    \qquad;\qquad 
    \rmd\phi_{\mathrm{inst}}\ =\ {\partial\phi_{\mathrm{inst}}\over\partial z^a}\rmd z^a
\ ,
\end{equation}
where $\rmd\phi_{\mathrm{inst}}$ is an instanton fluctuation.
Hence the metric on moduli space is
\begin{equation}
\label{Metric4Moduli}
    h_{ab}(z)\ =\ 
   \left({\partial\phi_{\mathrm{inst}}\over\partial 
    z^a}\right|\left.{\partial\phi_{\mathrm{inst}}\over\partial z^b}\right)
    \ = \ 
    {\mu_{0}^{2}\over 2\pi g}\,\int_{\rr}\, 
    {\partial\phi_{\mathrm{inst}}(\rr,z)\over\partial z^a}
   {\partial\phi_{\mathrm{inst}}(\rr,z)\over\partial z^b} \ ,
\end{equation}
and the corresponding measure is $\rmd \mu(z)=\rmd^m\!z\,\sqrt{\det (h)}$.
The contribution of the fluctuations orthogonal to the moduli space
$\CM_{z}$ is evaluated by the saddle-point method. The final result
for the contribution of the instanton to the partition function is
\begin{equation}
\label{ZfromInstanton}
    Z\quad\stackrel{\text{~~~instanton}}{\llongleftarrow}\quad
    \int_{\CM_{z}} 
    \rmd^{m}\! z_a \ 
    \sqrt{\mathrm{det}(h_{ab}(z))}\ 
    \rme^{-{1\over g}S[\phi_{\mathrm{inst}}]}\ 
    \Bigl(\mathrm{det'}\left[S''[\phi_{\mathrm{inst}}]/\mu_{0}^{2}\right]\Bigr)^{-1/2}
\ ,
\end{equation}
where $\mathrm{det'}\left[S''[\phi_{\mathrm{inst}}]\right]$ is the
product of the non-zero eigenvalues of $S''[\phi_{\mathrm{inst}}]$.
The $\det(h)$ gives a power of the coupling constant $g^{-m/2}$, where
$m$ is the number of instanton zero-modes.

Similarly, let us now consider the expectation value for an observable
$O[\phi]$, for instance a product of fields
$O=\phi(\rr_1)\cdots\phi(\rr_n)$. The expectation value is given by
\begin{equation}
    \langle O\rangle\ =\ 
    {1\over Z}\,\int\CD[\phi]\,O[\phi]\,\rme^{-{1\over g}S[\phi]}
    \label{eq:4.10}
   \end{equation}
The contribution of the (translationally invariant) classical vacuum 
to $\langle O\rangle$ is simply
\begin{equation}
    \langle O\rangle \quad\stackrel{\text{~~~classical
vacuum}}\llongleftarrow\quad O[\phi_{0}] \label{OfromVac} \ .
\end{equation}
The contribution to $\langle O\rangle$ of the instanton
$\phi_{\mathrm{inst}}$, is obtained from
\begin{equation}
\label{3.11.new}
\left< O \right> \approx \frac{\int {\cal D}[\varphi] \left( {O}
[\phi_{0}+\varphi] \rme^{-{1\over g}S[\phi_{0} +\varphi ]} +
{O} [\phi_{\mathrm{inst}}+\varphi]\rme^{-{1\over
g}S[\phi_{\mathrm{inst}} +\varphi ]}\right) }{\int {\cal D}[\varphi]\left( 
\rme^{-{1\over g}S[\phi_{0} +\varphi ]} +\rme^{-{1\over
g}S[\phi_{\mathrm{inst}} +\varphi ]}\right) }\ .
\end{equation}
This expression is rather symbolic, since we have not written the
integral over the 0-mode of the instanton. Since $S_{0}<S_{\mathrm{inst}}$, we
have $S_{0}/g\ll S_{\mathrm{inst}}/g$, for $g\to 0$. Thus the leading term of
(\ref{3.11.new}) is given by (\ref{OfromVac}), and the subleading one
is the contribution of the instanton, which (up to exponentially
small terms) reads
\begin{eqnarray}
\label{OfromInst}
   \langle O\rangle
\quad&\stackrel{\text{~~~instanton}}{\llongleftarrow}& \quad 
    \int_{\CM_{z}} \rmd^mz_a\,\sqrt{\mathrm{det}(h(z))}\,
    \Big(O[\phi_{\mathrm{inst}}[z]]-O[\phi_{0}]\Big)
    \,
    \rme^{-{1\over g}(S[\phi_{\mathrm{inst}}]-S[\phi_{0}])}
    \, \nonumber \\
&&\qquad \qquad \times 
    \left(
          {\mathrm{Det'}\left[S''[\phi_{\mathrm{inst}}]/\mu_{0}^{2}\right]
	  \over
	  \mathrm{Det}\left[S''[\phi_{0}]/\mu_{0}^{2}\right]}
    \right)^{-1/2}
\ .
\end{eqnarray}
One can check that the $\mu_{0}$ dependence disappears (remember that
the moduli metric $h$ depends on $\mu_0$) as long as there is no scale
anomaly coming from UV-divergences in the ratio of the two determinants
of the Hessians.

\subsubsection{Large orders of perturbation theory and instantons}
\label{sss:loi}

We now recall the basic argument which shows how the large orders of
perturbative series obtained by functional integrals are related to
instantons.

We assume that the observable $\Zfrak(g)$ has a series expansion for
small positive coupling constant $g$ and is in fact an analytic
function of the coupling constant $g$ in a complex neighborhood of the
origin away from the negative real axis (i.e.\ for $|g|$ small enough,
$|\mathrm{Arg}(g)|<\pi$),
but with a discontinuity along the negative real axis
($|\mathrm{Arg}(g)|=\pi$).

\begin{figure}
\begin{center}
\includegraphics[scale=1]{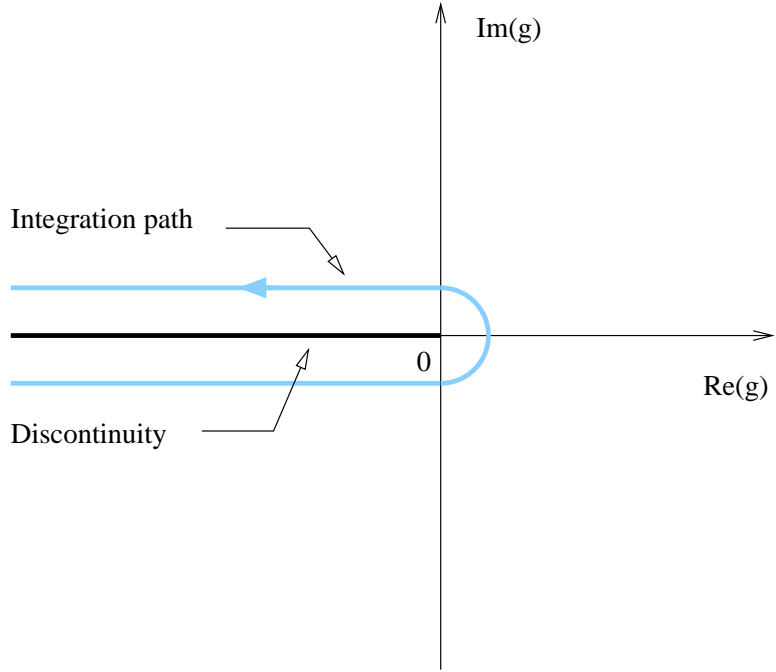}
\caption{Contour integration in the complex coupling constant plane for the large orders asymptotics}
\label{z-contour}
\end{center}
\end{figure}

Its asymptotic series expansion is written as
\begin{equation}
\label{pertexpz}
\Zfrak(g)\ =\ \sum_{k=0}^{\infty} \Zfrak_k\, g^k \ .
\end{equation}
The large order (large $k$) asymptotic behavior of the coefficients
$\Zfrak_k$ can be estimated by semi-classical methods.  Indeed, using a
classical dispersion relation, $\Zfrak_k$ can be written as a
Mellin-Barnes integral transform of the discontinuity of $\Zfrak(g)$
along the cut (see figure \ref{z-contour})
\begin{eqnarray}
\label{3.13}
\Zfrak_k\ =\ \int_{\mathcal{C}} {\rmd g\over 2\rmi\pi}\,
g^{-k-1}\,\Zfrak(g) \ &=&\ (-1)^{k} \int_0^{+\infty}{\rmd g\over
2\rmi\pi}\,
g^{-k-1}\,\left.\left[\Zfrak(-g+\rmi\epsilon)-\Zfrak(-g-\rmi\epsilon)\right]
\right|_{\epsilon\to 0_+}
\nonumber\\ 
\ &=&\ (-1)^{k} \int_0^{+\infty}{\rmd g\over \pi}\, 
g^{-k-1}\,\mathrm{Im}[\Zfrak(-g+\rmi 0_+)]
\end{eqnarray}
with $\mathcal{C}$ a counterclockwise contour around the cut ($\Zfrak$
is assumed to be real for real $g>0$).

For large positive $k$ this integral is dominated by the small $g$
behavior of the discontinuity, where semi-classical methods are
expected to be applicable.  Indeed, it turns out that for small real
negative $g$, the discontinuity of $\Zfrak(g)$ is dominated by the
contribution of a complex instanton (the real part of $\Zfrak(g)$
is still given by the contribution of the real classical solution).
Therefore the small $g$ behavior of $\mathrm{Im}[\Zfrak(g)]$ is of the
form
\begin{equation}\label{3.14}
\mathrm{Im}[\Zfrak(-g+0_+)]
\ =\ 
C\, |g|^\beta\,\rme^{- A |g|^{-\alpha}}\,\left[1+o(|g|^*)\right]
\end{equation}
with $A$ a number (corresponding to the instanton action
$S_{\mathrm{inst}}$), $\alpha$ a (positive) constant given by power
counting (in QM and standard local field theories $\alpha=1$), $C$ a
number related to the determinant of the Hessian operator of
fluctuations around the instanton, and $\beta$ a constant related to
the number $m_0$ of zero-modes of the Hessian (in standard local field
theories $\beta=1+m_0/2$).  (See appendix \ref{sec:o(n)vspolymer} for 
details on the polymer case.)

Given (\ref{3.14}), the integral (\ref{3.13}) can be calculated:
Changing variables from $g$ to $x:=g^{-\alpha}$, we obtain
\begin{eqnarray}\label{3.16}
\Zfrak_k &=& {(-1)^{k}}\frac C{\alpha \pi} \int_{0}^{\infty}\frac{\rmd x}{x} x^{(k-\beta)/\alpha} \rme^{-Ax}\nonumber \\
&=&(-1)^k\,{C\over
\pi\alpha}\,A^{{\beta-k\over\alpha}}\,\Gamma\left({k-\beta\over\alpha}\right)
\end{eqnarray}
$\Gamma$ is Euler's gamma function.  Note that since (\ref{3.14}) is
valid for small $g$ only, this result is valid for large $k$. Using
Stirling's formula $\Gamma(n)\simeq n^n\rme^{-n}\sqrt{{2\pi/ n}}$, this
amounts to
\begin{equation}
\label{lf19} \Zfrak_k\ =\ (-1)^k\,\left[{k\over
\alpha}\right]^{{k\over \alpha}}\,
\left[A\rme\right]^{\beta -k\over\alpha}\,\left[{k\over\alpha}\right]
^{-{\beta\over\alpha}}\,{C\over \pi\alpha}\, \sqrt{{2\pi\alpha\over
k}}
\,\left[ 
1+o(1/k^{*})\right]
\end{equation}
and for $\alpha=1$
\begin{equation}
\label{lf20}
\Zfrak_k\ =\ (-1)^k\,{k}^{{k}}\,
\left[{A \rme}\right]^{\beta -k}\,{k}^{-{\beta}}\,{C\over 
\pi}\, \sqrt{{2\pi\over k}}
\,\left[ 
1+o(1/k^{*})\right]
\ .
\end{equation}
It is an alternating asymptotic series with a Borel transform with
non-zero radius of convergence.

\subsection{Instanton for the SAM} \label{s:Instanton for the SAM} We
are thus interested in the analytic structure of the partition
function and the correlators of the SAM model for small negative
coupling constant $b$
\begin{equation}
\label{lf21}
b<0\quad,\qquad b\to 0
\ .
\end{equation}
In particular we are interested in the discontinuity along the
negative real axis.  As shown in \cite{DavidWiese1998} this can be
done more easily by first rescaling the fields and the size of the
manifold with $b$ in an adequate way.

\subsubsection{Complex rotation and rescalings for coupling constant
and fields}

We consider a finite manifold $\CM$ with internal size $L$ defined as
\begin{equation}
\label{VolL}
\mathrm{Vol}(\CM)\ =\ L^D
\ .
\end{equation}
We are interested in the model for small complex coupling constant
$b$, and more precisely in the discontinuity of the observables along
the negative real axis ($b<0$ real), where there is a cut.

We denote the argument of the coupling constant $b$  by $\theta$
\begin{equation}
\label{defTheta}
\theta = \mathrm{Arg}\left({b}\right)
\ .
\end{equation}
To reach the cut at negative $b$ from above or below amounts to taking
the limit
\begin{equation}
\label{Theta2pmPi}
\theta\ \to\ \pm\pi\qquad,\qquad |b|\ \mathrm{fixed}
\ .
\end{equation}
We now rescale
the internal coordinate of the membrane $\vx$ and the field $\rr$
with the size $L$ of the manifold and the modulus $|b|$ of the
coupling constant
\begin{equation}
\label{rescaleXR}
\vx\,\to\, |b|^{{1\over D-\epsilon}} L^{{D\over D-\epsilon}}\vx
\qquad,\qquad
\rr\,\to\,|b|^{{2-D\over 2(D-\epsilon)}} L^{{(2-D)D\over 2(D-\epsilon)}}\rr 
 \end{equation}
so that we now deal with a rescaled manifold $\CM_s$ with internal
size and internal volume
\begin{equation}
\label{rescaleSizeVol}
\mathrm{size}(\CM_s)\,=\,|b|^{-{1\over D-\epsilon}} L^{-{\epsilon\over
D-\epsilon}}  
\qquad,\qquad
\mathrm{Vol}(\CM_s)\,=\,|b|^{-{D\over D-\epsilon}} L^{-{\epsilon
D\over D-\epsilon}}  
\ .
\end{equation}
Similarly we must rescale the auxiliary fields $\rho$ and $V$ as
\begin{equation}
\label{rescaleRhoV}
\rho\ \to \ |b|^{-1} L^{-D} \rho
\qquad,\qquad
V\ \to\ |b|^{-{D\over D-\epsilon}} L^{-{D^2\over D-\epsilon}} V
\ .
\end{equation}
The purpose of these rescalings is that as the original coupling
constant $b$ goes to $0$, the effective theory for the auxiliary field
$V$ becomes simple. Indeed it appears that both terms in the effective
action $S[V]$ now scale in the same way, as will be detailed now.

\paragraph{Coupling constant:} 
Let us denote by $g$ the inverse of the volume of the rescaled manifold
\begin{equation}
\label{lf22}
g\ =\ {1\over \mathrm{Vol}(\CM_s)}\ =\ |b|^{{D\over
D-\epsilon}}\,L^{{D\epsilon\over D-\epsilon}} 
\end{equation}
$g$ is the (dimensionless) effective coupling constant of the theory,
which is real and positive and goes to 0 with $|b|$ as long as
$\epsilon<D$.

\paragraph{Partition function:}
The original partition function (for the manifold $\CM$) becomes for
the rescaled theory involving the manifold $\CM_s$
\begin{eqnarray}
\label{rescaledZ}
Z_{\CM_{s}}(b) & = & 
\int \CD[{\rr}]\,\CD[{V}]\, 
\exp\left(
-\int_{\CM_s} \left(\half (\nabla_{\!\vx} \rr)^2 +V(\rr)\right)
\,+\, \, {\rme^{-\rmi\theta}\over 2 g} \int_{\rr} V(\rr)^2 \right) \nonumber \\
 & = & \int \CD[V]\, \exp \left( -\,F_{\CM_s}[{V}] \,+\, \,
{\rme^{-\rmi\theta}\over 2 g} \int_{{\rr}} {V}({\rr})^2 \right) \ .
\end{eqnarray}
Due to (\ref{lf22}) both terms in the exponential scale as
$\mathrm{Vol}(\CM_s)=1/g$.

\paragraph{Functional measure:}
The functional measure over the rescaled field ${V}$ is now normalized
so that
\begin{equation}
\label{rescaledDV} \int \CD[{V}]\,\exp \left( {\rme^{-\rmi\theta}\over
2 g} \int_{{\rr}} {V}({\rr})^2 \right)\ =\ 1 \ .
\end{equation}

\paragraph{Correlation functions:}
The moments for the gyration radius of the manifold become in the
rescaled effective theory
\begin{eqnarray*}
\label{lf23} \mathsf{R_{gyr}^{(k)}}&=& b^{-2}\,L^{-2D}\,
\left(L\,g^{{1\over D-\epsilon}}\right)^{{(2-D)(2d+k)\over 2}-2D} \,
\int_{\rr_1}\int_{\rr_2} |\rr_1-\rr_2|^k\,V(\rr_1)\,V(\rr_2)
\\
&=&
\rme^{-2\rmi\theta}\,\left(|b|L^D\right)^{{2-D\over 2(D-\epsilon)}\cdot k}
\,\int_{\rr_1}\int_{\rr_2} |\rr_1-\rr_2|^k\,V(\rr_1)\,V(\rr_2)
\end{eqnarray*}
Hence
\begin{eqnarray}
\label{lf24}
\mathsf{R_{gyr}^{(k)}}
&=&
\rme^{-2\rmi\theta}\,L^{{2-D\over 2}k}\ 
g ^{\,{2-D\over 2D}k}\,
\int_{\rr_1}\int_{\rr_2} |\rr_1-\rr_2|^k\,V(\rr_1)\,V(\rr_2)
\ .
\end{eqnarray}
$L$ is the internal extension of the original manifold $\CM$.
This has the correct dimension $L^{\nu_0 k}$ with $\nu_0=(2-D)/2$, since 
$[ R^{(k)}]=[\rr]^k$ and $[\rr]=[x]^{{2-D}\over 2}$.
Note that there is no additional phase for $\theta=\pm\pi$.

\subsubsection{Semiclassical limit and the effective action $\CS[V]$}
Now come the crucial points:
\begin{enumerate}

\item As long as
\begin{equation*}
\label{lf25}
\epsilon<D
\end{equation*}
taking the semiclassical limit $b\to 0$ amounts to taking both the
small coupling limit $g\to 0$ in the rescaled theory and the
thermodynamic limit (infinite volume) for the rescaled manifold
\begin{equation*}
\label{lf26} b \to 0\qquad\Leftrightarrow \qquad g\to 0\ \
\mbox{and}\ \
\mathrm{Vol}\left({\CM_s}\right) \to \infty
\end{equation*}
for the rescaled manifold.

\item In this thermodynamic limit the free energy $F_{\CM_s}[V]$
 becomes proportional to the volume of the manifold
\begin{equation}
\label{FrEnAsymp}
F_{\CM_s}[{V}]\ =\ \mathrm{Vol}\left({\CM_s}\right)\,\CE[{V}]\,+\,\cdots
\ .
\end{equation} 
The free energy density $\CE[{V}]$ is defined as
\begin{equation}
\label{FrEnDeE}
\CE[{V}]\ :=\ \lim_{\mathrm{Vol}(\CM_{s})\to\infty} {1\over
{\mathrm{Vol}}(\CM_{s})} \,F_{\CM_{s}}[{V}] 
\end{equation}
in the limit where the size of the manifold $\CM_s$ is rescaled to
$\infty$, and its shape kept fixed. In this limit, the manifold
$\CM_s$ becomes locally a flat $D$-dimensional Euclidean space
$\mathbb{R}^D$:
\begin{equation} \label{lf27} {\CM_s}\ \to\
\mathbb{R}^D \ .
\end{equation}
The free energy density $\CE[V]$ is independent of the size \emph{and
of the shape} of the manifold and it is enough to compute it for the
infinite flat manifold.

\item Moreover -- and this is an important point -- as long as we are
interested in the contribution of potentials $V$ such that the
manifold is ``trapped'' in $V$ (namely such that the free energy
density $\CE[V]<0$ is negative, i.e.\ such that there is a ``bound
state'' in $V$) the neglected terms $+\cdots$
are expected to be exponentially small in $1/g$.
\item
Finally, since 
\begin{equation}
\label{lf28}
\mathrm{Vol}(\CM_s)\ =\ {1\over g}
\end{equation}
in the limit $g\to 0$
the functional integral takes the standard form
\begin{equation}
Z(b)\  \stackrel{g\to 0}{=} \  
\int \CD[V]\,\exp\left[-\,{1\over g} \mathcal{S}[V]
\right]
\ ,
\end{equation}
where $g$ is given by (\ref{lf22}), the measure is given by
(\ref{rescaledDV}) and the effective action $\CS[V]$ for the field $V$
is given by
\begin{equation}\label{3.35}
\mathcal{S}[V]\  = \ \mathcal{E}[V]-{\rme^{-\rmi\theta}\over 2}\int V^2
\end{equation}
$\CE[V]$ is the free energy density for an infinite flat manifold
trapped in the potential $V$, and is given by (\ref{FrEnDeE}).
\end{enumerate}

\subsubsection{The functional integral for negative $b$ and the instanton} 
We are interested in the imaginary part of the partition
function $Z(b)$ for $b$ along the negative real axis, that is for
\begin{equation}
\label{lf31}
\theta\ \to\ \pm\,\pi
\ .
\end{equation}
In this limit the effective action $\CS[V]$ for the rescaled theory is real
\begin{equation}
\label{lf32}
\mathcal{S}[V]\ =\ 
\mathcal{E}[V]\,+\,{1\over 2}\int V^2
\end{equation}
and the measure over $V$ is also real, since it is normalized such that
\begin{equation}
\label{lf33}
\int \mathcal{D}[V]\,\exp\left[-\,{1\over 2 g}\,\int V^2\right]\ =\ 1
\ .
\end{equation}
It is now the standard measure for a real white noise with variance
$g$:
\begin{equation}
\label{lf34}
\langle V(\rr_1)V(\rr_2)\rangle\ =\ {g}\,\delta(\rr_1-\rr_2)
\ .
\end{equation}
Thus we can chose for integration measure over $V$ the standard
measure over real $V(\rr)$ 
\begin{equation}
\label{lf35} \int\CD[V]_{\theta=\pm\pi}
\ =\ \int_{-\infty}^{\infty}\prod_\rr\, {\rmd V(\rr) \over \sqrt{2\pi
g \delta^d(0)}} \ .
\end{equation}

The instanton $V^{\mathrm{inst}}$ is a non-trivial finite action
extremum of the action $\mathcal{S}[V]$ and was found in
\cite{DavidWiese1998}. The saddle-point equation is
\begin{equation}
\label{lf36} 0 = \frac{\delta {\cal S}[V]}{\delta V (\rr)} =
V(\rr)\,+\,\langle\rho(\rr)\rangle_V \ ,
\end{equation}
where $\rho(\rr)$ is the manifold density at point $\rr$
\begin{equation}
\label{lf37}
\rho(\rr)\ =\ {1\over \mathrm{Vol}(\mathcal{M})}\int
\rmd^D\vx\,\delta(\rr-\rr(\vx)) 
\end{equation} and from now on we drop the index at $\CM_s$. 
$\langle\cdots\rangle_V$ refers to the expectation value for the
phantom manifold trapped in the external potential $V(\rr)$, that is
with the action
\begin{equation}
\label{lf38} \int_{\mathcal{M}} \rmd^D\vx\,\left({1\over
2}\left(\nabla_\vx\rr\right)^2+V(\rr)\right) \ .
\end{equation}
The ``classical vacuum'' is $V=0$ (free manifold).  The instanton
$V^{\mathrm{inst}}$ is a configuration of potential which is negative
(potential well $V(\rr)<0$), spherically symmetric, with $V\to 0$ as
$|\rr|\to\infty$.
The solution of the instanton equation and its properties have been
studied in \cite{DavidWiese1998}.

\subsection{Contribution of fluctuations around the instanton}
\label{s:Contribution of fluctuations around the instanton}
\subsubsection{Instanton zero modes} The Hessian matrix (second
derivative of the action) is
\begin{equation}
\label{lf39} {{\CS}''[V]}_{\rr_1 \rr_2}\ =\ {\delta^2 \CE[V]\over\delta
V(\rr_1)\delta V(\rr_2)} \ =\ \delta^d(\rr_1-\rr_2)\,+\,
\langle\rho(\rr_1)\rho(\rr_2)\rangle_V^{\mathrm{conn}}  \ .
\end{equation}
The instanton has $d$ translational zero modes, corresponding to the
position of the center of gravity $\rr_0$ of the instanton.  Thus the
Hessian has $d$ zero modes
\begin{equation}
\label{lf40}
V^{\mathrm{zero}}_a\ =\ \nabla_a V^{\mathrm{inst}}
\qquad;\qquad
{\CS}''[V]\cdot V^{\mathrm{zero}}_a\ =\ 0
\ .
\end{equation}
According to the previous analysis, see Eq.~(\ref{Metric4Phi}), the
metric on the instanton moduli space $\mathcal{M}=\mathbb{R}^d$, $
\rmd s^2\ =\ h_{ab}\,\rmd\rr_0^a \rmd\rr_0^b$, is
\begin{equation}
\label{metricmodr0} h_{ab}\ =\ {1 \over 2\pi g}\int
\rmd^d\rr\,V^{\mathrm{zero}}_a V^{\mathrm{zero}}_b \ =\
\delta_{ab}\,{1\over 2\pi\, g\,d}\int
\rmd^d\rr\,\left(\overrightarrow{\nabla} V^{\mathrm{inst}}\right)^2
\end{equation}
(using rotational invariance).
Therefore the measure over the instanton position $\rr_0$ is 
$$\rmd^d \rr_0\,\left[{1\over 2\pi\, g\,d}\int
\rmd^d\rr\,\left(\overrightarrow{\nabla}
V^{\mathrm{inst}}\right)^2\right]^{d/2}\ .$$ Hence the contribution of
the instanton to the partition function will be (depending on whether
$\theta=\mathrm{Arg}(b)=\pm\pi$)
\begin{equation}
\label{ImZMem0} Z(b)\ \
\stackrel{\text{instanton}}{\llongleftarrow}\ \ %\mp\,\half
\mathbf{C}_\pm \int \rmd^d\rr_0\,\left[{1\over 2\pi
d\,g}\int_\rr(\vec\nabla V)^2\right]^{d/2} \,\rme^{-\frac{1}{g}\CS[V]} \,
\left[{\det}'\bigl({\CS}''[V]\bigr)\right]^{-\half}
\end{equation}
with $\mathbf{C}_\pm$ a simple 
factor (usually 1 or an integer for a real instanton) giving the
weight of the instanton in the functional integral.

One might also expect zero-modes associated to the rotational
invariance of the theory. Such modes would indeed appear for a
non-rotationally invariant instanton solution. As it will turn out,
the instanton is rotationally invariant, such no such zero-modes
exist.

\subsubsection{Unstable eigenmode} \label{unsteigmode} However, as
expected for a theory with the wrong sign of the coupling and as shown
in \cite{DavidWiese1998}, the instanton has one unstable eigenmode
$V^-(\rr)$.
Thus the Hessian has one negative eigenvalue $\lambda^-$
and its determinant is real but negative: ${\det}'(S'')<0$. 
Therefore we expect that the factor $\mathbf{C}_\pm$ will be complex.

In fact, as this is the case for 
the instanton in the local $\phi^4$ field theory, the real part of
$\mathbf{C}_\pm$ is not unambiguously defined, but depends on the
resummation procedure used to define the contribution of the classical
saddle-point $V=0$ in the functional integral (this is known as the
Stokes phenomenon).  However, the instanton gives the dominant
contribution to the imaginary part of the functional integral, and one
can show that
\begin{equation}
\label{ImZMem1}  \mathrm{Im}\bigl[Z(b)\bigr]\ \
\stackrel{\text{instanton}}{\llongleftarrow}\ \ \mathbf{D}_\pm \int
\rmd^d\rr_0\,\left[{1\over 2\pi d\,g}\int_\rr(\vec\nabla
V)^2\right]^{d/2} \,\rme^{-\CV\,\CS[V]} \,
\left|{\det}'\bigl({\CS}''[V]\bigr)\right|^{-\half}
\end{equation}
with the weight factor $D_\pm$
\begin{equation}
\label{lf44}
  \mathbf{D}_\pm\ =\ \mp\,{\mathrm{i}\over 2}\ .
\end{equation}
This result can be obtained by a more precise analysis of the
respective position of the integration path and of the instanton
solution in the space of complex potentials $V(\rr)\in\mathbb{C}$ as
$\theta$ is rotated from $0$ to $\pm\pi$, using the steepest descent
method.  This is shown in Appendix \ref{section-contours}.

\subsection{Final result for the instanton contribution}
The final result for the imaginary part of the partition function at
negative coupling is
\begin{equation}
\label{ImZMem}
    \mathrm{Im}\,Z(b)\ =\ \mp\,\half \int \rmd^d\rr_0\,\left[{1\over
    2\pi d\,g}\int_\rr(\vec\nabla V)^2\right]^{d/2}
\,\rme^{-{1\over g}\,\CS[V]}
\, \bigl|{\det}'\left({\CS}''[V]\right)\bigr|^{-\half}   
\ .
\end{equation}
depending on whether $\mathrm{Arg}(b)=\pm\pi$.
The infinite bulk volume factor $\int \rmd^d\rr_0$ disappears (as it
should) in the normalized partition function $\Zfrak=Z/Z_0$
\begin{equation}
\label{ImZfrakMem}
\mathrm{Im}\,\Zfrak(b)\ =\ \mp\,\half \,g^{-d/D}\,
\left[{\rme^{-\bar\zeta'(0)}\over d}
\int_\rr(\vec\nabla V)^2\right]^{d/2}
\,\rme^{-{1\over g}\,\CS[V]}
\, \left|{\det}'\left({\CS}''[V]\right)\right|^{-\half}   
\ ,
\end{equation}
where $\tilde \zeta' (0)$ was defined in (\ref{zetadef}). 
One must remember that
\begin{equation}
\label{lf63}
g\ =\ \big(|b|L^{-\epsilon}\big)^{{D\over D-\epsilon}}
\end{equation}
and that $\rr$ is in fact the dimensionless rescaled field
$\widetilde{\rr}=\rr\left(|b|L^D\right)^{-{2-D\over
2(D-\epsilon)}}=\rr\left(g\,L^D\right)^{-{2-D\over 2D}}$ defined in
(\ref{rescaleXR}).  We thus obtain for the discontinuity of the
partition function $\CZ(b)$ for a marked manifold with a fixed point
(as defined by Eq.(\ref{Z4r}))
\begin{equation}
\label{ImCZMem}
    \mathrm{Im}\,\CZ(b)\ =\ \mp\,\half \,L^{-d{2-D\over 2}}
\,g^{\ -{d\over D}}\,\left[{1\over 2\pi d}
\int_\rr(\vec\nabla V)^2\right]^{d/2}
\,\rme^{-{1\over g}\,\CS[V]}
\, \left|{\det}'\left({\CS}''[V]\right)\right|^{-\half}   
\ .
\end{equation}
For the $N$-point correlators ${\cal R}^{(N)}(\rr_1,\cdots,\rr_N;b)$ defined
by \eq{defCRN} the result is more complicated since the $\rr_i$'s are
rescaled in the process $b\to g$. However he result takes a simple
form for global quantities such as the moments of the radius of
gyration of the manifold
$\mathcal{R}_{\mathrm{gyr}}^{(k)}=\langle\mathsf{R_{gyr}^{(k)}}\rangle$
defined by \eq{defGyrMom}
\begin{equation}
\label{ImRkMem}
\begin{split}
  \mathrm{Im}\,
  \mathcal{R}_{\mathrm{gyr}}^{(k)}
  \ =\   &  \mp\,\half 
\,L^{(k-d){2-D\over 2}}
\,g^{-{d\over D}+k{2-D\over 2D}}
\,\rme^{-{1\over g}\,\CS[V]}\times 
\\
  \   &  \ \times
\,\left[{1\over 2\pi d}
\int_\rr(\vec\nabla V)^2\right]^{d/2}
\, \big|{\det}'\left({\CS}''[V]\right)\big|^{-\half} 
\, \left[\int_{\rr_1}\int_{\rr_2}|\rr_1-\rr_2|^k\, V(\rr_1)V(\rr_2)\right]\ .
\end{split} 
\end{equation}

\subsection{Large orders}
In the rest of this article, we shall denote for simplicity
\begin{equation}
\label{SDLWdef}
\mathfrak{S}=\CS[V^{\mathrm{inst}}] \ ,\quad
\mathfrak{D}\,=\,{\det}'\left(\CS[V^{\mathrm{inst}}]\right) \ ,\quad
\mathfrak{L}\,=\,\log(\mathfrak{D})
\ ,\quad \mathfrak{W}\,=\,\left[{1\over 2\pi d}\int_\rr(\vec\nabla
V^{\mathrm{inst}})^2\right]^{d/2}\ .
\end{equation}
If no UV divergences were present at $\epsilon=0$, the final result at
$\epsilon =0$ would be
\begin{equation}
\label{ImCZsc} \mathrm{Im}\,\CZ(b)\ =\ \mp\,\half\,
L^{-2D}\,|b|^{{4\over 2-D}}\,\rme^{-{1\over
|b|}\mathfrak{S}}\,\mathfrak{W}\,|\mathfrak{D}|^{-{1\over 2}} \ .
\end{equation}
Using the the arguments of Sect.~\ref{sss:loi}, in particular the
dispersion relation (\ref{3.13}) and (\ref{lf20}), the large-order
asymptotics for the perturbative expansion of $\CZ(b)$
\begin{equation}
\label{CZbexp}
\CZ(b)\ =\ \sum_{k=0}^\infty \CZ_k\,b^k
\end{equation}
would be ($\epsilon =0$)
\begin{equation}
\label{CZklo1} \CZ_k\ \simeq\ (-1)^k \,\Gamma\left(k-{4\over
2-D}\right)\,{1\over
2\pi}\,L^{-2D}\,\mathfrak{W}\,|\mathfrak{D}|^{-{1\over
2}}\,\mathfrak{S}^{{4\over 2-D}-k}
\end{equation}
or equivalently  ($\epsilon =0$)
\begin{equation}
\label{CZklo2} \CZ_k\ \simeq\ (-1)^k \,\Gamma\left(k-2-{d\over
2}\right)\,{1\over 2\pi}\,L^{-{4d\over
4+d}}\,\mathfrak{W}\,|\mathfrak{D}|^{-{1\over
2}}\,\mathfrak{S}^{2+{d\over 2}-k}\ ,
\end{equation}
indicating that the Borel transform of $\CZ(b)$ has a finite radius of
convergence $\mathfrak{S}$. Of course the instanton normalization
$\mathfrak{W}\,|\mathfrak{D}|^{-{1\over 2}}$ depends also on $d$.

\section{UV Divergences and renormalization}
\label{s:renormalization}
We now discuss the UV divergences in the
determinant factor for the instanton, and how they are
renormalized. We remind the reader that at one loop in perturbation
theory, for $0<\epsilon\le D$ there is a divergence associated to the
operator $\BBI$ (super-renormalizable case); for $\epsilon=0$ two
divergences associated to the operators $(\nabla\rr)^2$ and
$\delta^{d}\bigl(\rr(\xx)-\rr(\yy)\bigr)$ (renormalizable case). For
$\epsilon<0$ the theory is not renormalizable.  The model is always
considered for $D<2$ and $\epsilon$ is given by
\begin{equation}
\label{ren1}
\epsilon=2D-{d\over 2}(2-D)\ .  
\end{equation} 

\subsection{Series representation of the determinant for the fluctuations}
\label{s:ren:series}
The Hessian matrix $\CS''$ is given by (\ref{lf39}). We rewrite it as
\begin{equation}
\label{S2O}
\CS''_{\rr_1\rr_2}\ =\ \JU _{\rr_1\rr_2}\,-\,\mathbb{O}_{\rr_1\rr_2}
\quad,\quad
\JU _{\rr_1\rr_2}\ =\ \delta^d(\rr_1-\rr_2)
\end{equation}
\begin{equation}
\label{Odef}
\mathbb{O}_{\rr_1\rr_2}\ =\ 
\lim_{
\CM\to\mathbb{R}^D }{1\over
\text{Vol}(\CM)}\int_{\xx_1}\int_{\xx_2}{\bigl\langle
\delta^d\bigl(\rr_1-\rr(\xx_1)\bigr)\delta^d\bigl(\rr_2
-\rr(\xx_2)\bigr)\bigr\rangle}_V^{\text{conn}}  \ ,
\end{equation}
where $V$ is the instanton potential $V^{\text{inst}}$.  $\mathbb{O}$
can be rewritten, using translational invariance $\xx\to\xx+\xx_0$
when $\CM\to\mathbb{R}^D$, and the saddle point equation for the
instanton potential $V$
\begin{eqnarray}\label{Odef2}
\mathbb{O}_{\rr_1\rr_2} &=&
\int_{\mathbb{R}^D}\!\mathrm{d}^D\xx\,{\bigl\langle
\delta^d\bigl(\rr_1-\rr(0)\bigr)\delta^d\bigl(\rr_2
-\rr(\xx)\bigr)\bigr\rangle}_V^{\text{conn}}
\nonumber \\
&=& \int_{\mathbb{R}^D}\!\mathrm{d}^D\xx\,\left[{\bigl\langle
\delta^d\bigl(\rr_1-\rr(0)\bigr)\delta^d\bigl(\rr_2-\rr(\xx)
\bigr)\bigr\rangle}_V
\,-\,{\bigl\langle \delta^d\bigl(\rr_1-\rr(0)\bigr)\big\rangle}_V
{\big\langle\delta^d\bigl(\rr_2-\rr(\xx)\bigr)\bigr\rangle}_V\right]\nonumber 
\\
&=& \int_{\mathbb{R}^D}\!\mathrm{d}^D\xx\,\left[{\bigl\langle
\delta^d\bigl(\rr_1-\rr(0)\bigr)\delta^d\bigl(\rr_2-\rr(\xx)\bigr)
\bigr\rangle}_V\,-V(\rr_1)V(\rr_2)\right]
\ .
\end{eqnarray}
Let us already note that such an integral is IR finite, since from
clustering we expect that at large distances
\begin{equation}
\label{clustering} {\bigl\langle
\delta^d\bigl(\rr_1-\rr(\xx)\bigr)\delta^d\bigl(\rr_2-\rr(\yy)\bigr)
\bigr\rangle}_V^{\text{conn}}  \ =\
\mathcal{O}\bigl(\exp(-|\xx-\yy|m)\bigr)\quad\text{when}\quad
|\xx-\yy|\,\to\,\infty \ ,
\end{equation}
where $m$ is the ``mass gap'' of the excitations for the manifold
trapped in the instanton potential $V$.

We have seen that the operator $\CS''$ has $d$ zero modes
$V_a^{\text{zero}}\propto\nabla_a V^{\text{inst}}$, which, as
discussed in section (\ref{s:large orders.a}), are eigenvectors of
$\mathbb{O}$ with eigenvalue $\lambda_0=1$, and one unstable eigenmode
$V^-$, which is an eigenvector of $\mathbb{O}$ with eigenvalue
$\lambda_-$ larger than $1$. For convenience, we normalize its
$\mathrm{L}_{2}$ norm to 1.  Let us denote $\BBP^{0}$ the projector
on the zero-modes, and $\BBP^{-}$ the projector on the unstable mode
\begin{equation}
\label{BBP0} {\BBP^0}_{\,\rr_1\rr_2}\ =\ {\sum_a
V_a^{\text{zero}}(\rr_1)V_a^{\text{zero}}(\rr_2)
} \ =\
{\overrightarrow{\nabla}V\otimes\overrightarrow{\nabla}V\over\int_\rr
\left(\overrightarrow{\nabla}V\right)^2} \quad,\quad
{\BBP^-}_{\,\rr_1\rr_2}\ =\ { V_-(\rr_1)V_-(\rr_2)
}
\end{equation}
and $\BBP$ the sum 
\begin{equation}
\label{BBP-}
\BBP\ =\ \BBP^0\,+\,\BBP^-
\ .
\end{equation}
Apart from these eigenvalues, is is easy to see that all other
eigenvalues of $\BBO$ are smaller than $1$, but positive. Indeed, from
Eq.(\ref{Odef}), $\BBO$ is a positive operator, since for any $f(\rr)$
\begin{equation}
\label{O>0}
f\cdot\BBO\cdot f\ =\ {1\over\text{Vol}(\CM)}\, \left<  
\left[\int_\xx f\bigl(\rr(\xx)\bigr) \right]^2
\right>_V^{\text{conn}}\ > \ 0 \ .
\end{equation}
To compute the determinant of the fluctuations we treat separately the
negative and zero modes from the rest. We write the logarithm of
${\det}'[\mathcal{S}'']$
\begin{equation}
\label{logmode} \mathfrak{L}\ =\
\log\big({{\det}'\left[\CS''\right]}\big)\ =\
\log(1-\lambda_-)\,+\,\tr\left[(\BBI-\BBP)\log(\BBI-\BBO)\right] \ .
\end{equation}
The first term is the contribution of the unstable mode (it has an imaginary part), the second term is 
the contribution of all other modes with $0<\lambda<1$. In this last
term we can expand the $\log$
and obtain a convergent series
\begin{equation}
\label{logser} \mathfrak{L}\ =\
\log(1-\lambda_-)\,-\,\sum_{k=1}^\infty {1\over k}\,L_k\quad,\quad
L_k\ =\ \tr\bigl[(\BBI-\BBP)\BBO^k\bigr]\ =\ \tr\left[
\BBO^k\right]-d-\lambda_-^k
 \ ,
\end{equation}
provided that each term is UV finite (that is the trace is well defined).

We now show that only the first two terms $k=1$ and $k=2$ are UV
divergent, and require renormalization.

\subsection{UV divergences}
\label{s:ren:UVdiv}
\subsubsection{UV divergences in $\rr$ and in $\xx$ space} 
UV divergences in the determinant are expected to come
from the high momentum eigenmodes of $\CS''$. If we consider a
potential $V=V^{\text{inst}}+V_>$, with $V_>$ a high momentum
fluctuation, we expect that a phantom manifold trapped in $V$ will
feel only weakly the small wavelength variations of $V$, so its free
energy $\CE[V]$ will depend only weakly on $V_>$. The other term $\int
\mathrm{d}^d\rr\,V^2(\rr)$ will be dominant in the variation of the
effective action $\CS[V]$. As a consequence, high momentum eigenmodes
of $\CS''$ will have eigenvalues close to $1$, that is will be
eigenmodes of $\BBO$ with very small eigenvalues $\lambda\to 0$.

Therefore UV divergences will come from the contribution of the
numerous eigenvalues of $\BBO$ close to $0$, that is from the
divergence of the spectral density $\rho_\BBO(\lambda)$ of the
operator $\BBO$ at $\lambda=0$.  We shall show that
$\rho_\BBO(\lambda)$ diverges as $\lambda^{\epsilon/D-3}$, and that
\begin{equation}
\label{trOkdiv}
\tr[\BBO]\quad\text{is UV divergent if}\ \epsilon\le D\quad,\quad
\tr[\BBO^2]\quad\text{UV divergent if}\ \epsilon= 0
\end{equation}
and that higher powers $\tr[\BBO^k]$ ($k\ge 3)$ are UV convergent.

The $\tr[\cdot]$ amounts in our representation to an integral over
$\rr$ in bulk space $\mathbb{R}^d$.  UV divergences will occur as
short-distance singularities in $\rr$ space. We shall also see that to
analyze the UV divergences it is more convenient to come back to the
equivalent representation of $\BBO$ in $\xx$ space (internal
manifold).
\subsubsection{$\tr[\BBO]$:}
This term is 
given by 
\begin{equation}
\label{trOfirst}
\tr[\BBO]\ =\ \int \mathrm{d}^d\rr\ \BBO_{\rr\rr}
\end{equation}
and is UV divergent for $\epsilon\le D$ because we expect that
\begin{equation}
\label{Oshort}
\BBO_{\rr\rr'}\ \simeq\ {|\rr-\rr'|}^{-d+{2(\epsilon-D)\over 2-D}}\quad\text{as}\quad \rr-\rr'\,\to\, 0
\ .
\end{equation}
The crucial point (to be proven later) is that the short-distance
behavior of $\BBO_{\rr\rr'}$ for a manifold in the background
potential $V(\rr)$ does not depend on the details of the potential $V$,
and is given (at leading order) by that of a free manifold in a
constant potential ($V(\rr)=V_0$).  We can compute explicitly
$\BBO_{\rr\rr'}$ in that case and find Eq.~(\ref{Oshort}).

Using (\ref{Odef2}) we can rewrite $\tr[\BBO]$ as an $\xx$-integral
over the manifold $\CM$, and integrate explicitly over $\rr$, with
the result
\begin{equation}
\label{trO}
\begin{split}
\tr[\BBO]\ 
&=\ 
 \int \mathrm{d}^d\rr\ \int_\CM \!\!\!\!\rmd^D\xx\, \Bigl[{\bigl\langle \delta^d\bigl(\rr-\rr(\xx_0)\bigr)\delta^d\bigl(\rr-\rr(\xx)\bigr)\bigr\rangle_V-V(\rr)^2}\Bigr]
\\
&=\ \int_\CM\!\!\!\!\mathrm{d}^D\xx\,\Bigl[{\bigl\langle \delta^d\bigl(\rr(\xx)-\rr(\xx_0)\bigr)\bigr\rangle_V\,-\,\int_\rr V(\rr)^2}\Bigr]\ .
\end{split}
\end{equation}
It contains the integral of the correlation function
\begin{equation}
\label{2ptV}
\langle{\delta^d\bigl(\rr(\xx)-\rr(\xx_0)\bigr)\rangle}_V
\end{equation}
for a phantom manifold (i.e.~without self-interaction) trapped in the
instanton potential $V(\rr)$.  The choice of the ``origin'' $\xx_0$ is
arbitrary, since (\ref{2ptV}) depends only on $\xx-\xx_0$
(translational invariance in $\CM=\mathbb{R}^D$).

This integral is IR convergent, as can be seen from
Eq.~(\ref{clustering}).
UV divergences occur if the $\xx$-integral is divergent at short
distances on the manifold, i.e.~for $|\xx-\xx_0|\to 0$.  The
correlation function (\ref{2ptV}) is very similar to the 2-point
correlation function which appears at first order in the perturbative
expansion of the self-avoiding manifold model, and more precisely for
the normalized partition function $\mathfrak{Z}(b)$
\begin{equation}
\label{ZfrakV0} \mathfrak{Z}(b) = 1- {b\over 2}\,\int_{\xx,\yy}\!
{\bigl\langle\delta^d\bigl(\rr(\xx)-\rr(\yy)\bigr)\bigr\rangle}_0+
{b^2\over 8} \,\int_{\xx,\yy,\xx',\yy'}\!
{\bigl\langle\delta^d\bigl(\rr(\xx)-\rr(\yy)\bigr)
\,\delta^d\bigl(\rr(\xx')-\rr(\yy')\bigr)\bigr\rangle}_0
+ \mathcal{O}(b^3)\ .
\end{equation}
One therefore expects that the renormalization group counter-terms at
leading order, which subtract the leading order UV-divergences in
(\ref{ZfrakV0}) are also sufficient to render (\ref{trO}) finite. That
this is indeed the case will be shown below.

\subsubsection{$\tr[\BBO^2]$:} 
Similarly, starting from (\ref{Odef2}),
we can rewrite $\tr[\BBO^2]$ in term of two ``replicas'' of the
manifold, labeled $\CM_1$ and $\CM_2$, fluctuating independently in
the same instanton potential $V$ (without interactions). If we denote
$\rr_1(\xx)$ and $\rr_2(\xx)$ the $\rr$-fields for the two replicas,
we have
\begin{equation}
\label{trO2}
\begin{split}
\tr[\BBO^2]\ 
=\ &
\int\! \mathrm{d}^d\rr \int\! \rmd^d{\rr '}
\ \BBO_{\rr\rr'}\BBO_{\rr '\rr}\ \\
=\ &\int\! \mathrm{d}^d\rr \int\! \mathrm{d}^d{\rr
'}\,\int_{\CM_1}\!\!\!\mathrm{d}^D\xx
\int_{\CM_2}\!\!\!\mathrm{d}^D\yy \, \Bigl[{\bigl\langle
\delta^d(\rr-\rr_1(\xx_0))\delta^d(\rr'-\rr_1(\xx))\bigr
\rangle_V-V(\rr)V(\rr')}\Bigr]
\times\\
&\hskip 4.7cm\times \bigl[{\bigl\langle
\delta^d(\rr'-\rr_2(\yy_0))\delta^d(\rr-\rr_2(\yy))\bigr\rangle_V
-V(\rr')V(\rr)}\bigr]
\\
=\ &\int_{\CM_1} \!\!\mathrm{d}^D\xx \int_{\CM_2} \!\!\mathrm{d}^D\yy
\, \left\{{ \Bigl\langle
\delta^d\bigl(\rr_1(\xx_0)-\rr_2(\yy)\bigr)\,\delta^d\bigl(
\rr_1(\xx)-\rr_2(\yy_0)\bigr)\Bigr\rangle_V
\vphantom{\left[\int_{\rr_1}V(\rr)^2 \right]^2} } \right.
\\
&\hskip 1.0cm\left.
\,-\,\bigl\langle V(\rr_1(\xx_0))V(\rr_1(\xx))\bigr\rangle_V
\,-\,\bigl\langle V(\rr_2(\yy_0))V(\rr_2(\yy))\bigr\rangle_V
\,+\,\left[\int_{\rr}V(\rr)^2 \right]^2
\right\} \ .
\end{split}
\end{equation}
The choice of the origins $\xx_0$ and $\yy_0$ on the two manifolds
$\CM_1$ and $\CM_2$ is arbitrary.

This integral is IR finite by the same arguments as those for
$\tr[\BBO]$.  UV divergences are only present in the first correlation
function
\begin{equation}
\label{ren3} \Bigl\langle \delta^d\bigl(\rr_1(\xx_0)-\rr_2(\yy)\bigr)\,
\delta^d\bigl(\rr_1(\xx)-\rr_2(\yy_0)\bigr)\Bigr\rangle_V
\end{equation}
very similar to the correlation function which appears at second order
in $\mathfrak{Z}(b)$, see Eq.~(\ref{ZfrakV0}).  We shall see that UV
divergences occur when
\begin{equation}
\label{ren4}
\xx\to\xx_0\quad,\quad \yy\to \yy_0\qquad\text{simultaneously}
\end{equation}
while the other terms $\bigl\langle
V\bigl(\rr(\xx)\bigr)V\bigl(\rr(\xx_0)\bigr)\bigr\rangle_V$ are not
singular.

\subsubsection{$\tr[\BBO^k]$, $k\ge 3$:} 
We can similarly write the higher order
terms.  At order $k$ we need $k$ copies $\CM_{\alpha}$ of the manifold
$\CM$, fluctuating in the same instanton potential $V(\rr)$.  The most
UV singular term in the $\xx$-representation of $\tr[\BBO^k]$ is
\begin{equation}
\label{ren5} \int_{\otimes\CM_\alpha} \hskip -1.ex
\mathrm{d}^D\xx
\Bigl<\prod_{\alpha=1}^k\delta^d\bigl(\rr_{\alpha-1}(\mathbf{o}_{\alpha-1})
-\rr_{\alpha}(\xx_{\alpha}\bigr)\Bigr>_V 
\end{equation}
(where we identify $\alpha=0$ with $\alpha=k$), that we can represent
graphically as a ``necklace of $k$ manifolds''.  The reference points
$\mathbf{o}_{\alpha}$ on each $\CM_\alpha$ can be chosen arbitrarily,
for instance fixed to the origin.  UV divergences occur when all pairs
of points $(\mathbf{o}_{\alpha},\xx_\alpha)$ collapse simultaneously
on each $\CM_\alpha$.  These terms are in fact UV finite for
$\epsilon=0$.

\begin{figure}[h]
\begin{center}
\includegraphics[width=2in]{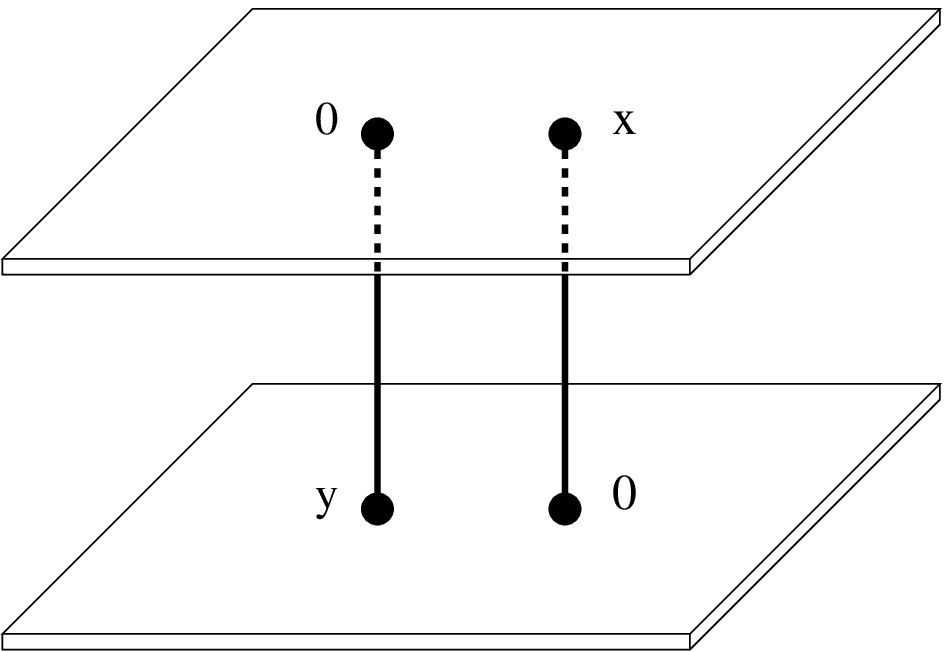}~~~
\includegraphics[width=2in]{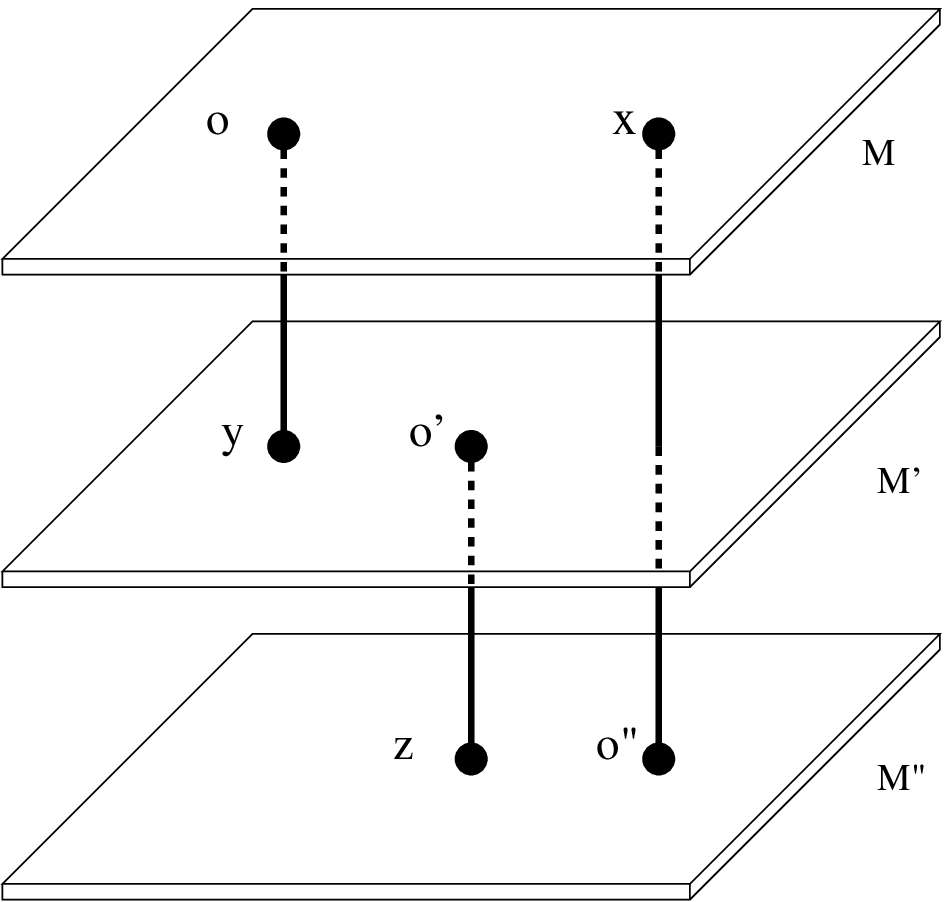} \caption{Diagrammatic
representation of the UV divergent correlation functions at order
$k=1$ (one manifold), $k=2$ (2 manifolds), and $k=3$ (3 manifolds).}
\label{neckfig}
\end{center}
\end{figure}

\subsection{MOPE for manifold(s) in a background potential}
\label{s:ren:mope} In \cite{DDG3,GBU} the UV divergences of the
self-avoiding manifold model have been analyzed using a Multilocal
Operator Product Expansion (MOPE). This formalism was developed to
study the correlation function of multilocal operators of the form
(\ref{lf3}),
\begin{equation}
\label{ren6}
\bigl<\prod_{i}\delta^d\bigl(\rr(\xx_i)-\rr(\yy_i)\bigr)\bigr>_0
\ ,
\end{equation}
where the expectation values $\bigl<\cdots\bigr>_0$ are calculated for
a free manifold model ($V=0$).  We show here how this formalism can be
adapted to deal with expectation values $\bigl<\cdots\bigr>_V$ for
manifolds trapped in a non-zero background potential $V(\rr)$.

\subsubsection{Normal product decomposition of the potential $V$}
\label{s:ren:mope:npV} In order to compute easily expectation values
of operators in the background potential $V$, 
we shall use the normal product formalism already developed in
\cite{DavidWiese1998}.

For simplicity we consider a potential $V(\rr)$ spherically symmetric
(as the instanton potential) with its minimum at $\rr=0$, of the form
\begin{equation}
\label{Vrdevv}
V(\rr)\ =\ \sum_{n=0}^\infty {v_n\over {2}^n n!} \left(\rr^2\right)^n
\quad,\quad v_1\,=\,m_0^2\,>\,0\ \quad m_0\, =\, \text{``bare mass''}
\ .
\end{equation}
We may (at least formally), compute expectation values of operators
$\bigl<\cdots\bigr>_V$ in perturbation theory, starting from the
Gaussian potential $V(\rr)={m_0^2\over 2}\rr^2$ and expanding in
powers of the non-linear couplings $\{v_k,\,k\ge 2\}$. This
perturbation theory involves Feynman diagrams with massive propagators
$1/(p^2+m_0^2)$.  It is more convenient to resum all tadpole diagrams
and to deal with an expansion of the potential $V(\rr)$ in terms of
normal products ${:\!\left(\rr^2\right)^n\!:}_m$.  The normal product
${:\! [\ ]\!:}_\mu$ with the subtraction mass scale $\mu$ is defined
by the global formula (expanded in $\kk$, it generates all operators
which are local powers of $\rr$)
\begin{equation}
\label{ren7} {:\!\mathrm{e}^{\mathrm{i} \kk \rr}\!:}_\mu\ =\
\mathrm{e}^{\kk^2 G_\mu/2}\,\mathrm{e}^{\mathrm{i} \kk \rr}\ \ ,
\end{equation}
where $G_\mu$ is the tadpole amplitude evaluated with the propagator
of mass $\mu$,
\begin{equation}
\label{ren8}
G_\mu\ =\  
\int {\mathrm{d}^D\pp\over (2\pi)^D}\,{1\over \pp^2+\mu^2}\ =\
{\Gamma\left({2-D\over 2}\right)\over (4\pi)^{D/2}}\,\mu^{D-2} \ .
\end{equation}
Thus we rewrite  the potential $V(\rr)$ given in (\ref{Vrdevv}) as
\begin{equation}
\label{Vnprdevm}
V(\rr)\ =\ 
\sum_{n=0}^\infty {g_n\over {2}^n n!}:\!{(\rr^2)}^n\!:_m \ .
\end{equation} 
The mass scale $m$ used to define the normal product ${:\!\cdots\!:}_m
$ is defined self-consistently from $V$ so that it coincides with the
``renormalized mass'' in (\ref{Vnprdevm})
\begin{equation}
\label{mnpcond}
g_1\ =\ m^2
\ .
\end{equation}
This gives a self-consistent equation for $m$ in terms of $V(\rr)$ (or
its Fourier transform $\hat V(\kk)$)
\begin{equation}
\label{mscequ}
\begin{split}
m^2\, &=\, -\,{1\over d}\int {\mathrm{d}^d\kk\over
(2\pi)^d}\,\kk^2\,\widehat{V}(\kk)\,\mathrm{e}^{-{\kk^2\over 2} G_m}
\\
&=\ -\, {1\over G_m}\, (2\pi G_m)^{-d/2}\ \int
\mathrm{d}^d\rr\,V(\rr)\left(1-{\rr^2\over
d\,G_m}\right)\,\mathrm{e}^{-{\rr^2\over 2\,G_m}} \ .
\end{split}
\end{equation}
All other couplings $g_0$, $g_2$, $g_3$, etc.\ in (\ref{Vnprdevm})
are then uniquely defined from the potential $V$. 
We rewrite $V$ as
\begin{equation}
\label{Vr2U} V(\rr) = g_0+{m^2\over 2}:\!\rr^2\!:+\,U(\rr)\quad;\quad
U(\rr)\,=\,{g_2\over 2^{2}2!}:\!(\rr^2)^2\!:+{g_3\over
2^{3}3!}:\!(\rr^2)^3\!:+ \dots 
\end{equation}
and we shall treat the non-linear terms $U(\rr)$ as perturbation.  The
expectation value of a (multilocal) operator $O(\xx_1,\dots ,\xx_K)$ can be
expanded as
\begin{equation}
\label{OVOU} {\langle O(\xx_1,\dots
,\xx_K)\rangle}_V=\sum_{N=0}^\infty {(-1)^N\over N!}\int \prod_{i=1}^N
\rmd^D\zz_i {\langle O(\xx_1,\dots ,\xx_K)\,U(\zz_1)\cdots
U(\zz_N)\rangle}_m^{\mathrm{connected}} \ ,
\end{equation}
where $\langle\cdots\rangle$ is the expectation value in the massive
free theory ($U=0$).

In this new perturbative expansion there are no tadpole diagrams. This
makes the diagramatics much simpler.  In addition many simplifications
occur in the limit $d\to\infty$, as already noted in
\cite{DavidWiese1998}.

\subsubsection{MOPE in a harmonic potential}
\label{s:ren:mope:harm}
First we consider the
case of a potential quadratic in $\rr$, which is especially simple.
The potential reads
\begin{equation}
\label{ren9}
V(\rr)\ =\ v_0\,+\,{m_0^2\over 2}\rr^2\ =\ g_0\,+\,{m^2\over 2}{:\!\rr^2\!:}_m
\quad,\quad
m_0\,=\,m
\quad;\quad
v_0\ =\ g_0\,-\,d\, {m^2\over 2}\,G_m
\ .
\end{equation}
 The field $\rr$ is still free but massive
with mass $m$ and the propagator is
\begin{equation}
\label{ren10}
G_m(\xx-\yy)\ =\ 
\int{\mathrm{d}^D\pp\over (2\pi)^D}\,{\rme^{i\pp(\xx-\yy)}\over
\pp^2+m^2}\ =\ {1\over 2\pi}\,\left[{m\over
2\pi|\xx-\yy|}\right]^{{D-2\over 2}}\,K_{{D-2\over 2}}(m|\xx-\yy|)
\ ,
\end{equation}
where $K_\nu$ is the modified Bessel Function.

It is simple to study the short-distance limit of products of local
and multilocal operators in this massive Gaussian theory, using
exactly the same ideas and techniques as for the free massless case
($m=0$) developed in \cite{GBU}.

\paragraph{OPE for the massive propagator $G_m$:}
We express the short-distance expansion of multilocal operators in terms of 
the expansion for the massive propagator%
%%%%%% BEGIN FOOTNOTE %%%%%%%%%%%%%%
\footnote{The expansion is easily obtained from the proper-time
integral representation of $G_{m} (\xx)$, by expanding the integrand
in $m^{2}$ to get the analytic terms in $m^2$, and in $\xx^{2}$ to get
the analytic terms in $\xx^2$:
\begin{subequations}
\begin{eqnarray}
\label{manalytic}
G_{m} (\xx) &=& \int\frac{\rmd^D\pp}{(2\pi)^{D}} \int_{0}^{\infty} \rmd
s\, \rme^{i\pp\xx} \rme^{- (\pp^2+m^{2})s} = \frac{1}{(4\pi)^{{D/2}}}
\int_{0}^{\infty} \rmd s\, \rme^{-m^{2}s} s^{-D/2} \rme^{-x^{2}/ (4s)}
\\
 &=& \frac{\Gamma (\frac{-2 + D}{2})}{4{\pi }^{\frac{D}{2}}}
|\xx|^{2-D}- \frac{\Gamma(\frac{-4 + D}{2}){{m}}^2}{16{\pi
}^{\frac{D}{2}}}|\xx|^{4 - D} + O (m^{4}) + \mbox{non-analytic terms
in $m^{2}$} \\
&=&  \frac{m^{D-2}}{
{( 4 \pi) }^{\frac{D}{2}}}\Gamma \left(1 - \frac{D}{2}\right) -
\frac{1}{4} \frac{m^D}{
{( 4 \pi) }^{\frac{D}{2}}} \Gamma \left(\frac{-D}{2}\right) |\xx|^{2}
+O (\xx^{4}) + \mbox{non-analytic terms
in $|\xx|^{2}$} \qquad 
\end{eqnarray}  
\end{subequations}
}
%%%%%% END FOOTNOTE %%%%%%%%%%%%%%%%%
\begin{equation}
\label{Gmexp} 
G_m(\xx-\yy)\ =\
c_0(D)\,m^{D-2}\,-\,d_0(D)\,|\xx-\yy|^{2-D}\,+\,c_1(D)\,m^D\,|\xx-\yy|^{2}
\,-\, d_1(D)\,m^2\,|\xx-\yy|^{4-D}\,+\,\cdots
\ .
\end{equation}
The coefficients $c_0$, $c_1$, $d_0$, $d_1$, are finite as long as
$D<2$ and are given by
\begin{equation}
\label{OPEGcoeffs}
c_0(D)={\Gamma\!\left({2-D\over 2}\right)\over (4\pi)^{D/2}}
\ ,\quad
c_1(D)={c_0(D)\over 2D}
\ ,\quad
d_0(D)=-{\Gamma\!\left({D-2\over 2}\right)\over 4\ \pi^{D/2}}
\ ,\quad
d_1(D)={d_0(D)\over 2(4-D)}
\ .
\end{equation}
Note that 
\begin{equation}
\label{d0volsphere} d_0(D)={1\over (2-D)\CS_D}
\quad\text{with}\quad\CS_D={2\,\pi^{{D\over 2}}\over\Gamma(D/2)}\ =\
\text{volume of the unit sphere in}\ \mathbb{R}^D \ .
\end{equation}
This expansion follows itself from the OPE for the product of two
$\rr$ fields in the massive theory, which reads
\begin{equation}
\label{OPErr} \rr^a(\xx)\rr^b(\yy)\,=\,-|\xx-\yy|^{2-D}\,d(|\xx-\yy|^2
m^2)\,\delta^{ab}\, \JU +\sum_{p_1,p_1}{\xx^{p_1}\over
p_1!}{\yy^{p_2}\over p_2!}:\!\nabla^{p_1}\rr^a\nabla^{p_2}\rr^b\!:_0 \
,
\end{equation}
where the coefficient $d(|\xx-\yy|^2 m^2)$ has an (asymptotic) series
expansion in $|\xx-\yy|^2m^2$
$$d(|\xx-\yy|^2 m^2)=d_0+d_1 |\xx-\yy|^2 m^2+d_2 |\xx-\yy|^4
m^4+\cdots$$ and where the normal products $:\!\cdots\!:_0$ with
respect to the zero mass means that the operators
$\nabla^\bullet\rr\nabla^\bullet\rr$ are defined through dimensional
regularization (see below).

\paragraph{MOPE for $\delta^d(\rr_1-\rr_1')$ and $\mbox{${\tr}$}[\mathbb{O}]$:}
We first consider the short-distance expansion for the operator 
$\delta^d(\rr(\xx)-\rr(\yy))$, which enters in $\mathrm{\tr}[\mathbb{O}]$. 
Using the definition (\ref{ren7}) for the normal product we can
write it as
\begin{equation}
\label{deltaxyexp}
\begin{split}
\delta^d(\rr(\xx)-\rr(\yy)) \ &=\ \int{\mathrm{d}^d\kk\over
(2\pi)^d}\,\mathrm{e}^{\mathrm{i}\kk(\rr(\xx)-\rr(\yy))}
\\
&=\ \int{\rmd^d\kk\over
(2\pi)^d}\,\mathrm{e}^{-\kk^2\bigl(G_m(0)-G_m(\xx-\yy)\bigr)}
\,:\!\mathrm{e}^{\mathrm{i}\kk(\rr(\xx)-\rr(\yy))}\!:_m \ .
\end{split}
\end{equation}
The last bilocal operator is regular at short distance (when
$\xx\to\yy$) and can be expanded in $\xx-\yy$ as
\begin{equation}
\label{OPEexp} :\!\mathrm{e}^{\mathrm{i}k(\rr(x)-\rr(\yy))}\!:_m\ =\
\JU (\zz)\,-\, {1\over 2}
\kk_a\kk_b\,(\xx^\mu-\yy^\mu)(\xx^\nu-\yy^\nu)
\,:\nabla_\mu\rr^a\nabla_\nu\rr^b(\zz):_m\, +\,\cdots \ ,
\end{equation}
where $\zz={\xx+\yy\over 2}$ and the subdominant terms are of order
$\mathcal{O}(|\xx-\yy|^4)$ with higher derivative operators.  We
insert (\ref{OPEexp}) into (\ref{deltaxyexp}) and integrate over $\kk$
to obtain
\begin{equation}
\label{deltaexp1}
\begin{split}
\delta^d(\rr(\xx)-\rr(\yy))\ 
=\ (4\pi)^{-{d\over 2}}&\left[G_m(0)-G_m(\xx-\yy)\right]^{-{d\over 2}}
\ \times
\\
\times & \left[\JU (\zz)- {\delta_{ab}\over 4}
{(\xx^\mu-\yy^\mu)(\xx^\nu-\yy^\nu) \over
\left[G_m(0)-G_m(\xx-\yy)\right]}
:\!\nabla_\mu\rr^a\nabla_\nu\rr^b(\zz)\!:_m\, +\,\cdots \right] \ .
\end{split}
\end{equation}
We now use the short-distance expansion (\ref{Gmexp}) of the massive
propagator $G_m(\xx-\yy)$
and insert it into (\ref{deltaexp1}) to obtain
\begin{equation}\label{deltaexp2}
\begin{split}
\delta^d(\rr(\xx{-}\rr(\yy)) = (4\pi\,d_0)^{-{d\over
2}}|\xx{-}\yy|^{\epsilon-2D}\,\left[ \! \left(1+{d\over 2}{c_1\over d_0}
m^D |\xx{-}\yy|^{D}-{d\over 2}{d_1\over d_0}m^2
|\xx{-}\yy|^2+\cdots\right) \JU (\zz) \right.
\\
 - \left.{\delta_{ab}\over 4} {(\xx^\mu-\yy^\mu)(\xx^\nu-\yy^\nu)
\over d_0\,|\xx{-}\yy|^{2-D} }
{:\!\nabla_\mu\rr^a\nabla_\nu\rr^b(\zz)\!:}_m\, +\,\cdots \right]
\ .
\end{split} 
\end{equation}
In (\ref{deltaexp2}) we can regroup the two terms of order
$|\xx-\yy|^{\epsilon-D}$ as
\begin{equation}
\label{4.36} (4\pi\,d_0)^{-d/2}\,|\xx-\yy|^{\epsilon-D}\,\left[
{d\over 2}{c_1\over d_0} m^D\,\JU (\zz)\,- {1\over 4
d_0}{(\xx^\mu-\yy^\mu)(\xx^\nu-\yy^\nu)\over |\xx-\yy|^2}\,
{:\!\nabla_\mu\rr\nabla_\nu\rr(\zz)\!:}_m \right]\ .
\end{equation}
Note that the OPE (\ref{deltaexp2}) is a relation between operators,
and is valid for any choice of the mass $m$ used to define the normal
product. Thus the term (\ref{4.36}) can
 be rewritten  as the normal ordered operator
${:\!\nabla\rr\nabla\rr\!}:_0$ with subtraction mass $\mu=0$
\begin{equation} \label{4.37}
(4\pi\,d_0)^{-d/2}\,|\xx-\yy|^{\epsilon-D}\,\left[ - {1\over 4
d_0}{(\xx^\mu-\yy^\mu)(\xx^\nu-\yy^\nu)\over |\xx-\yy|^2}\,
{:\!\nabla_\mu\rr\nabla_\nu\rr(\zz)\!:}_0 \right]
\ .
\end{equation}
Indeed we have the relation
\begin{equation}
\label{npmvs0}
{:\!\nabla_\mu\rr^a\nabla_\nu^b\rr\!:}_m\ =\ {:\!\nabla_\mu\rr^a\nabla_\nu^b\rr\!:}_0\ -\ 
2\,\delta^{ab}\,\delta_{\mu\nu}\,m^D\ c_1(D) \,\JU 
\ .
\end{equation}
Since this relation will be crucial to prove renormalizability, let us
show it explicitly.  From the definition of the normal product we have
\begin{equation}
\label{4.38} {:\!\nabla_\mu\rr^a\nabla_\nu^b\rr\!:}_m\ =\
\nabla_\mu\rr^a\nabla_\nu\rr^b\ -\
{\langle\nabla_\mu\rr^a\nabla_\nu\rr^b\rangle}_m\,\JU 
\end{equation}
for any $m$, hence
\begin{equation}
\label{4.39} {:\!\nabla_\mu\rr^a\nabla_\nu\rr^b\!:}_m-
:\!\nabla_\mu\rr^a\nabla_\nu\rr^b\!:_0 \ =\ -\
\bigl({\langle\nabla_\mu\rr^a\nabla_\nu\rr^b\rangle}_m-{\langle\nabla_\mu\rr^a\nabla_\nu\rr^b\rangle}_0\bigr)\,\JU 
\ .
\end{equation}
The r.h.s.\ is easily calculated using the OPE (\ref{Gmexp}) for the
propagator $G_m$ itself, since
\begin{equation}
\label{4.39b}
\langle\rr^a(x)\rr^b(y)\rangle_m\ =\ \delta^{ab}\,G_m(\xx-\yy)\ .
\end{equation}
This yields 
\begin{eqnarray}\label{4.40}
{\bigl< \nabla_\mu\rr^a\nabla_\nu\rr^b \bigr>}_m
-{\bigl< \nabla_\mu\rr^a\nabla_\nu\rr^b \bigr>}_0
&=&
\delta^{ab}\,
\left.{\partial\over\partial\xx^\mu}{\partial\over
\partial\yy^\nu}
\left[G_m(\xx-\yy)-G_0(\xx-\yy)\right]
\right|_{\xx=\yy}\nn\\
&=& -2\,\delta^{ab}\,\delta_{\mu\nu}\,m^D\ c_1(D) 
\ .
\end{eqnarray}
Note that the massless propagator $G_0(\xx-\yy)$ is IR divergent but
the IR divergent term is constant (independent of $\xx-\yy$) and
disappears in (\ref{4.40}) because of the $\xx$ derivatives.  Hence we
obtain (\ref{npmvs0}).

Thus we have obtained the first three terms of the MOPE for the
$\delta$ operator in the $U=0$ background
\begin{equation}
\label{deltaMOPE3}
\begin{split}
\delta^d(\rr(\xx)-\rr(\yy))\ =\ &  (4\pi d_0(D))^{-{d\over 2}}|\xx-\yy|^{\epsilon-2D}\left[1-
{d\over 4(4-D)}m^2|\xx-\yy|^2+\cdots\right]
\JU (\zz)
\\   
-\,\pi\,(4\pi &d_0(D))^{-\left(1+{d\over 2}\right)} |\xx-\yy|^{\epsilon-D-2}(\xx^\mu-\yy^\mu)(\xx^\nu-\yy^\nu)\,{:\!\nabla_\mu\rr\nabla_\nu\rr\!:}_0
\ +\ \cdots \ .
\end{split}
\end{equation}

The same argument can be used to construct the higher orders of the
MOPE. They involve higher dimensional operators of the form
$O_{p}={:\!\nabla^{p_1}\rr\nabla^{p_2}\rr\nabla^{p_3}\rr\cdots\!:}_0$
(since the operator $\delta^d(\rr(\xx)-\rr(\yy))$ is invariant by
translation $\rr\to\rr+\rr_0$ the $O_{p}$ must contain only
derivatives $\nabla\rr$, that is $p_j>0$, and by parity in $\rr$ the
$O_{p}$ must be even in $\rr$). They give subdominant powers of
$|\xx-\yy|$ of the form
$m^{2k}|\xx-\yy|^{\epsilon-2D+2k+\sum_{j}{(p_j-1+D/2)}}$.

Finally let us stress that the two first terms of the MOPE (for $D<2$)
are the terms of order $|\xx-\yy|^{\epsilon-2D}$ and
$|\xx-\yy|^{\epsilon-D}$ and that they are \emph{the same} as for the
MOPE for the free membrane, that is for $m=0$.  This will imply that
the (one-loop) UV divergences (single poles at $\epsilon=D$ and
$\epsilon=0$) due to this MOPE in the massive theory (self-avoiding
manifold in a harmonic confining external potential) are canceled by
the same counterterms as for the free theory (self-avoiding manifold
with no confining potential). These counterterms are proportional to
the operators $\JU $ and $(\nabla\rr)^2$.

\paragraph{MOPE for $\delta^d (\rr_1-\rr_2)\delta^d (\rr_1-\rr_2)$ 
and $\mbox{\rm\tr}[\mathbb{O}^2]$:}%\ \\
The reader familiar with the techniques of \cite{GBU} will see that
the same arguments can be used to construct the MOPE for general
products of local and multilocal operators in the $m\neq 0$, $U=0$
background.

Let us concentrate on the MOPE for two $\delta$ operators, which
enters in $\mathrm{\tr}[\mathbb{O}^2]$.
We are interested in the short-distance expansion ($\xx\to\xx_0$, $\yy\to\yy_0$) for the product of two bilocal operators
\begin{equation}
\label{ren23}
\delta^{d} (\rr_{1} (\xx_0)-\rr_2(\yy)) \delta^{d} (\rr_{1}
(\xx)-\rr_2(\yy_{0}))
\ ,
\end{equation}
where $\rr_1$ and $\rr_2$ belong to two independent manifolds $\CM_1$
and $\CM_2$.  As above, we write the $\delta$'s as a Fourier transform
of an exponential and reexpress it in terms of normal products
\begin{equation*}
\label{ren14}
\begin{split}& \delta^{d} (\rr_{1} (\xx_0)-\rr_2(\yy)) \delta^{d}
(\rr_{1} (\xx)-\rr_2(\yy_{0})) = \int \frac{\rmd^d \kk_{1}\rmd^d
\kk_{2}} {(2\pi)^{2d}} \rme^{\rmi\left( \kk_1
[\rr_{1}(\xx_0)-\rr_2(\yy)]+ \kk_2[\rr_{1}
(\xx)-\rr_2(\yy_{0})]\right) }
 \\
 &
 = \int \frac{\rmd^d \kk_{1}\rmd^d
\kk_{2}} {(2\pi)^{2d}} \, :\!\rme^{\rmi \left({\kk_1 [\rr_{1}
(\xx_0)-\rr_2(\yy)]+ \kk_2[\rr_{1} (\xx)-\rr_2(\yy_{0})]
}\right)}\!:\,
\rme^{-\kk_{1}\kk_2
\left[ G_{m}{(\xx_0-\xx)}+G_{m} (\yy-\yy_0)\right]-
(\kk_1^2+\kk_2^{2}) G_{m} (0)}
\end{split}
\end{equation*}
Note that there are no cross-terms, as those proportional to $G
(\xx-\yy)$, since $\xx$ and $\yy$ belong to different manifolds, thus
$\rr_1 (\xx)$ and $\rr_2 (\yy)$ are uncorrelated.
We now keep the dominant term for the OPE when $\xx\to\xx_0$ and
$\yy\to\yy_0$
\begin{equation*}
\label{ren24} :\!\rme^{\rmi \left({\kk_1 [\rr_{1}
(\xx_0)-\rr_2(\yy_0)]+ \kk_2[\rr_{1} (\xx)-\rr_2(\yy_{0})]
}\right)}\!: \ =\ :\!\rme^{\rmi \left({(\kk_1+\kk_2) [\rr_{1}
(\xx_0)-\rr_2(\yy_0)]}\right)}\!: \,+\,\cdots
\end{equation*}
(the neglected terms contain subdominant $\nabla^p\rr$'s), rewrite
this term as
\begin{equation*}
\label{ren25}
:\!\rme^{\rmi \left({(\kk_1+\kk_2) [\rr_{1} (\xx_0)-\rr_2(\yy_0)]}\right)}\!:
\ =\ 
\rme^{\rmi \left({(\kk_1+\kk_2) [\rr_{1} (\xx_0)-\rr_2(\yy_0)]}\right)}\,
\rme^{(\kk_1+\kk_2)^2 G_m(0)}
\end{equation*}
 and integrate over $\kk_1$ and $\kk_2$ to obtain
\begin{equation} 
\begin{split}
 \int &\frac{\rmd^d \kk_1\,\rmd^d\kk_2}{(2\pi)^{2d}} \,
:\!\rme^{\rmi (\kk_1+\kk_2) [\rr_{1} (\xx_0)-\rr_2(\yy_0)}\!:\,
\rme^{\kk_1\kk_2\left[ 2 G_m(0)-G_m(\xx_0-\xx)-G_m(\yy-\yy_0)\right]}
\\
&=
 \int \frac{\rmd^d \kk\,\rmd^d\kk'} {(2\pi)^{2d}} \, 
\rme^{i \kk [\rr_1 (\xx_0)-\rr_2(\yy_0)]} 
\rme^{ (k^2/4- {k'}^{2}) [2G_{m} (0)-G_m (\xx_0-\xx)-G_m (\yy-\yy_0)]}
\end{split}
\end{equation}
with $\kk=\kk_1+\kk_2$ and $\kk'=(\kk_1-\kk_2)/2$.  The leading term
is obtained by dropping the factor of $\kk^{2}/4$ in the second
exponential (the neglected terms give subdominant
$\delta^{(n)}(\rr_1-\rr_2)$ terms).
This allows to do the integrations explicitly
\begin{eqnarray}\label{ren15}
\delta^{d} (\rr_{1} (\xx_0)-\rr_2(\yy)) \delta^{d} (\rr_{1}
(\xx)-\rr_2(\yy_{0})) &\simeq& 
\left(4\pi \right)^{-d/2} \left[2G_{m} (0)-G_m (\xx-\xx_0)-G_{m} (\yy-\yy_0)
\right]^{-d/2}\nn\\
&&\times  \, 
\delta^{d} \left(\rr_{1}(\xx_0)-\rr_2(\yy_0)\right)
\ .
\end{eqnarray}
From the short-distance expansion (\ref{Gmexp}) for $G_m(0)-G_m(\xx)$
the most singular term
when both $\xx \to \xx_0$ and $\yy \to \yy_{0}$ is 
\begin{equation}
\label{dd2dMOPE}
\delta^{d} (\rr_{1} (\xx_0)-\rr_2(\yy)) \delta^{d} (\rr_{1}
(\xx)-\rr_2(\yy_{0})) 
=
\frac{\left[|\xx-\xx_0|^{2-D}+|\yy-\yy_0|^{2-D} \right]^{-d/2}}{(4 \pi d_{0}(D))^{d/2}}
\,
\delta^{d} (\rr_{1} (\xx_0)-\rr_2(\yy_0)) 
\,+\,\cdots
\ .
\end{equation}
Thus we have obtained the leading term for the MOPE in the harmonic
background $U=0$, $m\neq 0$.

This leading coefficient given by (\ref{dd2dMOPE}) is the same as for
the free membrane ($V=0$).  The same calculation can be done for the
MOPE of two $\delta$'s on the same membrane, and we  get (at
leading order) a MOPE with the same coefficient
\begin{equation}
\label{dd2d1MOPE} \delta^{d} (\rr (\xx_0)-\rr(\yy)) \delta^{d}
(\rr(\xx)-\rr(\yy_{0})) =
\frac{\left[|\xx-\xx_0|^{2-D}+|\yy-\yy_0|^{2-D} \right]^{-d/2}}{(4 \pi
d_{0}(D))^{d/2}}
\,
\delta^{d} (\rr (\xx_0)-\rr(\yy_0)) \,+\,\cdots
\ .
\end{equation}
This implies in particular that the (one-loop) UV divergence (single
pole at $\epsilon=0$) due to this MOPE in the massive theory
(self-avoiding manifold in a harmonic confining external potential) is
canceled by the same counterterm as for the free theory
(self-avoiding manifold with no confining potential). This counterterm
is proportional to the bilocal operator $\delta(\rr-\rr')$.

\paragraph{MOPE for higher order terms and \rm$\mathrm{\tr}[\mathbb{O}^k]$:}
The same analysis can be performed for the product of three
$\delta$'s, in particular
$\delta^d[\rr_1(\xx_0)-\rr_2(\yy)]\delta^d[\rr_2(\yy_0)-\rr_3(\zz)]\delta^d[\rr_3(\zz_0)-\rr_1(\xx)]$,
which has to be considered for the quantity
$\mathrm{\tr}[\mathbb{O}^3]$.  It shows that the leading singularity
when $\xx\to\xx_0$, $\yy\to\yy_0$, $\zz\to\zz_0$ is given by the same
MOPE as in the free theory, with the same leading coefficient. No
additional UV divergences arise. The same result holds for higher
order products of $\delta$'s.

\subsubsection{MOPE in the anharmonic potential} 
\label{sss:mopeanhar}
We now generalize
this analysis to the SAM model in an anharmonic confining potential.

\paragraph{General discussion:}
The perturbative expansion involves now interaction vertices given by
the expansion of the local potential $U(\rr)$.  For $D<2$ and as long as
no bilocal $\delta(\rr-\rr')$ operators are inserted this perturbation
theory is UV finite.  The only UV divergences that occur when $D\to 2$
are given by the tadpole amplitudes $G_m$, but they are subtracted by
the normal product prescription $:\!\cdots\!:_m$.  Thus as long as the
normal ordered potential $V$ is finite (i.e. its coefficients
$g_1=m^2$, $g_2$, $g_3$, etc.\ are UV finite) the ``vacuum diagrams''
are UV finite. Since we deal with a massive theory no IR divergences
are expected.

Now we have to consider insertions of the bilocal $\delta^{d}(\rr-\rr')$
operators, and thus to look for instance at
\begin{equation}
\label{ren27} \int \rmd^D\yy\,\rmd^D\yy\,\prod_{i=1}^N
\rmd^D\zz_i\,{\langle\delta^d(\rr(\xx)-\rr(\yy))U(\zz_1)\cdots
U(\zz_N)\rangle}_m^{\mathrm{connected}}
\ .
\end{equation}
The UV divergences which may occur when $|\xx-\yy|\to 0$, while the
other distances remain finite, have already been analyzed with the
MOPE in the harmonic case.  We have seen there that when some $\zz_i$
come close, no UV divergences occur.  The only dangerous case is when
some $\zz$'s, $\xx$ and $\yy$ come close at the same rate. Thus we
must study the short-distance expansion of a product of local operators
(the $U$'s) and of multilocal operators (the $\delta$'s), in the
massive theory.  This short-distance expansion can be studied by the
same MOPE techniques as above. Let us first give a simple explicit
example.

\paragraph{Example:}
To be explicit, we first regard as an example the simple case of the
contribution to $\tr (\JO)$ given by one of the terms of (\ref{ren27})
with only one $U(\zz)$, and more precisely one quartic term
$:\!\rr(\zz)^{4}\!:$.  The arguments for higher powers in $\rr$ or
higher orders in perturbation theory will be identical.
\begin{figure}
\fig{14cm}{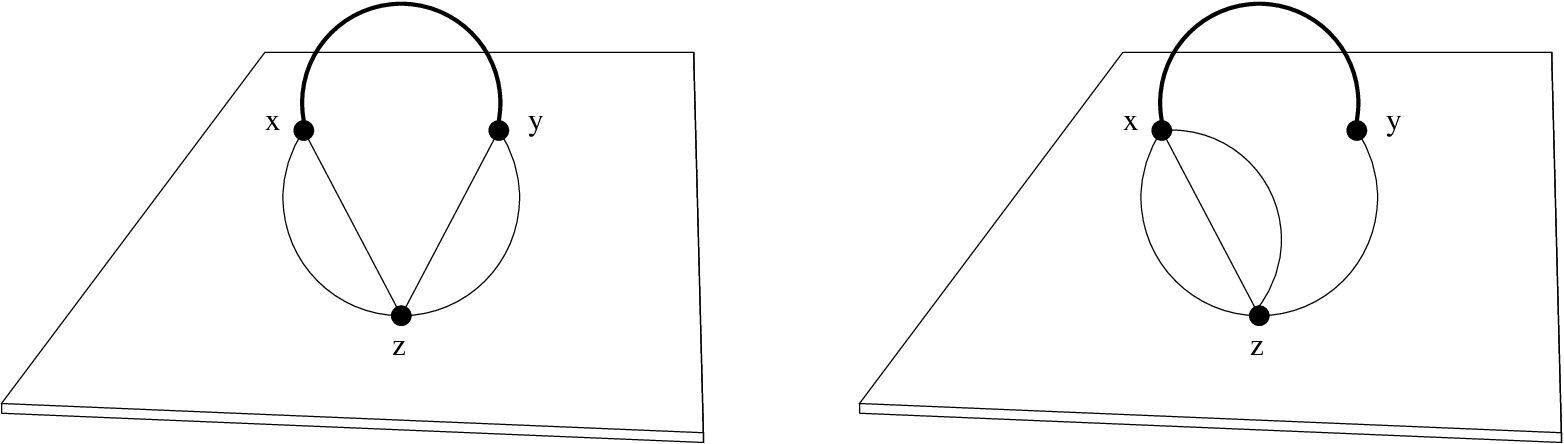}
\caption{Two
contributions to $\tr (\JO)$ at first order in perturbation theory, associated with the insertion of one $:\!\rr(\zz)^{4}\!:$}
\label{TRO1perturb}
\end{figure}
Following (\ref{trO}) the crucial term to calculate is
\begin{equation}
\label{trOp1} \int_\CM
\mathrm{d}^D\xx\, \int_\CM \mathrm{d}^D\yy\,
\bigl\langle
\delta^d\bigl(\rr(\xx)-\rr(\yy)\bigr)\,
:\!\rr(\zz)^{4}\!:
\bigr\rangle_m \ .
\end{equation}
Applying Wick's theorem we can decompose it in terms of multilocal
diagrams such as those depicted on Fig.~\ref{TRO1perturb}. More
explicitly this term can be written as
\begin{eqnarray}\label{ren17}
&& \!\!\!\!\int_\CM \mathrm{d}^D\xx\,
\int_\CM \mathrm{d}^D\yy\, \int
\frac{\rmd^D\kk}{(2\pi)^{d}} \,\left<
\rme^{i\kk [\rr (\xx)-\rr
(\yy)]} \, :\!\rr(\zz)^{4}\!:\right>_{m}
\nonumber \\
&&=\int_\CM \mathrm{d}^D\xx\,
\int_\CM
\mathrm{d}^D\yy\, \int \frac{\rmd^D\kk}{(2\pi)^{d}}
\,\left<
\rme^{i\kk [\rr (\xx)-\rr (\yy)]} \right>_{m} \left( \left<
\kk\left[\rr
(\xx)-\rr (\yy) \right]
\rr(\zz)\right>_{m}^{2}\right)^2\nonumber \\
&&\sim
\frac{1}{\mbox{Vol} ({\cal M})}\int_\CM \mathrm{d}^D\xx\,
\int_\CM
\mathrm{d}^D\yy\, \int_\CM \mathrm{d}^D\zz\,\int \left<
\delta^{d}
(\rr (\xx)-\rr (\yy)) \right> \times \frac{ \left[ G_m
(\xx-\zz)-G_m
(\yy-\zz)\right]^{4} }{\left[G_{m} (0)- G_{m}
(\xx-\yy) \right]^{2}
}\ ,\nn\\
&&=: \int_{\xx ,\yy\in {\cal M}} {\mathfrak F} (\xx,\yy)
\int_{\zz\in
\CM} \frac{\left[ G_m (\xx-\zz)-G_m (\yy-\zz)\right]^{4}
}{\left[
G_{m} (0)- G_{m} (\xx-\yy)\right]^{2}}
\ .
\end{eqnarray}
We now derive an important bound.  First of all, due to the triangular
inequality
\begin{equation}\label{triang}
\left(\rr(\yy)-\rr(\zz)
\right)^{2} \le \left(\rr(\xx)-\rr(\yy)
\right)^{2}
+\left(\rr(\xx)-\rr(\zz)
\right)^{2}
\end{equation}
\begin{equation}\label{ren18}
G_{m}
(0)-G_m(\yy-\zz) \le 2 G_{m} (0) - G_m(\xx-\yy)-
G_m(\xx-\zz)
\end{equation}
leading to
\begin{equation}\label{ren19}
 G_m(\xx-\zz) - G_{m} (\yy -\zz) \le
G_{m} (0)- G_m(\xx-\yy)\ .
\end{equation}
An analog relation is valid with $\xx$ and $\yy$ exchanged, resulting
in a bound for the absolute value. The r.h.s.\ is thus also positive
and we get the bound for the ratio
\begin{equation}\label{ren20}
\left|\frac{ G_m
(\xx-\zz)-G_m (\yy-\zz) }{G_{m} (0)-G_{m} (\xx-\yy)}
\right| \le1
\ .
\end{equation}
We now want to show that the counter-terms
remains the same.  Using (\ref{ren20}), we can write the bound
\begin{eqnarray}\label{renbound}
&&
\!\!\!\!\left|\int_\CM \mathrm{d}^D\xx\, \int_\CM
\mathrm{d}^D\yy\,
\int \frac{\rmd^D\kk}{(2\pi)^{d}} \,\left<
\rme^{i\kk [\rr (\xx)-\rr
(\yy)]} \, :\!\rr(\zz)^{4}\!:\right>_{m}
\right|\nonumber \\
&&\qquad \le \left|\int_{\xx
,\yy\in {\cal M}}
{\mathfrak F} (\xx,\yy)\left[ G_{m} (0)-G_{m}
(x-y)\right]^{2}
\right|\times \int_{\zz\in
\CM} \left[\frac{ G_m (\xx-\zz)-G_m
(\yy-\zz) }{G_{m} (0)-G_{m} (\xx-\yy)}
\right]^{4} \nonumber
\\
&&\qquad \le \left|\int_{\xx
,\yy\in {\cal M}} {\mathfrak F}
(\xx,\yy)\left[ G_{m} (0)-G_{m}
(x-y)\right]^{2} \right| \times
\mbox{Vol} (\mathcal{M})
\ .
\end{eqnarray}
The latter bound is already enough to show that no additional
counter-terms proportional to the elastic energy are necessary. It
would also be sufficient for the perturbation expansion of $\tr
(\JO^2)$. However, we can do better and show that there is no
divergence at all.  To do so, we now estimate the integral over
$\zz$. Two domains of integration have to be distinguished:
\begin{enumerate}
\item[{$\mathcal S$}]:
$\left|\zz-\frac{\xx+\yy}{2}\right| \le \alpha
|\xx-\yy|
$
\item[{$\mathcal L$}]: $\left|\zz-\frac{\xx+\yy}{2}\right| > \alpha
|\xx-\yy| $
\end{enumerate}
$\alpha$ is chosen large (to be specified below), but finite.  The
integrals over $\zz$ are bounded by
\begin{eqnarray}\label{termS}
 \int_{\zz\in \CM}\left[ \frac{
G_m (\xx-\zz)-G_m (\yy-\zz) }{G_{m}
(0)- G_{m} (\xx-\yy)}\right]^{4}
&\le&
\int_{\zz \in {\cal S}}\left[ \frac{ G_m (\xx-\zz)-G_m
(\yy-\zz)
}{G_{m} (0)-
G_{m} (\xx-\yy)}\right]^{4} \nonumber \\
&&+
\int_{\zz
\in {\cal L}}\left[ \frac{ G_m (\xx-\zz)-G_m (\yy-\zz)
}{G_{m} (0)-
G_{m} (\xx-\yy)}\right]^{4}
\ .
\end{eqnarray}
Using
(\ref{ren20}), the first term, is bounded
by
\begin{equation}\label{ren21}
\int_{\zz \in {\cal S}}\left[
\frac{ G_m (\xx-\zz)-G_m (\yy-\zz)
}{G_{m} (0)- G_{m}
(\xx-\yy)}\right]^{4} \le \int_{\zz\in \CS}\, 1 \le
\left(\alpha
|\xx-\yy| \right)^{D}
\ .
\end{equation}
In domain ${\cal L}$,
analyticity of the propagator allows the bound
\begin{equation}\label{ren22}
\left|\frac{ G_m (\xx-\zz)-G_m
(\yy-\zz) }{G_{m} (0)-G_{m} (\xx-\yy)}
\right| \le a_{1} \left|
\frac{(x-y)\nabla G_m (\xx-\zz) }{G_{m}
(0)-G_{m} (\xx-\yy)} \right|
\le a_{2} \left( m |\xx-\yy|\right)^{D-1}
\ .
\end{equation}
We do not give a rigorous proof here, but it is clear that $\alpha$
should be sufficiently larger than 1 (say 10), which allows to
establish a value for $a_1$, itself depending on $\alpha$, but
saturating for large $\alpha$.  The constant $a_{2}$ is chosen in
order to bound $\nabla G_{m} (x-z)$ by its maximal value on $\cal M$,
which has to scale with $m$ by power-counting in the way given above.

We are now in a position, to put everything together.

The integration over the distance $\mathsf{s}:=\xx-\yy$ (which
contains the possible UV-divergence) can now be written for small
$\mathsf{s}$ as follows (we drop all constants for simplicity of
notations)
\begin{equation}\label{ren28}
\int \frac{\rmd
\mathsf{s}}{\mathsf{s}} \, \mathsf{s}^{D}
\times
\mathsf{s}^{-\frac{2-D}{2}d} \times\mathsf{s} ^{2 (2-D)}
\times
\left\{{ \mathsf{s}^{D}\ , \atop \mathsf{s}^{4 (D-1)}\ ,
}
\right.~ {\mbox{for } \mathcal{S} \atop \mbox{for }\mathcal{L}
}
\ .
\end{equation}
The factor of $\mathsf{s}^{D}$ comes from the integration measure;
$\mathsf{s}^{-\frac{2-D}{2}d}$ is the leading UV-divergence in
${\mathfrak{F} (\xx,\yy)}$; the next factor $\mathsf{s}^{2 (2-D)}$ is
the short-distance scaling of $\left[G_{m} (0)-G_{m} (\xx-\yy)
\right]^{2}$, and the remaining factors have been established in
(\ref{ren21}) and (\ref{ren22}) respectively.  Using $\epsilon
=2D-\frac{2-D}{2}d$, this can be rewritten as
\begin{equation}\label{ren29}
\int \frac{\rmd
\mathsf{s}}{\mathsf{s}} \, \mathsf{s}^{\epsilon }
\times \left\{{
\mathsf{s}^{2 (2-D)}\ , \atop \mathsf{s}^{D}\ , }
\right.~ {\mbox{for
} \mathcal{S} \atop \mbox{for } \mathcal{L} }
\ .
\end{equation}
As long as $D<2$, all integrals are UV-convergent in the limit of
$\epsilon \to 0$. Thus no additional counter-terms are needed. The
only possible UV-divergence is when first taking $D\to 2$ before
$\epsilon \to 0$. Note however, that this divergence only effects the
contribution to the free energy (proportional to the counter-term 1),
but cancels in all properly normalized observables.

\paragraph{General analysis:}
We now consider the MOPE for the operator with one $\delta^{d}
(\rr-\rr')$ and $P=\sum\limits_i p_i$ fields $\rr$
\begin{equation}
\label{deltaUUU} O(\xx,\yy,\zz_i)\ =\
\delta^d(\rr(\xx)-\rr(\yy))\,\prod_{i=1}^N:\!\rr^{p_i}(\zz_i)\!:_m
\end{equation}
when the $N+2$ points $\xx$, $\yy$, $\zz_i\,\to \oo$.  The generating
functional for these operators is
\begin{equation}
\label{deltaVVV}
\begin{split}
\delta^d(\rr(\xx)&-\rr(\yy))\,\prod_{i=1}^N:\!\rme^{\qq_i\rr(\zz_i)}\!:_m
= \int
{\rmd^d\kk\over(2\pi)^d}\,\rme^{\rmi\kk(\rr(\xx)-\rr(\yy))}\prod_{i=1}^N:\!\rme^{\qq_i\rr(\zz_i)}\!:_m
\\
&=\int {\rmd^d\kk\over(2\pi)^d}\,
:\!\rme^{\rmi\kk(\rr(\xx)-\rr(\yy))+\sum\limits_i\qq_i\rr(\zz_i)}:_m \\
&\qquad \qquad \quad \times \rme^{-\kk^2[G_m(0)-G_m(\xx-\yy)]+{1\over
2}\sum\limits_{i\neq j}\qq_i\qq_j G_m(\zz_i-\zz_j)+\rmi\sum\limits_i
\kk\qq_i [G_m(\xx-\zz_i)-G_m(\yy-\zz_i)]} \ .
\end{split}
\end{equation}
Expanding the normal ordered operator in $\xx$, $\yy$ and $\zz$, using
the short-distance expansion for the propagator $G_m$ and integrating
over $\kk$ we get the MOPE.  We see that in this MOPE for
(\ref{deltaUUU}) local operators appear, of the form
\begin{equation}
\label{rMnablarQ}
A\ =\ \rr^M\nabla^{r_1}\rr \nabla^{r_2}\rr \cdots \nabla^{r_Q}\rr
\ ,\quad
0\le M\le P\ ,\quad r_j> 0
\ .
\end{equation}
The dimension of the operator (\ref{deltaUUU}) is
$\mathrm{dim}[O]=\epsilon-2D+P(2-D)/2$, while the dimension of
(\ref{rMnablarQ}) is $\mathrm{dim}[A]=(M+Q)(2-D)/2-R$ where
$R=\sum\limits_j r_j\ge Q$.  Hence the coefficients in the MOPE
\begin{equation}
\label{OCA}
O(\xx,\yy,\zz_i)\ =\ \sum_A C^O_A(\xx,\yy,\zz_i ;m)\,A(\oo)
\end{equation}
scale as
\begin{equation}
\label{COAscale}
C^O_A(S\xx,S\yy,S\zz_i ;m)\ \sim\ S^
\omega
\,
C^0_A(S\xx,S\yy,S\zz_i ;m)+\cdots
\ ,
\end{equation}
where
\begin{equation}
\label{omegadivcond} \omega={\mathrm{dim}[O]-\mathrm{dim}[A]}=
\epsilon-2D+(P-M)(2-D)/2+Q(4-D)/2+(R-Q)
\ .
\end{equation}
There will be short-distance UV divergences if the integration over
the $N+1$ independent positions $\xx$, $\zz_i$ is not convergent. This
occurs if
\begin{equation}
\label{ren30}
D(N+1)+\omega\le 0
\quad\Rightarrow\qquad
\epsilon+D(N-1)+(P-M){2-D\over 2}+Q{4-D\over 2}+{R-Q}\,\le\,0
\ .
\end{equation}
The case $N=0$ has already been studied.  When $N\ge 1$, since $P\ge
M$ and $R\ge Q$ we see that as long as $D<2$ the condition
(\ref{omegadivcond}) is satisfied only if $\epsilon=0$, $N=1$ and
$P=M$.  Let us look at what this last condition means.  $P=M$ means
that all the $\rr^{p_i}$ in $O$ appear in $A$, namely that no
combination of propagators of the form $[\prod
G_m(\xx-\zz_i)-G_m(\yy-\zz_i)]\prod G(\zz_j-\zz_k)$ appear in the
coefficient $C$ of the MOPE, which therefore depends only on
$\xx-\yy$.  In other words, this particular coefficient comes from the
product of two independent expansions
\begin{enumerate}
  \item The $N=0$ MOPE\ \ \ \  
  $\delta^{d}(\rr(\xx)-\rr(\yy))\ \to\  |\xx-\yy|^{\epsilon-2D}\JU $
  \item The trivial OPE\ \ \ \  
  $\prod\limits_i :\!\rr^{p_i}(\zz_i)\!: \ \to\ :\!\rr^P\!:$\ \ \ with
  coefficient $1$.
\end{enumerate}
and contains {\textbf{no connected diagram} with propagators
connecting \textbf{any} of the $\zz_i$'s \textbf{to} $\xx$ \textbf{or}
$\yy$}.  Thus this apparent divergence is not real. It is part of the
$N=0$ leading divergence at $\epsilon=D$ and disappears in the
connected expectation value $\langle\cdots\delta(\rr(\xx)-\rr(\yy))
U(\rr(\zz))\cdots\rangle_m^{\mathrm{connected}}$.

All other coefficients of the MOPE have scaling dimension $\omega$,
which satisfy the inequality (\ref{omegadivcond}).  No additional UV
divergences occur beyond those which appear already for the free
manifold and the manifold in a harmonic potential, even when
$\epsilon=0$.

The same argument can be developed when there are two $\delta$
operators and several $U$'s. One can show that when considering the
short-distance expansion of
\begin{equation}
\label{ren31}
\delta^{d}(\rr(\xx_0)-\rr(\yy_0))\delta^{d}(\rr(\xx_1)-\rr(\yy_1))\prod_i
U(\xx_i)\prod_j U(\yy_j) \quad,\qquad \xx_1,\,\xx_i\,\to\,\xx_0\quad
\text{and}\quad \yy_1,\,\yy_i\,\to\,\yy_0
\end{equation}
bilocal operators are generated by the MOPE. Power counting shows
nevertheless that no additional UV divergence appears beyond those
already studied for $\xx_1\to\xx_0$, $\yy_1\to\yy_0$ while all other
distances remain finite.

This is sufficient to prove (at least at one loop) that the
counterterms which make the SAM model UV finite at $\epsilon=0$
also render the SAM in a confining potential UV-finite, as long as
$D<2$.

\paragraph{The limit $D\to 2$:}
It is interesting to notice that there is a potentially divergent term
when $\epsilon=0$ and $D\to 2$ which corresponds to
\begin{equation}
\label{ren32}
 N=1\,,\ Q=R=0\,,\ 0\le M<P\ \ \text{arbitrary}\ .
\end{equation}
(We have already seen that the case $M=P$ is not relevant).  This fact
is not unrelated to the following observation.  In the MOPE
(\ref{deltaMOPE3}) for the single bilocal operator $\delta
(\rr-\rr')$, the third term, which is a subdominant term in the MOPE
of the form
$$\delta^{d}(\rr(\xx)-\rr(\yy))\quad \to\quad 
m^2|\xx-\yy|^{\epsilon-2D+2}\,\JU $$ is not UV divergent if
$D<2$, but when $D\to 2$ it becomes of the same order as the
divergent term
$$\delta^{d}(\rr(\xx)-\rr(\yy))\quad \to\quad 
|\xx-\yy|^{\epsilon-D}\,:\!(\nabla\rr)^2\!:$$ and is potentially
dangerous when $D\to 2$.  Since this term depends on $m$, it depends
linearly on the potential $V(\rr)$, like the $N=1$ terms that we
consider here. It would be interesting to study this more.

Since $Q=R=0$, this means that there are no $\nabla\rr$ involved in the
MOPE and we are only interested in the terms of the MOPE of the form
\begin{equation}
\label{ren33} \delta^{d}(\rr(\xx)-\rr(\yy))\,:\!\rr^p(\zz)\!:\quad
\to\quad :\!\rr^m(\oo)\!:\quad,\qquad 0\le m< p
\ .
\end{equation}
It is quite easy to compute the corresponding coefficients.  We find
\begin{equation}
\label{ren34} \delta^{d}(\rr(\xx)-\rr(\yy))\,:\!\rme^{\alpha\rr(\zz)}\!:_m\
\to\ (4\pi)^{-d/2} \, \left[G_m(0)-G_m(\xx-\yy)\right]^{-d/2} \,
\rme^{-{\alpha^2\over4}{(G_m(\xx-\zz)-G_m(\yy-\zz))^2\over
G_m(0)-G_m(\xx-\yy) }} \, :\!\rme^{\alpha\rr(\zz)}\!:_m
\end{equation}
hence at short distances 
\begin{equation}
\label{ren35}
\delta^d(\rr(\xx)-\rr(\yy))\,:\!\rme^{\alpha\rr(\zz)}\!:_m\ \to\ 
(4\pi\,d_0)^{-d/2}
\,
\,|\xx-\yy|^{\epsilon-2D}\,
\rme^{-{\alpha^2 d_0\over 4}
H(\xx,\yy,\zz)
}
\,
:\!\rme^{\alpha\rr(\zz)}\!:_m
\end{equation}
with the function
$H(\xx,\yy,\zz)$ defined as
\begin{equation}
\label{ren36} H(\xx,\yy,\zz) \ =\
{\left({|\xx-\zz|^{2-D}-|\yy-\zz|^{2-D}}\right)^2\over |\xx-\yy|^{2-D}
}
\end{equation}
or, after averaging with weight $\exp(\alpha^2 J/4)$ 
\begin{eqnarray}\label{ren37}
&&\!\!\!\delta^d(\rr(\xx)-\rr(\yy))\,:\!\rme^{J\rr^2(\zz)}\!:_m\nonumber \\
&&\qquad \to\ 
(4\pi\,d_0)^{-d/2}
\,
\,|\xx-\yy|^{\epsilon-2D}\,
\left[1-J d_0 H(\xx,\yy,\zz)
\right]^{-d/2}
\,
:\!\rme^{J \rr^2(\zz) \left[1-J d_0 
H(\xx,\yy,\zz)
 \right]^{-1}}\!:_m\qquad \qquad 
\end{eqnarray}

\subsection{Renormalization} 
\label{s:ren:ren} 

\subsubsection{Explicit form of the UV divergences for the determinant ${\det}'(\CS''[V])$}
\label{s:ren:ren:div} 

From the definition (\ref{trO}) of $\tr[\JO]$ as
an $\xx$ integral and the MOPE (\ref{deltaMOPE3}) for $\delta^d(\rr-\rr')$,
we see that the $\xx$ integral (\ref{trO}) has short-distance UV
divergences if $\epsilon\le D$. The usual rule of dimensional
regularization 
\begin{equation}
\label{fill1}
\int \rmd^D\xx\ |\xx|^{-a}\ =\ \CS_D\,{1\over D-a}\,+\,\text{finite terms}
\end{equation}
implies that $\tr[\JO]$ has an UV pole at $\epsilon=D$,
proportional to the insertion of the identity operator $\JU$, i.e.
\begin{equation}
\label{trOpoleD}
\tr[\JO]\ =\ 
\mathtt{C}_0
\,{1\over \epsilon-D}\,{\langle\JU\rangle}_V\,+\,\text{regular terms
at $\epsilon=D$}
\end{equation}
(of course ${\langle\JU\rangle}_V=1$), with the residue $\mathtt{C}_0$
given by
\begin{equation}
\label{C0tt}
\mathtt{C}_0=\mathtt{C}_0(D,d)=\CS_D\left[4\pi d_0(D)\right]^{-{d\over 2}}\quad .
\end{equation}
$\CS_D$ is the volume of the unit sphere in $\mathbb{R}^D$ and
$d_0(D)=1/(2-D)\CS_D$ the coefficient of the first subleading term in
the OPE of $G(\xx)$; they are given in (\ref{OPEGcoeffs}).

Using dimensional regularization, $ \tr[\JO]$ is analytically
continued to $0<\epsilon<D$. The next term in the MOPE gives the UV
divergence at $\epsilon=0$, hence a pole given from (\ref{deltaMOPE3})
by
\begin{equation}
\label{trOpole0}
\tr[\JO]\ =\ 
\mathtt{C}_1
\,{1\over\epsilon}\,{\langle
:\!(\nabla\rr)^2\!:_0\rangle}_V\,+\,\text{regular terms
at}\ \epsilon=0
\end{equation}
with residue
\begin{equation}
\label{C1tt} 
\mathtt{C}_1=\mathtt{C}_1(D,d)=-\CS_D{1 \over 4D d_0(D)}\left[4\pi
d_0(D)\right]^{-d/2}\ .
\end{equation}
Similarly, $ \tr[\JO^2]$ has an UV pole at $\epsilon=0$, given from
(\ref{dd2dMOPE}) by
\begin{equation}
\label{trO2pole0}
\tr[\JO^2]\ =\ \mathtt{C}_2\,{1\over\epsilon}\,
{\langle
\delta^d(\rr_1(\xx_0)-\rr_2(\yy_0))\rangle}_V\,+\,\text{regular terms
at $\epsilon=0$}
\end{equation}
with residue
\begin{equation}
\label{C2tt} \mathtt{C}_2= \mathtt{C}_2(D,d)= \CS_D^2{1 \over
2-D}{\Gamma\bigl(D/(2-D)\bigr)^2\over\Gamma\bigl(2D/(2-D)\bigr)}\bigl[4\pi
d_0(D)\bigr]^{-d/2}\ .
\end{equation}
Here $\rr_1$ and $\rr_2$ are associated to two independent copies
$\CM_1$ and $\CM_2$ of the infinite flat manifold $\CM$. Thus we have
\begin{equation*}
\label{ren38} {\delta^d(\rr_1(\xx_0)-\rr_2(\yy_0))\rangle}_V=\int
\rmd^d\rr\,{\langle\delta^d(\rr_1(\xx_0)-\rr)\rangle}_V
{\langle\delta^d(\rr_2(\yy_0)-\rr)\rangle}_V =\int \rmd^d\rr
\left[{{\left<\rho(\rr)\right>}_V}\right]^2
\end{equation*}
where $\rho(\rr)$ is the manifold density in bulk space.  Using
(\ref{logser}) and the discussion of section \ref{s:ren:UVdiv}, we see
that the logarithm of the determinant of the instanton fluctuations
$\mathfrak{L}=\log(\mathfrak{D})$ has a UV pole at $\epsilon=0$ given
by
\begin{equation}
\label{Lpoles0}
\mathfrak{L}\ =\ \log\left({\det}'[\CS'']\right)\ =\ {1\over \epsilon}\,\left(
-\,{\mathtt{C}_1}\left<(\nabla\rr)^2\right>_V
\,-\,{}{\mathtt{C}_2\over 2}\,
\int \rmd^d\rr \left[{\left<\rho(\rr)\right>_V}\right]^2
\right)
\,+\,
\LfMS
\end{equation}
where $\LfMS$ is the UV finite part of $\Lfrak$, obtained by
subtracting the UV pole of $\Lfrak$ at $\epsilon=0$; hence the ``MS"
(for minimal subtraction) subscript.

\subsubsection{Renormalized effective action} 
\label{sss:renSeff} 

We now study how the perturbative counterterms modify the effective
action $\CS[V]$ used in the instanton calculus. For this purpose, we
now repeat for the renormalized  theory the transformation
$S[\rr]\to\CS[V]$ and the rescalings performed for the bare theory in
Sect. \ref{ss:effectiveaction} and \ref{s:Instanton for the SAM}.

\paragraph{Renormalized original action $S_{\mathrm{ren}}[\rr]$:}\ \\
The renormalized action for the SAM model is according to \cite{GBU}
\begin{equation}
\label{Srenr1} \Sren[\rr]\ =\ {{Z(\bren)\over
2}}\int_{\xx\in\CM}\hskip-1em(\nabla\rr(\xx))^2+{\bren
Z_b(\bren)\mu^{\epsilon}\over 2}\iint_{\xx,\yy\in\CM} \hskip
-1.5em\delta^d(\rr(\xx)-\rr(\yy))
\end{equation} 
$\bren$ is the dimensionless renormalized coupling constant and $\mu$
is the renormalization mass scale.  At one loop the counterterms
$Z(\bren)$ and $Z_b(\bren)$ are found to be
\begin{equation}
\label{ZZ1r1l}
Z(\bren)\,=\,1-\bren{\mathtt{C_1}\over\epsilon}\ ,\quad 
Z_b(\bren)\,=\,1+\bren{1\over 2}{\mathtt{C_2}\over \epsilon}
\end{equation}
with $\mathtt{C_1}$ and $\mathtt{C_2}$ the same residues as those
obtained above in (\ref{C1tt}) and (\ref{C2tt}).  We first rewrite the
renormalized action as the bare action $S[\rr]$ plus the ``one-loop
counterterm $\Delta_1 S[\rr]$.
\begin{equation}
\label{ren39}
\begin{split}
\Sren[\rr]\ =\ &
S[\rr]\,+\,\Delta_1S[\rr]
\quad;\quad
S[\rr] \ =\ {{1\over
2}}\int_{\xx\in\CM}\hskip-1em(\nabla\rr(\xx))^2+{\bren\mu^{\epsilon}\over
2}\iint_{\xx,\yy\in\CM} \hskip -1.5em\delta^d(\rr(\xx)-\rr(\yy))
\\
\Delta_1S[\rr]\ =\ &-\,\bren\,{\mathtt{C}_1\over\epsilon}\, {{1\over
2}}\int_{\xx\in\CM}\hskip-.5em(\nabla\rr(\xx))^2
\,+\,{\bren^2\mu^{\epsilon}\over
4}{\mathtt{C}_2\over\epsilon}\iint_{\xx,\yy\in\CM}
\delta^d(\rr(\xx)-\rr(\yy))
\end{split}
\end{equation}
Note that
$(\nabla\rr)^2=:\!(\nabla\rr)^2\!:_0+d\,\delta^D(0)\JU$ and that in
dimensional regularization $\delta^D(0)=0$.

\paragraph{Renormalized effective action $\CS_{\mathrm{ren}}[V]$:}\ \\
We repeat the transformation of Sect.~\ref{ss:effectiveaction} to pass
from the action $S[\rr]$ to the effective action $\CS[V]$ for the
effective field $V(\rr)$, keeping $\Delta_1S[\rr]$ as a perturbation.
We thus arrive at the representation for the renormalized partition
function $Z_{\mathrm{ren}}(\bren)$
\begin{equation}
\label{ren40}
\begin{split}
\int \CD[\rr]\,\exp(-\Sren[\rr])\ &=\ \int
\CD[\rr]\CD[V]\,\exp\left(-\int_\xx\left[{1\over
2}(\nabla\rr)^2+V(\rr)\right]+{1\over 2\bren\mu^\epsilon}\int_\rr
V^2-\Delta_1[\rr]\right)
\\
&=\ \int \CD[V]\,\exp\left(-F_\CM[V]+{1\over
\bren\mu^\epsilon}\int_\rr V^2\right)\,
\left<\exp{(-\Delta_1S[\rr])}\right>_V
\end{split}
\ .
\end{equation}
We now perform the same rescalings and the same rotation in the
complex coupling-constant plane as for the bare theory (see section
\ref{s:Instanton for the SAM}):
\begin{equation}
\label{ren41}
\xx\to\left(|\bren|\mu^\epsilon L^D\right)^{{1\over D-\epsilon}}\xx
\quad,\qquad
\rr\to \left(|\bren|\mu^\epsilon L^D\right)^{{2-D\over 2(D-\epsilon)}}\xx
\quad,\qquad
\theta=\mathrm{Arg}(\bren)\to\pm\pi
\ .
\end{equation}
Starting from a finite manifold $\CM$ with size $L$ (volume $L^D$), we
end up with a rescaled manifold $\CM_s$ with volume $\mathrm{Vol}(\CM_s)$
and renormalized effective coupling $\gren$
\begin{equation}
\label{ren42} 
g_{\mathrm{r}}={1\over\mathrm{Vol}(\CM_s)} \quad,\qquad
\mathrm{Vol}(\CM_s)=|\bren|^{-{D\over
D-\epsilon}}\left[L\mu\right]^{-{D\epsilon\over D-\epsilon}}
\ .
\end{equation}
The functional integral becomes
\begin{equation}
\label{ren43}
\begin{split}
Z_{\mathrm{ren}}(\bren)\ &=\ \int \CD[V]\,\exp\left(-F_{\CM_s}[V]+{\rme^{-\rmi\theta}\over
2\gren}\int_\rr V^2\right)\, \left<\exp{(-\Delta_1' S[\rr])}\right>_V
\end{split}
\end{equation}
\begin{equation}
\label{ren44} 
\Delta_1' S[\rr]\ =\
\bren\left[-\,{\mathtt{C}_1\over\epsilon}{1\over
2}\int_{\CM_s}\!\!\!(\nabla\rr)^2+{\gren\rme^{\rmi\theta}\over
4}{\mathtt{C}_1\over\epsilon}\iint\delta^d[(\rr-\rr')\right]
\ .
\end{equation}
As in section \ref{s:Instanton for the SAM},
$\theta=\mathrm{Arg}(\bren)$.  We are interested in the semiclassical
limit $\bren\to 0$. Since this limit is a thermodynamic limit, where
the volume of the manifold
$\text{Vol}(\CM_s)=g_{\mathrm{r}}^{-1}\to\infty$, it is natural to
assume that clustering takes place (since for the instanton
configuration the manifold is confined in the potential $V$).  We may
thus approximate the contribution of the counterterm by
\begin{equation}
\label{ren45}
\left<\exp{(-\Delta_1' S[\rr])}\right>_V\ =\ 
\exp\left(-\left<\Delta_1' S[\rr]\right>_V\right)
\end{equation}
up to terms exponentially small in $g_{\mathrm{r}}$. 
The last expectation value is
\begin{equation}
\label{Delta1SV}
\left<\Delta_1' S[\rr]\right>_V\ =\ {\bren}\mathrm{Vol}(\CM_s)\left(
-\,{1\over 2}{\mathtt{C}_1\over\epsilon}\left<(\nabla\rr(\oo))^2\right>_V+
{\rme^{\rmi\theta}\gren \over 4}
{\mathtt{C}_2\over\epsilon}
\int_\xx\left<\delta^d(\rr(\oo)-\rr(\xx)\right>_V
\right)\ .
\end{equation}
Now we easily check that
\begin{equation}
\label{brenV}
\bren\mathrm{Vol}(\CM_s)=\bren/\gren=\rme^{\rmi\theta}
\left(\gren^{1/D}\mu L\right)^{-\epsilon}
\end{equation}
and that when $\epsilon=0$ it reduces to  $\rme^{\rmi\theta}=\CO(1)$.
The first expectation value in (\ref{Delta1SV})
$\left<(\nabla\rr)^2\right>_V$ is of order $\CO(1)$.  
The study of the second expectation value is slightly more subtle.  We write
\begin{equation}
\label{intdeltaV} \int_{\CM_s}\rmd^D\xx
\left<\delta^d(\rr(\oo)-\rr(\xx)\right>_V=\int {\rmd^{d}\kk\over (2\pi)^d}
\int_{\CM_s}\rmd^D\xx \left<\rme^{\rmi\kk(\rr(\oo)-\rr(\xx))}\right>_V
\ .
\end{equation}
From clustering we expect that what dominates is the large-$|\xx|$
regime where
\begin{equation}
\label{ren46} \left<\rme^{\rmi\kk(\rr(\oo)-\rr(\xx))}\right>_V\ =\
\left<\rme^{\rmi\kk\rr(\oo)}\right>_V\left<\rme^{-\rmi\kk\rr(\xx)}\right>_V
\ =\ \left<\hat\rho(\kk)\right>_V\,\left<\hat\rho(-\kk)\right>_V\ ,
\end{equation} 
and where $\hat\rho(\kk)$ is the Fourier transform of the manifold
density $\rho(\rr)$, see (\ref{lf37}).  So we finally obtain
\begin{equation}
\label{ren47} \gren \int_\xx\left<\delta^d(\rr(\oo)-\rr(\xx)\right>_V \
\simeq\ \gren\mathrm{Vol}(\CM_s)\int {\rmd^{d}\kk\over
(2\pi)^d}\left<\hat\rho(\kk)\right>_V\,\left<\hat\rho(-\kk)\right>_V \
=\ \int_\rr \left[{\left<\rho(\rr)\right>_V}\right]^2
\end{equation}
also of order $\CO(1)$.  (\ref{intdeltaV}) contains an UV-divergence
when $\xx\to 0$ and this will give a double pole when $\epsilon\to 0$
in (\ref{Delta1SV}), but this divergence is of order
$\bren\mathrm{Vol}(\CM_s)\gren\simeq\gren$. This is in fact a two-loop
divergence that we do not have to consider here.

The final result is that we can rewrite the renormalized functional
integral (at one loop) as
\begin{equation}
\label{ren48} 
Z_{\mathrm{ren}}(\bren)\ =\ \int\CD[V]\,\exp\left(-{1\over
\gren}\CS[V]-\rme^{\rmi\theta}\gren^{-\epsilon\over D} (\mu L)^{-\epsilon}
\Delta_1\CS[V]\right)
\end{equation}
with $\CS[V]$ the bare effective action (\ref{3.35}) and $\Delta_1\CS[V]$
the one-loop counterterm for the effective action
\begin{equation}
\label{DSV}
 \Delta_1\CS[V]\ =\ -\,{\mathtt{C_1}\over\epsilon}{1\over
 2}\left<(\nabla\rr)^2\right>_V+{\mathtt{C_2}\over\epsilon}
 {\rme^{\rmi\theta}\over 4}\int_\rr {\left<\rho(\rr)\right>_V}^2
\ .
\end{equation}
This amounts to state that the renormalised effective action
$\CS_{\mathrm{ren}}[V]$ at one loop is
\begin{equation}
\label{Seffren} \CS_{\mathrm{ren}}[V]\ =\
\CS[V]+\rme^{\rmi\theta}\gren^{{D-\epsilon\over D}}(\mu L)^{-\epsilon}
\Delta_1\CS[V]\ ,
\end{equation}
with $\CS[V]$ the original bare effective action (\ref{3.35}), and
$\Delta_1\CS[V]$ given by (\ref{DSV}). 

\subsubsection{1-loop renormalizability} \label{sss:ren1loop} It is
now easy to show that the renormalized action for the SAM model which
makes perturbation theory finite at one loop makes also the
determinant factor for the instanton
$\mathfrak{D}={\det}'(\CS''[V^{\mathrm{inst}}])$ UV finite at
$\epsilon=0$.

\paragraph{Instanton contribution in the renormalized theory}\ \\

If we evaluate the renormalized functional integral around the
instanton saddle point $V^{\mathrm{inst}}$ by the saddle-point method,
we see that the contribution at one loop of the instanton in the bare
theory (in (\ref{ImCZMem}) and (\ref{ImRkMem}))
\begin{equation}
\label{1loopbare}
\rme^{-{1\over
g}\CS[V]}\left|{\det}'(\CS''[V])\right|^{-{1\over 2}}\ =\ \rme^{-{1\over
g}\CS[V]-{1\over 2}\mathrm{Re}(\mathfrak{L})}
\end{equation}
is replaced in the renormalized theory by
\begin{equation}
\label{ESD2ESDR} 
\rme^{-{1\over \gren}\CS[V]}\left|{\det}'(\CS''[V])\right|^{-{1\over
2}}\rme^{\gren^{-{\epsilon\over D}} (\mu L)^{-\epsilon}
\Delta_1\CS[V]}
\ =\  \rme^{-{1\over
\gren}\CS[V]-{1\over 2}\mathrm{Re}(\mathfrak{L}_{\mathrm{ren}})}
\ ,
\end{equation}
where the ''renormalized trace-log'' of the instanton-fluctuations'
determinant
$\mathfrak{L}_{\mathrm{ren}}=\log(\mathfrak{D}_{\mathrm{ren}})$ is
simply (from now on we set $\theta=\pm\pi$)
\begin{equation}
\label{Lfrakren}
\mathfrak{L}_{\mathrm{ren}}\ =\ \mathfrak{L}-2
\left(\gren^{{1\over D}}\mu L\right)^{-\epsilon}\Delta_1\CS[V]
\ .
\end{equation}

\paragraph{Limit $\epsilon\to 0$ and UV finiteness.}\ \\

From Eq.~(\ref{DSV}) for the counterterm and Eq.~(\ref{Lpoles0}) which
gives the UV poles of $\mathfrak{L}$, one easily checks that
$\mathfrak{L}_{\mathrm{ren}}$ is UV finite when $\epsilon\to 0$.  It
is given in this limit by
\begin{equation}
\label{LfrakRen0} \mathfrak{L}_{\mathrm{ren}}\ = \ \LfMS -
\left({1\over D}\log \gren+\log(\mu L)\right) \,\mathtt{B}
\quad\text{when}\quad\epsilon=0 \ ,
\end{equation}
where $\LfMS$ is the UV-finite part of $\mathfrak{L}$, as defined in
Eq.~(\ref{Lpoles0}), and the coefficient $\mathtt{B}$ is (minus) the
residue in (\ref{Lpoles0})
\begin{equation}
\label{Bdef}
\mathtt{B}\,=\,
\mathtt{C}_1\left<(\nabla\rr)^2\right>_V+{\mathtt{C}_2\over 2}\int_\rr
V(\rr)^2\ .
\end{equation}
(We used the instanton equation $\langle\rho(\rr)\rangle_V+V(\rr)=0$
to simplify the last term).

Finally it is shown in Appendix \ref{condensates} that for the
instanton potential $V$ we have
\begin{equation}
\label{2Iconden} \langle
(\nabla\rr)^2\rangle_V\,=\,-\,d\left(1-{\epsilon\over
D}\right)^{-1}\,\mathfrak{S} \quad ,\qquad \int_\rr V(\rr)^2\,=\ 2
\left(1-{\epsilon\over D}\right)^{-1}\,\mathfrak{S} \ ,
\end{equation}
where $\mathfrak{S}=\CS[V]$ is the instanton action. Hence for
$\epsilon=0$ we have
\begin{equation}
\label{Bexplgen}
\mathtt{B}\,=\,\left(-d\,\mathtt{C_1} +\mathtt{C_2}\right)\mathfrak{S}
\ .
\end{equation}

\paragraph{UV pole at $\epsilon=D$}\ \\
A similar calculation shows that the counterterm which subtracts the
perturbative UV pole in ${\mathtt{C}_0\over\epsilon-D}$ also subtracts
the leading divergence for the instanton.  This justifies our use of
dimensional regularization to deal with this divergence.

\subsection{Large orders for the renormalized theory}
\subsubsection{Asymptotics} From these results we can easily obtain
the large-orders asymptotics for the renormalized theory at
$\epsilon=0$.  The semiclassical estimate (\ref{ImCZsc}) for the
discontinuity of the partition function $\CZ(b)$ becomes for the
renormalized partition function $\CZ_{\mathrm{ren}}(\bren)$
\begin{equation}
\label{CZrendisc}
\begin{split}
\mathrm{Im}\,\CZ_{\mathrm{ren}}(\bren)\ &=\ \mp\,{1\over 2}\,L^{-2D}
\,\left|\bren\right|^{{4\over 2-D}}\,\rme^{-{1\over |\bren
|}\mathfrak{S}}\,\mathfrak{W}\,\left|\mathfrak{D}_{\mathrm{ren}}\right]
^{-{1\over 2}}
\\
&=\ \mp\,{1\over 2}\,L^{-2D} \,\left|\bren\right|^{{4\over
2-D}+{\mathtt{B}\over 2D}} \,(\mu L)^{{\mathtt{B}\over
2}}\,\rme^{-{1\over |\bren |}\mathfrak{S}}\,
\mathfrak{W}\,\left|\mathfrak{D}_{\mathrm{MS}}\right]^{-{1\over 2}}
\end{split}
\end{equation}
with $\mathfrak{D}_{\mathrm{MS}}=\exp(\mathfrak{L}_{\mathrm{MS}})$.
The large order asymptotics for the renormalized partition function
\begin{equation}
\label{CZrenpert}
\CZ_{\mathrm{ren}}(\bren)\ =\ \sum_{k=0}^\infty \CZ^{\mathrm{ren}}_k\,\bren^k
\end{equation}
are
\begin{equation}
\label{CZrenklo1} \CZ^{\mathrm{ren}}_k\ \simeq\
(-1)^k\,\Gamma\left[k-{4\over 2-D}-{\mathtt{B}\over
2D}\right]\,{1\over 2\pi}\,L^{-2D}\,(\mu L)^{{\mathtt{B}\over
2}}\,\mathfrak{W}\,\left|\mathfrak{D}_{\mathrm{MS}}\right]^{-{1\over
2}}\, \mathfrak{S}^{{4\over 2-D}+{\mathtt{B}\over 2D}-k}
\end{equation}
and the analog of (\ref{CZklo2}) obtained by using $d/2=4/(2-D)-2$ at
$\epsilon=0$.

\subsubsection{Discussion}
\label{s:ren:ren:disc}

From these semiclassical estimates we expect that the Borel transform
of the renormalized theory still has a finite radius of convergence,
given by the instanton effective action $\mathfrak{S}$.  We also see
that as in ordinary QFT, renormalization at $\epsilon=0$ implies a
dependence on the renormalization scale $\mu$, an anomalous dependence
on the size $L$ of the manifold (anomalous dimension) and an anomalous
power dependence in the renormalized coupling constant $\gren$.  These
anomalous dimensions are given by the factor $\mathtt{B}$, which
combines the perturbative anomalous dimensions $\mathtt{C_1}$ and
$\mathtt{C_2}$ with the instanton action $\mathfrak{S}$.

\section{Variational calculation} 
\label{s:varcal}
\newcommand{\Svi}{{\mathcal{S}_{\mathrm{var}}^{\mathrm{inst}}}}
\newcommand{\Vvi}{{V_{\mathrm{var}}^{\mathrm{inst}}}}
\newcommand{\Vi}{{V^{\mathrm{inst}}}}
\newcommand{\JPo}{{\mathbb{P}_{\!\scriptscriptstyle{0}}}}
\newcommand{\JMvi}{{\mathbb{M}_{\mathrm{var}}^{\mathrm{inst}}}}
\newcommand{\JMv}{{\mathbb{M}_{\mathrm{var}}}}
\newcommand{\JGvi}{{\mathbb{G}_{\mathrm{var}}^{\mathrm{inst}}}}
\newcommand{\JGv}{{\mathbb{G}_{\mathrm{var}}}}
\newcommand{\Mv}{{{M}_{\mathrm{var}}}}
\newcommand{\Gv}{{{G}_{\mathrm{var}}}}

In \cite{DavidWiese1998} we used a Gaussian variational approximation
to compute the instanton $\Vvi$ and its action $\Svi$.  Moreover we
showed that the variational method was a good approximation for the
instanton in the limit $d\to\infty$ (for fixed $\epsilon$), and the
0-th order of a systematic $1/d$ expansion. We computed explicitly
the first correction in the $1/d$ expansion, and showed that for the
instanton action $\Svi$ it was finite when $\epsilon\to 0$.

We apply the same strategy here to compute the fluctuations around
the instanton, namely the determinant factor
\begin{equation}
\label{5.14}
\mathfrak{D}={\det}'(\CS")={\det}'(\JU-\JO)
\quad,\quad
\JO_{\rr_1\rr_2}=
-{\delta^2\CE[V]\over\delta V(\rr_1)\delta V(\rr_2)}
\end{equation}
We first recall briefly the principle of the variational method.  Then
we present a direct calculation of $\mathfrak{D}$ using a variational
estimate for $\JO$. We show that this method does not treat properly the
fluctuations and thus the UV divergences.  We then present a calculation
of $\mathfrak{D}$ based on the variational method and the
reorganization of the perturbative expansion at large $d$ already used
in \cite{DavidWiese1998} and in section \ref{s:ren:mope}.

\subsection{Variational approximation for the instanton}
\label{var approx inst}
\newcommand{\Htrial}{{H_{\mathrm{trial}}}} We first briefly recall the
variational approximation developed in \cite{DavidWiese1998}.  We
use a trial Gaussian Hamiltonian $H_{\mathrm{trial}}[\rr]$ of
the form
\begin{equation}
\label{lf64}
\Htrial[\rr]\ =\ \int_\CM \rmd^D\vx\,\left[{1\over
2}(\nabla\rr)^2\,+\,{1\over 2}(\rr-\rr_0){\mathbb{M}}(\rr-\rr_0)\right] 
\ ,
\end{equation}
where the variational parameters are the position of the instanton
$\rr_0$ and the variational mass matrix $\mathbb{M}=(\JM_{ab})$ (a {\em
symmetric} real $d\times d$ matrix).  The variational approximation
for the free energy of the manifold $\CM$ in the potential $V$ is 
\begin{equation}
\label{lf65}
F_{\mathrm{var}}[V]\ =\
\min_{\rr_0,\mathbb{M}}\bigl[F_{\mathrm{var}}[V;\mathbb{M},\rr_0] \bigr] 
\quad,\quad
F_{\mathrm{var}}[V;\mathbb{M},\rr_0]\ =\ 
F_{\mathrm{trial}}+\langle H - \Htrial
\rangle_{\Htrial}\ .
\end{equation}
$F_{\mathrm{trial}}=-\,\ln\left[\int
\mathcal{D}[\rr]\,\exp\left(-H_{\mathrm{trial}}[\rr]\right)\right] 
$ is the free energy for the trial Hamiltonian, and
is a function of ${\mathbb{M}}$ only (translational invariance).
We are interested in the limit of the infinite flat manifold
$\CM\to\mathbb{R}^D$, and we consider the free energy\emph{ densities}
\begin{equation}
\label{lf69-bis}
\CE_{\mathrm{var}}[V]\ =\ 
{1\over \mathrm{Vol}(\CM)}\,F_{\mathrm{var}}[V]
\quad,\quad
\CE_{\mathrm{var}}[V;\JM,\rr_0]\ =\ {1\over
\mathrm{Vol}(\CM)}\,F_{\mathrm{var}}[V;\JM,\rr_0] \ .
\end{equation}
Obviously
\begin{equation}
\label{lf69-ter}
\CE_{\mathrm{var}}[V] \ =\ 
\min_{\rr_0,\mathbb{M}} \bigl[\CE_{\mathrm{var}}[V;\JM,\rr_0] \bigr]\ .
\end{equation}
$\CE_{\mathrm{var}}[V;\JM,\rr_0]$ can be written in terms of the
Fourier transform of the potential $V(\rr)$
\begin{equation}
\label{lf68} \widetilde{V}(\mathsf{p})\ =\ \int
\rmd^d\rr\,\mathrm{e}^{-\mathrm{i}\mathsf{p}\mathsf{r}}\,V(\rr) \ .
\end{equation}
and in \cite{DavidWiese1998} is given as 
\begin{equation}
\label{lf70-0}
\CE_{\mathrm{var}}[V;\JM,\rr_0] =
{1\over D}{\Gamma\left(2-{D\over
2}\right)\over(4\pi)^{D/2}}\,\mathrm{tr}\left(\JM^{D/2}\right) \,+\,
\int {\rmd^d\mathsf{p}\over (2\pi)^d}\,\widetilde{V}(\mathsf{p})\,
\mathrm{e}^{\mathrm{i}\mathsf{p}\rr_0-{1\over
2}\mathsf{p}\mathbb{G}\mathsf{p}}
\ ,
\end{equation}
where
$\mathbb{G}=(\JG^{ab})$ is the ``variational tadpole'' matrix, defined as
\begin{equation}
\label{lf71}
\JG\,=\,\mathbb{G}(\JM)\,=\,\int {\rmd^D k\over (2\pi)^D}\,{1\over k^2+\JM}\ =\
{\Gamma\left(1-{D\over 2}\right)\over (4\pi)^{D/2}}\,\JM^{{D\over
2}-1} 
\ . 
\end{equation}
Extremization of (\ref{lf70-0}) with respect to the variational
parameters $\JM$ and $\rr_0$ for fixed $V$ gives the two equations for
the the variational parameters $\JM=\JM[V]$ and $\rr_0=\rr_0[V]$ as a
function of the potential $V$
\begin{eqnarray}
\label{varfromV}
\JM_{ab}  &=&   - \int {\rmd^d\mathsf{p}\over
(2\pi)^d}\,\mathsf{p}_a \mathsf{p}_b\,\widetilde{V}(\mathsf{p})\,
\mathrm{e}^{\mathrm{i}\mathsf{p}\rr_0-{1\over
2}\mathsf{p}\mathbb{G}\mathsf{p}} \\
\label{varfromr}
0 & =&  \int {\rmd^d\mathsf{p}\over
(2\pi)^d}\,\mathsf{p}_a\,\widetilde{V}(\mathsf{p})\,
\mathrm{e}^{\mathrm{i}\mathsf{p}\rr_0-{1\over
2}\mathsf{p}\mathbb{G}\mathsf{p}} \ .
\end{eqnarray}
Inserting these solutions in \eq{lf70-0} gives
$\CE_{\mathrm{var}}[V]=\CE_{\mathrm{var}}\left[V,\JM[V],\rr_0[V]\right]$.
Now, extremization of the variational effective action
\begin{equation}
\label{lf72}
\CS_{\mathrm{var}}[V]\ =\ \CE_{\mathrm{var}}[V]\,+\,{1\over 2}\int V^2
\end{equation}
with respect to variations of $V(\rr)$ 
leads to the equation for the variational instanton
$V^{\mathrm{inst}}_{\mathrm{var}}$, 
\begin{equation}
\label{lf72-2}
V_{\mathrm{var}}^{\mathrm{inst}}(\rr)\,+\,{\langle\delta^d
\bigl(\rr-\rr(x_0)\bigr)\rangle}_{H_{\mathrm{trial}}}\ =\ 0\ .
\end{equation}
The variational instanton is rotationally invariant (as expected), so
the associated mass matrix $\JMv=\JM[\Vvi]$ and the tadpole matrix
$\JGv=\JG(\JM[\Vvi])$ are constants times the unit matrix $\JU$,
\begin{equation}
\label{lf73} \JMv \, = \, \Mv\,\JU \quad,\quad\JGv \, = \, \Gv\,\JU
\quad,\quad \Gv\,=\,{\Gamma\left(1-{D\over 2}\right)\over
(4\pi)^{D/2}}\,\Mv^{{D\over 2}-1} \ .
\end{equation}
\eq{lf72-2} implies that the variational instanton has Gaussian
profile, and \eq{varfromV} gives $\Mv$ as the solution of
\begin{equation}
\label{lf74} 
2\,\Mv\,(4\pi)^{d/2}\,\Gv^{1+{d\over 2}}\ =\ 1\ .
\end{equation}
The variational instanton is a Gaussian well (centered at $\rr_0$),
its width is given by $\sqrt{\Gv}$
\begin{equation}
\label{lf75} 
\widehat{V}^{\mathrm{inst}}_{\mathrm{var}}(\mathsf{p})\
=\ -\, \mathrm{e}^{-\mathrm{i}\mathsf{p}\rr_0-{\Gv\over 2} \mathsf{p}^2}
\quad,\qquad
{V}^{\mathrm{inst}}_{\mathrm{var}}(\rr)\
=\ -\, (2\pi \Gv)^{-d/2}\mathrm{e}^{-{1\over 2\Gv}(\rr-\rr_0)^2}
\ .
\end{equation}
The variational instanton action was found to be \cite{DavidWiese1998} 
\begin{equation}
\label{varactDavWie98}
\Svi\,=\, \CS_{\mathrm{var}}[\Vvi]\ =\ \Gv\Mv\left(1-{\epsilon\over
D}\right)\ . 
\end{equation}

\subsection{A poor man's direct variational calculation of the
instanton determinant $\mathfrak{D}$}
\label{wrongfluctuations} \subsubsection{The approximation} We have to
compute the determinant of the fluctuations around the instanton
solution $V^{\mathrm{inst}}$
\begin{equation}\label{5.3}
\mathfrak{D}\ =\ \left. \mathrm{det}'_{V} \left[\frac{S[V]}{\rmd
V[\mathsf{r}]\rmd V[\mathsf{r}']} \right]
\right|_{V=V^{\mathrm{\mathrm{inst}}}}\ .
\end{equation}
In section \ref{var approx inst}, we have calculated the instanton
solution in the variational approximation
$V^{\mathrm{inst}}_{\mathrm{var}}$.  A first approximation for
$\mathfrak{D}$ is to replace it by
\begin{equation}\label{5.2} 
\mathfrak{D}_{\mathrm{var}}\ =\ \left. \mathrm{det}'_{V}
\left[\frac{\rmd^{2}S[V]}{\rmd V[\mathsf{r}]\rmd V[\mathsf{r}']}
\right] \right|_{V=V^{\mathrm{inst}}_{\mathrm{var}}}\ ,
\end{equation}
but this is still difficult to compute. 
A further approximation is to replace this by 
\begin{equation}\label{deftocalculate}
\mathfrak{D}_{\mathrm{var'}}\ =\ \left.\mathrm{det}'_{V}
\left[\frac{\rmd^{2} S_{\mathrm{var}}[V]}{\rmd V[\mathsf{r}]\rmd
V[\mathsf{r}']} \right]\right|_{V=V^{\mathrm{inst}}_{\mathrm{var}}} \
.
\end{equation}
since we have seen that $S_{\mathrm{var}}[V]$ for a general potential
$V$ is easy to calculate.

This first and simple approximation (\ref{deftocalculate}) is
presented in details in this section. We shall see from the result
that it
misses important features of the true result, especially the
UV-divergences due to the fluctuations, which are expected as we have
discussed in section \ref{s:renormalization}. In the following section
\ref{s:direct}, we will therefore calculate (\ref{5.2}), which seems
to be more appropriate.

\subsubsection{Reduction to a finite dimensional determinant in
variational space} In order to calculate (\ref{deftocalculate}), we
start from (\ref{lf70-0}),  and we need
\begin{equation}\label{wf1}
\frac{\rmd}{\rmd V(\mathsf{r})} \frac{\rmd}{\rmd V(\mathsf{r}')}
{\mathcal E}_{\ind{var}}[V] \ .
\end{equation}
We use
\begin{equation} \label{totalVderivative}
\frac{\rmd}{\rmd V(\mathsf{r})} = \frac{\p}{\p V(\mathsf{r})} +
\frac{\rmd r_0}{\rmd V(\mathsf{r})}\frac{\p}{\p \mathsf{r}_0} +
\frac{\rmd \JM}{\rmd V(\mathsf{r})}\frac{\p}{\p \JM} \ .
\end{equation}
Thus
\begin{equation}\label{wf2}
\frac{\rmd {\mathcal E}_{\ind{var}}}{\rmd V(\mathsf{r})} = \frac{\p
{\mathcal E}_{\ind{var}}}{\p V(\mathsf{r})} + \frac{\rmd r_0}{\rmd
V(\mathsf{r})}\frac{\p {\mathcal E}_{\ind{var}}}{\p \mathsf{r}_0} +
\frac{\rmd \JM}{\rmd V(\mathsf{r})}\frac{\p {\mathcal
E}_{\ind{var}}}{\p \JM} =\frac{\p {\mathcal E}_{\ind{var}}}{\p
V(\mathsf{r})} \ ,
\end{equation}
since due to the saddle-point equations
\begin{eqnarray} \label{SP-equ}
\frac{\p {\mathcal E}_{\ind{var}}}{\p \mathsf{r}_0} =0
\qquad\mbox{and}\qquad \frac{\p {\mathcal E}_{\ind{var}}}{\p \JM} =0 \
.
\end{eqnarray}
The second derivative is
\begin{eqnarray}
\frac{\rmd^2 {\mathcal E}_{\ind{var}}}{\rmd V(\mathsf{r})\rmd
V(\mathsf{r}')}&=& \frac{\p^2 {\mathcal E}_{\ind{var}}}{\p
V(\mathsf{r})\p V(\mathsf{r}')} +\frac{\rmd r_0}{\rmd
V(\mathsf{r}')}\frac{\p^2 {\mathcal E}_{\ind{var}}}{\p V(\mathsf{r})\p
\mathsf{r}_0} + \frac{\rmd \JM}{\rmd V(\mathsf{r}')}\frac{\p^2
{\mathcal E}_{\ind{var}}}{\p
V(\mathsf{r}) \p \JM}\nn\\
	&=& \frac{\rmd r_0}{\rmd V(\mathsf{r}')}\frac{\p^2 {\mathcal
E}_{\ind{var}}}{\p V(\mathsf{r})\p \mathsf{r}_0} + \frac{\rmd
\JM}{\rmd V(\mathsf{r}')}\frac{\p^2 {\mathcal E}_{\ind{var}}}{\p
V(\mathsf{r}) \p \JM} \ , \label{eq19}
\end{eqnarray}
since the explicit dependence of ${\mathcal E}_{\ind{var}}$ on $V$ is
linear.  Using the saddle-point equations \eq{SP-equ} we obtain
\begin{eqnarray}
\frac{\rmd}{\rmd V(\mathsf{r})} \frac{\p {\mathcal E}_{\ind{var}}}{\p
\mathsf{r}_0}&=&0 = \frac{\p^2 {\mathcal E}_{\ind{var}}}{\p
V(\mathsf{r}) \p \mathsf{r}_0} +\frac{\rmd r_0}{\rmd
V(\mathsf{r})}\frac{\p^2 {\mathcal E}_{\ind{var}}}{\p \mathsf{r}_0 \p
\mathsf{r}_0} + \frac{\rmd \JM}{\rmd V(\mathsf{r})}\frac{\p^2
{\mathcal E}_{\ind{var}}}{\p \mathsf{r}_0 \p \JM}
\\
\frac{\rmd}{\rmd V(\mathsf{r})} \frac{\p {\mathcal E}_{\ind{var}}}{\p
\JM}&=&0= \frac{\p^2 {\mathcal E}_{\ind{var}}}{\p V(\mathsf{r}) \p
\JM} +\frac{\rmd r_0}{\rmd V(\mathsf{r})}\frac{\p^2 {\mathcal
E}_{\ind{var}}}{\p \mathsf{r}_0 \p \JM} + \frac{\rmd \JM}{\rmd
V(\mathsf{r})}\frac{\p^2 {\mathcal E}_{\ind{var}}}{\p \JM \p \JM} \ .
\label{eq21}
\end{eqnarray}
\Eqs{eq19} to \eq{eq21} lead to  (attention to the counter-intuitive sign)
{\renewcommand{\arraystretch}{2.3}
\begin{equation}
\frac{\rmd}{\rmd V(\mathsf{r})}
\frac{\rmd}{\rmd V(\mathsf{r}')} {\mathcal E}_{\ind{var}}[V] =
- \left(
\begin{array}{c}
\ds \frac{\rmd \JM}{\rmd V(\mathsf{r}) }\\ \ds \frac{\rmd r_0}{\rmd
V(\mathsf{r}) }
\end{array}
\right)
\left(
\begin{array}{cc}\ds
\ds \frac{\p^2 {\mathcal E}_{\ind{var}}}{\p \JM\p \JM } &\ds \frac{\p^2 
{\mathcal E}_{\ind{var}}}{\p \JM\p \mathsf{r}_0 }\\
\ds \frac{\p^2 {\mathcal E}_{\ind{var}}}{\p \mathsf{r}_0\p \JM } &\ds \frac{\p^2 
{\mathcal E}_{\ind{var}}}{\p \mathsf{r}_0 \p \mathsf{r}_0 }
\end{array}
\right)
\left(
\begin{array}{c}
\ds \frac{\rmd \JM}{\rmd V(\mathsf{r}') }\\ \ds \frac{\rmd r_0}{\rmd V(\mathsf{r}') }
\end{array}
\right)
\label{fluctuations about Evar}
\end{equation}}
with (remind that everything is evaluated at the saddle-point)
\begin{equation}
\label{EvarVMr0}
{\mathcal E}_{\ind{var}} \left[\Vvi,\JM,r_0\right] = \frac 1 D
\frac{\Gamma\left(2-\frac{D}2\right)}{(4\pi)^{D/2}} \mbox{tr} \left (
\JM^{D/2}
 \right) -\frac1{(2\pi)^{d/2}} \det( A\JU + \JG)^{-1/2}
\rme^{-\half \mathsf r_0 \frac1{A\JU+\JG}\mathsf r_0}\ .
\end{equation}
The quantity $A$ is defined as follows:
\begin{equation}
\label{A def}
A:= G_{\mathrm{var}}^{\mathrm{inst}}\ ,
\end{equation}
i.e.\ it is the same as $G$, defined in (\ref{lf73}), but always taken at the
variational instanton. Thus when varying $V$, and thus $\JM$ and
$\JG$, only $\JG$ changes, but {\em not} $A$. 

The determinant to be calculated is  (the prime indicating that
the zero-modes are omitted)
\begin{eqnarray}
\label{Dvar'expl}
 \mathfrak{D}_{\mathrm{ver'}} &=& {\det}'_{V}\left[ \frac{\rmd^2
 S_{\ind{var}}[V]}{\rmd
V(\mathsf{r})\rmd V(\mathsf{r}')}  \right]
={\det}'_{V}\left[ \delta^d(\mathsf{r}-\mathsf r')+
\frac{\rmd^2 {\mathcal E}_{\ind{var}}[V]}{\rmd V(\mathsf{r})\rmd
V(\mathsf{r}')}  \right]  \nn\\
&=&{\det}'_{V}\left[ \delta^d(\mathsf{r}-\mathsf r')
{\renewcommand{\arraystretch}{2.3}
- \left(
\begin{array}{c}
\ds \frac{\rmd \JM}{\rmd V(\mathsf{r}) }\\ \ds \frac{\rmd \mathsf
r_0}{\rmd V(\mathsf{r}) }
\end{array}
\right)
\left(
\begin{array}{cc}\ds
\ds \frac{\p^2 {\mathcal E}_{\ind{var}}}{\p \JM\p \JM } &\ds \frac{\p^2
{\mathcal E}_{\ind{var}}}{\p \JM\p \mathsf{r}_0 }\\
\ds \frac{\p^2 {\mathcal E}_{\ind{var}}}{\p \mathsf{r}_0\p \JM } &\ds \frac{\p^2
{\mathcal E}_{\ind{var}}}{\p \mathsf{r}_0 \p \mathsf{r}_0 }
\end{array}
\right)
\left(
\begin{array}{c}
\ds \frac{\rmd \JM}{\rmd V(\mathsf{r}') }\\ \ds \frac{\rmd \mathsf  r_0}{\rmd V(\mathsf{r}') }
\end{array}
\right) } \right] 
\end{eqnarray}
Now we use the cyclic invariance of the determinant\footnote{If $X$ is
a $n\times m$ matrix and $Y$ a $m\times n$ matrix, and $\det '$
denotes the product over non-zero eigenvalues, we have the general
identity $\det '[1-XY]=\det '[1-YX]$, although the first determinant
is the determinant of a $n\times n$ matrix, and the second one the
determinant of a $m\times m$ matrix.} to reduce the above expression
(\ref{Dvar'expl}), which is the determinant of an integral kernel
operator over $\mathbb{R}^d$, to the determinant of a finite
dimensional matrix, acting on the space of the variational parameters
$\mathsf r_0$
($d$ dimensional) and $\JM$ ($d\times d$-dimensional):  
\begin{eqnarray}
&&={\det}'_{\mathsf r_0,\JM}\left[ \JU
{\renewcommand{\arraystretch}{2.3}
- \left\{ \int \rmd^d r
\left(
\begin{array}{c}
\ds \frac{\rmd \JM}{\rmd V(\mathsf{r}) }\\ \ds \frac{\rmd \mathsf r_0}{\rmd V(\mathsf{r}) }
\end{array}
\right)\otimes
\left(
\begin{array}{c}
\ds \frac{\rmd \JM}{\rmd V(\mathsf{r}) }\\ \ds \frac{\rmd \mathsf r_0}{\rmd V(\mathsf{r}) }
\end{array}
\right)\right\}
\left(
\begin{array}{cc}\ds
\ds \frac{\p^2 {\mathcal E}_{\ind{var}}}{\p \JM\p \JM } &\ds \frac{\p^2 
{\mathcal E}_{\ind{var}}}{\p \JM\p \mathsf{r}_0 }\\
\ds \frac{\p^2 {\mathcal E}_{\ind{var}}}{\p \mathsf{r}_0\p \JM } &\ds \frac{\p^2 
{\mathcal E}_{\ind{var}}}{\p \mathsf{r}_0 \p \mathsf{r}_0 }
\end{array}
\right)
} \right]\nn\\
&&=\ 
{\det}'_{\mathsf r_0,\JM}\left[ \JU
{\renewcommand{\arraystretch}{2.3}
- \left\{ \int \frac{\rmd^d \mathsf{p}}{(2\pi)^d}
\left(
\begin{array}{c}
\ds \frac{\rmd \JM}{\rmd \tilde V(\mathsf{p}) }\\ \ds \frac{\rmd \mathsf r_0}{\rmd \tilde V(\mathsf{p}) }
\end{array}
\right)\otimes
\left(
\begin{array}{c}
\ds \frac{\rmd \JM}{\rmd \tilde V(-\mathsf p) }\\ \ds \frac{\rmd \mathsf r_0}{\rmd 
\tilde V(-\mathsf p) }
\end{array}
\right)\right\}
\left(
\begin{array}{cc}\ds
\ds \frac{\p^2 {\mathcal E}_{\ind{var}}}{\p \JM\p \JM } &\ds \frac{\p^2 
{\mathcal E}_{\ind{var}}}{\p \JM\p \mathsf{r}_0 }\\
\ds \frac{\p^2 {\mathcal E}_{\ind{var}}}{\p \mathsf{r}_0\p \JM } &\ds
\frac{\p^2 {\mathcal E}_{\ind{var}}}{\p \mathsf{r}_0 \p \mathsf{r}_0 }
\end{array}
\right)
} \right] 
 \qquad \quad
\label{det'_r_0_M}
\end{eqnarray}
$\JU$ is the corresponding $d(d+1)$-dimensional unit-matrix.  In fact
the variational mass matrix parameter space is $d(d+1)/2$ dimensional,
since one has to consider only symmetric mass matrices
$\mathbb{M}$. However in our calculation is is simpler to consider the
$d^2$-dimensional variational space of all real matrices $\mathbb{M}$.

\subsubsection{The calculation} We now evaluate the elements of the
matrix.  First of all, due to rotational invariance and parity of the
instanton, the off-diagonal blocks of the two matrices $\{\square \}$
and $(\square )$ vanish
\begin{eqnarray}\label{0 from parity}
\frac{\p^2 {\mathcal{E}}_{\ind var}}{\p \mathsf{r}_0 \p \JM} &=&0\\
  \int \frac{\rmd^d \mathsf{p}}{(2\pi)^d} \frac{\p \JM}{\p \tilde
V(\mathsf{p})} \frac{\p \mathsf{r}_0}{\p \tilde V(-\mathsf p)} &=&0
\label{0 from parity 2} \ .
\end{eqnarray}
The second relation will be  explicitly checked below. 
As a consequence
(\ref{det'_r_0_M}) takes block-diagonal form, leading to the
factorization of the determinant as the product of the determinants
over each diagonal block
\begin{equation}
\label{DD1D2var'}
\mathfrak{D}_{\mathrm{var'}}\  = \ 
\mathfrak{D}_{\mathrm{var'}}^{(1)}\,
\mathfrak{D}_{\mathrm{var'}}^{(2)}
\end{equation}
\begin{equation}
\label{D1var'} \mathfrak{D}_{\mathrm{var'}}^{(1)}\ = \
{\det}'\left[\JU-\int_\pp\, {\rmd\JM\over\rmd\tilde V(\pp)}\otimes
{\rmd\JM\over\rmd\tilde
V(-\pp)}\,{\partial^2\CE_{\mathrm{var}}\over\partial\JM\partial\JM}\right]
\end{equation}
\begin{equation}
\label{D2var'} \mathfrak{D}_{\mathrm{var'}}^{(2)}\ = \
{\det}'\left[\JU-\int_\pp\, {\rmd\rr_0\over\rmd\tilde V(\pp)}\otimes
{\rmd\rr_0\over\rmd\tilde
V(-\pp)}\,{\partial^2\CE_{\mathrm{var}}\over\partial\rr_0\partial\rr_0}\right]\
.
\end{equation} 
Second, we shall see that the second block, relative to the zero-mode
collective coordinate $\rr_0$, is also 0. Indeed, we shall show that
\begin{equation}
\label{2ndmat=1} \int_\pp\, {\rmd\rr_0\over\rmd\tilde V(\pp)}\otimes
{\rmd\rr_0\over\rmd\tilde
V(-\pp)}\,{\partial^2\CE_{\mathrm{var}}\over\partial\rr_0\partial\rr_0}\
=\ \JU
\end{equation}
so that
\begin{equation}
\label{D2var=1} \mathfrak{D}_{\mathrm{var'}}^{(2)}\ = \
{\det}'\left[0\right]\ =\ 1\ .
\end{equation}
Thus it remains to compute the determinant of the first block,
involving only dependencies on the variational mass $\JM$.  Using
(\ref{EvarVMr0}) and the matrix derivative rules gathered in Appendix
\ref{a:matrixform}, we find
\begin{equation}
\label{d2Evar} \frac{\p^2 {\mathcal E}_{\ind{var}}}{\p \JM\p \JM } =
\frac A M \frac{2-D}{32} \left[ 2(2+D) \JE - d(2-D)\JP \right] \ ,
\end{equation}
with $\JE$ the projector on symmetric matrices and $\JP$ the projector
on the unity matrices
\begin{eqnarray}\label{wf3}
\JE_{ij,kl} &=&
	\half\left( \delta_{ik}\delta_{jl}+\delta_{il}\delta_{jk}\right)\\
\JP_{ij,kl} &=& \frac1d\delta_{ij}\delta_{kl}\ .\label{wf4}
\end{eqnarray}
Next, we calculate $\delta \JM_{ij}\over \delta V(\mathsf{r})$.  Using
\Eq{varfromV} and varying $V$ yields
\begin{eqnarray}\label{wf6}
\delta \JM^{ij}[V] &=& -\int \frac{\rmd^d \mathsf{p}}{(2\pi)^d}\,
\mathsf{p}^i \mathsf{p}^j\, \delta \tilde V(\mathsf{p})\, \rme^{i
\mathsf{p}\mathsf{r}_0} \,
\rme^{-\half \mathsf{p}^i\mathsf{p}^j \JG_{ij}} \nn\\
&&+\half \int \frac{\rmd^d \mathsf{p}}{(2\pi)^d}\, \mathsf{p}^i
\mathsf{p}^j\, \tilde V(\mathsf{p})\, \rme^{i \mathsf{p}\mathsf{r}_0}
\, \rme^{-\half \mathsf{p}^i\mathsf{p}^j \JG_{ij}}\, \mathsf{p}^k \mathsf{p}^l
\delta\JG_{kl} \ .
\end{eqnarray}
Using that at the saddle-point $\delta \JG= \frac{D-2}2 \frac AM \delta \JM$
and $\tilde V(\mathsf{p})$ from \Eq{lf75}, we obtain
\begin{eqnarray}\label{wf7}
\delta \JM^{ij}[V] &=& -\int \frac{\rmd^d \mathsf{p}}{(2\pi)^d}\,
\mathsf{p}^i \mathsf{p}^j\, \delta \tilde V(\mathsf{p})\, \rme^{i
\mathsf{p}\mathsf{r}_0} \, \rme^{-\half \mathsf{p}^2 A} -\half\int
\frac{\rmd^d \mathsf{p}}{(2\pi)^d}\, \mathsf{p}^i \mathsf{p}^j\,
\rme^{- \mathsf{p}^2 A}\,
\mathsf{p}^k \mathsf{p}^l \frac{D-2}2 \frac AM \delta \JM_{kl}\nonumber \\
&=& -\int \frac{\rmd^d \mathsf{p}}{(2\pi)^d}\, \mathsf{p}^i
\mathsf{p}^j\, \delta \tilde V(\mathsf{p})\, \rme^{i
\mathsf{p}\mathsf{r}_0} \, \rme^{-\half \mathsf{p}^2 A} +\frac{2-D}8
\delta \JM_{kl} \left( d \JP_{ij,kl}+2 \JE_{ij,kl}\right) \ .
\end{eqnarray}
This leads to
\begin{eqnarray}\label{wf8}
\frac{2-D}8d \JP \delta \JM -\frac{2+D}4 \delta \JM = \int
\frac{\rmd^d \mathsf{p}}{(2\pi)^d}\, \mathsf{p}^i \mathsf{p}^j\,
\delta \tilde V(\mathsf{p})\, \rme^{i \mathsf{p}\mathsf{r}_0} \,
\rme^{-\half \mathsf{p}^2 A}
\end{eqnarray}
and finally upon varying $\delta V$
\begin{eqnarray}
\label{delMdelVhelp} \frac{\delta \JM}{\delta \tilde V(\mathsf{p})}
\left( \frac{2-D}8d \JP -\frac{2+D}4 \JE \right) = \mathsf p \otimes
\mathsf p \, \rme^{i \mathsf{p}\mathsf{r}_0} \, \rme^{-\half
\mathsf{p}^2 A} \ .
\end{eqnarray}
This can be inverted (in the subspace of symmetric matrices) as 
\begin{eqnarray}\label{wf9}
\frac{\delta \JM}{\delta \tilde V(\mathsf{p})} = \left( -\frac4{2+D}
\JE - \frac{4d(2-D)}{(2+D)(4-2d+2D+Dd)} \JP \right) \mathsf p \otimes
\mathsf p \, \rme^{i \mathsf{p}\mathsf{r}_0} \, \rme^{-\half
\mathsf{p}^2 A} \label{dM dV(p)} \ .
\end{eqnarray}
Next, we need ${\delta \JM\over \delta V(\mathsf{r})} \otimes {\delta
\JM\over \delta V(\mathsf{r})} \frac{\p^2 {\mathcal E}_{\ind{var}}}{\p
\JM\p \JM }$.  Due to the saddle-point equations, or more explicitly
looking at \Eqs{delMdelVhelp} and \eq{d2Evar}, the following
combination is relatively simple:
\begin{eqnarray}\label{wf10}
{\delta \JM\over \delta \tilde V(\mathsf{p})} \frac{\p^2 {\mathcal
E}_{\ind{var}}}{\p \JM\p \JM } = -\frac{A}M \frac{2-D}4 \, \mathsf
p\otimes \mathsf p \, \rme^{i \mathsf{p}\mathsf{r}_0} \, \rme^{-\half
\mathsf{p}^2 A} \ ,
\end{eqnarray}
and after (Gaussian) integration over $\pp$ we obtain finally 
\begin{eqnarray}\label{wf11}
\int\frac{\rmd^d \mathsf{p}}{(2\pi)^d}\,
{\delta \JM\over \delta \tilde V(-\mathsf p)}
\otimes {\delta \JM\over \delta \tilde V(\mathsf{p})}
\frac{\p^2 {\mathcal E}_{\ind{var}}}\ {\p \JM\p \JM } =
\frac{2-D}{2+D} \JE + \frac{2d(2-D)}{(2+D)(\E+2-D)} \JP
\ .
\end{eqnarray}
The first block determinant (\ref{D1var'}) is therefore the
determinant of the following operator acting on the $d(d+1)/2$
dimensional space of $d\times d$ symmetric matrices
\begin{eqnarray}
\label{wf12}
\mathfrak{D}_{\mathrm{var'}}^{(1)}
&&\  =\ {\det}'\left(  \frac{2D}{2+D} \JE - 
\frac{2d(2-D)}{(2+D)(\E+2-D)} \JP\right)\ .
\end{eqnarray}
Since in this space the projector $\JE$ reduces to the identity, while $\JP$ is the projector on the 1-dimensional subspace generated by the identity, it is easy to see that the operator has
$d(d+1)/2-1$ eigenvalues equal to $2D/(2+D)$, plus one eigenvalue equal to 
$2D/(2+D)-2d(2-D)/(2+D)(\epsilon+2-D)=-2(D-\epsilon)/(\epsilon+2-D)$. 
Hence the final result is
\begin{equation}
\label{Dvar'fin}
\mathfrak{D}_{\mathrm{var'}}^{(1)}\ =\ 
\mathfrak{D}_{\mathrm{var'}}\ =\  - \frac{2(D-\E)}{\E+2-D}\, \left( \frac{2D}{2+D}
\right)^{{d(d+1)\over 2}-1} \ .
\end{equation}

\subsubsection{Terms associated with the zero modes}
Before discussing this result, we  calculate the other entries of the matrix
\eq{fluctuations about Evar}, associated with the 0-modes.
First we vary \Eq{varfromr} with respect to $\delta \mathsf r_0$ and the
corresponding $\delta \tilde V(\mathsf{p})$:
\begin{equation}\label{wf15}
\int\frac{\rmd^d \mathsf{p}}{(2\pi)^d} \delta \tilde V(\mathsf{p})\,
i\mathsf p \, \rme^{-\half \mathsf p \JG \mathsf p} \rme^{i
\mathsf{p}\mathsf{r}_0} = \int\frac{\rmd^d \mathsf{p}}{(2\pi)^d}
\tilde V(\mathsf{p})\, \mathsf p \, \rme^{-\half \mathsf p \JG \mathsf
p} \rme^{i \mathsf{p}\mathsf{r}_0} (\mathsf p \delta \mathsf r_0) \ .
\end{equation}
Deriving with respect to $\delta \tilde V(\mathsf{p})$ and evaluating at
$V_{\ind{inst}}$ yields
\begin{eqnarray}\label{wf16}
  i\mathsf p_i \, \rme^{-\half A \mathsf{\mathsf p}^2 } \rme^{i
\mathsf{p}\mathsf{r}_0} &=& \int\frac{\rmd^d \mathsf{p}}{(2\pi)^d}
\tilde V(\mathsf{p})\, p_i \, \rme^{-\half \mathsf p \JG\mathsf p}
\rme^{i \mathsf{p}\mathsf{r}_0} \left(\mathsf p \frac{\rmd \mathsf r_0}{\rmd \tilde 
  V(\mathsf{p})}\right) \nn\\ 
&=& -\JM_{ij} \frac{\rmd \mathsf r_0^j}{\rmd \tilde  V(\mathsf{p})}
=
-M \delta_{ij} \frac{\rmd \mathsf r_0^j}{\rmd \tilde V(\mathsf{p})} \ .
\end{eqnarray}
This gives
\begin{equation}
\frac{\rmd \mathsf r_0^i}{\rmd \tilde V(\mathsf{p})}\lts_{V_{\ind{inst}}} =
-\frac 1 M i\mathsf p_i \, \rme^{-\half A \mathsf{p}^2 } \rme^{i
\mathsf{p}\mathsf{r}_0} \label{dr_0 dV(p)} \ .
\end{equation}
Combining \Eqs{dM dV(p)} and \eq{dr_0 dV(p)} checks (\ref{0 from parity 2}).

We now calculate the determinant of the lower block, for which we need
\begin{equation}\label{wf18}
  \int \frac{\rmd^d \mathsf{p}}{(2\pi)^d} \frac{\rmd \mathsf r_0^j}{\rmd
\tilde V(\mathsf{p}) } \frac{\rmd \mathsf r_0^k}{\rmd \tilde V(-\mathsf p) }
=\frac1{M^2} \int \frac{\rmd^d \mathsf p}{(2\pi)^d} \mathsf{p}^j \mathsf{p}^k
\rme^{-A\mathsf p^2} =\frac1{M}\delta^{jk} \ ,
\end{equation}
as well as
\begin{equation}\label{wf19}
\frac{\p^2 {\mathcal E}_{\ind{var}}}{\p \mathsf{r}_0^i\p \mathsf
r_0^j}\lts_{V_{\ind{inst}}} =\frac{\p}{\p \mathsf{r}_0^i}\frac{\p}{\p
\mathsf{r}_0^j} \int\frac{\rmd^d \mathsf{p}}{(2\pi)^d} \tilde
V(\mathsf{p}) \rme^{i\mathsf p \mathsf r_0} \rme^{-\half \mathsf
p\cdot\JG\cdot \mathsf p} = \int\frac{\rmd^d \mathsf{p}}{(2\pi)^d}
\mathsf{p}^i \mathsf{p}^j \rme^{-A \mathsf{p}^2} = M \delta^{ij} \ ,
\end{equation}
where we used that the first term of ${\mathcal E}_{\ind{var}}$ in
\eq{lf70-0} does not depend on $\mathsf r_0$, as well as the instanton
at the saddle-point from \Eq{lf75} and the mass from \Eq{lf74}.  Hence
the second block matrix, relative to the zero mode $\rr_0$, is
identically zero. This is not surprising. Therefore
\begin{equation}\label{wf20}
\left\{\int \frac{\rmd^d \mathsf{p}}{(2\pi)^d} \frac{\rmd \mathsf
r_0^j}{\rmd \tilde V(\mathsf{p}) } \frac{\rmd \mathsf r_0^k}{\rmd
\tilde V(-\mathsf p) }\right\} \frac{\p^2 {\mathcal E}_{\ind{var}}}{\p
\mathsf{r}_0^i \p\mathsf{r}_0^j} = \delta^{ik} \ ,
\end{equation}
and indeed the determinant \eq{det'_r_0_M} is the contribution of the
$d$ translational instanton zero-modes.
\begin{equation}
\label{det'0=1}
\mathfrak{D}^{(2)}_{\mathrm{var'}}\ =\ {\det}'_{\rr_0}\left[0\right]\ =\ 1\ .
\end{equation}

\subsubsection{Discussion} We now discuss our result (\ref{Dvar'fin})
for $\mathfrak{D}_{\mathrm{var'}}$ in our simple variational
approximation.  We see that $\mathfrak{D}_{\mathrm{var'}}$ is finite
and negative for $\epsilon<D$, thus we recover the unstable mode with
a negative eigenvalue for $\CS''$.  However we see that for
$\epsilon=0$, $\mathfrak{D}_{\mathrm{var'}}$ is still finite, while we
expect from our general argument that $\mathfrak{D}$ will have UV
divergences.  Thus our approximation does not properly take into
account the short-wavelength fluctuations around the instanton, and
renormalization, which is important when $\epsilon\to 0$.

Finally it is interesting to look at the behavior of
$\mathfrak{D}_{\mathrm{var'}}$ in the limit $d\to\infty$, $\epsilon$
fixed.  We find for the logarithm of $\mathfrak{D}_{\mathrm{var'}}$,
\begin{equation}
\label{fill2}
\mathfrak{L}_{\mathrm{var'}}=\log(\mathfrak{D}_{\mathrm{var'}})\simeq
-{d\over 2}\left(1-{\epsilon\over 4}\right)=\CO(d)
\end{equation}
as expected from the variational approximation.  However, as we shall
see later, the better approximation $\mathfrak{L}_{\mathrm{var}}$ and
the exact solution $\mathfrak{L}$ behaves respectively at large $d$ as
\begin{equation}
\label{fill3}
\mathfrak{L}_{\mathrm{var}}\simeq {d\over \epsilon^2}\quad,\qquad
\mathfrak{L}\simeq {d^2\over \epsilon}\ .
\end{equation}

\subsection{Expansion around the variational approximation and $1/d$ expansion}
\label{s:direct}

\subsubsection{The large-$d$ limit} 
\label{ss:5.C}
A better method to compute
$\mathfrak{D}$ is to start from \eq{Odef}
\begin{equation}
\label{Orrdd} \JO_{\rr_1\rr_2}=
\int_{\xx}{\left<\delta^d(\rr_1-\rr(\oo))
\delta^d(\rr_2-\rr(\xx)))\right>}_\Vi^{\mathrm{conn}}
\end{equation}
($\oo$ is an arbitrary point on $\CM=\mathbb{R}^D$) and to make a perturbation
expansion around the variational Gaussian Hamiltonian
$H_{\mathrm{trial}}$.
Since the problem is invariant under translations, we chose for $V$
the instanton centered at the origin ($\rr_0=0$).  $m$ will denote the
variational mass ($M=m^2$) and $G_m$ the variational tadpole
$G_m=(4\pi)^{-D/2}\Gamma((2-D)/2)m^{2-D}$. $m$ is solution of
\eq{lf74}, that we rewrite
\begin{equation}
\label{varmassequ}
2m^2 G_m=(4\pi G_m)^{-d/2}
\ .
\end{equation}

\begin{figure}[h]
\fig{0.4\textwidth}{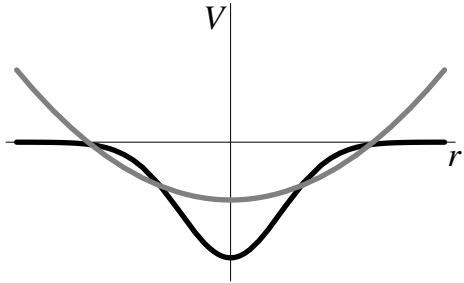}
\caption{The variational instanton (black) and its approximation by a
harmonic potential (grey) (here for $D=1$, $d=4$).
Note that the
curvature of $V (r)$ is the quadratic term before normal-ordering,
whereas in the variational approximation the quadratic term after
normal-ordering appears.}
\label{approx}
\end{figure}

\paragraph{Large-$d$ limit and the variational approximation:}\ \\
The first crucial point used in \cite{DavidWiese1998} is that when the
variational instanton potential \eq{lf75} is written in terms of
normal products relative to the variational mass $m$, it takes the
simple form
\begin{equation}
\label{dir1}
\Vvi(\rr)
=-(4\pi G_m)^{-d/2}\,{:\!\mathrm{e}^{-{\rr^2\over 4G_m}}\!:}_m
\,=\, -2\,m^2\,G_m\,{:\!\mathrm{e}^{-{\rr^2\over 4G_m}}\!:}_m
\end{equation}
that we rewrite as the variational trial potential
$\frac{1}{2}m^2\rr^2$ plus a perturbation $U(\rr)$ as in section
\ref{s:ren:mope} (see \eq{Vr2U})
\begin{equation}
\label{dir2}
\Vvi(\rr)\ =\ -2m^2G_m\,\JU\,+\,{m^2\over 2}\,{:\!\rr^2\!:}_m\,+\,U(\rr)
\quad,\quad
U(\rr)\,=\,-\,2m^2G_m\sum_{n=2}^\infty {1\over n!}\left({-1\over
4G_m}\right)^{n}\,{:\!\left(\rr^2\right)^n\!:}_m 
\end{equation}
and to treat $U(\rr)$ as perturbation, see
\eq{OVOU} and Fig.~\ref{approx}.

The second point is that in the limit when
\begin{equation}
\label{dir3}
d\,\to\,\infty\quad,\qquad\epsilon\ \text{fixed}
\end{equation}
these perturbative terms are subdominant (of order $1/d$) with respect
to the leading term obtained by replacing $V$ by the trial harmonic
potential ${m^2\over 2}:\!\rr^2\!:_m$.  This is seen by rescaling
$\xx$ and $\rr$ in units of the variational mass $m$ (as described in
 detail in Appendix \ref{appendixnormvar}), so that  $m\to
1$, and the propagator $G_m(\xx)$ becomes $G(\xx)=G_1(x)$
\begin{equation}
\label{Gm2G}
G_m(\xx)\to G(\xx)=G_{m=1}(\xx)= (2\pi)^{-{D\over 2}}K_{{D-2\over 2}}(|\xx|)
\end{equation}
and the tadpole amplitude $G_m$ becomes
\begin{equation}
\label{dir4} G_m\to c_0(D)=G(0)=(4\pi)^{-D/2}\Gamma\left({2-D\over
2}\right)\,\simeq\,{1\over4 \pi}{d\over
4-\epsilon}\qquad\text{when}\quad d\to\infty\ ,\ \ \epsilon\
\text{fixed}\ .
\end{equation}
$c_0(D)$ is noted $\mathbb{C}$ in \cite{DavidWiese1998}. When we shall
not deal with the explicit dependence on $D$ of $c_0(D)$ we shall
denote it simply by $c_0$.

The variational instanton potential becomes (see Appendix
\ref{appendixnormvar})
\begin{equation}
\label{dir2res}
\Vvi(\rr)\ =\ -2c_0\,\JU\,+\,{1\over 2}\,{:\!\rr^2\!:}\,+\,U(\rr)
\quad,\quad
U(\rr)\,=\,-\,2c_0\sum_{n=2}^\infty {1\over n!}\left({-1\over
4c_0}\right)^{n}\,{:\!\left(\rr^2\right)^n\!:} \ ,
\end{equation}
where the normal product $:\cdots :$ refers to the normal product with
respect to the unit mass $m=1$, i.e. $:\cdots :=:\cdots :_{m=1}$.

Since $c_0\sim d$, in perturbation theory, the $2n$-leg vertices carry
a weight $d^{1-n}$ and closed loops carry a weight $d$ (summation over
bulk space indices). Counting the resulting factors of $d$ for each
graph, as in the large-$N$ expansion for vector models, only ``cactus
diagrams'' with tadpoles survive in the large-$d$ limit. However
within our normal product scheme, there are no tadpoles.  Therefore
for any observable at large $d$ we can replace
\begin{equation}
\label{dir5}
\left< \text{Observable} \right>_{\Vvi}\ =\ \left< \text{Observable}
\right>_m\,+\, \text{subdominant terms in}\ {1\over d} 
\ ,
\end{equation}
where $\left<\cdots\right>_m$ refers to the expectation value with
respect to the trial variational action
\begin{equation}
\label{dir6}
H_{\text{trial}}^{\text{var}}\ =\ \int_\xx\, {1\over
2}(\nabla\rr)^2+{m^2\over 2}\rr^2 
\ .
\end{equation}

For the same reason, as shown in \cite{DavidWiese1998}, at leading
order in $1/d$, the variational instanton, solution of
\begin{equation}
\label{dir7}
\Vvi(\rr)\,+\,\left<\delta^d(\rr-\rr(\oo))\right>_{m}=0
\end{equation}
is a good approximation for the exact instanton $\Vi$, solution of
\eq{lf36}:
\begin{equation}
\label{dir9}
\Vi(\rr)\ =\, \Vvi(\rr)\,\Bigl(1+\CO(1/d)\Bigr)
\ .
\end{equation}
The first correction of order $1/d$ was computed in \cite{DavidWiese1998}.  
Finally the action for the variational
instanton was found to be
\begin{equation}
\label{dir10} \Svi\ =\ m^D\,\left(1-{\epsilon\over D}\right) c_0(D)
\ .
\end{equation}
If we rescale the effective coupling constant $g$ (or equivalently the
initial coupling constant~$b$) in terms of the variational mass $m$,
\begin{equation}
\label{dir11}
g\ \to\ m^D\,\underline{g}\qquad\text{i.e.}\qquad b\ \to\ m^{D-\epsilon}\,\underline{b}
\end{equation}
the instanton action becomes
\begin{equation}
\label{dir12}
\underline{\Svi}\ =\ \left(1-{\epsilon\over D}\right) c_0(D)\ =\ \CO(d)
\ .
\end{equation}
This rescaling  will the done at the end, but for the time-being, we
keep the explicit mass dependence.

\paragraph{Large-$d$ limit for $\JO$:}\ \\
For our problem, in the large-$d$ limit, we shall firstly
approximate the Hessian $\JO$ in the exact instanton background, with
kernel $\JO_{\rr_1\rr_2}$ given by \eq{Orrdd}, by the Hessian
$\JO^{\text{var}}$ in the variational instanton background, with
kernel $\JO_{\rr_1\rr_2}^{\text{var}}$ given by
\begin{equation}
\label{JOvar}
\JO_{\rr_1\rr_2}^{\text{var}}\ = \ 
\int_{\xx}{\left<\delta^d(\rr_1-\rr(\oo))\delta^d(\rr_2-\rr(\xx))\right>}_\Vvi^{\mathrm{conn}} 
\end{equation}
and then approximate this $\JO^{\text{var}}$ by its large-$d$ limit
$\JO^{\mathrm{var'}}$, with kernel
\begin{equation}
\label{JOm}
\JO_{\rr_1\rr_2}^{\mathrm{var'}}\ = \ 
\int_{\xx}{\left<\delta^d(\rr_1-\rr(\oo))\delta^d(\rr_2-\rr(\xx))\right>}_m^{\mathrm{conn}}
\ .
\end{equation}
This  will be the leading term of a systematic ${1\over d}$
expansion, which can be performed along similar lines as in
\cite{DavidWiese1998}.

$\JO_{\rr_1\rr_2}^{\mathrm{var'}}$ can easily be computed, since we now
deal with a massive free theory. It is even easier to compute its
Fourier transform
\begin{equation}
\label{hatJOm}
\begin{split}
\widehat{\JO}_{\kk_1\kk_2}^{\mathrm{var'}}\,&=\,
\int_{\rr_1}\int_{\rr_2}\rme^{-\rmi(\kk_1\rr_1+\kk_2\rr_2)}\,\JO_{\rr_1\rr_2}^{\text{var}}
\,=\, \int_\xx
{\left<\rme^{\rmi\kk_1\rr(\oo)}\rme^{\rmi\kk_2\rr(\xx)}\right>}_m
-{\left<\rme^{\rmi\kk_1\rr(\oo)}\right>}_m
{\left<\rme^{\rmi\kk_2\rr(\xx)}\right>}_m \\ &=\,
\rme^{-{(\kk_1^2+\kk_2^2)}G_m(0)/2}\,\int_\xx\,
\left[\rme^{-\kk_1\kk_2 G_m(\xx)}-1\right]
\end{split}
\ ,
\end{equation}
where $G_m(\xx)$ is the massive scalar propagator (\ref{ren10}).
Note that we have
$G_m=G_m(0)$. 

\paragraph{Zero modes:}\ \\
In order to compute $\mathfrak{D}$, we must take into account the
translational zero modes of $\CS''\!\!=\JU-\JO$ and the projector
$\JPo$ onto the subspace of zero modes.  According to section
\ref{s:Contribution of fluctuations around the instanton}, these zero
modes are the partial derivatives of $\Vi$,
$V^{\mathrm{zero}}_a=\partial_a\Vi$, and from section
\ref{s:ren:series} (see \eq{BBP0}) the projector is
$$\JPo_{\rr_1\rr_2}=
\mathtt{c_0}\sum\limits_a\partial_a\Vi(\rr_1)\partial_a\Vi(\rr_2)$$
(with the constant $\mathtt{c_0}$ such that $\JP^2_{0}=\JP_{0}$).  In the large-$d$
limit we may approximate $\JPo$ by ${\JPo}^{\mathrm{var}}$
\begin{equation}
\label{JPovar} \JPo^{\mathrm{var}}_{\rr_1\rr_2}\ =\
\mathtt{c'_0}\sum\limits_a\partial_a\Vvi(\rr_1)\partial_a\Vvi(\rr_2)
\end{equation}
and since $\Vvi$ is a Gaussian function, $\JP^{\mathrm{var}}_{0}$ is easily
computed. We obtain for its Fourier transform
\begin{equation}
\label{hatJPovar} \widehat{\JPo}^{\mathrm{var}}_{\kk_1\kk_2}\ =\
-\,{\kk_1\kk_2\over m^2} \,\rme^{-{1\over 2}(\kk_1^2+\kk_2^2)G_m(0)}
\ .
\end{equation}
Finally, in the large-order formulas such as \eq{ImCZMem}, to the
instanton zero-modes is associated the weight factor
$$\mathfrak{W}\,=\,g^{-{d\over D}}\,\left[{1\over 2\pi
d}\int_\rr(\nabla \Vi)^2\right]^{d/2}\ .$$ In the large-$d$ limit this
gives
\begin{equation}
\label{Wvar} \mathfrak{W}^{\mathrm{var}}\,=\,g^{-{d\over
D}}\,\left[{1\over 2\pi d}\int_\rr(\nabla
\Vvi)^2\right]^{d/2}\,=\,g^{-{d\over D}}\,\left[{m^2\over
2\pi}\right]^{d/2} \ =\ \left[{{\overline{g}}^{-{2\over D}}\over
2\pi}\right]^{d/2} 
\end{equation}
so that
$$\log\left(\mathfrak{W}^{\mathrm{var}}\right)=\CO(d)\ .$$

\subsubsection{Large-$d$ calculation of $\mathfrak{L}$}
\label{ss:5.C.1}
\paragraph{Series representation for $\mathfrak{L}$:} \ \\
We now apply these results to the computation of the determinant, or
rather of its logarithm
\begin{equation}
\label{Ldef1}
\mathfrak{L}=\log\left({\det}'[\CS'']\right)
\ .
\end{equation}
Since $\CS''=\JU-\JO$ and since ${\det}'$ subtracts the zero modes, we
can write
\begin{equation}
\label{LandJQ}
\mathfrak{L}=\tr\bigl[\log\left(\JU-\JQ\right)\bigr]
\quad;\qquad
\JQ\ =\ \JO-\JPo
\ .
\end{equation}
Note that $\mathfrak{L}$ has an imaginary part since $\CS''$ has one negative eigenvalue 
$\lambda^-$.
We expand the log as
\begin{equation}
\label{LexpQ}
\mathfrak{L}=-\sum_{k=1}^\infty {1\over k}\,\tr\bigl[\JQ^k\bigr]
\ .
\end{equation}
As we shall see, further simplifications occur  in the large-$d$ limit.
In this limit we can approximate $\JQ$ by $\JQ^{\mathrm{var'}}$ 
given by 
\begin{equation}
\label{JQvar}
\JQ^{\mathrm{var'}}\ =\ \JO^{\mathrm{var'}}-\JPo^{\mathrm{var}}
\ ,
\end{equation}
where $ \JO^{\mathrm{var'}}$ defined by \eq{JOm} is the Hessian
$-\CE''$ at the variational instanton, computed in the variational
approximation, while $\JPo^{\mathrm{var}}$ defined by
(\ref{hatJPovar}) is the projector on the zero-modes of $\CS''$ in the
variational approximation.  Therefore we approximate $\mathfrak{L}$ by
\begin{equation}
\label{Lvar'sum} \mathfrak{L}^{\mathrm{var'}}
=-\sum_{k=1}^\infty {1\over
k}\,\tr\Bigl[{\bigl(\JQ^{\mathrm{var'}}\bigr)}^k\Bigr]
\ .
\end{equation}

\paragraph{``Beads'' and ``necklace'' diagrammatic representation:}\ \\
Starting from \eq{hatJOm}, \eq{hatJPovar} and using the fact that
$\int_\xx G_m(\xx)=1/m^2$, we can write the kernel of
$\JQ^{\mathrm{var'}}$ as
\begin{equation}
\label{hatJQx} \widehat{\JQ}^{\mathrm{var'}}_{\kk_1,\kk_2}=
\rme^{-{(\kk_1^2+\kk_2^2)}G_m(0)/2} \int_\xx \left[ \rme^{-\kk_1\kk_2
G_m(\xx)}-1+\kk_1\kk_2\,G_m(\xx) \right]\ ,
\end{equation}
and expanding in $\kk_1\kk_2$, we get a simple diagrammatic
representation for $\widehat{\JQ}^{\mathrm{var'}}_{\kk_1,\kk_2}$ as a
sum of ``watermelon'' diagrams
\begin{equation}
\label{hatObeads}
\begin{split}
\widehat{\JQ}^{\mathrm{var'}}_{\kk_1,\kk_2}&=
 \rme^{-{(\kk_1^2+\kk_2^2)}G_m(0)/2}\,
 \sum_{n=2}^{\infty} {(-1)^n\over n!}\,(\kk_1\kk_2)^n\,\int_\xx\,
G_m(\xx)^n
\\
&= \ \parbox{8cm}{\fig{8cm}{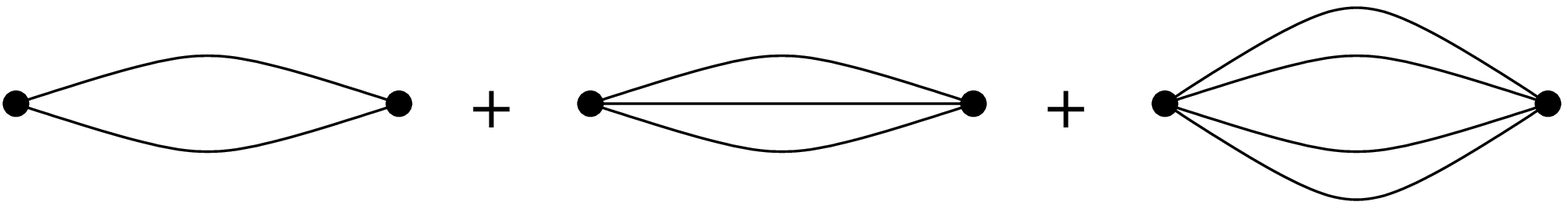}}
\end{split}
\ .
\end{equation}
Each line represents a propagator $G_m$. No internal $\CM$ momentum
flows in the diagram, the $\pp$'s are external momenta relative to the
embedding space $\mathbb{R}^d$.  In this series the term $n=0$ is
removed by the fact that $\JO$ is defined by a connected correlator in
\eq{Orrdd}; while the term $n=1$ is removed by the projector onto the
zero modes $\JPo$.

Now we consider the
$\tr [{(\JQ^{\mathrm{var'}})}^k]$ in
\eq{Lvar'sum}. Each trace is given by
\begin{equation}
\label{trOkint} \tr\Bigl[{\bigl(\JQ^{\mathrm{var'}}\bigr)}^k\Bigr]=
\int{\rmd^{d}\kk_1\over (2\pi)^d}\cdots{\rmd^{d}\kk_k\over (2\pi)^d} \
\widehat{\JQ}^{\mathrm{var'}}_{\kk_1,\,-\kk_2}
\widehat{\JQ}^{\mathrm{var'}}_{\kk_2,\,-\kk_3} \cdots
\widehat{\JQ}^{\mathrm{var'}}_{\kk_k,\,-\kk_1} \ .
\end{equation}
Thus $\mathfrak{L}^{\mathrm{var'}}$ can be represented as a sum over
``necklace'' diagrams made out of the ``beads'' of \eq{hatObeads}.  The
integration over the $\kk$'s can be done explicitly and gives a
decomposition of the form
\begin{equation}
\label{Oksum}
\tr\Bigl[{\bigl(\JQ^{\mathrm{var'}}\bigr)}^k\Bigr]=\sum_{n_1,\dots ,n_k\ge 2}
P_{n_1,\dots ,n_k}(d)\ \prod_{i=1}^{k}2m^2G_m(0){I_{n_i}\over 2^{n_i}n_i!}
\quad,\quad
I_n=\int_\xx \left[{G_m(\xx)\over G_m(0)}\right]^n
\ ,
\end{equation}
where $P_{n_1,\dots ,n_k}(d)$ is a polynomial in $d$ (the bulk space
dimension), with integer coefficients, given by the average
\begin{equation}
\label{Pndef} P_{n_1,\dots ,n_k}(d)=\overline{(-\kk_1\kk_2)^{n_1}
(-\kk_2\kk_3)^{n_2}\cdots(-\kk_k\kk_1)^{n_k}}
\end{equation}
with the normalized Gaussian independent variables 
$\kk_{i}\in\mathbb{R}^d$, i.e.
$\overline{\kk_i^a\kk_j^b}=\delta^{ab}\delta_{ij}$\ . The polynomial
$P$ can be computed by Wick's theorem. Typical configurations are:
$$
\includegraphics[width=5in]{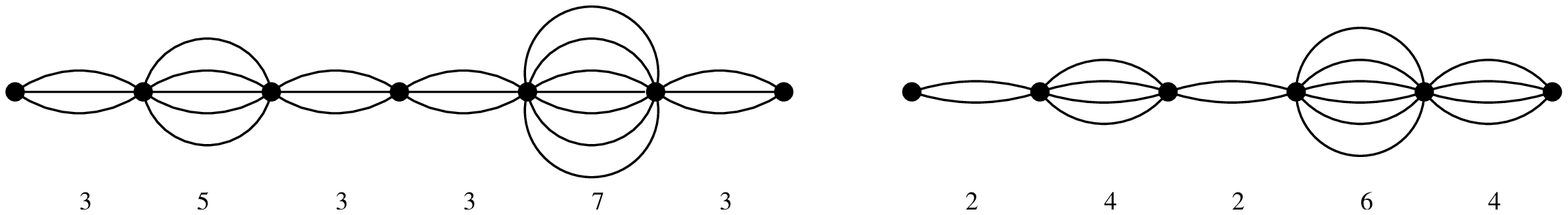}
$$
Note that the first and last points are identified. 
Let us denote by $N$ the total number of lines $N=\sum n_i$ in the
diagram.  From \eq{Pndef} the $P$'s are non zero if and only if the
$n_i$'s are either all even, or all odd.
\begin{itemize}
\item If $k=1$ this is always true and $P_{n}(d)$ is of degree $n$
in $d$.  
\item If $k>1$ and the $n_i$'s are even, $N\ge 2k$ and the
degree of $P(d)$ is $N/2\ge k$. 
\item If $k>1$ the $n_i$'s are odd, $N\ge 3k$ and the degree of $P(d)$
is $1+(N-k)/2 > k$.
\end{itemize}

\paragraph{Large-$d$ power counting:} \ \\
We now look at the behavior of these terms when $d\to\infty$,
$\epsilon$ fixed.  First we rescale everything in units of the
variational mass $m$,
\begin{equation}
\label{dir17} \xx\to\xx/m\quad,\qquad\pp\to\pp\, m^{{2-D\over 2}}
\quad,\qquad G_m(\xx)\to m^{2-D}G(\xx)\ ,
\end{equation}
that is set the variational mass $m$ to unity in our calculations,
since the $\tr[\JQ^k]$ are dimensionless quantities.  We refer to
Appendix \ref{appendixnormvar} for the details on this rescaling.
Then we note that the propagator $G(\xx)$ for $\xx \neq 0$ given by
\eq{ren10} is of order $\CO(1)$ when $d\to\infty$
$$G(\xx)=(2\pi)^{-{D\over 2}}|\xx|^{{2-D\over 2}}K_{{D-2\over 2}}(|\xx|)
\ \to\ {1\over 2\pi}K_0(|\xx|)\ =\ \CO(1)\ ,
$$
while $G(0)$ is of order $\CO(d)$ since
$$ G(0)=(4\pi)^{-{D\over 2}}\Gamma\left({\scriptstyle{2-D\over
2}}\right)\ \to\ {1\over 2\pi}{1\over 2-D}\ \simeq {d\over
4\pi(4-\epsilon)}\ =\ \CO(d)\ .$$ 
Thus the integrals $I_n$ given by
\eq{Oksum} are of order $d^{-n}$
$$I_n=\int_\xx \left[{G(\xx)\over G(0)}\right]^n=\CO(d^{-n})\ ,$$ and the
term associated to the $k$-bead necklace $[n_1,n_2,\cdots,n_k]$ in
the decomposition \eq{Oksum} is of order
$$[n_1,n_2,\cdots,n_k]\ \to\ \CO\left({d}^{\text{degree}[P]+k-N}\right)$$
where $N=\sum n_i$.
\begin{itemize}
  \item If $k=1$, we have seen that $\text{degree}[P]=N$, and all the
terms are of order $d$.  Therefore, if the series over the $n$'s
converges (we shall discuss this later)
\begin{equation}
\label{k=1Od}
\tr\left[{\JQ^{\mathrm{var'}}}\right]\ =\ \CO(d)
\ .
\end{equation}

  \item If $k>1$ we have seen that there are two cases. For even
necklaces the $n_i$'s are all even and $\text{degree}[P]=N/2\ge k$ so
we obtain a term of order ${d\,}^{k-{N\over 2}}\le d^{0}$ We note that the
most dominant terms are those with $N=2k$.  These are the
$[2,2,\cdots,2]$ necklaces whose beads contain 2 links (chains of
bubbles).
\begin{equation}
\label{dir18}
\includegraphics[width=2in]{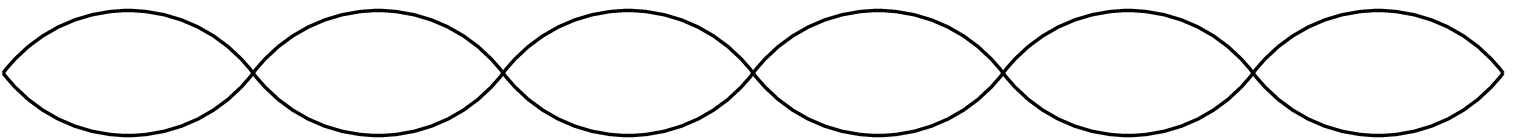}
\end{equation}
For odd necklaces, the $n_i$'s are all odd and
$\text{degree}[P]=1+(N-k)/2>k$, while $N\ge 3k$. This gives a term of
order ${d\,}^{1-{{N-k\over 2}}}\le d^{1-k}\ll 1$.
The conclusion is that (as long as we can sum the necklace series) the
$k>1$ terms are of order $\CO(1)$
\begin{equation}
\label{k>1O1} 
k>1\quad
\Rightarrow\quad\tr\left[\big({\JQ^{\mathrm{var'}}}\big)^k\right]\
=\CO(1)\ ,
\end{equation} 
and that the dominant contribution is given by the chain of bubbles.
\end{itemize}

\paragraph{Final result:}\ \\
All the $k>1$ terms in \eq{Oksum} are subdominant with respect to the
$k=1$ term.  In the large-$d$ limit, $\mathfrak{L}$ is of order
$\CO(d)$ and can be approximated by
\begin{equation}
\label{Ldinfinite}
\mathfrak{L}\ =\ -\,\tr\left[{\JQ^{\mathrm{var'}}}\right]\,+\,\CO(1)
\ .
\end{equation}
We shall check this result with explicit calculations. As we shall
see, the summation of the necklace series is not completely obvious,
and is impaired by the UV divergences of the theory.  Let us also note
that the imaginary part which comes from the unstable eigenmode of
$\CS''=\JU-\JO$ is an effect of order $\CO(1)$ (since it is associated
with one single eigenvalue).

\subsection{Explicit calculations at large $d$}
\label{s:expdl}

\subsubsection{$\tr\!\left[{\JQ^{\mathrm{var'}}}\right]$ and its
large-$d$ limit for $\epsilon>0$:} \label{ss:5.C.2} We first consider
the leading term $\tr\left[{\JQ^{\mathrm{var'}}}\right]$, given by
\begin{equation}
\label{ }
\tr\left[{\JQ^{\mathrm{var'}}}\right]\ =\
\tr\left[{\JO^{\mathrm{var'}}}\right]\ -\ d\ 
\end{equation}
\paragraph{$\xx$-integral representation:} 
$\tr\left[{\JO^{\mathrm{var'}}}\right]$ is easily
calculated from \eq{hatJOm} and \eq{trOkint}.
\begin{equation*}
\begin{split}
\tr\left[{\JO^{\mathrm{var'}}}\right]\ &=\ \int {\rmd^{d}\kk\over
(2\pi)^d}\,{\JO^{\mathrm{var'}}_{\kk,-\kk}} \ =\ \int {\rmd^{d}\kk\over
(2\pi)^d} \rme^{-\kk^2G_m(0)}\,\int_\xx\, \left[\rme^{\kk^2
G_m(\xx)}-1\right]
\end{split}
\end{equation*}
The $\kk$-integration is Gaussian and gives, using the equation for $m$ (\ref{varmassequ})
\begin{equation}
\label{trOexpl}
\begin{split}
\tr\left[{\JO^{\mathrm{var'}}}\right]\ &=\
2\,m^2\,G_m(0)\int_\xx\left(\left[1-{G_m(\xx)\over G_m(0)}\right]^{-{d\over
2}}-1\right)
\end{split}
\end{equation}
Since $\tr\left[{\JO^{\mathrm{var'}}}\right]$ is dimensionless we can set 
 the variational mass $m$ to unity $m=1$, in the r.h.s.\ of
 (\ref{trOexpl}) (see Appendix \ref{appendixnormvar}).
Using the explicit form (\ref{Gm2G}) for the
propagator $G(x)$ and integrating over the $\xx$ angular variables via $\int
\rmd^D\xx=\CS_D\int_0^\infty \rmd x\,x^{D-1}$ with $x=|x|$ we obtain
\begin{equation}
\label{trOexpl2}
\begin{split}
\tr\left[{\JO^{\mathrm{var'}}}\right]\ &=\
2^{2-D}{\Gamma\left({2-D\over 2}\right)\over\Gamma\left({D\over
2}\right)}\int_0^\infty \rmd x\,x^{D-1}\,\left(\left[1-2 \left[{x\over
2}\right]^{{2-D\over 2}}{K_{{D-2\over 2}}(x)\over
\Gamma\left({2-D\over 2}\right)}\right]^{-{d\over 2}}-1\right)
\end{split} 
\ .
\end{equation}
Let us first consider this integral for finite (but a priori large)
$d$, and study its convergence.  

\paragraph{IR convergence:}\ \\
 At large $\xx$ the integral is convergent. Indeed the massive
propagator is exponentially decreasing as
$G(\xx)\simeq\exp(-|\xx|)$. For finite $d$, and thanks to the $-1$
that comes from the subtraction of the disconnected part, the integrand in \eq{trOexpl}
is also exponentially decreasing at large $\xx$.

\paragraph{UV divergences:}\ \\ The small-$\xx$ behavior of the
integral \eq{trOexpl} has in fact already been studied in
Sect.\eq{s:ren:mope}. It was shown that this behavior is governed by
the MOPE \eq{deltaMOPE3} and is related to the UV divergences at one
loop of the model. The integrand in (\ref{trOexpl2}) behaves as
$\int_0^{\cdots}\rmd x\,x^{\epsilon-D-1}\,\mathtt{C_0}$ with
$\mathtt{C_0}$ given by (\ref{C0tt}).
We thus recover the expected UV divergence at $\epsilon\le D$, which
is proportional to the insertion of the operator $\JU$.  This UV
divergence appears in the series representation \eq{Oksum} of
$\tr\left[{\JO^{\mathrm{var'}}}\right]$ as the onset of the
non\-sum\-ability of the series\footnote{This non\-sum\-ability has of
course nothing to do with the large-order behavior we are
after.}. Indeed, this series is
\begin{equation}
\label{trOserexp}
\tr\left[{\JO^{\mathrm{var'}}}\right]=2G(0)\sum_{n=1}^\infty
{P_n(d)\over {{2}^n} n!} I_n \quad\text{with}\quad
P_n=d(d+2)\cdots(d+2n-2)\ ,
\end{equation}
and $I_n=\int_\xx [G(x)/G(0)]^n\sim\  n^{-{D\over 2-D}}$ at large $n$. 
It is easy to check that the series \eq{trOserexp} behaves as $\sum_n n^{-1+{d\over 2}-{D\over 2-d}}$
and is convergent only if $\epsilon>D$. 

Since the model is defined for $\epsilon<D$ by dimensional
regularization, the analytic continuation of the integral
\eq{trOexpl} is its finite part (in the sense of distribution
theory).  Therefore $\tr\left[{\JO^{\mathrm{var'}}}\right]$ is defined
for $\epsilon>0$ by 
\begin{equation}
\label{trOexplfp}
\begin{split}
\tr\left[{\JO^{\mathrm{var'}}}\right]\ &=\ 2\,G(0) \,\times\
\mbox{f.p.} \int_\xx\left(\left[1-{G(\xx)\over
G(0)}\right]^{-{d\over 2}}-1\right) \ ,
\end{split}
\end{equation}
or equivalently by the resummation of the series \eq{trOserexp} by a
zeta-function prescription.

For $\epsilon=0$ the integral has another UV divergence, which is
canceled by the $(\nabla\rr)^2$ counterterm of the renormalized
theory. We shall discuss this point later.

\paragraph{large-$d$ limit:}
We can now  take the limit of \eq{trOserexp} when
$$d\to\infty \quad,\qquad \epsilon>0\quad\text{fixed}\ .$$
Since in this limit
$$G(\xx)\to{1\over 2\pi}K_0(|\xx|)\quad,\qquad G(0)\to\ {1\over
4\pi}{d\over 4-\epsilon}$$ we obtain
\begin{equation}
\label{dir24}
\begin{split}
\tr\left[{\JO^{\mathrm{var'}}}\right]\ &=\ d\,{1\over 4-\epsilon}\
\mbox{f.p.}\int_{0}^\infty\hskip -1ex\rmd x\,
x\,\left[\rme^{{(4-\epsilon)} K_0(x)}-1\right]\ +\ \CO(1)
\end{split}
\ .
\end{equation}
From the short distance behavior of the 2-dimensional propagator
$K_0(x)\simeq \log(1/x)$, the last integral is
\begin{equation*}
T_1(\epsilon)=\text{f.p.}\int_{0}^\infty\hskip -1ex\rmd x\,
x\,\left[\rme^{{(4-\epsilon)} K_0(x)}-1\right] =\
\int_{0}^\infty\hskip -1ex\rmd x\, x\,\left[\rme^{{(4-\epsilon)}
K_0(x)}-1-x^{-4+{\epsilon}}\right]
\end{equation*}
and is UV finite for $0<\epsilon<2$. 
Thus we recover that 
$\tr\left[{\JO^{\mathrm{var'}}}\right]=\CO(d)$ in this case.

$T_1(\epsilon)$ has a single pole at $\epsilon=2$, as expected. It is
UV divergent when $\epsilon\to 0$. This will be studied later.

\subsubsection{$\tr\!\left[\left({\JQ^{\mathrm{var'}}}\right)^2\right]$
and its large-$d$ limit for $\epsilon>0$:}
 
\label{ss:5.C.3}
We now perform the same analysis for
$\tr\left[\JQ^2\right]$. We have, using (\ref{varmassequ})
\begin{equation}
\label{dir26}
\begin{split}
&\tr\!\left[\left({\JQ^{\mathrm{var'}}}\right)^2\right]  =\int{\rmd^{d}
\kk_1\over (2\pi)^{d}}{\rmd^{d} \kk_2\over (2\pi)^{d}}
\,\widehat{\JQ}^{\mathrm{var'}}_{\kk_1,-\kk_2}
\widehat{\JQ}^{\mathrm{var'}}_{\kk_2,-\kk_1} =\int{\rmd^{d} \kk_1\over
(2\pi)^{d}}{\rmd^{d} \kk_2\over (2\pi)^{d}}
\,\rme^{-(\kk_1^2+\kk_2^2)G_m(0)}\ \times
\\
&\qquad \qquad  \qquad \quad \times 
\int_{\xx_1}\int_{\xx_2}
\left[\rme^{-\kk_1\kk_2 G_m(\xx_1)}-1+\kk_1\kk_2 G_m(\xx_1)\right]
\left[\rme^{-\kk_1\kk_2 G_m(\xx_2)}-1+\kk_1\kk_2 G_m(\xx_2)\right]
\end{split}
\end{equation}
Setting $m=1$ and performing the $\kk$ integrations we get
\begin{equation}
\label{dir26a}
\begin{split}
& 4\,G(0)^2 \int\limits_{\xx_1}\int\limits_{\xx_2} \Bigg \{
{
\left[1-\left[{G(\xx_1)+G(\xx_2)\over 2 G(0)}\right]^2\right]^{-{d\over 2}}
\hskip -1ex -\left[1-\left[{G(\xx_1)\over 2 G(0)}\right]^2\right]^{-{d\over 2}}
\hskip -1ex-\left[1-\left[{G(\xx_2)\over 2 G(0)}\right]^2\right]^{-{d\over 2}}
\hskip -1ex+1}
\\
&\qquad \qquad  \qquad \qquad {-\ d\,{G(\xx_1)G(\xx_2)\over
4\,G(0)^2}\,\left[\left[1-\left[{G(\xx_1)\over
2G(0)}\right]^2\right]^{-{d\over 2}-1} 
+\left[1-\left[{G(\xx_2)\over 2G(0)}\right]^2\right]^{-{d\over
2}-1}-1\right]}
\Bigg \}
\end{split}
\end{equation}
This integral is IR and UV finite as long as $\epsilon>0$. When
$\epsilon=0$ we recover the UV divergence when both $\xx_1$ and $\xx_2
\to 0$.

Now in the large-$d$ limit, $\epsilon$ fixed, since $G(0)\sim d$ and
$G(x)\sim 1$ we can expand the $\left[\cdots\right]^{-d/2}$ and get
\begin{eqnarray}\label{dir27}
{3 d(d+2)\over 16\,G(0)^2}\int_{\xx_1}\int_{\xx_2} G(\xx_1)^2 \, G(\xx_2)^2
\ +\ \CO(d^{-1}) &=& \frac{3}{16}  \left(2D-\epsilon
\right)\left(2+D-\epsilon \right) + \ \CO(d^{-1})\nonumber \\
& \approx& 
{3}\left(1-{\epsilon\over 4}\right)^2\,+\ \CO(d^{-1})\
\end{eqnarray}
This expansion is not valid for $\xx_1$ and $\xx_2=0$ when
$\epsilon=0$ and this gives the $1/\epsilon$ UV pole (coupling
constant renormalization), but this is an effect exponentially small
in the large-$d$ limit.  We have thus checked the fact that
\begin{equation}
\label{dir28}
\tr\!\left[\left({\JQ^{\mathrm{var'}}}\right)^2\right]=\CO(1)
\end{equation}

\subsubsection{$\tr\!\left[\left({\JQ^{\mathrm{var'}}}\right)^k\right]$
and its large-$d$ limit for $\epsilon>0$:} Calculation of higher
powers can be done along the same line. We get
\begin{equation}
\label{dir30}
\tr\!\left[\left({\JQ^{\mathrm{var'}}}\right)^k\right]=
\left(1-{\epsilon\over 4}\right)^k\,+\,\CO(d^{-1}) 
\ .
\end{equation}

\section{$1/d$ corrections to the large $d$ limit} 
\label{s:1/dcor} 
In this section we study the first $1/d$ correction to the variational
solution, which was shown to be valid for large $d$, $\epsilon$ being
kept fixed.

\subsection{$1/d$ diagrammatic}
\label{ss:1/dexp} 
We first recall in
this subsection how is constructed and organized the $1/d$ expansion,
following the ideas of our first paper \cite{DavidWiese1998}.  We have
performed the rescaling (\ref{dir17}) so that the variational mass $m$
is set to unity. This rescaling is detailled in
Appendix~\ref{appendixnormvar}.  We denote by $c_0$ the normalized
tadpole amplitude\footnote{$c_0$ is denoted $\mathbb{C}$ in
\cite{DavidWiese1998}.} and the integration measure over $d$-momenta
$\kk$ is now normalized so that we have
\begin{equation}
\label{mc0int} m=1\quad,\quad c_0=(4\pi)^{-D/2}\Gamma((2-D)/2)=
\parbox{1cm}{\includegraphics[width=1.cm]{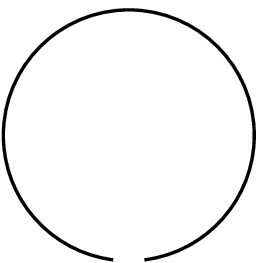}} \quad,\quad
\int_\kk\rme^{-\kk^2 c_0}=2c_0
\end{equation}
The exact instanton potential is in these units of the form
\begin{equation}
\label{Vinstnp2} \Vi(\rr)=2c_0\sum_{n=0}^\infty{1\over
2^nn!}\left({-1\over 2 c_0}\right)^n\mu_n\,{:\!{(\rr^2)}^n\!:}\ ,
\end{equation}
where now the normal products are defined with respect to the unit
variational mass $m=1$, i.e.  $$:\quad :\,=\,:\quad :_{m=1} $$
and the coefficients $\mu_n$ are of order $1$ in the large-$d$ limit,
and are found to be
\begin{equation}
\label{mun1/d}
\mu_n=-1+{\delta_n\over d}+\CO(d^{-2})
\end{equation}
in the large-$d$ limit, with $\delta_n=\delta_n(D,d)$ 
given by a self-consistent equation that we
recall later.  We remind the reader that if we set the $\mu_n=-1$ we
recover the variational instanton $\Vvi$.

\begin{figure}[h]
\begin{center}
$$
\begin{matrix}
\text{order\ }n&0&1&2&3&4&\cdots\\
&&&&&&\\
2n-\text{vertex}&\includegraphics[scale=.5]{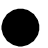}
&\includegraphics[scale=.5]{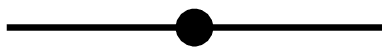}
& \parbox{42.5pt}{\includegraphics[scale=.5]{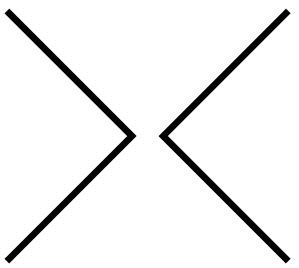}}
&\parbox{49.5pt}{\includegraphics[scale=.5]{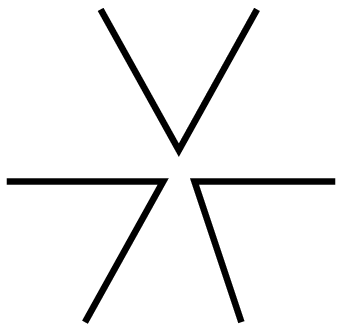}}
&\parbox{47pt}{\includegraphics[scale=.5]{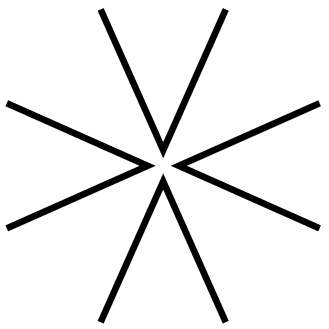}}
&\cdots\\
&&&&&&\\
\text{couplings}&c_0(d/2-2\mu_0)&\mu_1+1&{-\mu_2\over
2c_0}&{\mu_3\over (2c_0)^2}&{-\mu_4\over (2c_0)^3}&\cdots
\\
&&&&&&\\
\CO(1/d)&c_0(d/2+2-{2\delta_1/d})&{\delta_1\over d}&{1\over
2c_0}&{-1\over (2c_0)^2}&{-1\over (2c_0)^3}&\cdots
\end{matrix}
$$
\caption{Self energy ($n=0$), mass ($n=1$) and interaction ($n\ge2$)
vertices and couplings in the $U$ expansion (the symmetry factors
$1/(2^n n!)$ for the vertices are not written).}
\label{verticeslarged}
\end{center}
\end{figure}

The perturbative diagrammatics is obtained by writing
\begin{equation}
\label{Vr2U2}
\Vi(\rr)={1\over 2}\rr^2+U(\rr)
\end{equation}
and treating $U$ as a perturbation. The corresponding $2n$-vertices
and couplings are schematically depicted on Fig.~\ref{verticeslarged}.
The last line represents the couplings which have to be kept at order
$1/d$.  The propagator is the usual bosonic propagator with unit mass
$G(x)$.  The one-loop tadpole graph is absent since it is subtracted by the
normal-product prescription.  The external $\rr$-space  indices
$a=1,\cdots, d$ flow along the closed lines as in a standard O($n$)
model.

\begin{figure}[h]
\begin{center}
$ \parbox{74pt}{\includegraphics[scale=.5]{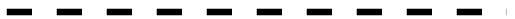}}=1+
\parbox{28.5pt}{\includegraphics[scale=.5]{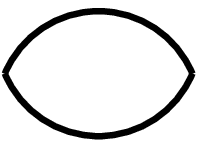}}
+\parbox{58pt}{\includegraphics[scale=.5]{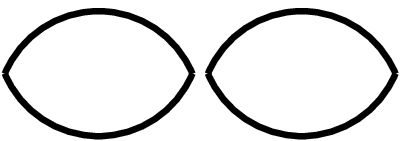}}+
\parbox{87pt}{\includegraphics[scale=.5]{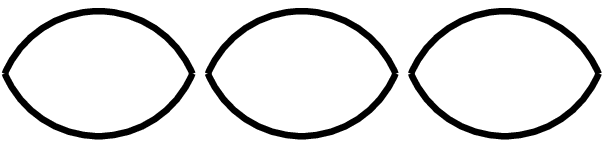}}+\cdots $
\caption{The chains of bubbles in the large-$d$ expansion}
\label{chainofbubble}
\end{center}
\end{figure}
It was shown in \cite{DavidWiese1998} that the diagrams can be
reorganized in a $1/d$ expansion by summing all the chains of bubbles,
as depicted in Fig.\ref{chainofbubble}.  More precisely, the
propagator for the chain is given by the geometric series (in Fourier
transform w.r.t. $\xx$ space)
\begin{equation}
\begin{split}
 \parbox{74pt}{\includegraphics[scale=.5]{bbline0}}=H(\pp)
=\left[1+\mu_2{d\over 4c_0}B(\pp)\right]^{-1} \ ,
\end{split}
\label{chaine0}
\end{equation}
$\pp$ being the $D$-momentum flowing through the chain, and $B(\pp)$
the bubble amplitude (one-loop diagram)
\begin{equation}
\label{B2pex}
\begin{split}
{\parbox{28.5pt}{\includegraphics[scale=.5]{onebubble}}}
=B(\pp)=&\int_\pp\rme^{\rmi\pp\xx}\,G(\xx)^2
\\ &={1\over \pi}{\mathrm{arcth}\left(p/\sqrt{4+p^2}\right)\over
p\sqrt{4+p^2}} = \frac{1}{\pi} \frac{\mbox{arcsinh} \left(p/2
\right)}{p \sqrt{4+p^{2}}} \qquad\text{when}\quad D=2
\end{split}
\end{equation}
For zero momentum, we have
\begin{equation}
\label{B20}
B (0) = \frac{2-D}{2} c_{0}\ . 
\end{equation}
In practice we also have to consider the chains with $n\ge 1$ or $n\ge
2$ bubbles.  They are depicted as follows, with the associated
amplitude $H^{(1)}(\pp)$ and $H^{(2)}(\pp)$
\begin{equation}
\label{chaine1}
\begin{split}
\parbox{74pt}{\includegraphics[scale=.5]{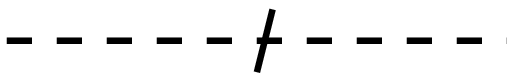}}&=
\parbox{28.5pt}{\includegraphics[scale=.5]{onebubble}}
+\parbox{58pt}{\includegraphics[scale=.5]{twobubble}}+
\parbox{87pt}{\includegraphics[scale=.5]{threebubble}}+\cdots
\\
&=H^{(1)}(\pp)=
\left[1+\mu_2{d\over 4c_0}B(\pp)\right]^{-1}-1
\end{split}
\end{equation}
\begin{equation}
\label{chaine2}
\begin{split}
\parbox{74pt}{\includegraphics[scale=.5]{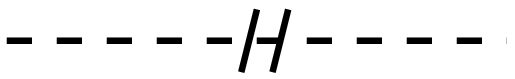}}&=
\parbox{58pt}{\includegraphics[scale=.5]{twobubble}}+
\parbox{87pt}{\includegraphics[scale=.5]{threebubble}}+\cdots
\\
&=H^{(2)}(\pp)=
\left[1+\mu_2{d\over 4c_0}B(\pp)\right]^{-1}-1+\mu_2{d\over 4c_0}B(\pp)
\end{split}
\end{equation}
At the diagrammatic level this reorganisation of perturbation theory
is very similar to what is done in the $1/N$ expansion for the (linear
or non-linear) sigma models, where the bubble chain is the propagator
for an auxiliary $\sigma$ field, and the interaction involves only
$\rr\rr\sigma$ and $\sigma^k$ ($k\ge 3$) terms. The analytic structure
of the perturbation theory is nevertheless quite different, in
particular for the UV and IR divergences of the theory, as already
discussed in \cite{DavidWiese1998}, and as we shall see below.

\begin{figure}[h]
\begin{center}
$$
\begin{matrix}
 \text{vertex}\quad &\quad \includegraphics[scale=.5]{2L-vertex} \quad
&\quad \parbox{58.5pt}{\includegraphics[scale=.5]{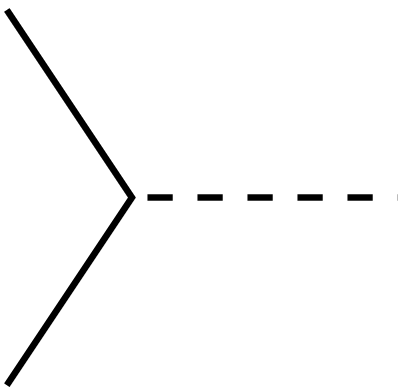}} \quad
& \quad
\parbox{56pt}{\includegraphics[scale=.5]{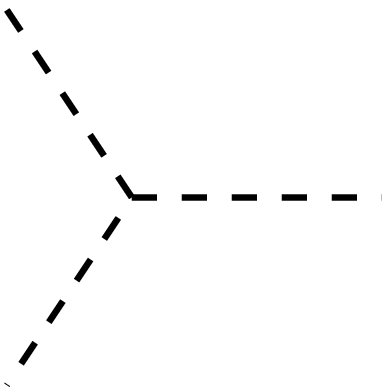}}
\quad&\quad\parbox{42.5pt}{\includegraphics[scale=.5]{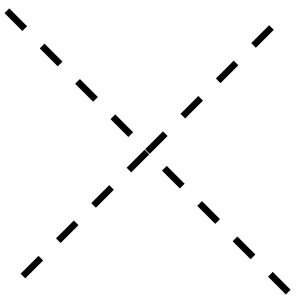}} 
\quad&\quad\cdots\\
\quad   &\quad      & & & & \\
\text{coupling} \quad &\quad \mu_1+1 & \sqrt{{-\mu_2\over 2c_0}} &
{1\over\sqrt{2c_0}}{\mu_3\over (-\mu_2)^{3/2}} & {1\over
2c_0}{-\mu_4\over (-\mu_2)^2} & \quad\cdots
 \\
 \quad   &\quad      & & & & \\
 \text{coupling at order}\ {1\over d}\quad & {\delta_1\over d} &
 {1\over \sqrt{2 c_0}}& -{1\over \sqrt{2 c_0}}& {1\over 2 c_0} &
\end{matrix}
$$
\caption{The vertices contributing to $-V^{\mathrm{inst}} (r)$ and
their couplings in the large-$d$-reorganized perturbative expansion.}
\label{R&Bcouplings}
\end{center}
\end{figure}
After this resummation the new vertices with their couplings are
depicted in Fig.\ref{R&Bcouplings}.  The crucial point is that in the
limit $d\to\infty$, $\epsilon$ fixed, since $D\to 2$ the tadpole
coefficient $c_0$ diverges as $d$ so that
\begin{equation}
\label{4vdlim}
\mu_2 {d\over 2 c_0}\ \to\ - 8\pi\left(1-{\epsilon\over 4}\right)\ =\ \CO(1)
\end{equation}
and the bubble propagator $H(\pp)$ is of order $\CO(1)$, while the
vertices are of order $1/\sqrt{d}$, $1/d$, etc. It was shown in
\cite{DavidWiese1998} that only a finite number of diagrams contribute
to a given order in $1/d$, and explicit calculations where done at the
first non-trivial order.

With these notations we have found in \cite{DavidWiese1998} that at
order $1/d$ the following diagrams contribute to the expectation value
of the exponential (or vertex) operator
\begin{equation}
\label{vopdiag}
{\langle\rme^{\rmi\kk\rr(0)}\rangle}_V=\rme^{-{\kk^2\over 2}c_0}\ \left[1\, -{\kk^2}\left(
\parbox{1cm}{\includegraphics[width=1.cm]{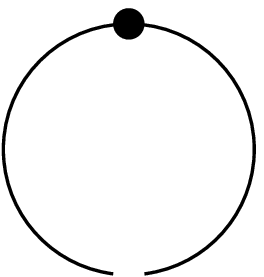}}+
\parbox{1cm}{\includegraphics[width=1.cm]{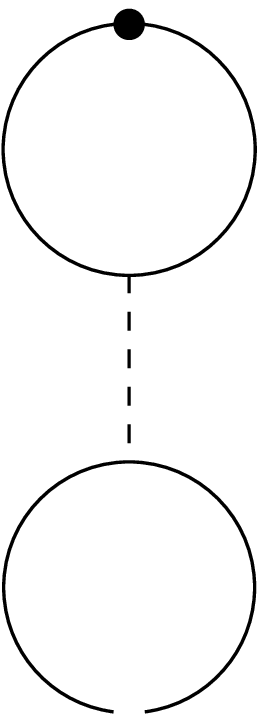}}+
\parbox{1cm}{\includegraphics[width=1.cm]{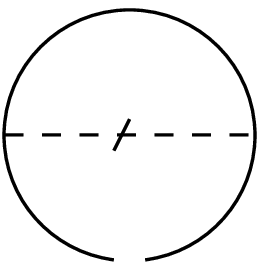}}+
\parbox{1cm}{\includegraphics[width=1.cm]{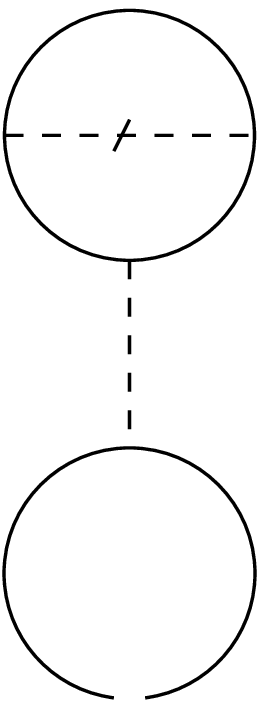}}+
\parbox{1cm}{\includegraphics[width=1.cm]{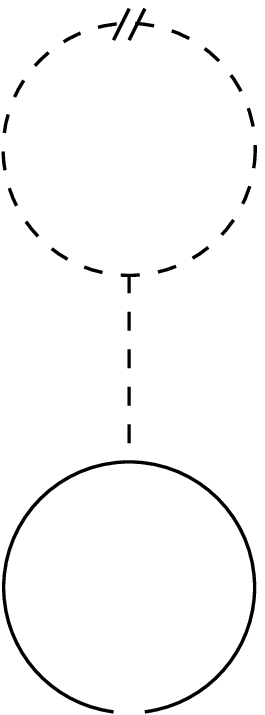}}
\right)+{(\kk^2)}^2\ 
\parbox{1cm}{\includegraphics[width=1.cm]{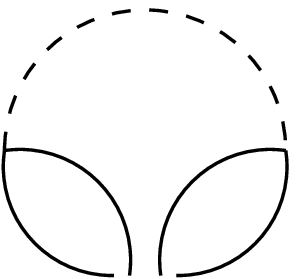}}
\right]
\end{equation}
The symmetry factors of the diagrams are not written, they are
respectively $1/2$, $1/4$, $1/2$, $1/4$, $1/4$ and $1/8$ for the
diagrams in (\ref{vopdiag}).  No $\rr$-space indices flow through the
unclosed line% 
\footnote{This is different from the following graph
considered in \cite{DavidWiese1998}, whose amplitude differs by a
factor of d $d$.
$$\parbox{1cm}{\includegraphics[width=1.cm]{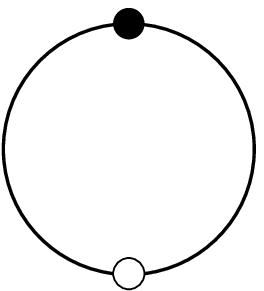}}\ =\ {d\over
2}\,({1+\mu_1})\,\int_\xx G(\xx)^2$$
}
diagram is (taking into account the couplings and the symmetry
factors)
\begin{equation*}
\label{G1ampl} \parbox{1cm}{\includegraphics[width=1.cm]{graph-1}}\ =\
{1\over 2}\,({1+\mu_1})\,\int_\xx G(\xx)^2\, =\ {2-D\over 4} (1+\mu_1)
c_0
\end{equation*}
Similarly for the last diagram
\begin{equation*}
\label{G6ampl} \parbox{1cm}{\includegraphics[width=1.cm]{graph-6}}\ =\
{1\over 8}{-\mu_{2}\over 2 c_0}\,\,\int_\xx\int_\yy G(\xx)^2 G(\yy)^2
H(\xx-\yy)\,
\end{equation*}
The exact instanton saddle-point equation, which is (once again)
\begin{equation}
\label{vhatinsta}
\widehat{V}(\kk)\,+\,{\langle\rme^{\rmi\kk\rr(0)}\rangle}_V\  =\ 0
\end{equation}
fixes the $\mu_n$'s.
In particular $\mu_1$ is given by
\begin{equation}
\label{mu1equ}
\mu_1=-{1\over d}\int_\kk \kk^2\, \widehat{V}(\kk)\, \rme^{-{\kk^2\over 2}c_0}
\end{equation}
and using (\ref{vhatinsta}) and (\ref{vopdiag}) at order $1/d$ we get
the equation for $\mu_1$ (i.e. $\delta_1$) which reads
diagrammatically 
\begin{equation}
\label{d1equdiag}
\includegraphics[scale=.25]{2L-vertex}+{2c_0\over d}\left[
-\overline{\kk^4}\left(
\parbox{1cm}{\includegraphics[width=1.cm]{graph-1}}+
\parbox{1cm}{\includegraphics[width=1.cm]{graph-4}}+
\parbox{1cm}{\includegraphics[width=1.cm]{graph-2}}+
\parbox{1cm}{\includegraphics[width=1.cm]{graph-3}}+
\parbox{1cm}{\includegraphics[width=1.cm]{graph-5}}
\right)+\overline{{\kk^6}}\ 
\parbox{1cm}{\includegraphics[width=1.cm]{graph-6}}
\right]=0\ ,
\end{equation}
where $\overline{\kk^4}$ and $\overline{\kk^6}$ mean the average value
of $\kk^4$ and $\kk^6$ respectively with the Gaussian weight
$\rme^{-\kk^2 c_0}$.  Since
\begin{equation}
\label{k4k6aver}
\overline{\kk^4}={d(d+2)\over (2c_0)^2}\simeq\left({d\over 2c_0}\right)^2
\quad,\quad
\overline{\kk^6}={d(d+2)(d+4)\over (2c_0)^3}\simeq\left({d\over 2c_0}\right)^3
\end{equation}
are of order $\CO(1)$ we recover that $\mu_1=-1+\CO(1/d)$.

\subsection{The Hessian $\mathbb{O}$} \label{ss:1:dO} We now show how
this method to construct a $1/d$ expansion can be applied to compute
the matrix elements 
of the Hessian $\CS''$ and of the associated operator $\JO$.  We start
from the expression for $\JO$ in momentum space
\begin{equation}
\label{Ohat2}
\widehat{\JO}_{\kk_1 \kk_2}=\int_\xx\left<\rme^{\rmi\kk_1\rr(\oo)}
\rme^{\rmi\kk_2\rr(\xx)}\right>_V^{\mathrm{conn}}
=\int_\xx\left<\rme^{\rmi\kk_1\rr(\oo)}
\rme^{\rmi\kk_2\rr(\xx)}\right>_V
-\left<\rme^{\rmi\kk_1\rr(\oo)}\right>_V
\left<\rme^{\rmi\kk_2\rr(\xx)}\right>_V
\end{equation}
and we use our perturbative rules to expand the
e.v. $\left<\cdots\right>_V$ in $1/d$.

\subsubsection{$\JO$ at order $1$:}\label{Ohlead}
At leading order $\CO(1)$, we get (\ref{hatJOm}) that we can represent
as a sum over diagrams with $n\ge 1$ propagators between $\oo$ and
$\xx$, integrated over $\xx$
\begin{equation}
\label{Ohat0}
\begin{split}
\widehat{\JO}_{\kk_1\kk_2}^{(0)}&=\int_\xx\rme^{-(\kk_1^2+\kk_2^2)c_0/2}
\left[\rme^{-\kk_1\kk_2 G(\xx)}-1\right]
\\
&=\sum_{n=1}^\infty\rme^{-\kk_1^2c_0/2}\rme^{-\kk_2^2c_0/2}(\rmi\kk_1\cdot
\rmi\kk_2)^n \ \raisebox{-2.5 ex}{\includegraphics[width=2.cm]{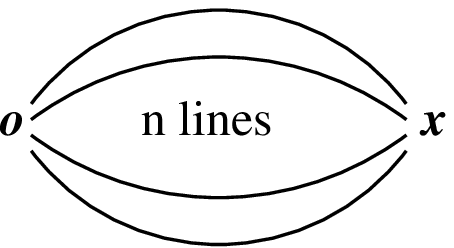}}
= \sum_{n=1}^\infty \ \raisebox{-2.5
ex}{\includegraphics[width=2.cm]{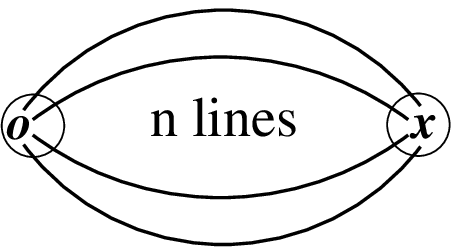}}
\end{split}
\end{equation}
where the integration over $\xx$ and the symmetry factor $1/n!$ of the
graphs are implicit.  We have introduced here an additional
diagrammatic notation, which will be very convenient in the following
discussion.

\subsubsection{A diagrammatic representation for the vertex
$\widehat{V}(\kk)$:}\label{cir4Vk} The circles in the last graph are a
symbol for the factors which depend respectively on $\kk_1$ and
$\kk_2$ and are attached to the vertices $\oo$ and $\xx$. More
precisely, the circle represents the exponential $\rme^{-\kk^2 c_0/2}$
and each line entering into the circle represents an additional
(multiplicative) factor $\rmi\kk$, with an external space index $a$
carried by the line. Thus the following picture, a circle with $n$
external lines, represents the factor
\begin{equation}
\label{circlestar} \raisebox{-2.5
ex}{\includegraphics[scale=.5]{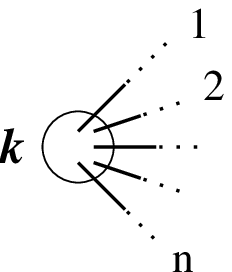}}\ =\ \rme^{-\kk^2
c_0/2}\,(\rmi\kk^{a_1})\cdots(\rmi\kk^{a_n}) \ .
\end{equation}

\subsubsection{$\JO$ at order $1/d$:} \label{Ohsubl} Now
we make the perturbative expansion and keep the diagrams which
contribute to $\JO$ at order $\CO(1/d)$ only. We find that only (!) 21
different (classes of) diagrams contribute
\begin{equation}
\label{Ohat1}
\begin{split}
\widehat{\JO}_{\kk_1,\kk_2}^{(1)}&=
\raisebox{-2.5 ex}{\includegraphics[scale=.33]{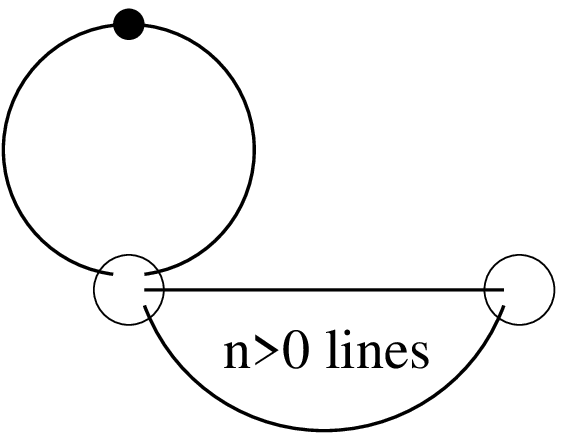}}
+
\raisebox{-2.5 ex}{\includegraphics[scale=.33]{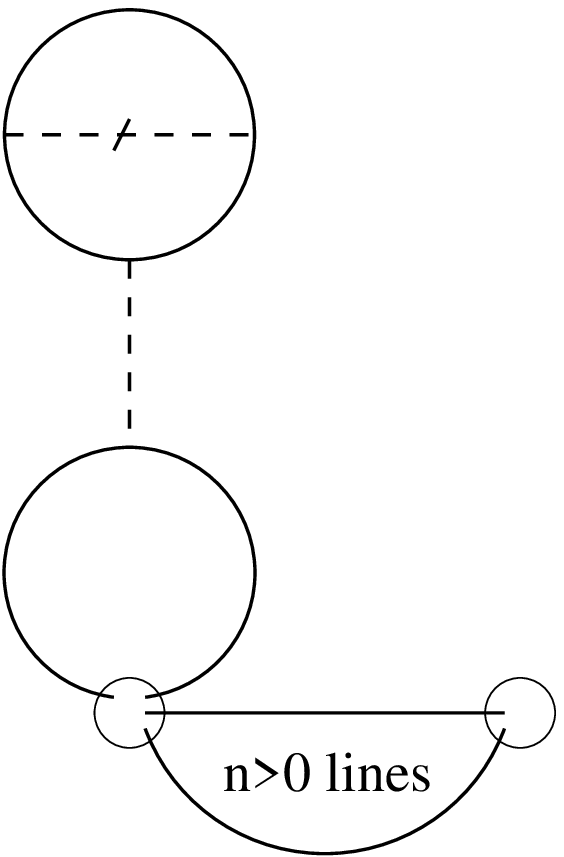}}
+
\raisebox{-2.5 ex}{\includegraphics[scale=.33]{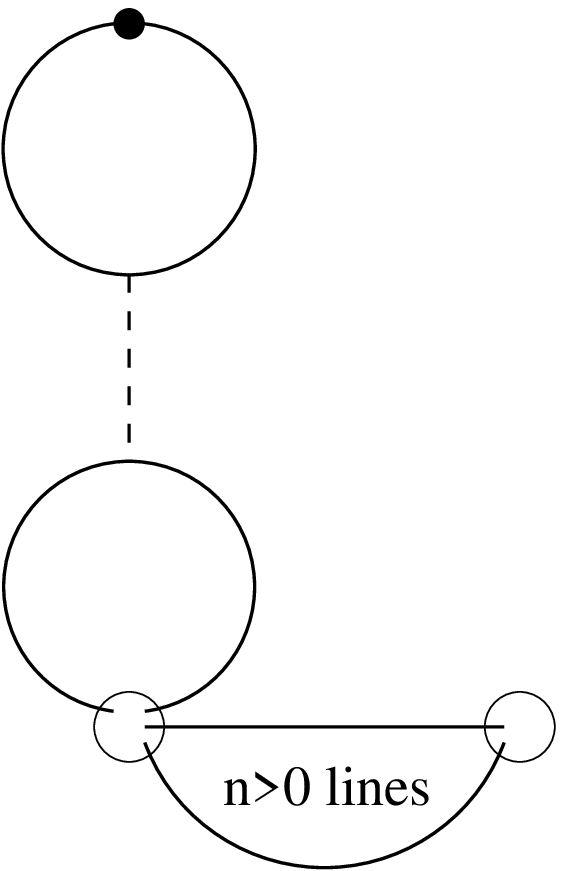}}
+
\raisebox{-2.5 ex}{\includegraphics[scale=.33]{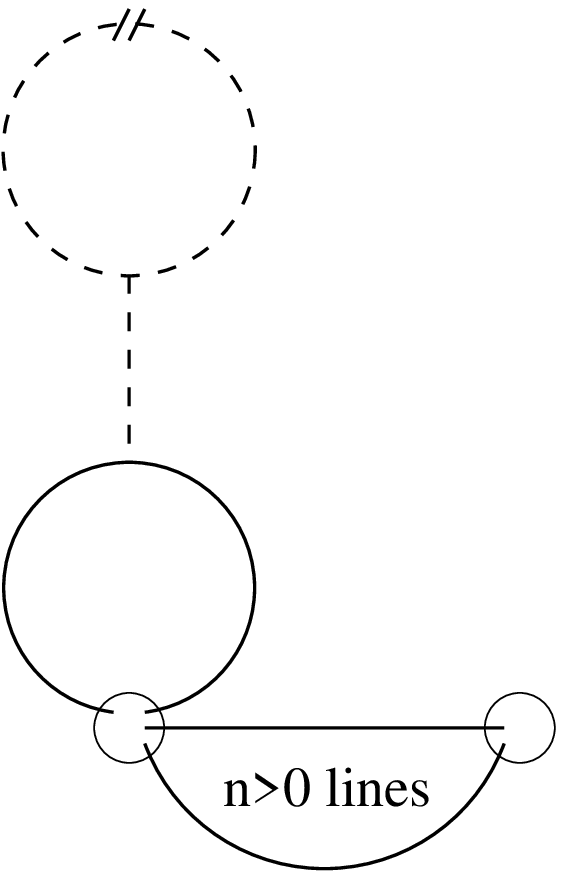}}
+
\raisebox{-2.5 ex}{\includegraphics[scale=.33]{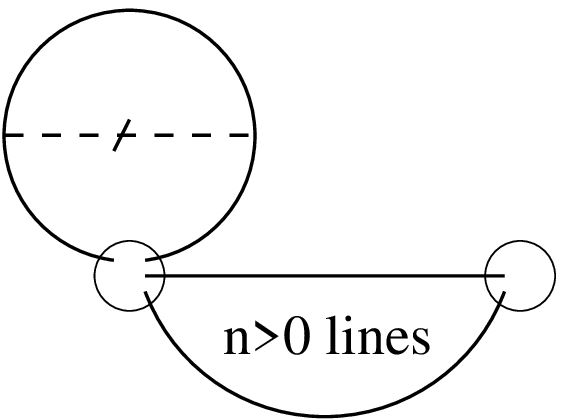}}
+
\raisebox{-2.5 ex}{\includegraphics[scale=.33]{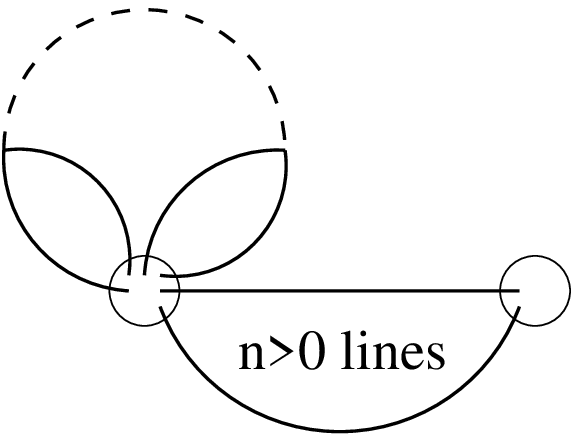}}
\\
&+
\raisebox{-2.5 ex}{\includegraphics[scale=.33]{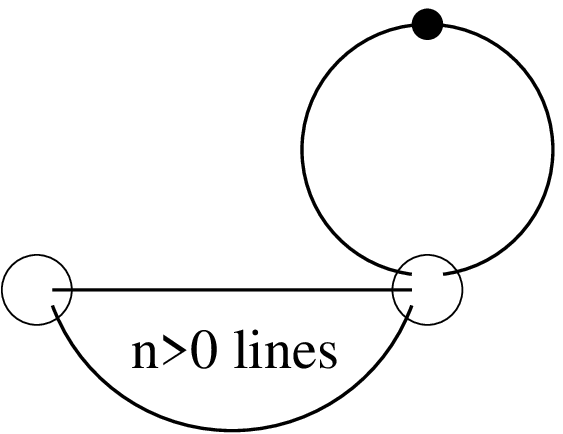}}
+
\raisebox{-2.5 ex}{\includegraphics[scale=.33]{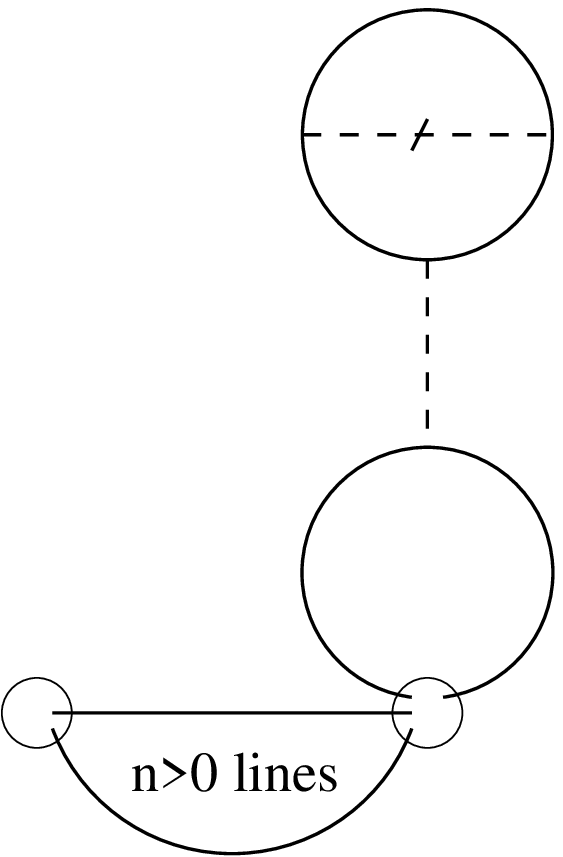}}
+
\raisebox{-2.5 ex}{\includegraphics[scale=.33]{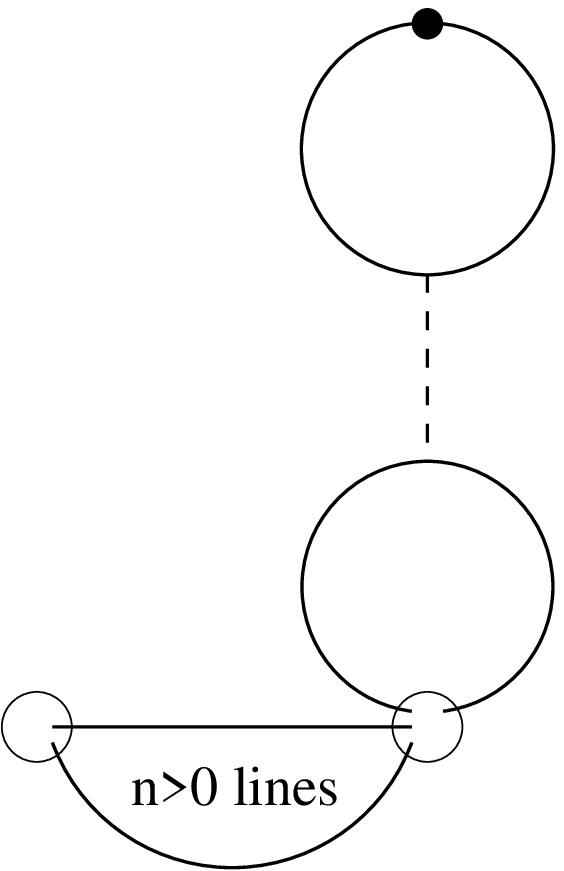}}
+
\raisebox{-2.5 ex}{\includegraphics[scale=.33]{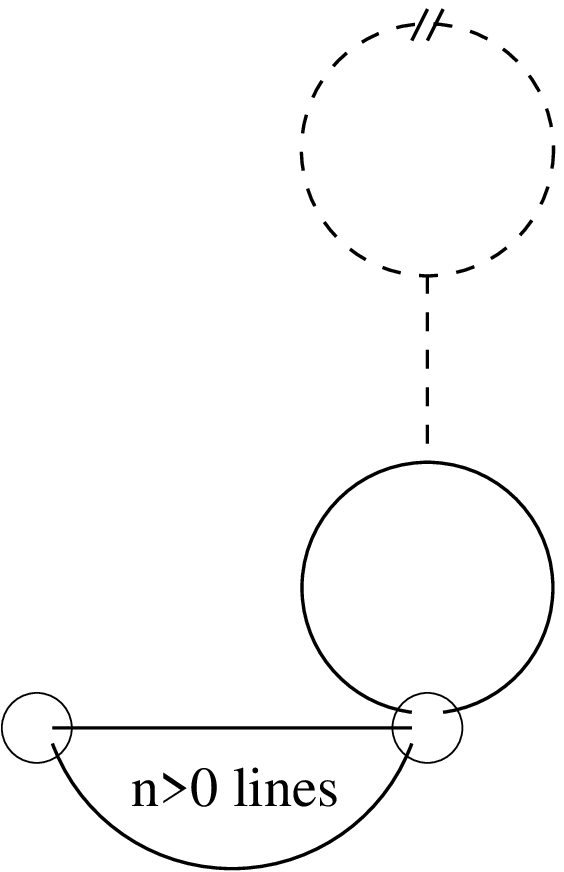}}
+
\raisebox{-2.5 ex}{\includegraphics[scale=.33]{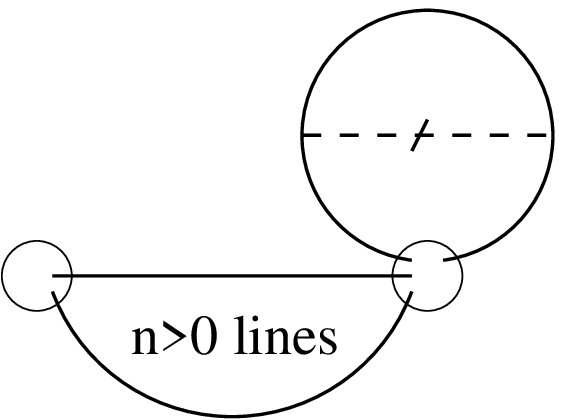}}
+
\raisebox{-2.5 ex}{\includegraphics[scale=.33]{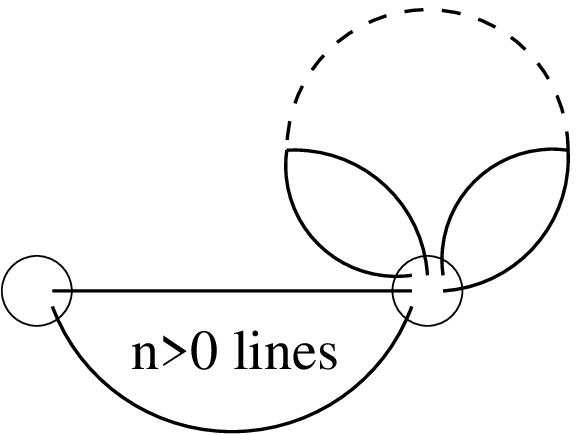}}
\\
&+
\raisebox{-2.5 ex}{\includegraphics[scale=.33]{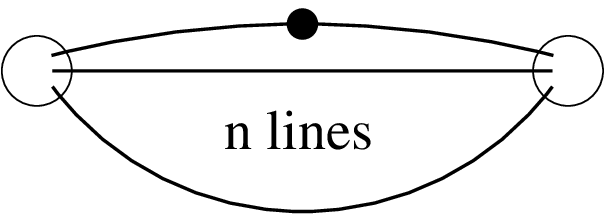}}
+
\raisebox{-2.5 ex}{\includegraphics[scale=.33]{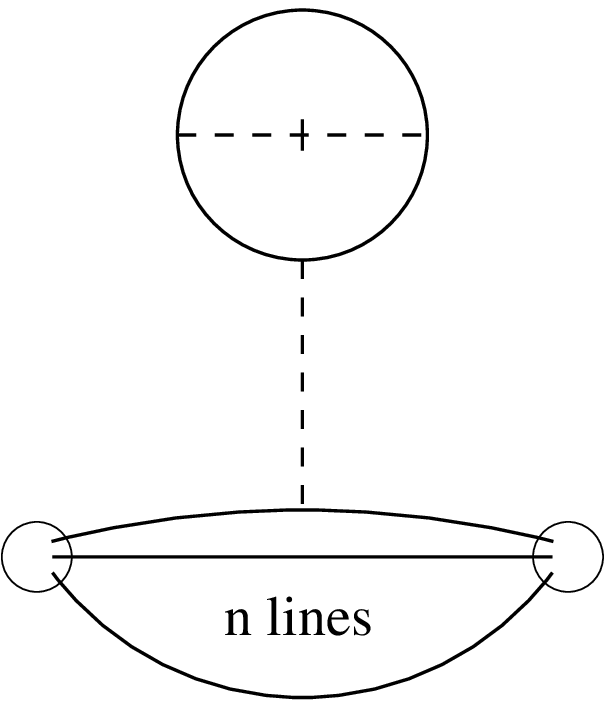}}
+
\raisebox{-2.5 ex}{\includegraphics[scale=.33]{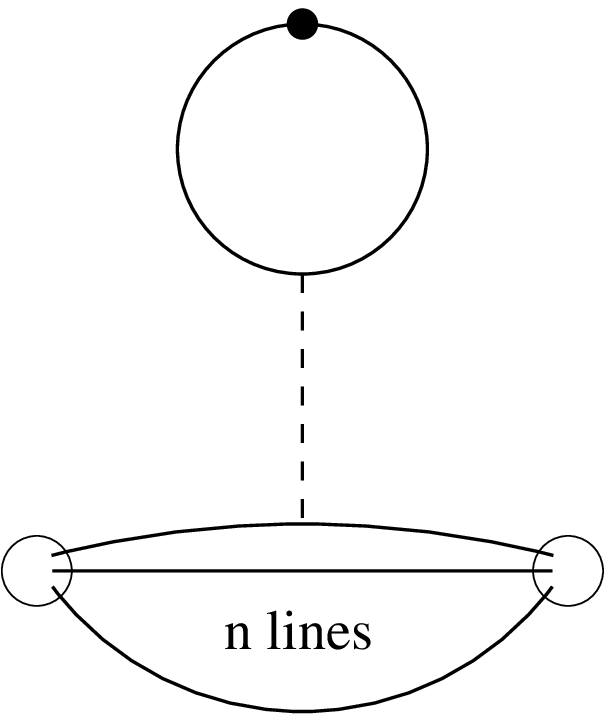}}
+
\raisebox{-2.5 ex}{\includegraphics[scale=.33]{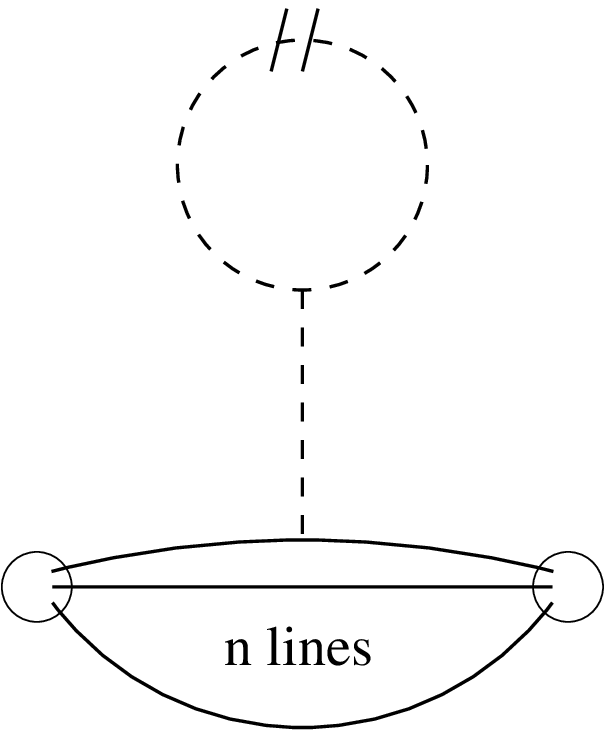}}
\\
&+
\raisebox{-2.5 ex}{\includegraphics[scale=.33]{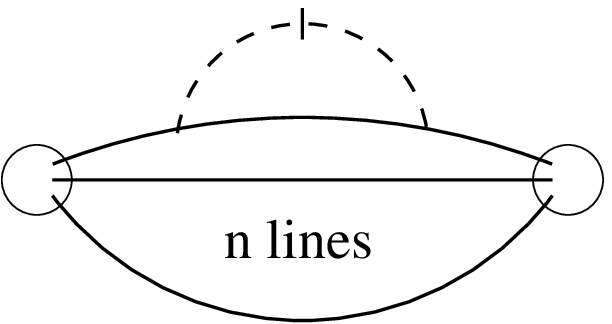}}
+
\raisebox{-2.5 ex}{\includegraphics[scale=.33]{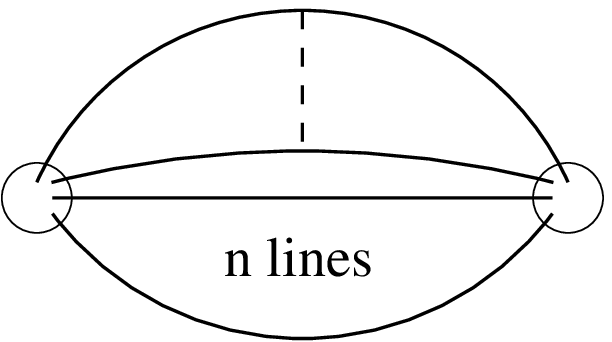}}
+
\raisebox{-2.5 ex}{\includegraphics[scale=.33]{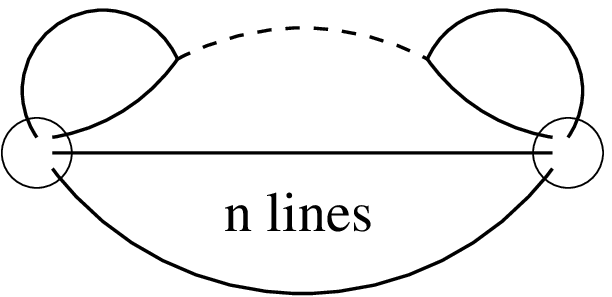}}
+
\raisebox{-2.5 ex}{\includegraphics[scale=.33]{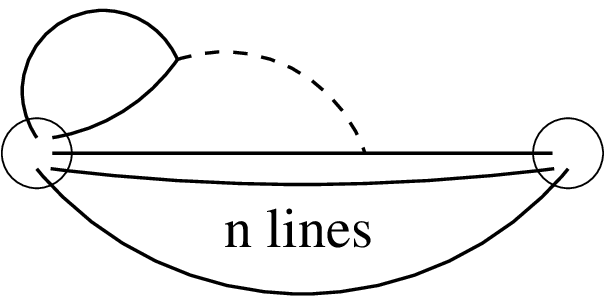}}
+
\raisebox{-2.5 ex}{\includegraphics[scale=.33]{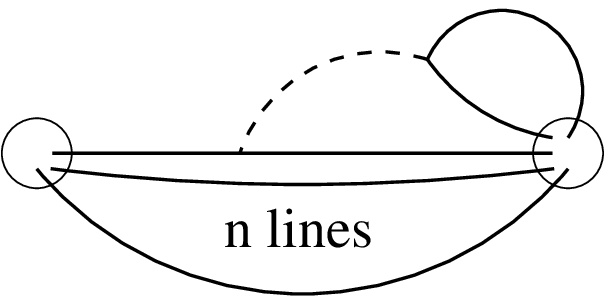}} \ .
\end{split}
\end{equation}

\subsubsection{Diagrammatic for the mass renormalization and $V$:}
\label{mu1Vdi}
Moreover, Eq.~(\ref{d1equdiag}) for the mass renormalization $\mu_1$
may be rewritten at order $\CO(1/d)$ with our notations as
\begin{equation}
\label{mu1eqdiag}
\raisebox{-.0 ex}{\includegraphics[scale=.33]{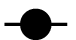}}\ =
\raisebox{-2.5 ex}{\includegraphics[scale=.33]{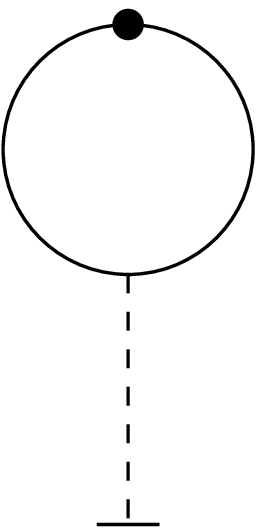}}+
\raisebox{-2.5 ex}{\includegraphics[scale=.33]{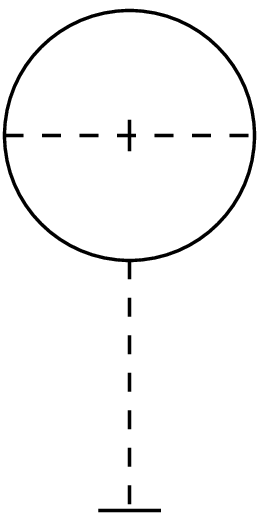}}+
\raisebox{-2.5 ex}{\includegraphics[scale=.33]{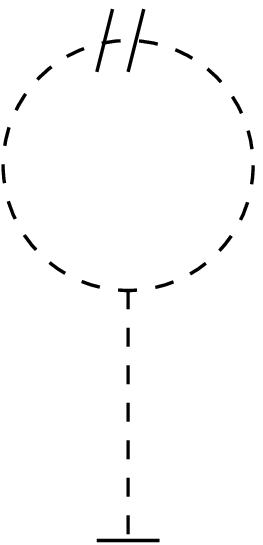}}
+\CO(1/d^2)\ .
\end{equation}
While Eq.(\ref{vopdiag}) for $V$ reads
\begin{equation}
\label{Vhatdiag}
-\,\widehat{V}(\kk)=
\raisebox{-.0 ex}{\includegraphics[scale=.33]{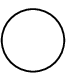}}+
\raisebox{-.0 ex}{\includegraphics[scale=.33]{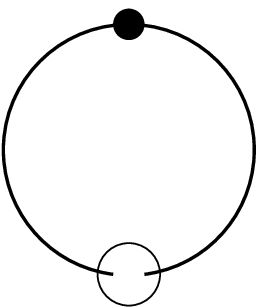}}+
\raisebox{-.0 ex}{\includegraphics[scale=.33]{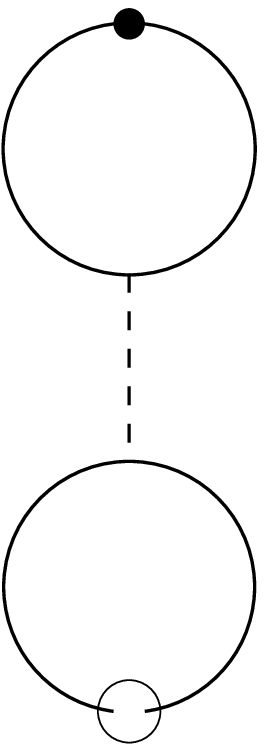}}+
\raisebox{-.0 ex}{\includegraphics[scale=.33]{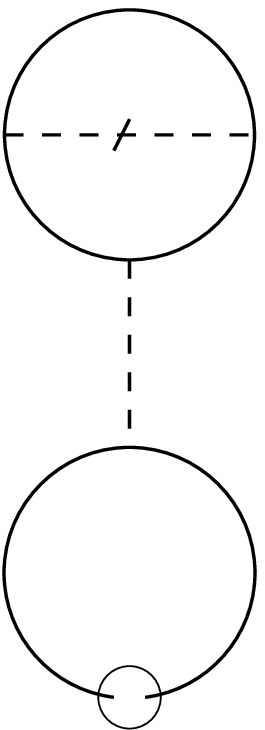}}+
\raisebox{-.0 ex}{\includegraphics[scale=.33]{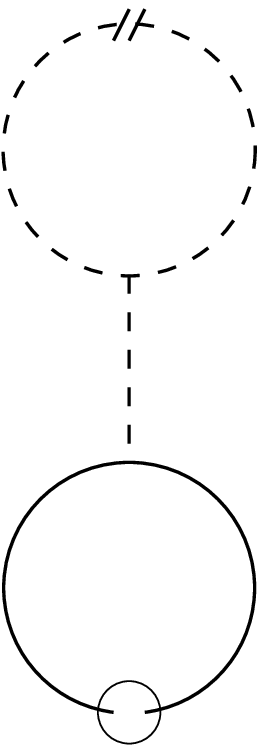}}+
\raisebox{-.0 ex}{\includegraphics[scale=.33]{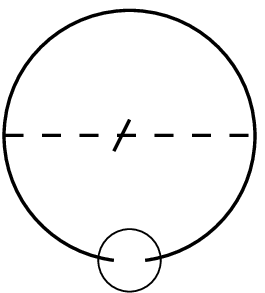}}+
\raisebox{-.0 ex}{\includegraphics[scale=.33]{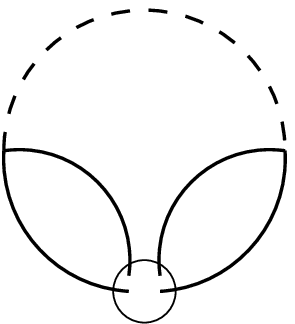}}
+\CO(1/d^2)
\end{equation}
Note that in Eq.~(\ref{Ohat1}) the first 4 diagrams of lines 1 and 2,
and the 4 diagrams of line 3 can be absorbed into a mass shift $m=1\
\to\ m=1-\ \raisebox{.2 ex}{\includegraphics[scale=.33]{graph-13a}}$
in the leading contribution represented in (\ref{Ohat0}).  The same mass
shift absorbs the diagrams 2--5 in Eq.(\ref{Vhatdiag}) for $V$.

\subsection{The zero-mode projector $\JPo$}
\label{ss:Po1/d}
We now compute the projector onto the zero-modes
\begin{equation}
\label{P0hatII} {\widehat{\JPo}}_{\kk_1\kk_2}
={\rmi\kk_1\widehat{V}(\kk_1)\cdot\rmi\kk_2\widehat{V}(\kk_2)\over{1\over
d}\int_\kk \kk^2 \widehat{V}(\kk)^2}
\end{equation}
\subsubsection{$\JPo$ at order 1}\label{P0lead} We have already seen
that at leading order in the $1/d$ expansion $\int_\kk \kk^2
\widehat{V}(\kk)^2=d$ and since $\int_\xx
G(\xx)=\widehat{G}(0)=\frac{1}{m^{2}}= 1$ so that with our
diagrammatic notations
\begin{equation}
\label{P0hat0}
\widehat{\JPo}_{\kk_1\kk_2}^{(0)}=\int_\xx(-\kk_1\kk_2)\rme^{-{\kk_1^2}c_0/2}
\rme^{-{\kk_2^2 }c_0/2} G(\xx)=\ \raisebox{-.0
ex}{\includegraphics[scale=.33]{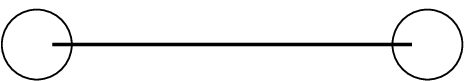}} \ .
\end{equation}
Thus the projector $\JPo$ subtracts the one-line diagram in $\JO$ (see
Eqs.(\ref{hatJQx})-(\ref{hatObeads})).  \subsubsection{$\JPo$ at order
$1/d$}\label{P0subl} We can now compute explicitely the first
correction in $1/d$ to $\JPo$, using (\ref{Vhatdiag}) for $V$.  It is
easy to see that the numerator in (\ref{P0hatII}) gives all the
diagrams in the lines (rows) 1 and 2 of (\ref{Ohat1}) with $n=1$ line
between the two points $\oo$ and $\xx$.
\begin{equation}
\label{numgraph}
\begin{split}
(-\kk_1&\cdot\kk_2)\widehat{V}(\kk_1)\widehat{V}(\kk_2)=\
\raisebox{-.0 ex}{\includegraphics[scale=.33]{1-La}} \ +
\\
&
\raisebox{-.0 ex}{\includegraphics[scale=.33]{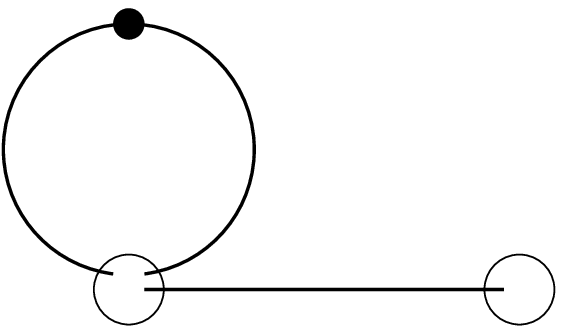}}
+
\raisebox{-.0 ex}{\includegraphics[scale=.33]{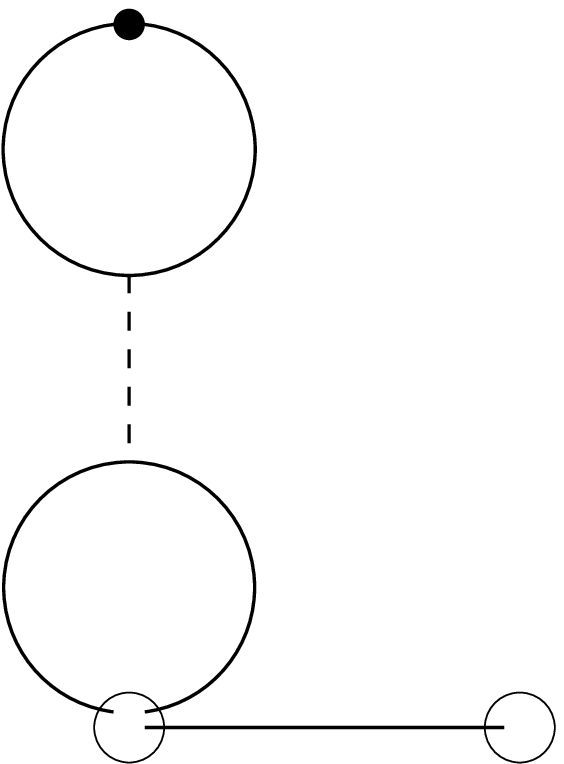}}
+
\raisebox{-.0 ex}{\includegraphics[scale=.33]{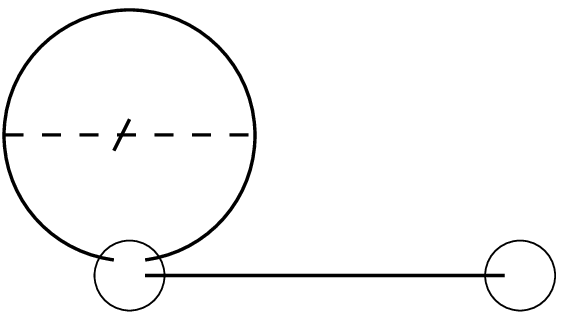}}
+
\raisebox{-.0 ex}{\includegraphics[scale=.33]{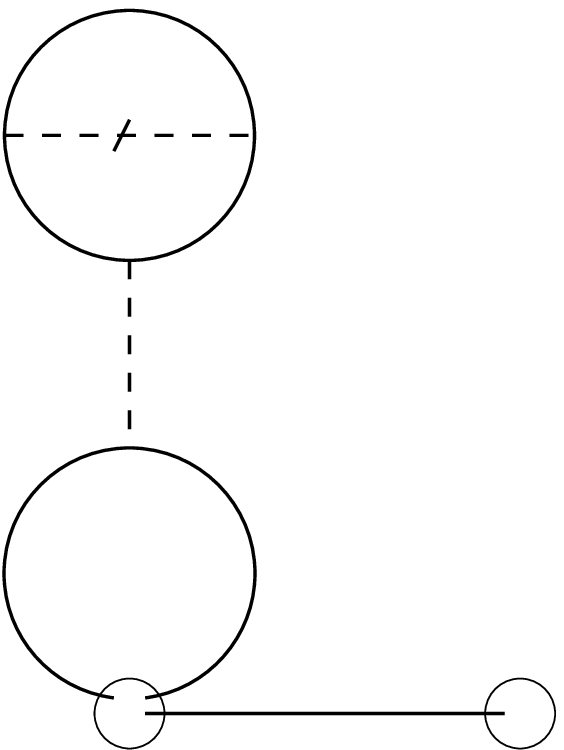}}
+
\raisebox{-.0 ex}{\includegraphics[scale=.33]{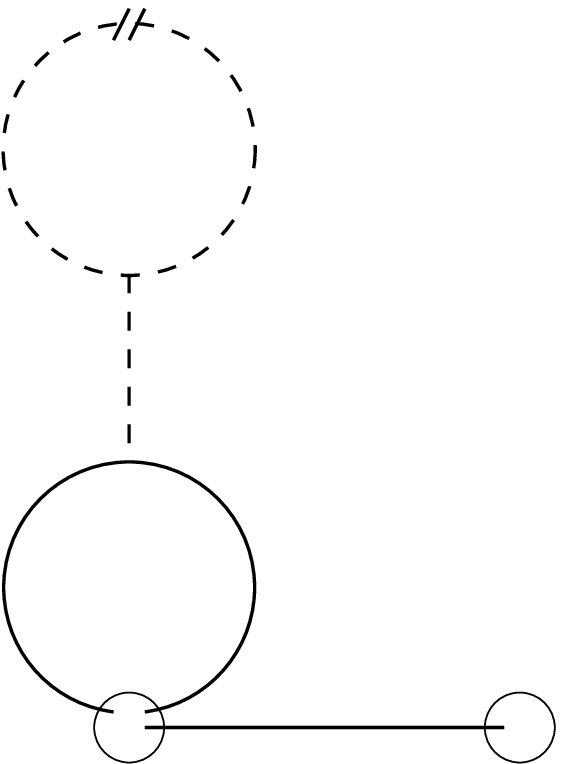}}
+
\raisebox{-.0 ex}{\includegraphics[scale=.33]{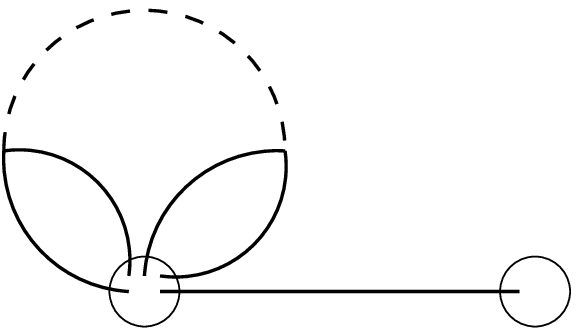}}
\\
+&
\raisebox{-.0 ex}{\includegraphics[scale=.33]{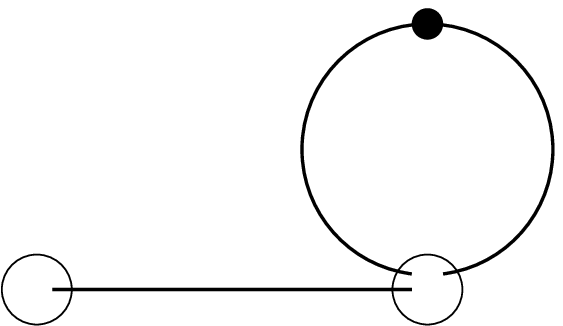}}
+
\raisebox{-.0 ex}{\includegraphics[scale=.33]{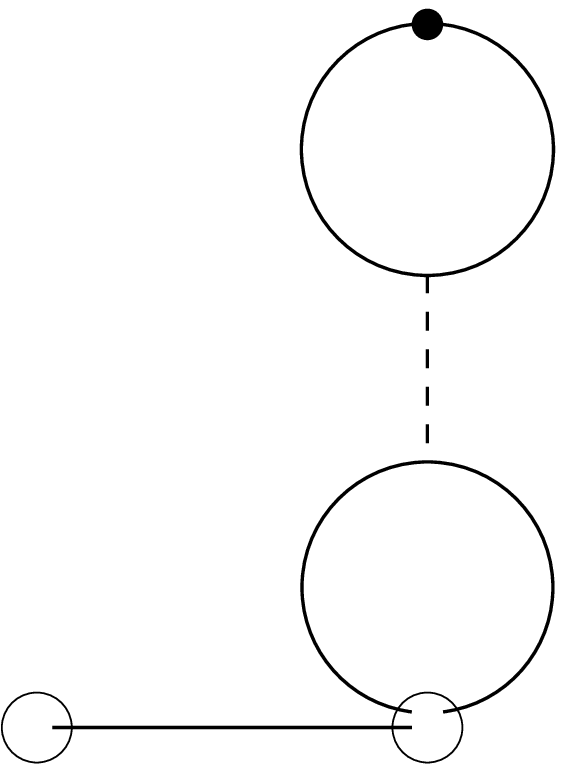}}
+
\raisebox{-.0 ex}{\includegraphics[scale=.33]{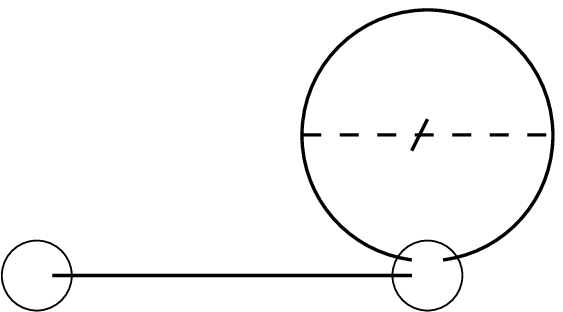}}
+
\raisebox{-.0 ex}{\includegraphics[scale=.33]{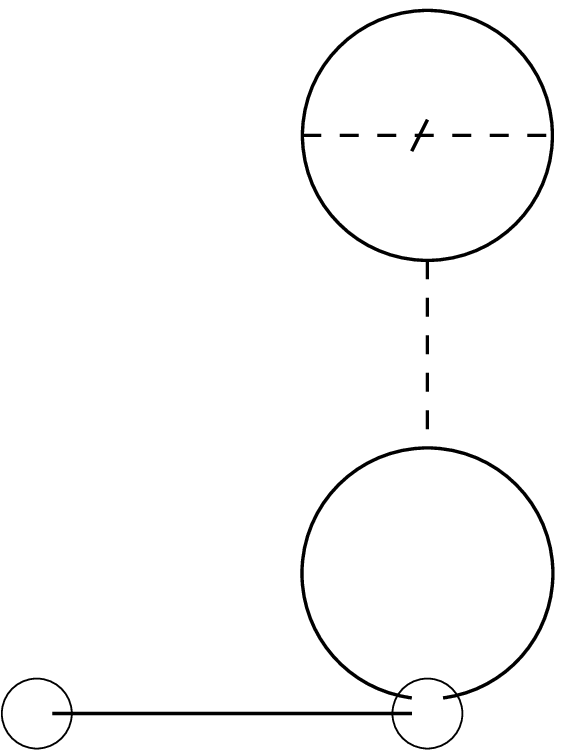}}
+
\raisebox{-.0 ex}{\includegraphics[scale=.33]{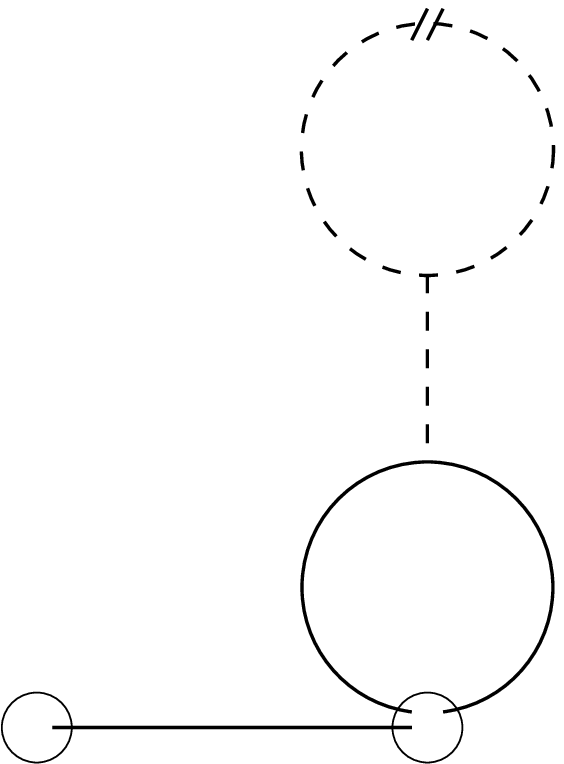}}
+
\raisebox{-.0 ex}{\includegraphics[scale=.33]{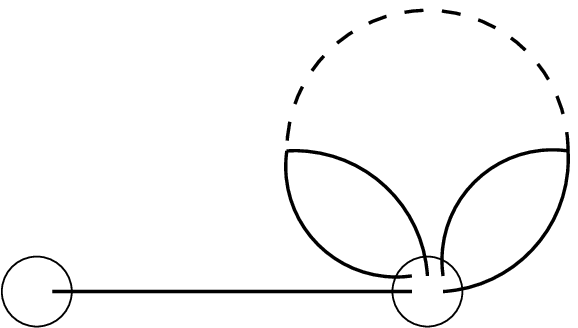}}
\end{split}
\end{equation}
The denominator is computed from the explicit form of $V$ given by
(\ref{mu1eqdiag})-(\ref{Vhatdiag}).  In fact it is easy to see, using
(\ref{vopdiag}), that
\begin{equation}
\label{denomgraph}
\begin{split}
{1\over d}\int_\kk \kk^2{\widehat{V}(\kk)}^2 &= 1-2\,{\delta_1\over
d}+\CO(1/d^2)= 1- \raisebox{-.0
ex}{\includegraphics[scale=.33]{graph-13a}}- \raisebox{-2.5
ex}{\includegraphics[scale=.33]{graph-12a}}- \raisebox{-2.5
ex}{\includegraphics[scale=.33]{graph-11a}}- \raisebox{-2.5
ex}{\includegraphics[scale=.33]{graph-10a}} +\CO(1/d^2)
\end{split}
\end{equation}
and since $\widehat{G}(0)=1$ we can write $\raisebox{-.2
ex}{\includegraphics[scale=.33]{1-La}} \times \raisebox{.2
ex}{\includegraphics[scale=.33]{graph-13a}} = \raisebox{-.0
ex}{\includegraphics[scale=.33]{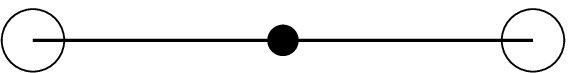}}$ , etc. 
We obtain that only (!) 16 diagrams contribute 
to $\JPo$; the final result is 
\begin{equation}
\label{Pohatgraph}
\begin{split}
&{\widehat{\JPo}}_{\kk_1\kk_2}\ =\ \raisebox{-.0 ex}{\includegraphics[scale=.33]{1-La}}
+
\raisebox{-.0 ex}{\includegraphics[scale=.33]{graph-13}}
+
\raisebox{-.0 ex}{\includegraphics[scale=.33]{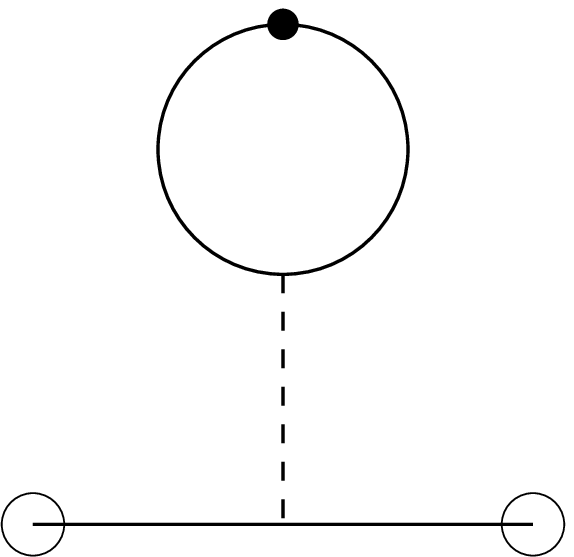}}
+
\raisebox{-.0 ex}{\includegraphics[scale=.33]{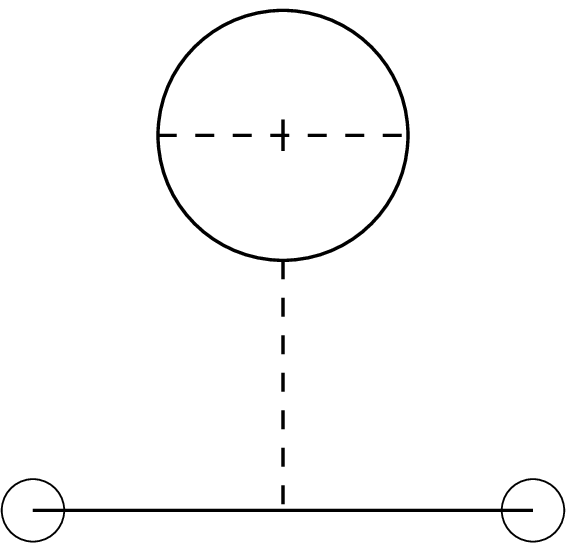}}
+
\raisebox{-.0 ex}{\includegraphics[scale=.33]{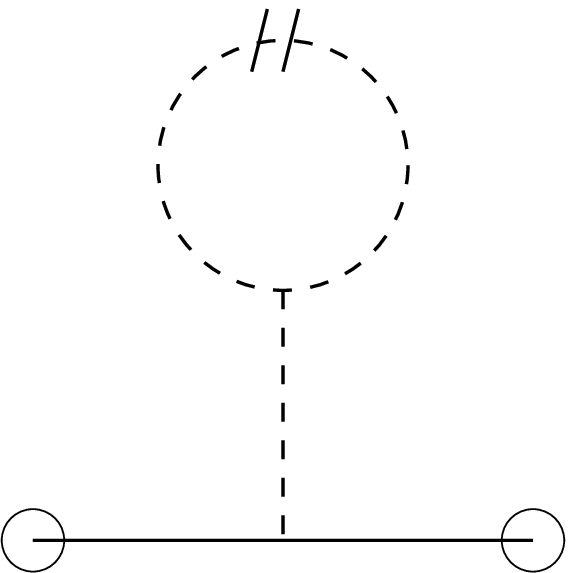}}
\\
&+
\raisebox{-.0 ex}{\includegraphics[scale=.33]{graph-O-1c}}
+
\raisebox{-.0 ex}{\includegraphics[scale=.33]{graph-O-4c}}
+
\raisebox{-.0 ex}{\includegraphics[scale=.33]{graph-O-2c}}
+
\raisebox{-.0 ex}{\includegraphics[scale=.33]{graph-O-3c}}
+
\raisebox{-.0 ex}{\includegraphics[scale=.33]{graph-O-5c}}
+
\raisebox{-.0 ex}{\includegraphics[scale=.33]{graph-O-6c}}
\\
+&
\raisebox{-.0 ex}{\includegraphics[scale=.33]{graph-O-1a}}
+
\raisebox{-.0 ex}{\includegraphics[scale=.33]{graph-O-4a}}
+
\raisebox{-.0 ex}{\includegraphics[scale=.33]{graph-O-2a}}
+
\raisebox{-.0 ex}{\includegraphics[scale=.33]{graph-O-3a}}
+
\raisebox{-.0 ex}{\includegraphics[scale=.33]{graph-O-5a}}
+
\raisebox{-.0 ex}{\includegraphics[scale=.33]{graph-O-6a}}
+\ \CO(1/d^2)
\end{split}
\end{equation}
We note that the denominator gives all the one-line reducible
diagrams (with only a single line joining $\oo$ and $\xx$) with a
tadpole-like graph attached to the line.  The diagrams $\raisebox{-.0
ex}{\includegraphics[scale=.33]{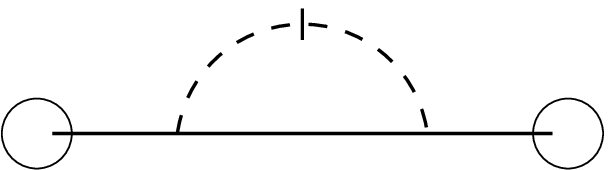}}$, $\raisebox{-.0
ex}{\includegraphics[scale=.33]{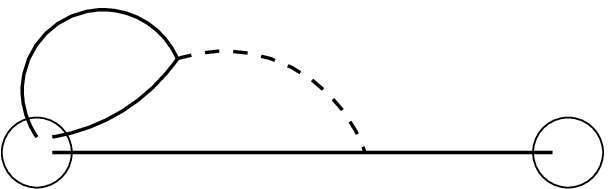}}$ and $\raisebox{-.0
ex}{\includegraphics[scale=.33]{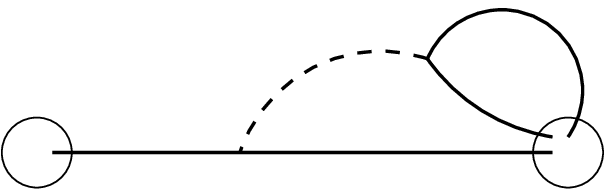}}$ although
one-line-reducible, are not contained in (\ref{Pohatgraph}).

\subsubsection{An all order argument relating $\JPo$ with tadpole graphs}\label{P0tadp}
A simple general argument shows that the denominator in
(\ref{P0hatII}), given at first order by the tadpole diagrams depicted
in (\ref{denomgraph}), is given at all orders by the tadpole diagrams
with two truncated external legs attached to the same vertex. In
diagrammatic language we shall show that
\begin{equation}
\label{tad2pattes}
1-\raisebox{-.0 ex}{\includegraphics[scale=.5]{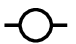}}
-\raisebox{-.0 ex}{\includegraphics[scale=.5]{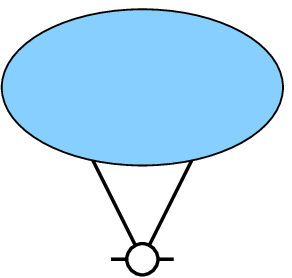}}
-\raisebox{-.0 ex}{\includegraphics[scale=.5]{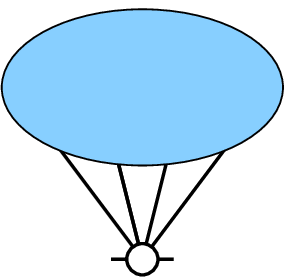}}
-\ \cdots
\ =\ {1\over d}\int_\kk\,\kk^2\,\widehat{V}(\kk)^2
\end{equation}
The l.h.s.\ of (\ref{tad2pattes}) is nothing but
\begin{equation}
\label{lhsoftad2pattes}
1-\raisebox{-.0 ex}{\includegraphics[scale=.33]{tadpole-0}}
-\raisebox{-.0 ex}{\includegraphics[scale=.33]{tadpole-2}}
-\raisebox{-.0 ex}{\includegraphics[scale=.33]{tadpole-4}}
-\ \cdots
\ =\ 
{1\over d}{\left<V''\bigl(\rr(\oo)\bigr)\right>}_V
\qquad\text{with}\qquad
V''(\rr)=\sum_a{\partial^2 V(\rr)\over\partial\rr^a\partial\rr^a}
\end{equation}
(the two $\rr$ derivatives pick two legs out of the vertex $V(\rr)$).
We can rewrite it as
\begin{equation}
\label{V''2}
V''\bigl(\rr(\oo)\bigr)=\int_\rr V''(\rr)\,\delta\bigl((\rr-\rr(\oo)\bigr)=\ \int_\kk(-\kk^2)\,\widehat{V}(\kk)\,\rme^{\rmi\kk\rr(\oo)}
\end{equation}
and using the exact equation (\ref{vhatinsta}) for the instanton
potential, we get
\begin{equation}
\label{V''3}
{\left<V''\bigl(\rr(\oo)\bigr)\right>}_V\ =\ \int_\kk (-\kk^2) \,\widehat{V}(\kk)\, \left<\rme^{\rmi\kk\rr(\oo)}\right>_V\ =\ \int_\kk \,\kk^2 \,\widehat{V}(\kk)^2
\end{equation}
Q.E.D.

(\ref{tad2pattes}) implies that $\JPo$ will contain all the tadpole
chains with tadpole graphs attached at the $\oo$ and $\xx$ end-points,
of the form
\begin{equation}
\label{P0all}
\JPo\ =\ \raisebox{-.0 ex}{\includegraphics[scale=.33]{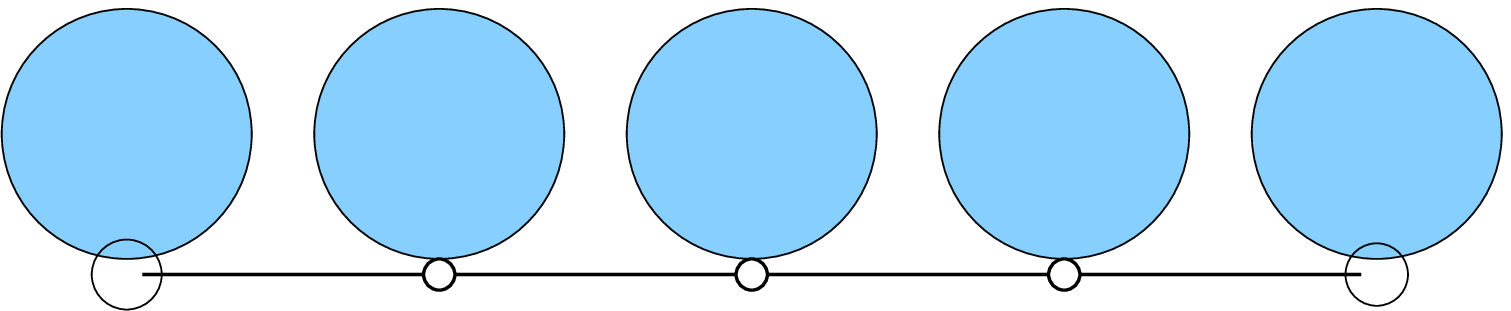}}
\end{equation}

\subsection{Final result for $\JQ=\JO-\JPo$}\label{ss:Q1/d} As a
consequence the subtracted operator $\JQ=\JO-\JPo$ is given at order
$\CO(1/d)$ by the same diagrams as those depicted in Eq.(\ref{Ohat1})
for $\JO$, with the simple restriction that the diagrams in the first
two lines of (\ref{Ohat1}) must have at least 2 lines joining the two
end-points ($n\ge 2$), and that the diagrams in the third line must
have at least one non-dressed line joining the two end-points ($n\ge
1$), while for the diagrams in the fourth line, there is no additional
restriction (no constraints on the number $n$ of simple lines, $n\ge
0$).  Using equation (\ref{mu1eqdiag}) for $\mu_1$ we can rewrite it
as a sum over only 12 graphs (instead of 21!)
\begin{equation}
\label{Qhat1}
\begin{split}
\widehat{\JQ}^{(1)}_{\kk_1,\kk_2}&=
2\raisebox{-2.5 ex}{\includegraphics[scale=.33]{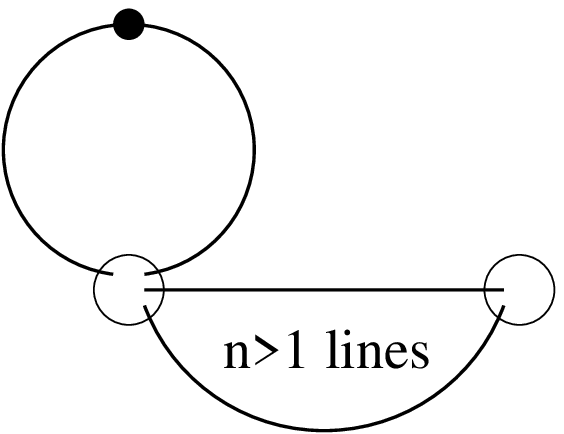}}
+
\raisebox{-2.5 ex}{\includegraphics[scale=.33]{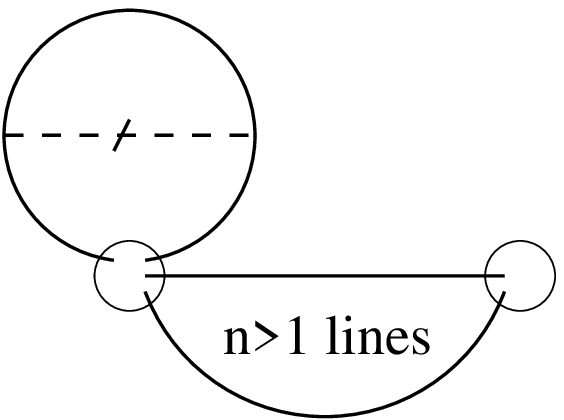}}
+
\raisebox{-2.5 ex}{\includegraphics[scale=.33]{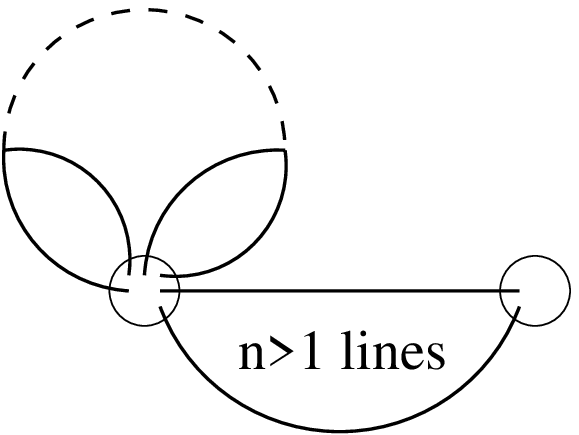}}
+
2\,\raisebox{-2.5 ex}{\includegraphics[scale=.33]{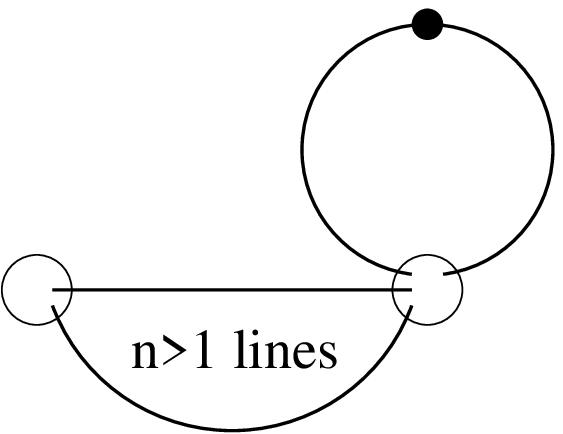}}
+
\raisebox{-2.5 ex}{\includegraphics[scale=.33]{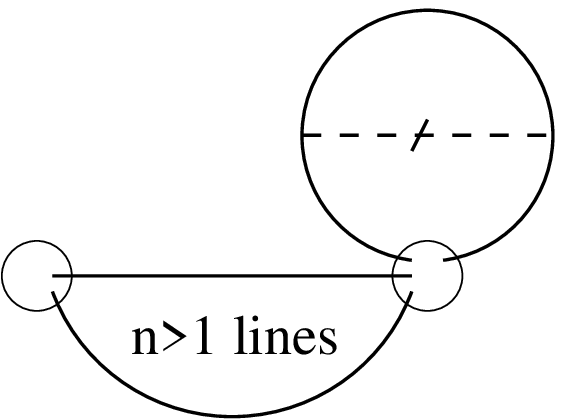}}
+
\raisebox{-2.5 ex}{\includegraphics[scale=.33]{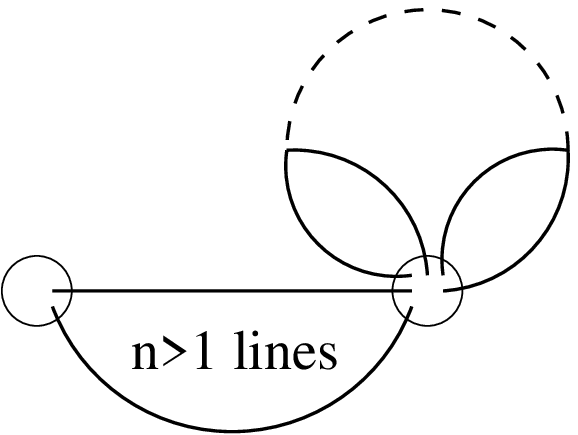}}
\\
&\hskip -2em +
2\,\raisebox{-2.5 ex}{\includegraphics[scale=.33]{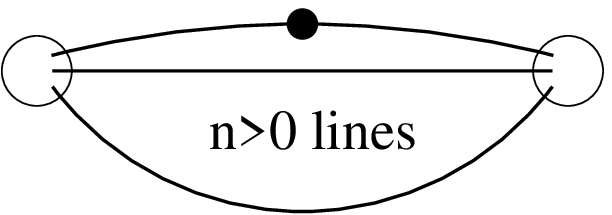}}
+
\raisebox{-2.5 ex}{\includegraphics[scale=.33]{graph-O-14}}
+
\raisebox{-2.5 ex}{\includegraphics[scale=.33]{graph-O-15}}
+
\raisebox{-2.5 ex}{\includegraphics[scale=.33]{graph-O-16}}
+
\raisebox{-2.5 ex}{\includegraphics[scale=.33]{graph-O-17}}
+
\raisebox{-2.5 ex}{\includegraphics[scale=.33]{graph-O-17b}} \ .
\end{split}
\end{equation}

\subsection{The determinant $\mathfrak{D}$}
\label{ss:thedet}

We now compute the $\log$ of the determinant of instanton fluctuations 
\begin{equation}
\label{Lseriesagain}
\mathfrak{L}=\log(\mathfrak{D})=\tr\log(\JU-\JQ)=-\sum_{k=1}^\infty {1\over
k}\,\tr(\JQ^k)
\end{equation}

\subsubsection{Diagrammatic representation of the trace}
\label{sss:trdiag}
With our rescalings (see Appendix \ref{appendixnormvar}) each trace
still reads
\begin{equation}
\label{trQknew} \mathfrak{T}_k=
\tr(\JQ^k)=\int_{\kk_1}\cdots\int_{\kk_k}\widehat{\JQ}_{\kk_1,-\kk_2}
\widehat{\JQ}_{\kk_2,-\kk_3}\cdots\widehat{\JQ}_{\kk_k,-\kk_1}
\end{equation}
and with the representation for the kernel $\JQ$, we have to compute
integrals over $\kk$ of the form, 
\begin{equation}
\label{intkwick} \int_\kk \rme^{-\kk^2
c_0}(-\rmi\kk^{a_1})\cdots(-\rmi\kk^{a_{m}})\,
(\rmi\kk^{b_1})\cdots(\rmi\kk^{b_{n}})\ =\ (-1)^{{m-n\over 2}} (2
c_0)^{1-{m+n\over 2}}\,\sum_{\mathrm{pairing}}
\delta^{..}\cdots\delta^{..}
\ .
\end{equation}
Using Wick's theorem we can represent each term by pairing of lines
between the left $\JQ$ and the right $\JQ$, as already discussed when
we introduced the diagrammatic necklace representation for
$\mathfrak{L}$. This is depicted below
\begin{equation}\label{}
\raisebox{-1.ex}{\includegraphics[scale=.5]{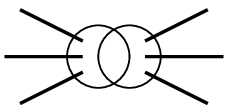}} = 
\mbox{\scriptsize $m$ lines entering}\left\{
\raisebox{-1.ex}{\includegraphics[scale=.5]{circle-star-2}}
\right\}\mbox{\scriptsize $n$ lines exiting}
\end{equation}
\begin{equation}
\label{intkvertex}
\int_\kk 
\raisebox{-1.ex}{\includegraphics[scale=.5]{circle-star-2}}
=(-1)^{{m-n\over 2}}(2c_0)^{1-{m+n\over
2}}\raisebox{-2.ex}{\includegraphics[scale=.5]{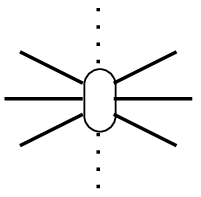}}\ .
\end{equation}
For instance for $m=n=2$
\begin{equation}
\label{vertex-2-2}
\raisebox{-2.ex}{\includegraphics[scale=.5]{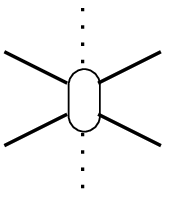}}
=
\raisebox{-2.ex}{\includegraphics[scale=.5]{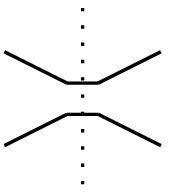}}
+
\raisebox{-2.ex}{\includegraphics[scale=.5]{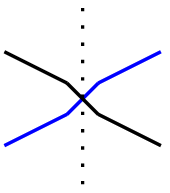}}
+
\raisebox{-2.ex}{\includegraphics[scale=.5]{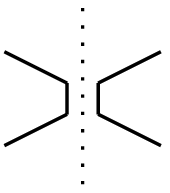}}
\end{equation}
and for $m=3$, $n=1$
\begin{equation}
\label{vertex-3-1}
\raisebox{-2.ex}{\includegraphics[scale=.5]{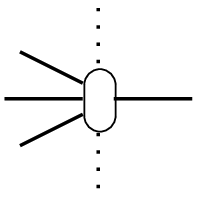}}
=
\raisebox{-2.ex}{\includegraphics[scale=.5]{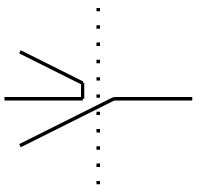}}
+
\raisebox{-2.ex}{\includegraphics[scale=.5]{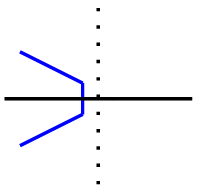}}
+
\raisebox{-2.ex}{\includegraphics[scale=.5]{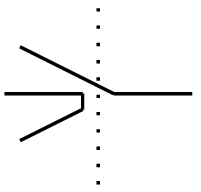}}
\ .
\end{equation}
The vertical dotted line indicates that no $\CM$-momenta $\pp$ flows
through the vertex, since each $\JQ$ is attached to a different
replica of the manifold $\CM$.  With these graphical notations, if we
represent the kernel $\JQ$ by the ``bead"
\begin{equation}
\label{Bbead}
\JQ\ =\ \raisebox{-2.5ex}{\includegraphics[scale=.5]{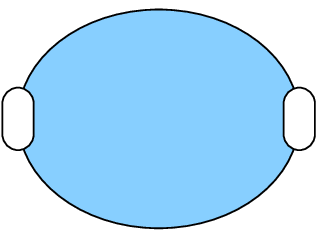}}
\end{equation}
$\tr[\JQ^k]$ is represented by the k-bead necklace (with periodic
boundary condition between the left and right dashed vertical lines)
\begin{equation}
\label{trsbeads}
\tr[\JQ]\,=\,\raisebox{-1.5ex}{\includegraphics[scale=.4]{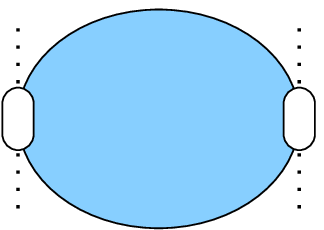}}
\quad,\quad
\tr[\JQ^2]\,=\,\raisebox{-1.5ex}{\includegraphics[scale=.4]{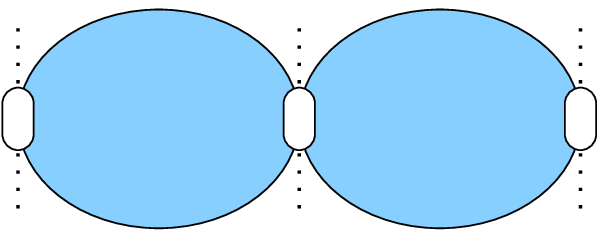}}
\quad,\quad
\tr[\JQ^k]\,=\,\raisebox{-1.5ex}{\includegraphics[scale=.4]{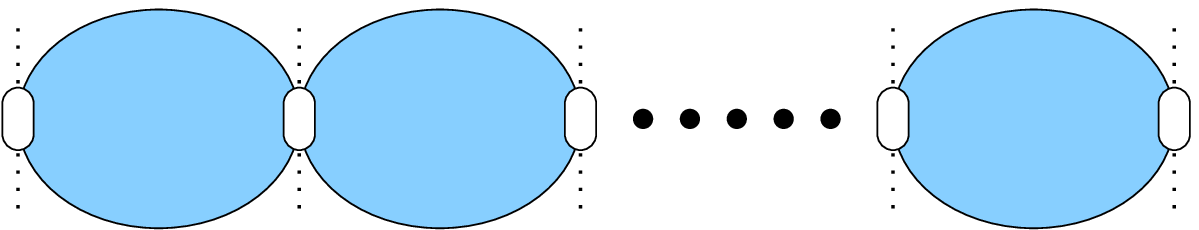}}
\ .
\end{equation}

\subsubsection{$\tr\left[\JQ\right]$} 
\label{sss:trQ} 
We first
consider the term $k=1$. We have already seen that at leading order in
$1/d$
\begin{equation}
\label{trQ0} \tr\left[\JQ^{(0)}\right]\ =\ \int_\kk
\widehat{\JQ}_{\kk,-\kk}^{(0)}\ =\ \int_\kk\int_\xx\rme^{-\kk^2
c_0}\left(\rme^{\kk^2G(\xx)}-1-\kk^2G(\xx)\right) \ =\ \CO(d)
\end{equation}
($\JQ^{(0)}=\JO^{(0)}-\JPo^{\!(0)}$).  This can be depicted
graphically as 
\begin{equation}
\label{ trQ0graph}
\begin{split}
\tr\left[\JQ^{(0)}\right]\ &=\ \sum_{n=2}^\infty \,(2 c_0)^{1-n} \
\left[  \raisebox{-3.ex}{\includegraphics[scale=.5]{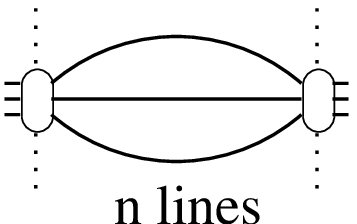}} \right]
\\
&=\ \sum_{n=2}^\infty \,(2 c_0)^{1-n}\,\left[ 
\raisebox{-2.5ex}{\includegraphics[scale=.5]{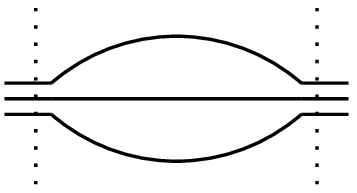}}
\,+\,
\raisebox{-2.5ex}{\includegraphics[scale=.5]{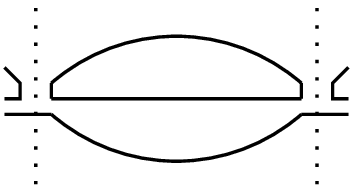}}
\,+\,
\raisebox{-2.5ex}{\includegraphics[scale=.5]{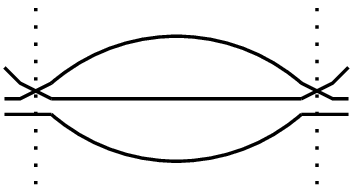}} \right]
\,+\,\CO(1/d)
\end{split}
\end{equation}
and one checks easily that the first graph is of order $\CO(d)$, the
second and the third of order $\CO(1)$, since each closed loop carries
a factor of $d$, and there are periodic boundary conditions between the
left and right vertical dashed lines.

It is easy to see that the trace of the first order correction is of
order $\CO(1)$
\begin{equation}
\label{trQ1}
\tr[\JQ^{(1)}]\,=\,\tr\bigl[\JO^{(1)}-\JPo^{\!(1))}\bigr]\,=\,\CO(1)
\end{equation}
and that at this order it is given by the following 12 diagrams
\begin{equation}
\label{trQ1gr}
\begin{split}
&
2\ \raisebox{-2.5ex}{\includegraphics[scale=.5]{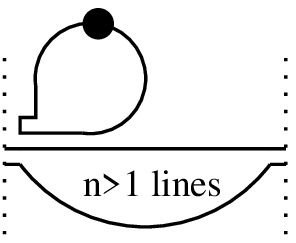}}
\,+\,
\raisebox{-2.5ex}{\includegraphics[scale=.5]{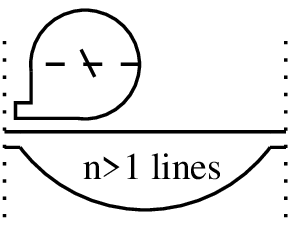}}
\,+\,
\raisebox{-2.5ex}{\includegraphics[scale=.5]{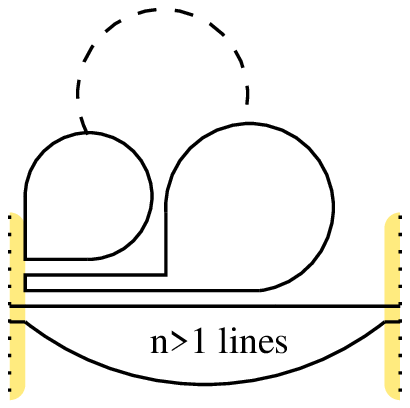}}
\,+\,
2\ \raisebox{-2.5ex}{\includegraphics[scale=.5]{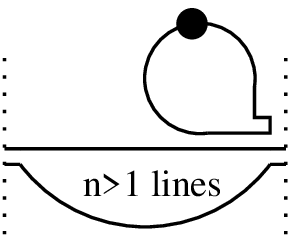}}
\,+\,
\raisebox{-2.5ex}{\includegraphics[scale=.5]{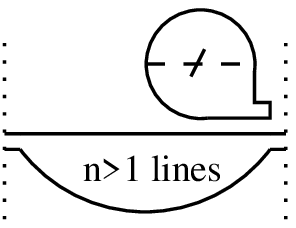}}
\,+\,
\raisebox{-2.5ex}{\includegraphics[scale=.5]{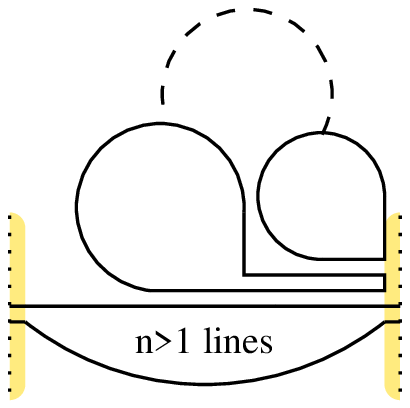}}
\\
+\,&
2\  \raisebox{-2.5ex}{\includegraphics[scale=.5]{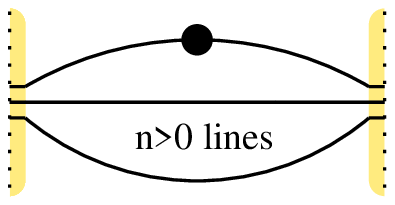}}
\,+\,
\raisebox{-2.5ex}{\includegraphics[scale=.5]{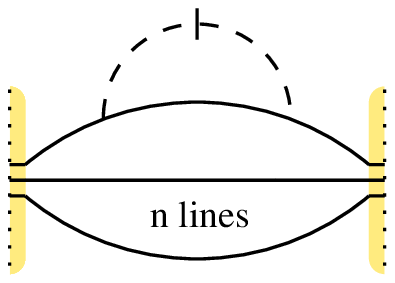}}
\,+\,
\raisebox{-2.5ex}{\includegraphics[scale=.5]{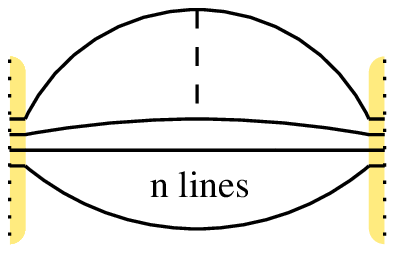}}
\,+\,
\raisebox{-2.5ex}{\includegraphics[scale=.5]{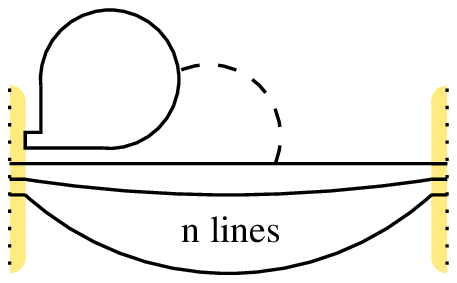}}
\,+\,
\raisebox{-2.5ex}{\includegraphics[scale=.5]{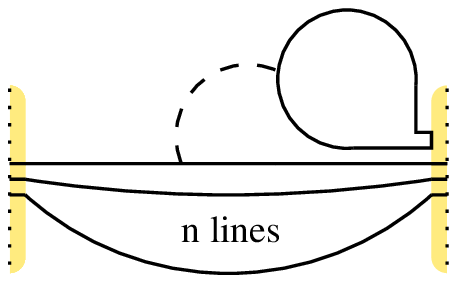}}\\
\,+\,&
\raisebox{-2.5ex}{\includegraphics[scale=.5]{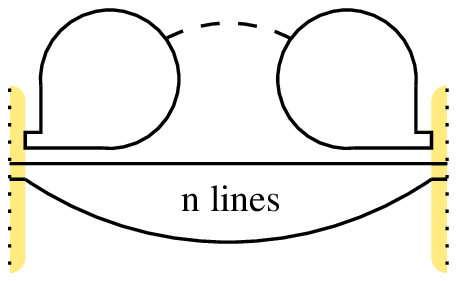}}
\end{split}
\ .
\end{equation}
This corresponds to a specific, but complicated analytical expression,
that we do not write here.

Finally it is quite easy to check that higher order diagrams that
contribute to the term of order $\CO(d^{-r})$ of $\JQ$, will
contribute to the terms of order $\CO(d^{1-r})$ of $\tr[\JQ]$.

\subsubsection{$\tr\left[\JQ^2\right]$} 
\label{ss:trQ2} 
We have seen
in Sect.~\ref{s:direct} that $\tr\left[{\JQ^{\mathrm{var'}}}^2\right]$
was of order $\CO(1)$.  Since $\JQ^{\mathrm{var'}}=\JQ^{(0)}$ we could
have expected that the next order correction $\JQ^{(1)}$ would
contribute by a term of order $\CO(1/d)$ to $\tr[\JQ^2]$. We shall see
that this is not exact, but that there are nevertheless a lot of
simplifications, and that a simple subclass of diagrams contributes at
order $\CO(1)$.

In fact there are simply two beads which contribute at leading order
to $\tr[\JQ^2]$. These are
\begin{equation}
\label{dombeads}
\raisebox{-2.5ex}{\includegraphics[scale=.5]{Q-graph-a}}\ \to \ 
\raisebox{-2.5ex}{\includegraphics[scale=.5]{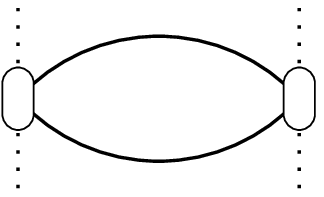}}\ +\, 
\raisebox{-2.5ex}{\includegraphics[scale=.5]{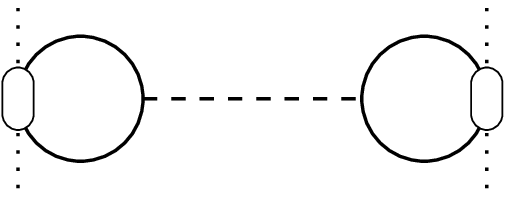}}
\ .
\end{equation}
More precisely, the only 2-bead necklaces which are of order $\CO(1)$
are
\begin{equation}
\label{domnecklace}
\raisebox{-1.3ex}{\includegraphics[scale=.5]{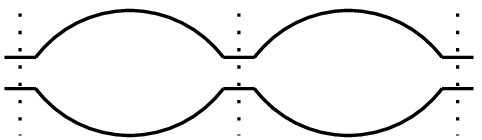}}+
\raisebox{-1.3ex}{\includegraphics[scale=.5]{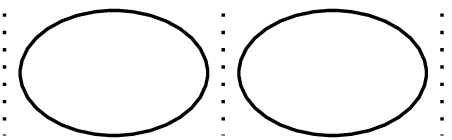}}+
\raisebox{-1.3ex}{\includegraphics[scale=.5]{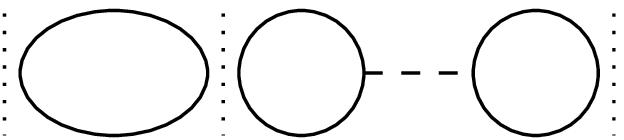}}+
\raisebox{-1.3ex}{\includegraphics[scale=.5]{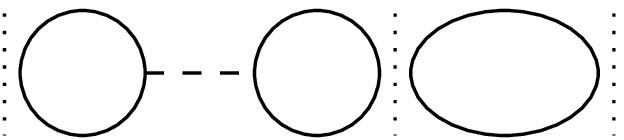}}+
\raisebox{-1.3ex}{\includegraphics[scale=.5]{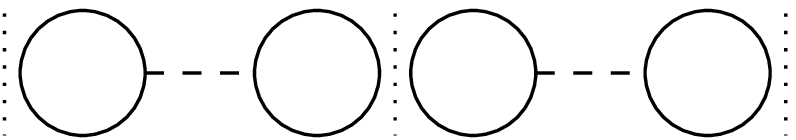}}
\ .
\end{equation}
A careful but not difficult analysis shows that each of these
diagrams is of order $\CO(1)$, and that all the other possible diagrams
are of order $\CO(1/d)$.

The first diagram contributes by 
\begin{equation}
\label{domdiag1}
\raisebox{-1.3ex}{\includegraphics[scale=.5]{Q2-a}}\,=\, {1\over
2}\,d^2\,(2 c_0)^{-2}\,B(0)^2\ =\ {2}\,\left[{d\over 4
c_0}B(0)\right]^2
\end{equation}
(we have taken into account the different contractions of the vertices of (\ref{vertex-2-2}) which give this diagram).
The four last one give the square opf a single bead amplitude
\begin{equation}
\label{domdiag2}
\left[\raisebox{-1.3ex}{\includegraphics[scale=.5]{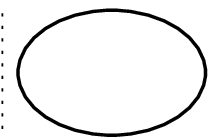}}+
\raisebox{-1.3ex}{\includegraphics[scale=.5]{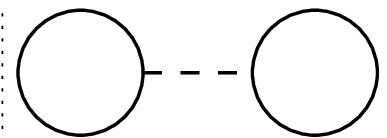}}\right]^2
\end{equation}
with the single bead amplitude
\begin{equation}
\label{domdiag3}
\raisebox{-1.3ex}{\includegraphics[scale=.5]{Q-dom-c}}+
\raisebox{-1.3ex}{\includegraphics[scale=.5]{Q-dom-d}}\ =\ {1\over 2
c_0}\left({1\over 2}\, d\,B(0)+{1\over 4}\,d^2\,{1\over 2
c_0}\,B(0)^2\,H(0)\right) \ =\ {d\,B(0)\over 4
c_0}\left[1+{{d\,B(0)\over 4 c_0}\over 1+\mu_2 {d\,B(0)\over 4
c_0}}\right]
\ ,
\end{equation}
where we used (\ref{chaine0}) and (\ref{B2pex}).

We thus see that the chain of bubbles contributes already to
$\tr[\JQ^2]$ at the leading order $\CO(1)$. In the large-$d$ limit, $\epsilon$ being fixed,
since $\mu_2\to -1$ and $dB(0)/4c_0\to 1-\epsilon/4$ we get 
\begin{equation}
\label{tQ2} \tr[\JQ^2]\ =\ 2\left(1-{\epsilon\over
4}\right)^2+\left({4\over\epsilon}-1\right)^2\,+\,\CO(1/d)\ .
\end{equation} 

\subsubsection{$\tr\left[\JQ^k\right]$, $k>2$} 
\label{sss:trQk} 
The same analysis can be done for the general term $\tr[\JQ^k]$. Here also
the only diagrams that contribute to order $\CO(1)$ are those of
(\ref{dombeads}), and more precisely those with the beads of
(\ref{domdiag3}). It follows that at leading order
\begin{equation}
\label{tQk}
\tr[\JQ^k]\ =\ \left[\raisebox{-1.3ex}{\includegraphics[scale=.5]{Q-dom-c}}+
\raisebox{-1.3ex}{\includegraphics[scale=.5]{Q-dom-d}}\right]^k
\ =\ 
\left[{4\over\epsilon}-1\right]^k\ +\ \CO(1/d)
\ .
\end{equation}
We shall comment later on the meaning of the pole in $1/\epsilon$.

\subsubsection {Summation of the $\log$ series} 
\label{sss:sumlog} 
We
see that, except for more complicated graphs coming from the $k=1$ and
$k=2$ terms, the whole series (\ref{Lseriesagain}) for
$\mathfrak{L}=\log(\mathfrak{D})$ contains the series $-\sum_{k>1}
{1\over k}(4/\epsilon-1)^k$ which can be resummed formally as a
logarithm, so that (the second term compensates for the missing term
in the sum giving the log) 
\begin{equation}
\label{Lresum1} \mathfrak{L}\ =\ -\tr\bigl[\JQ^{(0)}+\JQ^{(1)}\bigr]
\,+\,\left[{4\over\epsilon}-1\right]+\left[1-{\epsilon\over
4}\right]^2\,+\,\log\left[2-{4\over\epsilon}\right]\,+\,\CO(1/d)\ .
\end{equation}
This last series is not convergent if $\epsilon<2$ and the argument of
the logarithm is negative, hence $\mathfrak{L}$ has an imaginary part
$\pm\pi$.

In fact this is not surprising, and is a feature of the model, since
we have in fact recovered the unstable eigenvalue
$\lambda_{\mathrm{min}}=1-\lambda_-$ of $\CS"$ of the Hessian $\CS"$,
which indeed gives an imaginary part $\pm\pi$ to $\mathfrak{L}$. We
show this fact in the next section.

\subsection{The unstable mode} 
\label{ss:unstmode} 
It was shown in
\cite{DavidWiese1998} and in Sect.~\ref{unsteigmode} by general
arguments that as long as $0\le\epsilon<D$ the Hessian $\CS"[\Vi]$ has
one single negative eigenvalue $\lambda_{\mathrm{min}}<0$,
corresponding to the mode of unstable fluctuations around the
instanton configuration.

In appendix \ref{varbound} we derive a variational estimate for an
upper bound for this $\lambda_{\mathrm{min}}$. This estimates is
given by Eq.~(\ref{resultbound2}) and becomes in the large-$d$ limit
\begin{equation}
\label{lminvar1} \lambda_{\mathrm{min}}^{\mathrm{var}}\ =\
{-2\epsilon(D-\epsilon)\over(2-D)(2D-\epsilon)+\epsilon^2}\ \to\
2-{4\over\epsilon}\quad\text{when}\quad D\to 2,\ \epsilon\
\text{fixed}\ .
\end{equation}
This is precisely the argument of the $\log$ in (\ref{Lresum1}).

Here we show that this is not a coincidence, and that the variational
bound $\lambda_{\mathrm{min}}^{\mathrm{var}}$ is saturated in the
limit $d\to\infty$, $\epsilon$ finite, so that the infinite series of
necklace diagrams,  with beads made themselves out of  chains of bubbles
of (\ref{domdiag3}) reconstructs precisely the logarithm of the unstable
eigenvalue $\log(\lambda_{\mathrm{min}})$.
\begin{equation}
\label{lmin1} \lambda_{\mathrm{min}} \ =\
2-{4\over\epsilon}\quad\text{when}\quad d\to \infty,\ \epsilon\
\text{fixed}
\ .
\end{equation}
To obtain this result, we shall simply take the following ansatz
$\Psi_-$ for the unstable eigenmode
\begin{equation}
\label{Psiansatz}
\widehat{\Psi}_-(\kk)\ =\ {1\over 2}\,\kk^2\,\rme^{-\kk^2 c_0/2}
\end{equation}
and show that at leading order
\begin{equation}
\label{IdQPsi}
(\JU-\JQ)\,\Psi_-\ =\ (2-4/\epsilon)\,\Psi_- \ +\ \CO(1/d)
\ .
\end{equation}
Let us first compute $\JQ^{(0)}\Psi_-$ 
\begin{equation}
\label{Q0Psi}
\begin{split}
\widehat{\JQ}^{(0)}\widehat{\Psi}_-(\kk_1)&=
\int_{\kk_2}\widehat{\JQ}^{(0)}_{\kk_1,-\kk_2}\widehat{\Psi}_-(\kk_2)
={1\over 2}\,\rme^{-\kk_1^2 c_0/2}\int_{\kk_2}\rme^{-\kk_2^2 c_0} \kk_{2}^{2}
\int_\xx \left[\rme^{-\kk_1\kk_2 G(\xx)}-1+\kk_1\kk_2 G(\xx)\right]
\\
&= {1\over 2} 2c_0\,\rme^{-\kk_1^2 c_0/2}\int_\xx
\left(\rme^{{\kk_1^2\over 4}{G(\xx)^2\over c_0}}\left({\kk_1^2\over
4}{G(\xx)^2\over c_0^2}+{d\over 2 c_0}\right)-{d\over 2 c_0}\right) \ .
\end{split}
\end{equation}
In the limit $d\to\infty$ since $c_0\sim d$ the dominant term is
\begin{equation}
\label{Q0Psidom} {1\over 2}{d\over 4 c_0}\,\kk_1^2\,\rme^{-\kk_1^2
c_0/2}\,\int_\xx G(\xx)^2\ \simeq\ {d\over 4 c_0}{1\over
4\pi}\,\widehat{\Psi}_-(\kk_1)\ =\, \left(1-{\epsilon\over
4}\right)\,\widehat{\Psi}_-(\kk_1)
\ .
\end{equation}
Note that we may represent graphically $\widehat{\Psi}_-$ and
$\widehat{\JQ}\widehat{\Psi}_-$ by
\begin{equation}
\label{PsiQPsigraph} \widehat{\Psi}_-\ =\
\raisebox{-.2ex}{\includegraphics[scale=.5]{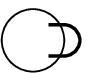}} \quad,\quad
\widehat{\JQ}\widehat{\Psi}_-\ =\
\raisebox{-3.ex}{\includegraphics[scale=.5]{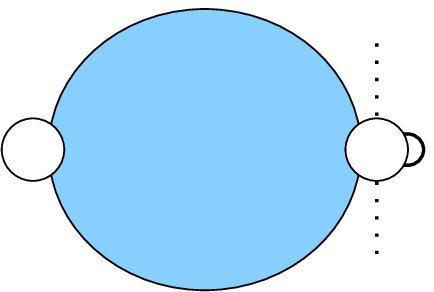}}
\end{equation}
(where the little handle represents $\delta_{ab}$) and that the
dominant contribution (\ref{Q0Psidom}) at large $d$ corresponds simply
to the diagram
\begin{equation}
\label{Q0Psigraph} \JQ^{(0)}\,\Psi_-\ \simeq\
\raisebox{-1.3ex}{\includegraphics[scale=.5]{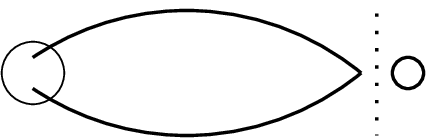}} =\
B(0){1\over 2c_0}{d\over 2}\,\Psi_- \ .
\end{equation}
Note that the rightmost little loop is just 
$\raisebox{-0ex}{\includegraphics[scale=.5]{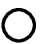}}=d/2$.

We can now compute in the large-$d$ limit the contribution of
$\JQ^{(1)}\Psi_-$. It is given a priori by all the diagrams of
(\ref{Ohat1}) inserted into (\ref{PsiQPsigraph}).  However a careful but
easy analysis shows that the only diagram which contributes finally at
leading order $\CO(1)$ is  the chain of bubbles, that
appears already in (\ref{domdiag3})
\begin{equation}
\label{Q1Psigraph}
\begin{split}
\JQ^{(1)}\Psi_-\ &\simeq\
\raisebox{-1.3ex}{\includegraphics[scale=.5]{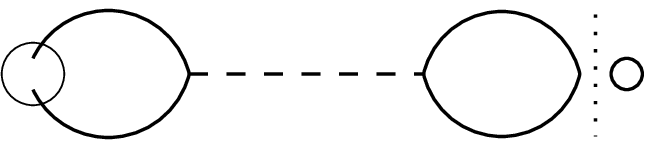}} \ =\
B(0){H(0)}{-\mu_2\over 2c_0}{d\over 2}\,B(0){1\over 2c_0}{d\over
2}\Psi_-
\\
&=\ {4\over \epsilon}\left(1-{\epsilon\over
4}\right)^2\Psi_-\quad\text{when}\quad d\to\infty
\end{split}
\ .
\end{equation}

Now if we consider the graphs that appear in the higher-order terms
$\JQ^{(r)}$ of the $1/d$ expansion of $\JQ$, one can see that when
applied to $\Psi_-$ they also give only terms of order at most
$\CO(1/d)$. Hence we have
\begin{equation}
\label{QQ1Q2Psi}
\JQ\Psi_-\ =\ \JQ^{(0)}\Psi_-+\JQ^{(1)}\Psi_-+\CO(1/d)
\end{equation}
and combining (\ref{Q0Psidom}) and (\ref{Q1Psigraph}) we obtain
(\ref{IdQPsi}). Q.E.D.

\subsection{The zero-mode measure} 
\label{ss:zeromodmeas} 
Finally, we
have to compute the $1/d$ correction to the weight $\mathfrak{W}$ for
the collective-coordinate measure for the instanton.  According to
(\ref{SDLWdef}), this weight is given by
\begin{equation}
\label{Wagain} \mathfrak{W}\ =\ g^{{d/ D}}\left[ {1\over
2\pi\,d}\int_\rr(\nabla V)^2\right]^{d/2} \ =\ g^{{d/ D}}\left[
{1\over 2\pi\,d}\int_\kk \kk^2 \widehat{V}(\kk)^2\right]^{d/2}\ ,
\end{equation}
and using the explicit form for $V$, and in particular
(\ref{denomgraph}) we get (for $D\to 2$) 
\begin{equation}
\label{W1/d} \mathfrak{W}\ =\ \left[{g\over
2\pi}\right]^{d/2}\,g^{{4-\epsilon\over
2}}\,\rme^{-\delta_1(\epsilon)}\bigl(1+\CO(1/d)\bigr)
\ ,
\end{equation}
where $\delta_1(\epsilon)$ is the coefficient for the mass correction at
order $1/d$ defined by (\ref{mun1/d}). $\delta_1(\epsilon)$ is of order
$\CO(1)$ in the large-$d$ limit, its exact value is given by the
self-consistent equation (\ref{mu1equ})-(\ref{mu1eqdiag}).
The large-$d$ limit for $\delta_1(\epsilon)$ was already obtained in
\cite{DavidWiese1998}. It is given by the integral
\begin{equation}
\label{delt1expl} \delta_1(\epsilon) = {1\over
4-\epsilon}\int_0^\infty \!\!\rmd p\
p\,\left[{-\log\left[{1-\left[1-{\epsilon\over
4}\right]J(p)}\right]-\left[1-{\epsilon\over 4}\right]J(p)}\right] \
,\quad J(p)={2\,\mathrm{arcsinh}(p/2)\over p\sqrt{1+p^2/4}}
\ ,
\end{equation}
which is convergent as long as $\epsilon>0$.

\section{The limit $\epsilon=0$  and the renormalized theory}
\label{s:ren1/d} 
We are interested in the renormalized theory in which
the UV divergences have been subtracted and the limit $\epsilon\to 0$
has been taken. We have already discussed in Section
\ref{s:renormalization} the UV divergences and how they are
renormalized. Here we discuss this limit in more detail and its
interplay with the large-$d$ limit. Our main result is that the $1/d$
expansion is plagued by IR divergences when $\epsilon=0$, so that the
limits $d\to\infty$ and $\epsilon\to 0$ do not simply commute. As we
shall see in our discussion, this does not mean that our instanton
calculus does not make sense at $\epsilon=0$, but rather that when
$\epsilon=0$ the large-$d$ limit is of a different nature and contains
non-analytic terms in $d$ such as logarithms of $d$.

\subsection{Minimal subtraction schemes} 
\label{ss:MSsch} 
To study the
renormalized theory at a given dimension $d$ we must first specify a
renormalization scheme.  We shall use the minimal subtraction scheme
(MS) such that the field and coupling-constant counterterms in the
original action subtract the poles at $\epsilon=0$ (see
Eqs.~\eq{Srenr1}-\eq{ZZ1r1l}).  In fact the definition of a MS scheme
requires some care. Indeed, since $\epsilon=2D-d(2-D)/2$ depends both
on $D$ and $d$ (the manifold and bulk space dimensions) the limit
$\epsilon\to 0$ to construct the renormalized theory of a $D$-manifold
in $d=d_c(D)=4D/(2-D)$ dimension can be taken in different ways. These
different limits corresponds to different renormalized theories which
differ by a finite renormalization of the field and the coupling
constant, i.e.\ these limits correspond to different renormalization
schemes.  \subsubsection{Definition of the MS-D and MS-d
schemes}\label{sss:schemes}
\begin{enumerate}
  \item \textbf{MS-D scheme:}\label{MSDs}
  A first scheme is to work at fixed manifold dimension $D=D_c$ and to
  take the limit $d\to d_c(D_c)=4D_c/(2-D_c)$. Then
  $\epsilon=(d_c-d)(2-D_c)/2$. This allows direct comparison with the
  field theoretical calculations for SAW and polymers, since for
$D_c=1$ and $\epsilon=\varepsilon/2$ where $\varepsilon=4-d$ is the
parameter of the standard Wilson-Fisher expansion.
  \item \textbf{MS-d scheme:}\label{MSds}
  Another scheme, more natural for 2-dimensional manifolds ($D=2$), is
  to fix $d=d_c$ and to take the limit $\epsilon\to 0$ by varying
  $D$. In this case
$\epsilon=(2+d_c/2)(D-D_c)$ with $D_c=D_c(d_c)=2d_c/(4+d_c)$.
 \end{enumerate}

\subsubsection{Relation between the schemes}
\label{sss:relsch} In both
schemes we take as counterterms
\begin{equation}
\label{ }
Z(\bren)=1-\bren{\mathtt{C_1}(D_c,d_c)\over\epsilon}\quad,\qquad
Z_b(\bren)=1+\bren{1\over 2}{\mathtt{C_2}(D_c,d_c)\over\epsilon}
\end{equation}
and the relation between the bare fields $\rr$ and coupling constant
$b$ and renormalized ones $\rr_{\mathrm{r}}$ and $\bren$ is
\begin{equation} 
\label{ }
\rr=Z^{1/2}\rr_{\mathrm{r}}\quad,\qquad b=\bren\mu^\epsilon Z_b Z^{d/2}
\ .
\end{equation}
We see that both $\epsilon$ and $d$ appear explicitly in the second
relation for $b$. At one loop it gives
\begin{equation}
\label{ } b=\mu^\epsilon \bren\left[1+\bren{1\over
2}{\mathtt{C_2}-d\mathtt{C_1}\over \epsilon}+\cdots\right]
\ .
\end{equation}
In the MS-d scheme the last term gives
\begin{equation}
\label{ }
\mathtt{C_2}-d\mathtt{C_1}=\mathtt{C_2}(D_c,d_c)-d_c\mathtt{C_1}(D_c,d_c)
\end{equation}
while in the MS-D scheme it gives
\begin{equation}
\label{ }
\mathtt{C_2}-d\mathtt{C_1}=\mathtt{C_2}(D_c,d_c)-d_c\mathtt{C_1}(D_c,d_c)
+\epsilon\,{2\mathtt{C_1}(D_c,d_c)\over 2-D_c}
\ .
\end{equation}
We see that renormalization in the MS-D and the MS-d schemes with the
same subtraction mass scale $\mu$ amounts to a finite coupling-constant
renormalization
\begin{equation}
\label{ }
b_{\mathrm{MS-d}}=b_{\mathrm{MS-D}}+b_{\mathrm{MS-D}}^2\,{\mathtt{C_1}\over
2-D_c}
\end{equation}
or equivalently that the MS-d subtraction scale $\mu_{\mathrm{MS-d}}$
and the MS-D subtraction scale $\mu_{\mathrm{MS-D}}$ are related by
\begin{equation}
\label{ } \log\left[{\mu_{\mathrm{MS-d}}\over
\mu_{\mathrm{MS-D}}}\right]={2\over
2-D}{\mathtt{C_1}\over\mathtt{C_2}-d\mathtt{C_1}} \ .
\end{equation}
Let us also note that we recover the combination of counterterms
$\mathtt{C_2}-d\mathtt{C_1}$ that appears in the result \eq{Bexplgen}
for the coefficient $\mathtt{B}$ (defined by Eq.~\eq{Bdef}) of the UV
pole in $1\over\epsilon$ for the effective action $\CS[V]$ (see
Eq.~\eq{DSV}) and for $\mathfrak{L}=\tr\log\CS''[V]$.

\subsection{Variational mass subtraction scale}
\label{ss:varsubm} Now
for simplicity and in order to study more easily the large-$d$ limit
of the renormalized theory we shall work with the normalizations of
Appendix \ref{appendixnormvar} where $\xx$ and $\rr$ are rescaled as
$\xx\to \mvar\xx$, $\rr\to\mvar^{(2-D)/2}$ and the coupling constant
$b$ is redefined by $b\to\mvar^{\epsilon-D} b$ so that the variational
mass is now set to unity ($m_{\mathrm{var}}=1$) in all the
calculations.  Since the rescaling of the coupling constant amounts to
$g\to\mvar^{-D} g$, this last rescaling amounts to choosing as
subtraction scale a multiple of the variational mass
($\mu\to\mu\mvar$) in the renormalized theory.

In this normalization the field and coupling-constant counterterms (as
defined in \eq{ren39}) $\mathtt{C_1}=\mathtt{C_1}(D_c,d_c)$ and
$\mathtt{C_2}=\mathtt{C_2}(D_c,d_c)$ in the action become (see
Appendix \ref{appendixnormvar} and in particular
Eqs. \eq{Cttresc2}-\eq{SDcsurd})
\begin{equation}
\label{C12res2} \mathtt{C_1}={-\CS_D\over 2D}\left[{c_0\over
d_0}\right]^{1+{d\over 2}} \quad,\quad \mathtt{C_2}={2 \CS_D^2\over
(2-D)^2}{\Gamma[D/(2-D)]^2\over \Gamma[2D/(2-D)]}\left[{c_0\over
d_0}\right]^{1+{d\over 2}} \ .
\end{equation}
The logarithm  of the renormalized instanton determinant $\mathfrak{L}_{\mathrm{r}}$ is still given by \eq{Lfrakren}
\begin{equation}
\label{Lfr2}
\mathfrak{L}_{\mathrm{r}}=\mathfrak{L}+\left(\gren^{1\over D}\mu
L\right)^{-\epsilon}\,
\left[{\mathtt{C_1}\over\epsilon}\langle{(\nabla\rr)^2\rangle_V
+{\mathtt{C_2}\over 2\epsilon}\int_\rr}V(\rr)^2\right] \ .
\end{equation}
We have seen that in the large-$d$ limit ($\epsilon$ fixed), the first
counterterm is of order one $\mathtt{C_1}=\CO(1)$ while the second one
is exponentially small, $\mathtt{C_2}\sim\CO(\exp(-d))$.  For our
discussion of the variational approximation and of the large-$d$ limit
we only have to consider the wave-function counterterm
$\mathtt{C_1}=\mathtt{C_1}(D,d_c(D))=\mathtt{C_1}(D_c(d),d)$ which is
given explicitly when $\epsilon=0$ by
\begin{equation}
\label{C1resexp}
\mathtt{C_1}=-{4\over D}{(4\pi)^{D/2}\over {\Gamma[D/2]}}\left[{\Gamma[(2-D)/2]\over -\Gamma[(D-2)/2]}\right]^{{2+D\over2-D}}
=
{-{4\over d}{(4\pi)^{{2d\over 4+d}}\over{\Gamma[d/(4+d)]}^2}\left[{-\Gamma[-4/(4+d)]\over \Gamma[4/(4+d)]}\right]^{-{d\over 2}}}
\end{equation}

\subsection{Renormalized theory in the variational approximation for finite $d$} 
\label{ss:rendlead} We first consider the renormalized
instanton determinant in the variational approximation, but for finite
embedding space dimension $d$, following the lines of
Sect.~\ref{s:expdl}.  We thus approximate $\mathfrak{L}=\log\det
'[\CS'']$ by $\mathfrak{L}^{(0)} =-\tr\left[\JQ^{(0)}\right]$ (as
defined by Eq.~\eq{trQ0}). This gives, after integration over $\kk$
and using \eq{gaussresc},
\begin{equation}
\label{L0exp1}
\mathfrak{L}^{(0)}=-\tr\left[\JQ^{(0)}\right]=-\int_\kk\rme^{-\kk^2c_0}\int_\xx\left[\rme^{\kk^2
G(\xx)}-1-\kk^2 G(\xx)\right]
=d-2c_0\,\int_\xx\,\left(\left[1-{G(\xx)\over c_0}\right]^{-{d\over 2}}-1\right)
\ .
\end{equation}
To renormalize consistently $\mathfrak{L}$ we must take for the condensate $\langle(\nabla\rr)^2\rangle_V$ in  the counterterm in \eq{Lfr2} its value in the variational approximation
\begin{equation}
\label{gradr2var}
\langle(\nabla\rr)^2\rangle_V\,\to\,\langle(\nabla\rr)^2\rangle_{m=1}\,=\ d\int_\pp{\pp^2\over\pp^2+1}\,=\,-d c_0
\end{equation}
(we use dimensional regularization), and neglect the coupling-constant
counterterm $\mathtt{C_2}$, since there is no coupling-constant
renormalization in the variational approximation.  Thus we obtain for
the renormalized log \emph{in the MS-D scheme}
\begin{equation}
\label{L0ren1}
\Lfren^{(0)}=\lim_{\epsilon\to 0,\,D\,\text{fixed}}
\left[\mathfrak{L}^{(0)}-d c_0
\left(\gren^{1\over D}\mu L\right)^{-\epsilon}\,{\mathtt{C_1}(D)\over\epsilon}\right]
=-\mathtt{B_{var}}\left[{1\over D}\log(\gren)+\log(\mu L)\right]+\mathfrak{L}^{(0)}_{\mathrm{MS-D}}
\end{equation}
with
\begin{equation}
\label{Bttvar}
\mathtt{B_{var}}=-d_c(D)\mathtt{C_1}(D)c_0(D)={32\over (2-D)^2}\left[{\Gamma[(2-D)/2]\over -\Gamma[(D-2)/2]}\right]^{{4\over 2-D}}
\end{equation}
and
\begin{equation}
\label{L0MSD}
\mathfrak{L}^{(0)}_{\mathrm{MS-D}}\ =\ 
\lim_{\epsilon\to 0,\,D\,\text{fixed}}
\left[\mathfrak{L}^{(0)}-d c_0
\,{\mathtt{C_1}(D)\over\epsilon}\right]
\ .
\end{equation}
Integrating over the angular degrees of freedom of $\xx$ we can
rewrite $\mathfrak{L}^{(0)}$ as
\begin{equation}
\label{L0intx}
\mathfrak{L}^{(0)}=d-2c_0(D)\CS_D\ \mathfrak{I}\quad,\qquad\mathfrak{I}=\text{f.p.}\int_0^\infty \rmd x\,x^{D-1}
\left(\left[1-{G(x)\over c_0}\right]^{-{d\over 2}}-1
\right)
\ ,
\end{equation}
where the finite part prescription ``f.p.'' deals with the short-distance divergence at $x=0$ still present when $\epsilon>0$.
The UV divergence of $\mathfrak{I}$
comes from the short-distance behavior of the propagator $G(\xx)$,
obtained from \eq{Gmexp}
\begin{equation}
\label{Gsdex} G(\xx)=c_0-d_0\, x^{2-D}+{c_0\over 2D}\,x^2-{d_0\over
2(4-D)} \,x^{4-D}+\CO(x^{4}) \ .
\end{equation}
This implies that the integrand in $\mathfrak{I}$ behaves at small $x$ as
\begin{equation}
\label{Lf0sdexp} \mathfrak{I}\ \simeq\ \int_0^{\cdots} \rmd x\
\left[{c_0\over d_0}\right]^{{d\over 2}}\,
\left[x^{\epsilon-D-1}+{d\over 4D}{c_0\over d_0}x^{\epsilon-1}-{d\over
4(4-D)}\,x^{\epsilon+1-D}+\CO(x^{\epsilon+1})\ \right]\ .
\end{equation}
The first term gives the UV pole at $\epsilon=D$, which is subtracted
by dimensional regularization, and is dealt with by the
f.p. prescription. The second term gives the UV pole at $\epsilon=0$.
The third one gives a non-singular pole at $\epsilon=D-2$, but will be
important in the large-$d$ limit.  Now we use the explicit result
\eq{C12res2} for $\mathtt{C_1}$, which implies that we can rewrite the
counterterm in \eq{L0MSD} as
\begin{equation}
\label{Lf0ct}
d c_0
\,{\mathtt{C_1}(D)\over\epsilon}
=
-2c_0\CS_D\left[{c_0\over d_0}\right]^{1+{d_c(D)\over 2}}{d\over 4D}\int_0^1\rmd x\,x^{\epsilon-1}
\ .
\end{equation}
Thus this counterterm cancels the pole at $\epsilon=0$, but we must
notice that since we use the MS-D scheme, there is a slight difference
between the coefficient of the $x^{\epsilon-1}$ term in \eq{Lf0sdexp}
and \eq{Lf0ct}: the first one contains $[c_0/d_0]^{1+d/2}$ and the
second one $[c_0/d_0]^{1+d_c(D)/2}$. Since $d=d_c(D)-2\epsilon/(2-D)$
this gives a difference of order $\CO(\epsilon)$ for the residue of
the poles at $\epsilon=0$, hence a term of order $\CO(1)$ in the limit 
$\epsilon\to 0$.  We carefully rewrite the expression \eq{L0MSD}
for $\mathfrak{L}^{(0)}_{\mathrm{MS-D}}$ as
\begin{equation}
\label{L0MS1}
\begin{split}
&\mathfrak{L}^{(0)}_{\mathrm{\tiny{MS-D}}}=\lim_{\epsilon\to 0,\,D\,\text{fixed}}
\left\{
d+{d c_0\over 4D}\left[{c_0\over d_0}\right]^{{2+D\over 2-D}}{1\over\epsilon}\left[1-\left[{c_0\over d_0}\right]^{{-\epsilon\over 2-D}}\right]
\,-2c_0\CS_D\,\mathfrak{I'}
\right\}
\\
&
\mathfrak{I'}=\int_0^\infty \!\rmd x\,
	\left[x^{D-1}
		\left[
			\left[1-{G(x)\over c_0}\right]^{-{d\over 2}}
			\!\!\!-1
			\right]
		-\left[{c_0\over d_0}\right]^{{d\over 2}}x^{\epsilon-D-1}
		-{d\over 4D}\left[{c_0\over d_0}\right]^{1+{d\over 2}}
\hskip -1ex x^{\epsilon-1}\theta(1-x)
	\right]
\end{split}
\end{equation}
$\theta(1-x)=1$ if $x<1$, $0$ if $x>1$ is the Heaviside step function.
This integral representation is a priori valid for $\epsilon>0$ but is
now convergent if we take the limit $\epsilon\to 0$. We can
interchange this limit and the small $x$ integration and obtain, using
$d_c(D)=4D/(2-D)$ and $\CS_D=1/(2-D)d_0$,
\begin{equation}
\label{L0MS2}
\begin{split}
&\mathfrak{L}^{(0)}_{\mathrm{MS-D}}=d+
{ c_0\over (2-D)^2}\left[{c_0\over d_0}\right]^{{2+D\over 2-D}}\log\left[{c_0\over d_0}\right]
\,-{2\over 2-D}{c_0\over d_0}\,\mathfrak{I'}
\\
&
\mathfrak{I'}=
\int_0^\infty \!\rmd x\,
	\left[x^{D-1}
		\left[
			\left[1-{G(x)\over c_0}\right]^{-{2D\over 2-D}}
			\!\!\!-1
			\right]
		-\left[{c_0\over d_0}\right]^{{2D\over 2-D}}x^{-D-1}
		-{1\over 2-D}\left[{c_0\over d_0}\right]^{{2+D\over 2-D}}x^{-1}\theta(1-x)
	\right]
\end{split}
\end{equation}
This last integral over $x$ is UV and IR convergent as long as $D<2$. It can be computed numerically.

For $D=1$ we have $c_0=1/2$, $d_0=1/2$ and $G(x)=\rme^{-|x|}/2$, and we obtain
\begin{equation}
\label{L0MSD1}
\mathfrak{L}=4-2\mathfrak{I'}\quad,\qquad
\mathfrak{I'}=\int_0^\infty \rmd x\,\left[\left[1-\rme^{-x}]\right]^{-2}-1-x^{-2}-x^{-1}\theta(1-x)\right]=\zeta(0)=-{1\over 2}
\end{equation}

{\footnotesize{
\noindent\textbf{Details of the calculation:} We compute the integral 
$$\mathfrak{L}=4-2\mathfrak{I'}\ ,\ \ \mathfrak{I'}=\int_0^\infty \rmd x\,\left[\left[1-\rme^{-x}]\right]^{-2}-1-x^{-2}-x^{-1}\theta(1-x)\right]\ .$$
First we put a regulator $\varepsilon$, and notice that the last term is here to subtract the pole in $\varepsilon$
$$\mathfrak{I'}=\lim_{\varepsilon\to 0} \mathfrak{J}(\varepsilon)-{1\over \varepsilon}\ ,\ \ \mathfrak{J}(\varepsilon)=\mathrm{f.p.}\int_0^\infty \rmd x\,x^{\varepsilon}\,\left[\left[1-\rme^{-x}]\right]^{-2}-1\right]\ .
$$
Now 
$$\mathfrak{J}(\varepsilon)=\sum_{n=1}^\infty (n+1)\int_0^\infty \rmd x\,x^{\varepsilon}\,\rme^{-nx}
=\sum_{n=1}^\infty (n+1)\Gamma(\varepsilon+1)\,n^{-\varepsilon-1}=\Gamma(\varepsilon+1)\bigl(\zeta(\varepsilon)+\zeta(\varepsilon+1)\bigr)
$$
Now we use $$\zeta(0)=-{1\over 2}\ ,\ \ \zeta(1+\varepsilon)={1\over\varepsilon}+\gamma_{\mathrm{\scriptscriptstyle{E}}}+\CO(\varepsilon)
\ ,\ \ \Gamma(1+\varepsilon)=1-\gamma_{\mathrm{\scriptscriptstyle{E}}}\varepsilon+\CO(\varepsilon^2)
$$ and obtain
$$\mathfrak{J}(\varepsilon)={1\over\varepsilon}-{1\over 2}+\CO(\varepsilon)\qquad\text{hence}\qquad I=-{1\over 2}$$
But note that $\text{Mathematica}^{\text{®}}$ 5 gives the result directly.
}}
\begin{table}[h]
  \centering 
$$
\begin{array}{|c|c|c|c|c|c|c|c|c|c|c|}
\hline
d&0&2&4&6&8&10&12&14&16&20\\
\hline
\mathfrak{L}^{(0)}_{\mathrm{MS-D}}&\ 2\ \ &4.01&\ 5\ \ &6.60&9.16&13.0&18.5&25.9&35.6&62.7\\
\hline
\end{array}
$$ 
 \caption{ }\label{tableLrend}
\end{table}
\begin{figure}[h]
\begin{center}
\includegraphics[width=3in]{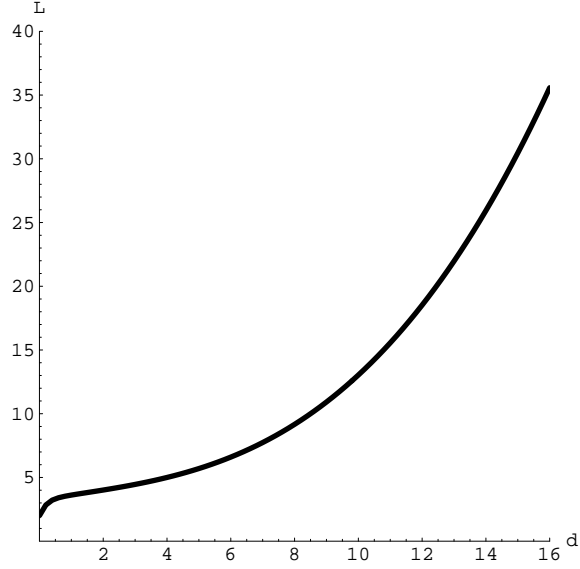}
\caption{$L=\mathfrak{L}^{(0)}_{\mathrm{MS-D}}$ as a function of the external
dimension $d$.}  
\label{figLrend}
\end{center}
\end{figure}
For $D\neq 1$ this integral representation allows easy numerical
integration. This gives the following results, presented on Table
\ref{tableLrend} and Figure \ref{figLrend}.  Finally, we see on the
numerical results that $\mathfrak{L}^{(0)}_{\mathrm{MS-D}}$ diverges
when $d\to\infty$ (i.e.\ when $D\to 2$). As we shall see later, it
behaves as
\begin{equation}
\label{L0MSdom}
\mathfrak{L}^{(0)}_{\mathrm{MS-D}}\ \simeq\ \left(2\rme^{\gamma_{\tiny{\text{\tiny{E}}}}}\right)^{-4} \,d^3
\end{equation}
and this asymptotic behavior is reached as soon as $d\simeq 20$,
as shown on Fig. \ref{figLrend2}.
\begin{figure}[h]
\begin{center}
\includegraphics[width=3in]{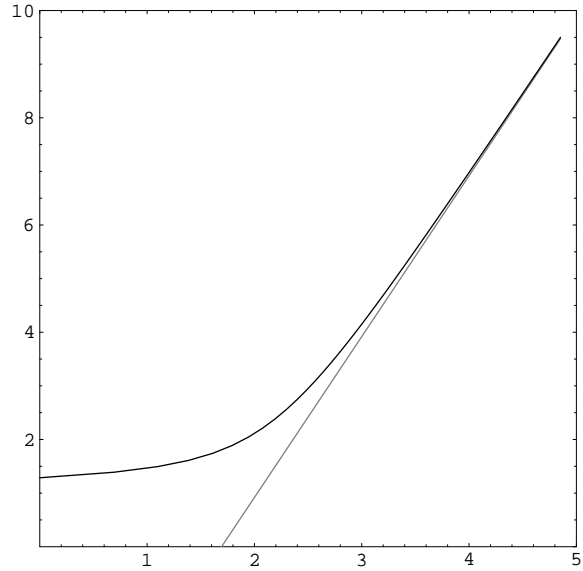}
\caption{$\mathfrak{L}^{(0)}(d)$ (black curve) in Log-Log coordinates
compared to its large-$d$ asymptotics (grey curve, straight line).} \label{figLrend2}
\end{center}
\end{figure}
\par
{\footnotesize{ \noindent{\textbf{Details of the calculation:}} To
compute the integral $\mathfrak{I'}$ numerically, it is more
convenient to separate the integral over $x\in ]0,1]$ and over $x\in
[1,\infty[$
\begin{equation}
\label{II1I2} \mathfrak{I'}=\mathfrak{I'_1}+\mathfrak{I'_2}\ ,\ \
\mathfrak{I'_1}=\int_0^1\rmd x\,\left[\cdots\right] \ ,\ \
\mathfrak{I'_2}=\int_1^\infty\rmd x\,\left[\cdots\right]
\ .
\end{equation}
For the second one we can integrate directly the first term, and
explicitly the counterterms and get
\begin{equation}
\label{I2} \mathfrak{I'_2}=-{1\over D}\left[{c_0\over
d_0}\right]^{{2D\over 2-D}}+\mathfrak{I''_2}\quad,\qquad
\mathfrak{I''_2}=\int_1^\infty\rmd x\,x^{D-1}\left[\left[1-{G(x)\over
c_0}\right]^{-{2D\over 2-D}}
			\!\!\!-1\right]
\ .
\end{equation}
For the first one it is better to over-subtract it, in order to improve
the integration at $x=0$ and the study of the large-$d$ limit, and
write
\begin{equation}
\label{Ip1}
\mathfrak{I'_1}=-{1\over D}-{D\over (2-D)^2(4-D)}\left[{c_0\over d_0}\right]^{{2D\over 2-D}}+\mathfrak{I''_1}
\end{equation}
\begin{equation}
\label{Is2} \mathfrak{I''_1}=\int_0^1\rmd
x\,\left[x^{D-1}\left[1-{G(x)\over c_0}\right]^{-{2D\over 2-D}}
			\!\!\!-\left[{c_0\over d_0}\right]^{{2D\over
			2-D}}x^{-D-1}
		-{1\over 2-D}\left[{c_0\over d_0}\right]^{{2+D\over
		2-D}}x^{-1}+{D\over (2-D)(4-D)}\left[{c_0\over
		d_0}\right]^{{2D\over 2-D}}x^{1-D}\right]
\ .
\end{equation}
Both integrals $\mathfrak{I''_1}$ and $\mathfrak{I''_2}$ are
convergent for any $D<2$ and have a smooth limit when $D\to 2$.  }}

\subsection{Do the limit $d\to\infty$ and the limit $\epsilon\to 0$
commute?}  
\label{ss:Dinfeps0} 
\subsubsection{An apparent paradox}
\label{sss:paradox} 
For $\epsilon>0$ the variational approximation
$L^{(0)}$ for $L$ has a regular large-$d$ limit. We have studied it
already in Sect.~\ref{ss:5.C.3}.  It is of order $\CO(d)$ and is given
by the convergent integral
\begin{equation}
\label{L0dinf} \mathfrak{L}^{(0)}=d\left(1-{1\over
4-\epsilon}\,\text{f.p.}\int_0^\infty\rmd
x\,x\,\left[\rme^{(4-\epsilon)K_0(x)}-1
\right]\right)
\ ,
\end{equation}
which is the large-$d$ limit of the integral \eq{L0intx} ($K_0(x)$ is
the Modified Bessel function of the 2nd kind).

This integral can be computed numerically.  To study its UV structure,
we use the small $x$ expansion for the 2D propagator
\begin{equation}
\label{ldr1} K_0(x)\ =\ -\log(x/2)-\gamma_E +{1\over
4}x^2\left(-\log(x/2) +1-\gamma_E\right)+\CO(x^4)
\end{equation}
The integrand in \eq{ldr1} behaves at small $x$ as
\begin{equation}
\label{ldr2} \left[{\rme^{\gamma_E}/
2}\right]^{\epsilon-4}\,\left(x^{\epsilon-3}+{4-\epsilon\over
4}x^{\epsilon-1}\left(-\log(x/2)-\gamma_E+1\right)+\CO(x^{1+\epsilon}\log
x)\right)\,-1
\ .
\end{equation}
The first term $x^{\epsilon-3}$ gives the UV pole at $\epsilon=2$, and
is subtracted by the f.p. prescription. The second term
$x^{\epsilon-1}$ gives the poles at $\epsilon=0$ but the $\log x$
gives in fact a double pole, so that
\begin{equation}
\label{ldr3} \mathfrak{L}^{(0)}\ \simeq\ -\,d\
4\rme^{-4\gamma_E}\left({1\over \epsilon^2}+{3\over 4
\,\epsilon}+\CO(1)\right)
\ .
\end{equation}
There seems to be a discrepancy between this calculation and the
results of the previous section:
\begin{itemize}
  \item Here we take the limit $d\to \infty$, then $\epsilon\to 0$;
        $\mathfrak{L^{(0)}}$
  has a UV pole $\propto {d/ \epsilon^2}$
  \item Previously we took the limit $\epsilon\to 0$, then
        $d\to\infty$; $\mathfrak{L^{(0)}}$ has a UV pole $\propto
        {d^2/ \epsilon}$
\end{itemize}
\textbf{Clearly the limits $\epsilon\to 0$ and $d\to\infty$ do not
simply commute.}

\subsubsection{Resolution of the paradox} \label{sss:respardx} This
apparent paradox can be understood if we use the results of the
previous section to study carefully how the bare quantity
$\mathfrak{L}^{(0)}=d-\tr[\JO^{(0)}]$ behaves when both $\epsilon\to
0$ and $d \to \infty$. $\mathfrak{L}^{(0)}$ is given by (\ref{L0intx})
and in that limit the dominant contribution is the integral
$\mathfrak{I}$, and more precisely the terms of order $x^{-1}$ when
$x\to 0$ in the integral \eq{Lf0sdexp} for $\mathfrak{I}$.  There are
two such terms, the dominant one of order $x^{\epsilon-1}$ (which will
give the UV pole at $\epsilon=0$), and the subdominant one of order
$x^{\epsilon+1-D}$. These two terms combine so that in the large-$d$,
small-$\epsilon$ limit, we have
\begin{equation}
\label{L0asympt} \mathfrak{L}^{(0)}\ \simeq\ -\,{d^2\over
\epsilon}\,+\, {d^2\over \epsilon+2-D}
\quad\text{with}\quad\epsilon=2D-{d\over 2}(2-D)\ .
\end{equation}

If we take the limit $\epsilon\to 0$ with $d$ finite and large (hence
$2-D$ small but non-zero) the first term is singular, and we recover the
standard single UV pole, while the second one stays finite.
Renormalization within the MS-scheme amounts to subtracting the first term
and we recover
\begin{equation}
\label{L0MSasy} \mathfrak{L}^{(0)}_{\text{\tiny{MS-D}}}\ \simeq\
{d^2\over 2-D}\ =\ {d^3\over 4D}
\ .
\end{equation} 
All the other terms contributing to $\mathfrak{L}^{(0)}_{\text{ren}}$
are at most of order $d^2$.  Thus we recover the fact that the
renormalized $\mathfrak{L}$ is of order $\CO(d^3)$.

If we now take the limit $d\to \infty$ with $\epsilon$ non-zero but
small, we rewrite \eq{L0asympt} as
\begin{equation}
\label{L0asy2} \mathfrak{L}^{(0)}\ \simeq\ -\,{d^2(2-D)\over
\epsilon(\epsilon+2-D)}\ =\ -{2d(2D-\epsilon)\over
\epsilon(\epsilon+2-D)}\ \simeq\ -\,{8d\over\epsilon^2}
\end{equation}
and we recover the fact that the bare $\mathfrak{L}$ is of order
$\CO(d)$ but with a double pole when $\epsilon\to 0$. Thus
\eq{L0asympt} contains both \eq{L0MSasy} and \eq{L0asy2}.

\subsubsection{Discussion} \label{sss:disc} Of course in the full
theory, it is the first limit ($\epsilon\to 0$ then $d\to\infty$)
which must be considered to study the large-$d$ behavior of the
renormalized theory.  At the level of the variational approximation
framework, from the previous calculations one can show that at
large-$d$ the variational renormalized $\log\det$
$\mathfrak{L}^{(0)}_{\text{\tiny{MS-D}}}$ has a regular large-$d$
asymptotic expansion in powers of $1/d$, 
\begin{equation}
\label{L0MSdexp} \mathfrak{L}^{(0)}_{\text{\tiny{MS-D}}}\ =\
\mathfrak{l^{0}_{0}}\,d^3+\mathfrak{l^{0}_{1}}\,d^2\,
+\,\mathfrak{l^{0}_{2}}\,d\,+\,
\mathfrak{l^{0}_{3}}\,d^0\,+\,\cdots
\end{equation}
with the $\mathfrak{l^{0}_{n}}$'s real and finite.

Indeed, setting $\epsilon=0$, starting from \eq{L0MS2}, using
equations \eq{II1I2}-\eq{Is2} and the fact that $c_0/d_0$ is analytic
in $1/d$ ($c_0/d_0=1+\CO(1/d)$), one sees that the only possible
non-analytic terms are the integrals $\mathfrak{I''_1}$ and
$\mathfrak{I''_2}$. Now the mass$=1$ propagator $G(x)$ is analytic in
$1/d\sim 2-D$, except at $x=0$, where it has a $\log(x)$ singularity
when $D=2$. At $x=\infty$ it behaves as $\exp(-x)$ for any $D$ and it
is then easy to see that $\mathfrak{I''_2}$ is analytic in $1/d$.  The
integral $\mathfrak{I''_1 }$ given by \eq{Is2} behaves at $D=2$ as
$\int_0\rmd x\,x\,\log x $ and its $n^{\text{th}}$ derivative with
respects to $D$ behaves as $\int_0 \rmd x\, x\,\log^n x$ and is
convergent for any $n$. Hence we deduce that $\mathfrak{I''_1 }$ too
has an (asymptotic) expansion in $2-D\sim 1/d$. Q.E.D.

Thus the variational renormalized theory at $\epsilon=0$ scales with
$d$ as $d^3$ (and not as $d$), but still is amenable by a $1/d$
expansion. As we shall see in the next section, the situation becomes
more complicated when we deal with the corrections to the variational
approximation. Indeed, the perturbative expansion studied in
Sect.~\ref{s:1/dcor} is plagued with infra-red (IR) divergences at
$\epsilon=0$, in addition to the UV divergences, and we shall argue
that this means that the renormalized theory contains non-analytic
terms such as $\log(d)$'s in the large-$d$ limit.

\subsection{Renormalized theory: first $1/d$ correction and IR divergences}
\label{ss:L1eps0}

\subsubsection{The IR divergences at $\epsilon=0$} \label{sss:IRd} In
section \ref{s:1/dcor} we have isolated the classes of diagrams in the
expansion of the kernel $\JQ$ which give a contribution of order
$\CO(1)$ in $\mathfrak{L}=\tr\log[\JU-\JQ]$.  This analysis is valid
provided that $\epsilon>0$. Indeed, as long as $\epsilon>0$ the
individual diagrams are IR and UV convergent, and the summation over
the diagrams in each class in also convergent.

If we now take the limit $\epsilon\to 0$, IR problems may occur when
summing these diagrams.

First we consider the diagrams that contribute to $\tr[\JQ^{(1)}]$,
depicted in \eq{trQ1gr}. In each of the 11 classes of diagrams in
\eq{trQ1gr} the sum over the $n$ lines joining the left to the right
contribute to a similar sum as the sum considered in
Sect.~\ref{ss:5.C.2} at leading order, i.e.\ to integrals of the form
\begin{equation*}
\label{ } \int_\xx \int_\kk \left(\rme^{-\kk^2
(c_0-G(\xx))}-\rme^{-\kk^2 (c_0)}\right)\,\times\cdots\ \simeq\
2c_0\int_\xx\left(\left[1-G(\xx)/c_0\right]^{-d/2}-1\right)\,\times\cdots
\end{equation*}
These integrals are UV and IR finite when $\epsilon>0$ (with a
finite-part prescription to deal with the singularity at $\xx=0$).
When $\epsilon=0$ they are still IR finite (the $\xx$-integration is
convergent at $|\xx|\to\infty$ since the propagator $G(\xx)$ is
massive, hence exponentially decaying at large distance).  On the
other hand, these integrals are UV divergent when $\epsilon=0$ (there
is a singularity at $\xx=0$ which gives a $1/\epsilon$ UV pole) but
this divergence is dealt by the renormalization procedure.

Now the second infinite sum in the 11 classes of \eq{trQ1gr} is given
by the ``chain-of-bubbles" propagator of the $1/d$ expansion
\begin{equation*}
\label{ }
H(\pp)\,=\, \includegraphics[scale=.5]{bbline0}\, 
=1+
\parbox{28.5pt}{\includegraphics[scale=.5]{onebubble}}
+\parbox{58pt}{\includegraphics[scale=.5]{twobubble}}+
\parbox{87pt}{\includegraphics[scale=.5]{threebubble}}+\cdots
\end{equation*}
given by \eq{chaine0} and depicted in
Fig.\ref{chainofbubble}. Combining the results of
sect.~\ref{ss:1/dexp}, in particular \eq{chaine0}, \eq{B2pex} and
\eq{4vdlim}, we easily obtain that in the limit $d\to\infty$,
$\epsilon$ finite, this propagator is
\begin{equation}
\label{Hexpl2}
H(\pp)\ =\ \left[1-\left(1-{\epsilon\over 4}\right)J(p)\right]^{-1}
\quad\text{with}\qquad J(p)\,=\,{2\over p}{\mathrm{arcsinh}(p/2)\over
\sqrt{1+p^2/4}}\,=\,\pi\,B(\pp)
\end{equation}
(we use the notations of \cite{DavidWiese1998} for $J(p)$, $B(\pp)$ is
the bubble amplitude \eq{B2pex} at $D=2$).  For large $p$ the UV
behavior of $H$ is
\begin{equation}
\label{Hplarge} H(p)\,\simeq\ 1+(4-\epsilon){\log p\over
p^2}+\cdots\quad\text{as}\quad p\to\infty
\end{equation}
and does not raise additional UV problems.  For small $p$ its IR
behavior is
\begin{equation}
\label{Hpsmall}
H(p)\,\simeq\ {1\over {p^2/ 6}+{\epsilon/ 4}}\quad\text{as}\quad p\to 0
\ .
\end{equation}
As long as $\epsilon>0$, the IR behavior of $H$ is that of a massive
scalar field with effective mass
\begin{equation}
\label{Hmeff}
m_{\text{eff}}=\sqrt{3\epsilon/2}\ \ ,
\end{equation}
to be compared with the variational mass $m_{\text{var}}=1$.  However,
when $\epsilon=0$, this propagator becomes massless $m_{\text{eff}}=0$
and since we are dealing with an effective theory in two dimensions
($D=2$), IR divergences occur! Indeed, in the diagrams of \eq{trQ1gr}
there are two sources of IR divergences:
\begin{enumerate}
  \item Firstly, the mass shift depicted by \raisebox{.3ex}
        {\includegraphics[scale=.33]{graph-13a}} is given by the
        solution of \eq{mu1eqdiag} which involves tadpole diagrams
        with the propagator $H(\pp)$ at zero momentum $\pp=0$, which
        gives potential powerlike IR divergences $\raisebox{.3
        ex}{\includegraphics[scale=.33]{graph-13a}}\propto H(0)$
        since $$H(0)={4\over\epsilon}\ .$$
  \item Secondly, both the tadpole diagrams in the r.h.s.\ of
        \eq{mu1eqdiag} and the other diagrams in \eq{trQ1gr} contain
        internal loops with the propagator $H(\pp)$. Integration over
        the internal loop momentum gives logarithmic IR divergences
        since
  $$\int_\pp H(\pp)\ =\ {3\over 2\pi}\,\log(1/\epsilon)\,+\,\cdots$$ 
\end{enumerate}

If we now consider the diagrams which contribute to $\tr[\JQ^{(k)}]$,
$k\ge 2$, depicted in \eq{domnecklace}, they also contain the
zero-momentum propagator $H(0)=4/\epsilon$. Their amplitudes at large
$d$ are given by \eq{tQk} and have a powerlike IR divergence in $1/
\epsilon^k$.

This IR problem was in fact first discovered by the authors in
\cite{DavidWiese1998}, and its significance for the calculation of the
instanton action studied.  It was shown in \cite{DavidWiese1998} that
these IR divergences exist for the instanton profile $V(\rr)$, but
cancel in the first $1/d$ correction to the instanton action
$\CS_{\mathrm{inst}}$.  As we shall see now, some partial
cancellations of IR divergences also occur in the contributions of the
fluctuations around the instanton, but the first $1/d$ correction
$\mathfrak{L}^{(1)}_{\mathrm{ren}}$ to the renormalized fluctuation
contribution $\mathfrak{L}_{\mathrm{ren}}$ is \textsl{still IR
divergent} at $\epsilon=0$.

\subsubsection{Cancellation of IR divergences in the mass shift
$\delta_1$} \label{d1IRdiv} We first look at the mass shift $\delta_1$
depicted by $\raisebox{.1ex}{\includegraphics[scale=.33]{graph-13a}}$
and solution of \eq{mu1eqdiag}. We have already computed $\delta_1$
in \cite{DavidWiese1998} and $\delta_1$ is in fact IR finite when
$\epsilon\to 0$.  We refer to sect.~6.5 and Appendix B of
\cite{DavidWiese1998} for the details of the calculation, the final
result being given by Eqs.~(135), (137)~\&~(138) of \cite{DavidWiese1998},
i.e.\ the integral
\begin{equation}
\label{d1intex}
\begin{split}
\delta_1\ &=\ {4-\epsilon\over 2-\epsilon}\,2\pi\int_\pp
\left(-\log\left[H(\pp)\right]-\left(1-{\epsilon\over
4}\right)J(\pp)\right)
\\
&=\ {4-\epsilon\over 2-\epsilon}\int_0^\infty \rmd p\,p\,
\left(-\log\left[1-\left(1-{\epsilon\over
4}\right)J(p)\right]-\left(1-{\epsilon\over 4}\right)J(p)\right)
\\
&=\ {4-\epsilon\over 2-\epsilon}\int_0^\infty \rmd v\,\sinh
v\,\left(-\log\left[1-\left(1-{\epsilon\over 4}\right){v \over \sinh
v}\right]-\left(1-{\epsilon\over 4}\right){v\over \sinh v}\right)
\\
&=\ 7.5583\dots  \quad\text{for}\quad\epsilon=0 \ .
\end{split}
\end{equation}
This integral is IR and UV convergent even for $\epsilon=0$, since it
behaves at small $\pp$ as $\int_\pp 1$ and at large $\pp$ as $\int_\pp
\pp^{-4}\log^2(\pp)$.

To prove the IR finiteness of $\delta_1$ it is sufficient to rewrite
\eq{mu1eqdiag} as
\begin{equation}
\label{d1eqrew} \raisebox{-.0
ex}{\includegraphics[scale=.33]{graph-13a}}\ =\raisebox{-2.5
ex}{\includegraphics[scale=.33]{graph-12a}}+ \raisebox{-2.5
ex}{\includegraphics[scale=.33]{graph-11a}}+ \raisebox{-2.5
ex}{\includegraphics[scale=.33]{graph-10a}} 
=
\raisebox{-.0 ex}{\includegraphics[scale=.33]{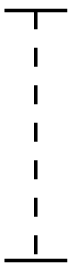}}\left[
\raisebox{-.0 ex}{\includegraphics[scale=.33]{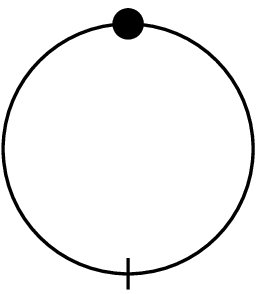}}
\,+\,
\raisebox{-.0 ex}{\includegraphics[scale=.33]{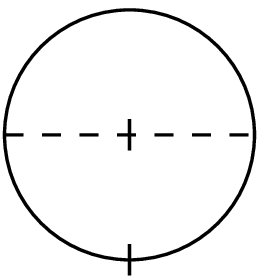}}
\,+\,
\raisebox{-.0 ex}{\includegraphics[scale=.33]{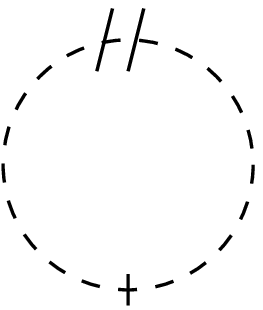}}
\right]
=
\raisebox{-.0 ex}{\includegraphics[scale=.33]{piece-1}}\left[
\raisebox{-.0 ex}{\includegraphics[scale=.33]{graph-13a}}
\ \times\raisebox{-.0 ex}{\includegraphics[scale=.33]{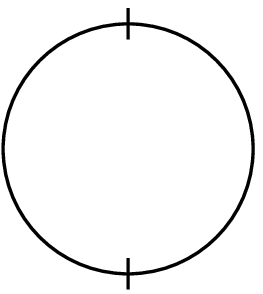}}
\,+\,
\raisebox{-.0 ex}{\includegraphics[scale=.33]{piece-3}}
\,+\,
\raisebox{-.0 ex}{\includegraphics[scale=.33]{piece-4}}
\right]
\ .
\end{equation}
Since no momentum flows through the first (vertical) $H$ line we have
\begin{equation*}
\label{d1eqII} \raisebox{-2
ex}{\includegraphics[scale=.33]{piece-1}}\ =\ {1\over 2
c_0}\,{4\over\epsilon} \ \text{, \qquad while}\qquad \raisebox{-2
ex}{\includegraphics[scale=.33]{piece-2}}\ =\ {d\over
2}B(0)\,=\,{d\over 2}{1\over 4\pi}
\end{equation*}
is finite, and the last two diagrams contain $\log(\epsilon)$ IR
divergences.  However, one can easily check that these IR divergences
cancel. Indeed the coefficient of the log is obtained by using that as long as $\int_{\pp} f(\pp)$ is not itself IR-divergent, 
\begin{eqnarray}
\int_{\pp} \raisebox{.5ex}{\includegraphics[scale=.33]{bbline0}} f(\pp) &=& \int_{\pp} H(\pp) \times f(0) + \mbox{infrared convergent terms} \nonumber\\
&=& \frac{3}{2\pi} \ln (1/\epsilon) f(0) + \mbox{infrared convergent terms} \ .
\end{eqnarray}
This means that any  $H$ line
$\raisebox{.5ex}{\includegraphics[scale=.33]{bbline0}}$ is to be replaced by $\frac3{2\pi} \log(1/\epsilon)$ and treated as if no momentum flows through it. We thus
obtain
\begin{equation*}
\label{d1eqIII} \raisebox{-2
ex}{\includegraphics[scale=.33]{piece-3}}\ =\ {3\over
2\pi}\,\log({1/\epsilon})\,{1\over 2c_0}\,\raisebox{-2
ex}{\includegraphics[scale=.33]{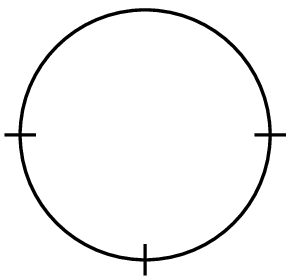}}\,+\,\CO(1)
\quad\text{with}\qquad\raisebox{-2
ex}{\includegraphics[scale=.33]{piece-6}}\,=\, {d\over 2}\,{1\over
8\pi}
\end{equation*}
while for the second graph
\begin{equation*}
\label{d1eqIV} \raisebox{-2
ex}{\includegraphics[scale=.33]{piece-4}}\ =\ -\,{3\over
2\pi}\,\log({1/\epsilon})\,{1\over 2}\,+\,\CO(1)\ .
\end{equation*}
The coefficient of the $\log(1/\epsilon)$ IR divergences 
is therefore zero, since
\begin{equation*} 
\label{d1eqV} \raisebox{-2
ex}{\includegraphics[scale=.33]{piece-3}}+\raisebox{-2
ex}{\includegraphics[scale=.33]{piece-4}}\ =\ {3\over
2\pi}\,\log({1/\epsilon})\,\left({d\over 4c_0}{1\over 8\pi}-{1\over
2}\right)\ \simeq\ -{3\over 16\pi}\,\epsilon\log({1/\epsilon})
\end{equation*}
Thus eq.~\eq{d1eqrew} for $\delta_1$ is of the form
\begin{equation}
\label{d1eqVI}
\delta_1
\ =\ {1\over\epsilon}\left[
\delta_1
+\CO(1)\right]\qquad\Rightarrow\qquad 
\delta_1
\,=\, \CO(1)\qquad\text{when}\qquad\epsilon=0
\ .
\end{equation}

\subsubsection{IR divergences in $\JQ^{(1)}$} \label{Q1IRdiv} We now
perform the same analysis for the first $1/d$ correction to the
Hessian $\JQ$, $\JQ^{(1)}$, calculated in Sect.\ref{s:1/dcor}.  The
Fourier transform of $\JQ^{(1)}$ is given by the graphs of
\eq{Qhat1}. For reasons that will become clear later, let us separate
$\JQ^{(1)}$ into 4 parts
\begin{equation}
\label{Qhat1sum} \widehat{\JQ}^{(1)}\ =\
\widehat{\JQ}^{(1a)}+\widehat{\JQ}^{(1b)}+\widehat{\JQ}^{(1c)}+\widehat{\JQ}^{(1d)}
\end{equation}
$\widehat{\JQ}^{(1a)}$ is the sum of the graphs which contain the mass
shift $\raisebox{-.0 ex}{\includegraphics[scale=.33]{graph-13a}}$
\begin{equation}
\label{Qhat1a} \widehat{\JQ}^{(1a)}\ =\ 2\raisebox{-2.5
ex}{\includegraphics[scale=.33]{graph-O-1e}}
+
2\,\raisebox{-2.5
ex}{\includegraphics[scale=.33]{graph-O-1f}}
+
2\,\raisebox{-2.5
ex}{\includegraphics[scale=.33]{graph-O-13b}}
\end{equation}
$\widehat{\JQ}^{(1b)}$ is the sum of the graphs which are ``really
irreducible"
\begin{equation}
\label{Qhat1b}
\begin{split}
\widehat{\JQ}^{(1b)}\ & =\ 
\raisebox{-2.5 ex}{\includegraphics[scale=.33]{graph-O-2e}}
+
\raisebox{-2.5 ex}{\includegraphics[scale=.33]{graph-O-2f}}
+
\raisebox{-2.5 ex}{\includegraphics[scale=.33]{graph-O-6e}}
+
\raisebox{-2.5 ex}{\includegraphics[scale=.33]{graph-O-6f}}
\\
&
+
\raisebox{-2.5 ex}{\includegraphics[scale=.33]{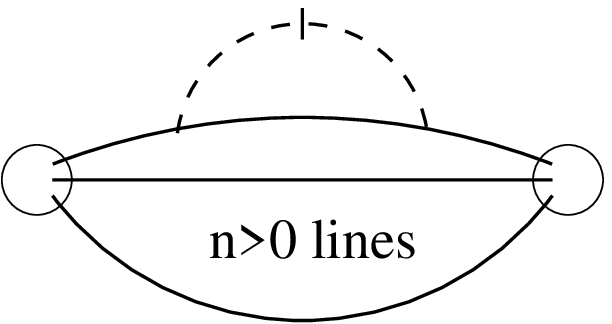}}
+
\raisebox{-2.5 ex}{\includegraphics[scale=.33]{graph-O-15}}
+
\raisebox{-2.5 ex}{\includegraphics[scale=.33]{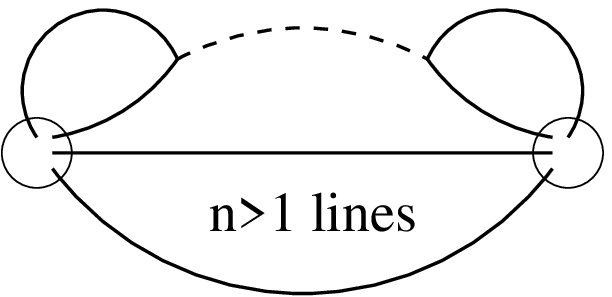}}
+
\raisebox{-2.5 ex}{\includegraphics[scale=.33]{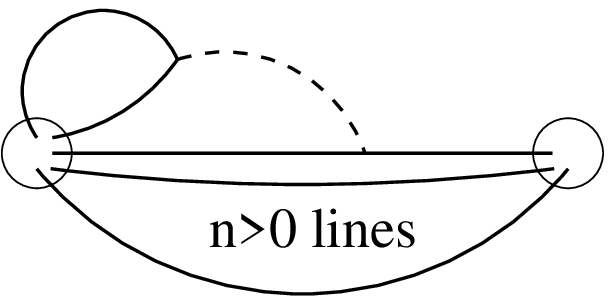}}
+
\raisebox{-2.5 ex}{\includegraphics[scale=.33]{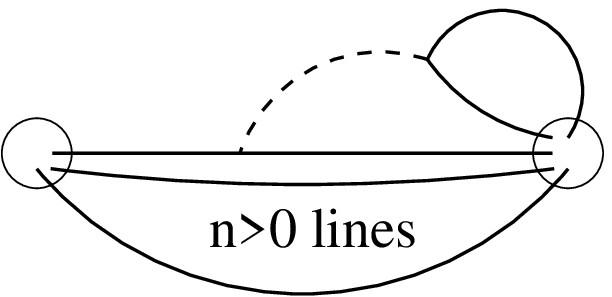}}
\end{split}
\end{equation}
$\widehat{\JQ}^{(1c)}$ is the sum of the 4 graphs 
\begin{equation}
\label{Qhat1c}
\begin{split}
\widehat{\JQ}^{(1c)}\ & =\ 
\raisebox{-.5 ex}{\includegraphics[scale=.33]{graph-O-14-A}}
+
\raisebox{-.5 ex}{\includegraphics[scale=.33]{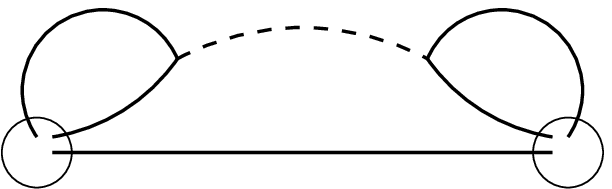}}
+
\raisebox{-.5 ex}{\includegraphics[scale=.33]{graph-O-17c}}
+
\raisebox{-.5 ex}{\includegraphics[scale=.33]{graph-O-17a}}
\end{split} 
\end{equation}
and $\widehat{\JQ}^{(1d)}$ is the single remaining graph.
\begin{equation}
\label{Qhat1d} \widehat{\JQ}^{(1d)}\ =\ \raisebox{-.5
ex}{\includegraphics[scale=.33]{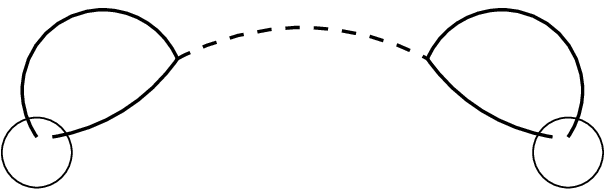}} \ .
\end{equation}

$\widehat{\JQ}^{(1a)}$ is IR finite since the mass shift
$\raisebox{-.0 ex}{\includegraphics[scale=.33]{graph-13a}}$ is IR
finite.
\begin{equation}
\label{Q1aDiv}
\widehat{\JQ}^{(1a)}\ =\ \CO(1)\ .
\end{equation}
$\widehat{\JQ}^{(1b)}$ and $\widehat{\JQ}^{(1c)}$ have a logarithmic
IR divergence in $\log(1/\epsilon)$. By the same argument as above,
the coefficient of the IR divergence is obtained by removing the $H$
propagator in the graph, so that
\begin{equation}
\label{Q1bDiv} \widehat{\JQ}^{(1b)}\ =\ {1\over 2 c_0}\,{3\over
2\pi}\,\log(1/\epsilon)\,\widehat{\mathbb{D}}^{(1b)}\,+\, \CO(1)
\end{equation}
with
\begin{equation}
\label{D1b}
\begin{split}
\widehat{\mathbb{D}}^{(1b)}\ & =\ 
\raisebox{-2.5 ex}{\includegraphics[scale=.33]{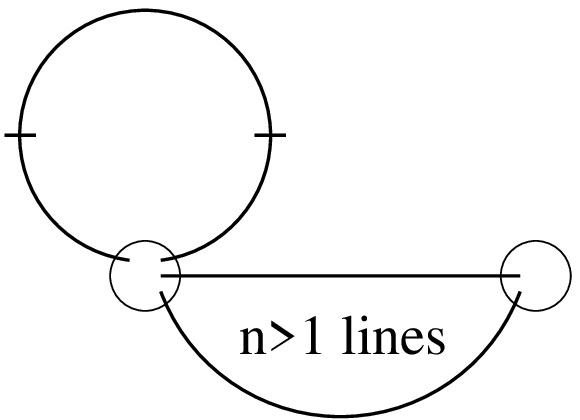}}
+
\raisebox{-2.5 ex}{\includegraphics[scale=.33]{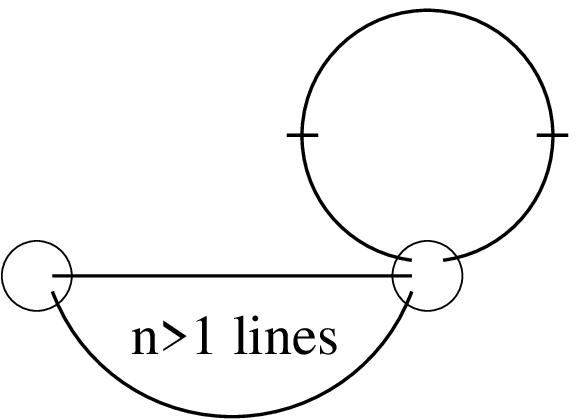}}
+
\raisebox{-2.5 ex}{\includegraphics[scale=.33]{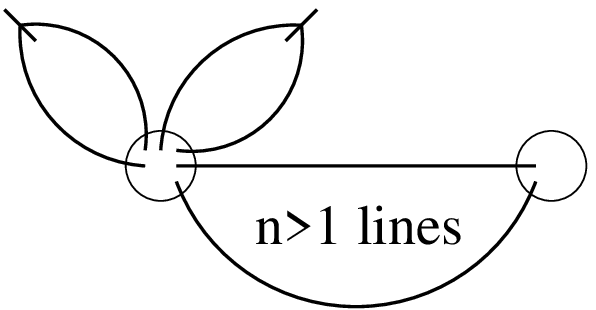}}
+
\raisebox{-2.5 ex}{\includegraphics[scale=.33]{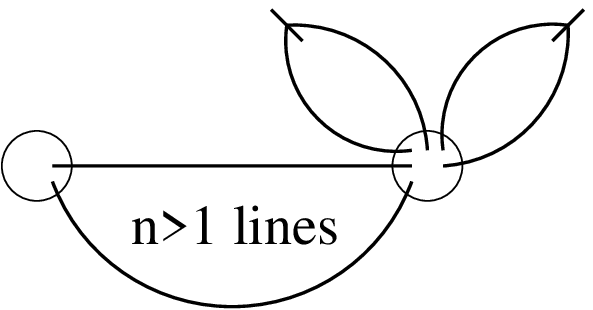}}
\\
&
+
\raisebox{-2.5 ex}{\includegraphics[scale=.33]{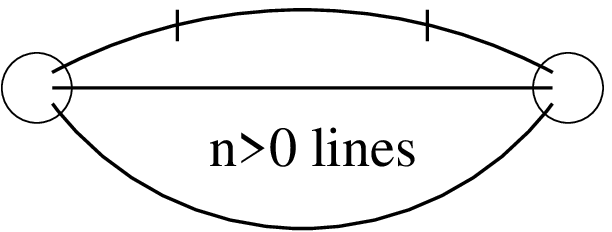}}
+
\raisebox{-2.5 ex}{\includegraphics[scale=.33]{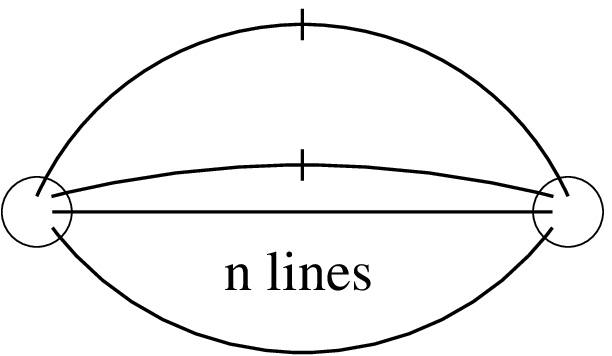}}
+
\raisebox{-2.5 ex}{\includegraphics[scale=.33]{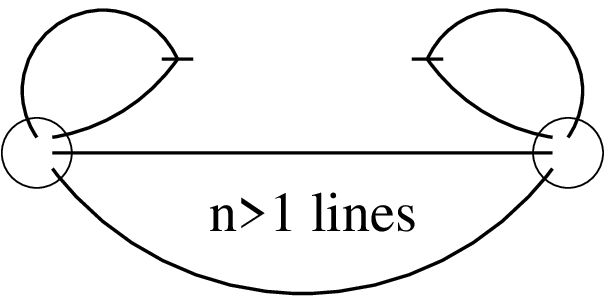}}
+
\raisebox{-2.5 ex}{\includegraphics[scale=.33]{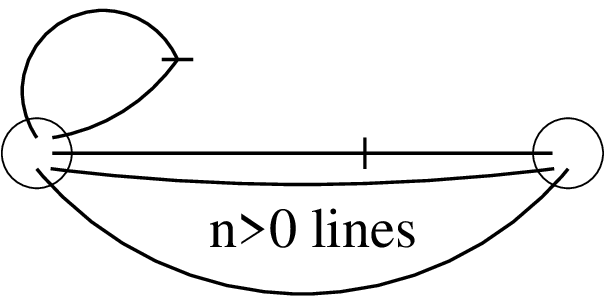}}
+
\raisebox{-2.5 ex}{\includegraphics[scale=.33]{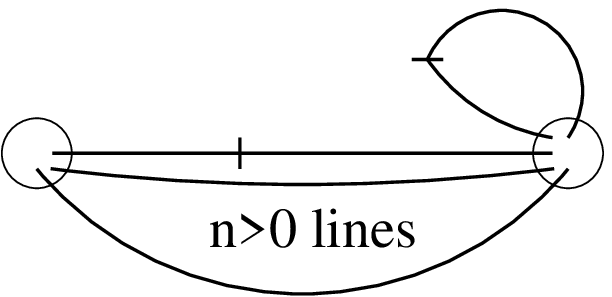}}
\end{split}
\ .
\end{equation}
Similarly for $\widehat{\JQ}^{(1c)}$ we have 
\begin{equation}
\label{Q1cDiv} \widehat{\JQ}^{(1c)}\ =\ {1\over 2 c_0}\,{3\over
2\pi}\,\log(1/\epsilon)\,\widehat{\mathbb{D}}^{(1c)}\,+\, \CO(1)
\end{equation}
with
\begin{equation}
\label{D1c}
\begin{split}
\widehat{\mathbb{D}}^{(1c)}\ & =\ 
\raisebox{-.5 ex}{\includegraphics[scale=.33]{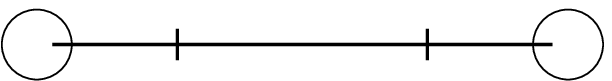}}
+
\raisebox{-.5 ex}{\includegraphics[scale=.33]{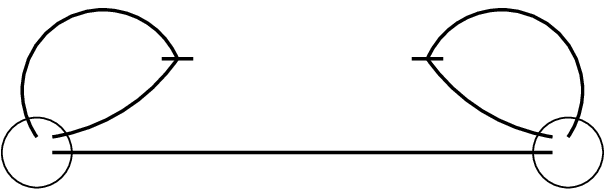}}
+
\raisebox{-.5 ex}{\includegraphics[scale=.33]{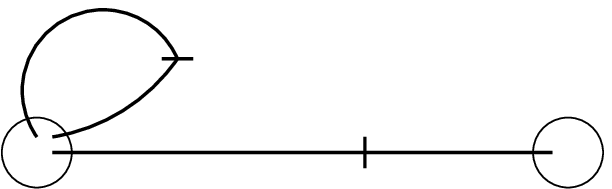}}
+
\raisebox{-.5 ex}{\includegraphics[scale=.33]{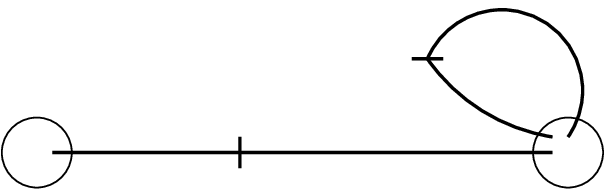}}
\end{split} 
\ .
\end{equation}
Finally $\widehat{\JQ}^{(1d)}$ has an IR pole in $1/\epsilon$, since
the bubble propagator carries zero momentum,
\begin{equation}
\label{Q1dDiv}
\widehat{\JQ}^{(1d)}\ =\ \CO(1/\epsilon)
\ .
\end{equation}

\subsubsection{Partial IR cancellations in $\tr[\JQ^{(1)}]$}
Now come the IR cancellations in $\tr[\JQ^{(1)}]$. We first consider
the second term.  We notice that the diagrams in \eq{D1b}, which
contribute in $\widehat{\mathbb{D}}^{(1b)}$, are  obtained by two
mass insertions in the diagrams which contribute to
$\widehat{\JQ}^{(0)}$, 
\begin{equation}
\label{Q0diag2} \widehat{\JQ}^{(0)}\ =\ \raisebox{-.5
ex}{\includegraphics[scale=.33]{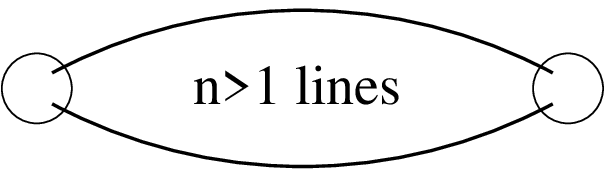}}
\end{equation}
and more explicitly, since a mass insertion corresponds to (minus) a
derivative with respect to $m^2$ (where $m$ is the variational mass in
the propagators)
\begin{equation}
\label{D1bQ0} \widehat{\mathbb{D}}^{(1b)}_{\kk_1 ,\kk_2}\ =\
\left.{{1\over 2}{\partial^2\over \partial {(m^2)}^2}\,
\left({\int_\xx \left<\rme^{\rmi\kk_1\rr(\oo)}\,
\rme^{\rmi\kk_2\rr(\xx)}\right>_m^{\mathrm{conn}}-d}\right)}\right|_{m=1}
\,=\, \left.{{1\over 2}{\partial^2\over \partial
{(m^2)}^2}\,\widehat{\mathbb{Q}}^{(0)}_{\kk_1 ,\kk_2} }\right|_{m=1} \
.
\end{equation}
The effect of the derivative on the propagators from $\oo$ to $\xx$ is
easy to understand. The tadpoles with one or two mass insertions are
generated by the derivative acting on the circle \eq{circlestar} at
$\oo$ or $\xx$.

The kernel $\widehat{\mathbb{D}}^{(1b)}$ is clearly non-zero, but it
is traceless for $\epsilon=0$. Indeed,
\begin{equation}
\label{trD1b} \tr\left[\widehat{\mathbb{D}}^{(1b)}\right]\ =\ \
\left.{{1\over 2}{\partial^2\over \partial {(m^2)}^2}\,
\tr\left[\JQ^{(0)}\right]}\right|_{m=1}
\end{equation}
and $\tr\left[\JQ^{(0)}\right]$ scales with the mass $m$ like
\begin{equation}
\label{trQ0m} \tr\left[\JQ^{(0)}\right]\ =\
m^{D-\epsilon}\,\left.{\tr\left[\JQ^{(0)}\right]}\right|_{m=1} \ .
\end{equation}
Therefore, since in the large-$d$ limit, $D=2$, we have
\begin{equation}
\label{trD1bfin} \tr\left[{\mathbb{D}}^{(1b)}\right]\ =\
-{\epsilon(2-\epsilon)\over 8}\,\tr\left[\JQ^{(0)}\right]
\end{equation}
and formally\footnote{One must be cautious since
$\tr\left[\JQ^{(0)}\right]$ has an UV pole at $\epsilon=0$, so there
is a mixture of IR and UV singularities, that we shall discuss later.}
$\tr\bigl[{\mathbb{D}}^{(1b)}\bigr] =0$ when $\epsilon=0$.

Similarly we can compute the IR coefficient
$\widehat{\mathbb{D}}^{(1c)}$ for the 3rd term and its trace.  We
obtain easily the explicit result
\begin{equation}
\label{trD1cfin} \tr\bigl[{\mathbb{D}}^{(1c)}\bigr]\ =\
4\pi\,c_0\,{\epsilon^2(1-\epsilon/4)}\ =\ d\,\epsilon^2 /4 \ .
\end{equation}
It is also zero when $\epsilon\to 0$.
\begin{description}
    \item[Details of the calculation:]
We have
\begin{equation*}
\label{D1cexpl}
\begin{split}
\raisebox{-.5 ex}{\includegraphics[scale=.33]{graph-O-16-f}}&= {1\over
4}\,\left({1\over 4\pi}\right)^2\,(-\kk_1^2) (-\kk_2^2)(-\kk_1\kk_2)
\,\rme^{-(\kk_1^2+\kk_2^2)c_0/2}
\\
\raisebox{-.5 ex}{\includegraphics[scale=.33]{graph-O-17-g}}&= {1\over
2}\,\left({1\over
4\pi}\right)\,(-\kk_1^2)(-\kk_1\kk_2)\,\rme^{-(\kk_1^2+\kk_2^2)c_0/2}
\\
\raisebox{-.5 ex}{\includegraphics[scale=.33]{graph-O-17-h}}&= {1\over
2}\,\left({1\over
4\pi}\right)\,(-\kk_2^2)(-\kk_1\kk_2)\,\rme^{-(\kk_1^2+\kk_2^2)c_0/2}
\\
\raisebox{-.5 ex}{\includegraphics[scale=.33]{graph-O-14-f}}&=
(-\kk_1\kk_2)\,\rme^{-(\kk_1^2+\kk_2^2)c_0/2}
\end{split} 
\end{equation*}
we integrate over $\kk_1=-\kk_2$ to obtain the trace and we use the
fact that $${d\over 4c_0}=4\pi(1-\epsilon/4)$$ Q.E.D.
\end{description}
The final result is therefore that the IR divergence of
$\tr[\JQ^{(1)}]$ comes only from the last single diagram in
\eq{Qhat1d}, which gives $\JQ^{(1d)}$! It is a single IR pole in
$1/\epsilon$, since
\begin{equation}
\label{Q1dexpl} \widehat{\JQ}^{(1d)}_{\kk_1 ,\kk_2}={1\over
4}\left({1\over 4\pi}\right)^2{1\over 2 c_0}{4\over\epsilon}
({-\kk_1^2})( {-\kk_2^2})\rme^{-(\kk_1^2+\kk_2^2)c_0/2}
\end{equation}
hence
\begin{equation}
\label{trQ1dexpl} \tr[\JQ^{(1d)}]\ =\
{4\over\epsilon}\left(1-{\epsilon\over 4}\right)^2 \ .
\end{equation}

\subsubsection{IR divergence and the unstable mode}
\label{ss:IRlambda-} We now look at the IR singularity in the other
terms of the expansion for $\mathfrak{L}$,
$$\mathfrak{L}=\tr\log[\JU-\JQ]=-\sum_{k=1}^\infty {1\over
k}\,\tr[\JQ^k]\ .$$
 We have shown in section \ref{ss:thedet} that the
next terms $\tr[\JQ^k]$ are of order $\CO(1)$ and can be computed
explicitly at that order, since they are given by the contribution of
two diagrams only (see \eq{dombeads}) in $\JQ$, namely the single
bubble diagram
$\raisebox{-2.5ex}{\includegraphics[scale=.5]{Q-dom-a}}$, which is IR
finite, and the diagram
$\raisebox{-2.5ex}{\includegraphics[scale=.5]{Q-dom-b}}$, which is
nothing but the IR divergent diagram of \eq{Qhat1d}.  Thus we see that
all the IR divergences of $\mathfrak{L}^{(1)}$, i.e.\ the term of
order $\CO(1)$ in the $1/d$ expansion of $\mathfrak{L}$ are contained
in the last term of \eq{Lresum1}, namely in the summation of the
$\log$ series
\begin{equation}
\label{L1IR} \mathfrak{L}^{(1)}\ \simeq\
\log\left(2-{4\over\epsilon}\right)\,+\,\text{IR finite (but UV
divergent) term when } \epsilon\to 0 \ .
\end{equation}
Now we have shown in Sect.~\ref{ss:unstmode} that this IR singular
$\log(2-4/\epsilon)$ is nothing but the contribution of the smallest
(and negative) eigenvalue of the Hessian $\CS''[V]$ associated to the
unstable eigenmode (dilation) for the instanton
\begin{equation}
\label{lmindinf2}
\lambda_{\mathrm{min}}=2-{4\over\epsilon}\,<\,0
\ .
\end{equation}
The conclusion of our analysis of the IR divergence of
$\mathfrak{L}=\log\det ' [\CS '']$ is that, at least at order $1/d$,
the IR divergence can be attributed entirely to the contribution of
the smallest eigenvalue. This is in fact quite natural, since IR
divergences must come from the large distance properties of the
fluctuations around the instanton configuration.

\subsubsection{A conjecture for the large-$d$ behavior of the unstable
mode} 
\label{sss:discIR}

This IR divergence in our large-$d$ estimate of the negative
eigenvalue $\lambda_{\mathrm{min}}$ for the instanton Hessian does not
mean that $\lambda_{\mathrm{min}}$ is IR singular when $\epsilon=0$,
but rather that $\lambda_{\mathrm{min}}$ does not behave in the same
way when $d\to\infty$, depending on whether $\epsilon>0$ or $\epsilon=0$.
\begin{itemize}
  \item If $\epsilon>0$, we have seen that
        $\lambda_{\mathrm{min}}=\CO(1)$ when $d\to\infty$.
  \item If $d$ is finite and we take the limit $\epsilon\to 0$, we
        have also $\lambda_{\mathrm{min}}=\CO(1)$, as can be checked
        explicitly for the case $d=4, \epsilon=0$, where we recover the
        classical $\phi^4_4$ instanton for the O($n=0$) model for SAW.
  \item Therefore we expect that the IR pole in \eq{lmindinf2} means
        simply that when we first take $\epsilon=0$, then
        $d\to\infty$, $\lambda_{\mathrm{min}}$ is no more of order
        $\CO(1)$, but becomes infinite
        ($\lambda_{\mathrm{min}}\to\infty$).
 \end{itemize}
  In fact we conjecture that for the renormalized theory,
  $\lambda_{\mathrm{min}}$ scales as $d$ \begin{equation}
\label{lminscd}
\lambda_{\mathrm{min}}(\epsilon=0,d)\,\simeq\,\CO(d)\quad\text{when}\quad
d\to\infty
\end{equation}
by analogy with the behavior of the leading term $\mathfrak{L^{(0)}}$
which is found to behave as
\begin{equation}
\label{ deps2d2eps}
\begin{split}
    {d\over\epsilon}\,\cdot\,{1\over \epsilon}  &\quad   \text{when}\quad d\to\infty\quad\text{then}\quad\epsilon\to 0 \\
  {d\over\epsilon}\,\cdot\,{d} &\quad \text{when}\quad \epsilon\to 0
  \quad\text{then}\quad d\to\infty
\end{split}
\end{equation}
The first ${d\over\epsilon}$ being an UV pole, and only the last
${1\over \epsilon}\sim d$ being IR.  Even if this form of the
conjecture is not correct, it is clear that once again for the
unstable mode the limits $\epsilon\to 0$ and $d\to\infty$ do not
commute.

\section{Conclusion}
In this paper we have shown how to compute at one loop the
fluctuations around the instanton in the self-avoiding manifold model,
and how this is related to the normalization for the large order
asymptotics for the SAM model.  We have shown that the perturbative
counterterms which make the SAM model UV finite in perturbation theory
do renormalize (at one loop) the instanton contribution.  We have
constructed a systematic $1/d$ expansion, and studied the first terms
of this expansion and the interplay between the $1/d$ expansion and
renormalization.

Although we have obtained many results in this article, and checked at
one loop the consistency of the instanton calculus for the SAM model,
several points deserve further studies:
\begin{itemize}
  \item It would be interesting to get a better understanding of how
        to resum the IR divergences present in the $1/d$ expansion for
        the renormalized theory at $\epsilon=0$, or to find another
        approximation scheme which does not suffer from IR divergences.
  \item We have checked that the instanton factor obtained by our
        method is for $D=1$ (self-avoiding walk) equal to the factor
        obtained by field theoretical methods. However, it would be
        interesting in this case to compare the approximate result
        that we obtain via the large-$d$ limit with the exact result
        (as was done for the instanton action in
        \cite{DavidWiese1998}).
    \item A practical application of the theoretical results obtained in
        this paper would be to compare our large-order asymptotics
        with our explicit calculations at 2-loop order for the scaling
        exponents for the SAM \cite{DavWie96,WieDav97}. Since the
        non-perturbative effects become small when $d$ is large, it is
        expected (and checked numerically) that the 2-loop estimates
        for the critical exponents are reliable for large $d$. Such a
        study would help our understanding of the domain of validity
        of the 2-loop calculations, and perhaps suggest better
        resummation procedures than those used previously.
    \item For renormalized local field theories, in addition to
        instantons, other contributions occur in the large-order
        asymptotics, denoted renormalons. They are associated both to
        the short-distance behavior of the theory (UV renormalons) and
        to its large-distance behavior (IR renormalons). We expect
        that such effects occur also for the SAM model at
        $\epsilon=0$, since for $D=1$ it is equivalent to the $\phi^4$
        theory, but it is not known how to treat these renormalon
        effects (if they are present) in the framework of the SAM
        model, which is a multi-local theory.
  \end{itemize}
%\acknowledgements{The authors are very indebted to themselves for their patience.}

\newpage
%\newpage

\newpage
\appendix
\section{Measure and normalizations for the functional integral}
\label{section-measures}

In this appendix we precise the normalization for the functional
integration over the fields and the treatment of the zero modes.
 
\subsection{DeWitt metric and measure for the functional integral} We
consider the free membrane model.  The functional measure
$\mathcal{D}[r]$ is normalized as follows. We start from the DeWitt
metric $G$ over the manifold configuration space
$\mathcal{C}=\{\rr(\vx)\}$
\begin{equation}\label{Metric4r} 
G(\delta\rr,\delta \rr) \ =\ {\mu_{0}^{2}\over 2\pi
}\,\int_{\mathcal{M}} \rmd^D{\vx}\,|\delta \rr(\vx) |^2 \ =\
{\mu_{0}^{2}\over 2\pi}\, \|\delta \rr\|^{2}_{\scriptscriptstyle 2}\ .
\end{equation} 
$\|\dots \|_{\scriptscriptstyle 2}$ is the $L_{2}$ norm
over $\mathcal{M}$.  This metric depends explicitly on an (arbitrary)
normalization mass scale $\mu_{0}$.

The corresponding measure is defined (formally)
by $\CD[\rr]=\prod_{\vx}\rm\rmd^d\rr(\vx)\,\sqrt{\det G}$.  This
corresponds to the normalization
\begin{equation}\label{measure4rr}    
\int \CD[\rr]\,\exp\left(-{\mu_{0}^{2}\over
2}\int_{\mathcal{M}\!\!\!}\rmd^D\vx\,\rr(x)^{2}\right)\ =\ 1\ \ .
\end{equation}
With this normalization, a quadratic form $\mathbf{A}$ with kernel
$A^a_{\ b}$, i.e.\ $(\mathbf{A}\rr(\vx))^{a}=\int \rmd^D\mathbf{\vy}A^a_{\
b}(\vx,\vy)\rr^b(\vy)$, yields
\begin{equation}\label{measure4rrb}    
\int \CD[\rr]\,\exp\left(-{1\over
2}\int_{\mathcal{M}}\!\!\!\rmd^D\vx\int_{\mathcal{M}}\!\!\!\rmd^D\vy\,
\rr(x)\mathbf{A}(\vx,\vy)\rr(\vy)\right)\ 
=\ \det\left[\mathbf{A}/\mu_0^2\right]^{-1/2}\ \ .
\end{equation} To evaluate the partition function for the free
membrane
$$Z_0\ =\ \int\CD[\rr]\,\exp\left(-{1\over
2}\int_\CM\!\!\rmd^D\vx\,\nabla\rr(\vx)^2\right)\ ,$$ we must treat
separately the zero modes $\rr_0(x)=\rr_0$ of the scalar Laplacian
$\Delta_\vx$ over $\CM$ and the fluctuations $\widetilde{\rr}$
orthogonal to the zero mode, $G(\rr_0,\widetilde{\rr})=0$.  Let
$G^{(0)}$ be the metric for the collective coordinate $\rr_0$ of the
zero mode induced on the ``moduli space" of minima of the action
$\rr(\vx)=\rr_0$ by the DeWitt metric
\begin{equation}
\label{app-measure1} G(\delta\rr_0,\delta\rr_0)\ =\ {\mu_0^2\over
2\pi}\int_\CM\!\! \rmd^D \vx\, |\delta\rr_0|^2 \ =:\
G_{ab}^{(0)}\delta\rr_0^a\delta\rr_0^b \quad\Rightarrow\quad
G_{ab}^{(0)}\,=\,{\mu_0^2\over 2\pi}\text{Vol}(\CM)\delta_{ab} \ .
\end{equation}
Hence the measure is 
\begin{equation}
\label{app-measure2} \rmd\mu(\rr_0)\ =\
\rmd^d\rr_0\,\sqrt{\det\bigl(G_{ab}^{(0)}\bigr)}\ =\ \rmd^d\rr_0\,
\left[{\mu_0^2\over 2\pi}\text{Vol}(\CM)\right]^{d/2} \ .
\end{equation}
The integration over the modes $\widetilde{\rr}$ orthogonal to the
zero modes gives
\begin{equation}
\label{app-measure3}
\left({\det}'\left[-\Delta/\mu_0^2\right]\right)^{-d/2}
\ ,
\end{equation}
where ${\det}'$ is the reduced determinant, that is the product over
the non-zero eigenvalues of the operator $-\Delta/\mu_0^2$. Hence
\begin{equation}
\label{app-measure4} Z_0\ =\ \int
\rmd\mu[\rr_0]\,\left({\det}'\left[{-\Delta/\mu_0^2}\right]\right)^{-d/2}
\ =\ \int \rmd^d\rr_0\ \mathcal{ Z}_0
\end{equation}
with the partition function for the marked manifold $\mathcal{ Z}_0$
\begin{equation}
\label{app-measure5} \mathcal{ Z}_0\ =\
\left({{\det}'\left[{-\Delta\over\mu_0^2}\right]{2\pi\over\mu_0^2\,
\text{Vol}(\CM)}}\right)^{-d/2}\ .
\end{equation}

\subsection{Zeta-function regularization} 
The ${\det}'$ requires UV
regularization for its definition. We use the standard zeta-function
regularization (see for instance \cite{GBU} and \cite{Dup87}).
\begin{equation}
\label{app-measure6}
\log({\det}'[-\Delta/\mu_0^2])\ =\ {\tr}'(\log[-\Delta/\mu^2_0])
\ =\ -\,\left.{\rmd\over \rmd s}\zeta(s)\right|_{s=0}
\ ,
\end{equation}
where the zeta-function $\zeta(s)$ for the operator
$A=-\Delta/\mu^2_0$ is defined by the sum over the non-zero
eigenvalues $\lambda_i$ 
\begin{equation}
\label{app-measure7}
\zeta_A(s)\ =\ \sum_{{\begin{smallmatrix}
\lambda_i\neq 0
\end{smallmatrix}
}}\lambda_i^{-s}
\end{equation}
for $\mathrm{Re}(s)$ large enough, and by analytic continuation down
to $s=0$.  ${\tr}'$ means the trace over the subspace orthogonal to
$\mathrm{Ker}(A)$ ( w.r.t.\ to the metric $G$).

The operator $-\Delta$ scales with the internal size $L$ of the
manifold $\CM$ as $L^{-2}$.  Therefore $\zeta(s)$ scales as
\begin{equation}
\label{app-measure8}
\zeta(s)\ =\ (L\mu_0)^{2s}\,\bar\zeta(s)
\ ,
\end{equation}
where $\bar\zeta(s)$ is a scale invariant zeta function which depends
on the shape of $\CM$ but not on its size $L$.

If there is  no global conformal anomaly, $\zeta(s)$ is analytic
around $s=0$ and $\zeta'(0)=\bar\zeta'(0)+2\log(L\mu_0)\zeta(0)$.
Moreover, for any such $A$, one has
\begin{equation}
\label{app-measure9}
\zeta_A(0)\ =\ -\,\mathrm{dim}(\mathrm{Ker}(A))\ =\ -\,\text{number of
zero modes of}\ A 
\ .
\end{equation}
Indeed, one can show that if $A$ has no zero-mode, for instance
$A=-\Delta+m^2$, then $\zeta_A(0)=\tr(1)=0$ {(this is analogous to the
celebrated rule $\delta(0)=0$ in dimensional regularization)}, and if
$A$ has $N$ zero modes, $\zeta_A(s)=\lim_{\epsilon\to
0}[\zeta_{A+\epsilon}(s)-N\epsilon^{-s}]$, therefore
$\zeta_A(0)=\lim_{\epsilon\to 0}[\zeta_{A+\epsilon}(0)-N]=-N$.  The
Laplacian $\Delta$ has one zero mode, and therefore
\begin{equation}
\label{app-measure10}
\zeta(0)\ =\ -1\quad,\quad
\zeta'(0)\ =\ \bar\zeta'(0)-2\log(L\mu_0)
\ .
\end{equation}
Using the fact that the size of the manifold is defined as
\begin{equation}
\label{app-measure11}
L\ =\ \mathrm{Vol}(\CM)^{1/D}
\end{equation}
we obtain for the partition function
\begin{equation}
\label{app-measure12}
\CZ_0\ =\ L^{d(2-D)/2}\left[{\mathrm{e}^{\bar\zeta'(0)}\over 2\pi}\right]^{d/2}
\ .
\end{equation}
The dependence on the mass scale $\mu_0$ used to define the measure
$\CD[\rr]$ has disappeared, as expected in the absence of a conformal
anomaly.  \subsection{Conformal anomaly} It is known that there is no
conformal anomaly if
\begin{enumerate}
\item $D=1$ and the manifold has no boundary (closed loop). This
corresponds to a ring polymer.  
\item $D=2$, the manifold has no boundary and has Euler
characteristics $\chi=0$. This corresponds to a closed membrane with
the topology of a torus (or a Klein bottle).
\item $D$ non-integer. The model is defined by dimensional
regularization, as detailed in \cite{GBU}. This is the relevant case
for the $\epsilon$-expansion.
\end{enumerate}
If there is a conformal anomaly, $\zeta(0)\neq -1$ and there is an
additional power of $L\mu_0$ in the partition function $\CZ_0$, which
depends explicitly on the scale $\mu_0$.  For instance for $D=2$
(membrane) it is known that
\begin{equation}
\label{app-measure13}
\zeta	(0)\ =\ -1\,+{c\over 6}\chi \quad,\quad\text{with} \ c=1\
\text{the central charge for the free boson} 
\ .
\end{equation}
Hence
\begin{equation}
\label{app-measure14}
\CZ_0\ \propto\ (L\mu_0)^{\chi\,d/6}\ .
\end{equation}

\section{Integration paths for the functional integration over $V[r]$}
\label{section-contours}
\renewcommand{\arg}{{\mathrm{Arg}}}
\newcommand{\vinst}{V^{\mathrm{inst}}}  In this appendix we discuss
in more detail via the steepest descent method the functional
integration over $V(\rr)$ and the relative position of the instanton
$V^{\mathrm{inst}}$ and of the integration contour over $V$, as the
argument $\theta$ of the coupling constant $b$ varies in $[-\pi,\pi]$.
This is required to treat properly the contribution of the unstable
eigenmode of the Hessian $\CS''[V]$ for the instanton.

For general $\theta\in[-\pi,\pi]$ we know from (\ref{rescaledDV}) that the
functional integral for the rescaled potential $V(\rr)$ is normalized
so that
\begin{equation}
\label{normDVb}
\int\CD[V]\,\exp\left({\rme^{-\rmi\theta}\over
2g}\int\rmd^d\rr\,V(\rr)^2\right)\ =\ 1 
\end{equation}
($g$ is real positive). The effective action for $V$ is given in (\ref{3.35})
\begin{equation}
\label{appcont1}
\CS_\theta[V]\ =\ \CE[V]\,-{\rme^{-\rmi\theta}\over 2}\,\int
\rmd^d\rr\,V(\rr)^2 
\end{equation}
and for large $V$ is dominated by the last term $\int V^2$.  The
steepest descent integration path for $V(\rr)$ in $\mathbb{C}$ is such
that (at least for large $|V|$)
\begin{equation}
\label{argVtheta}
\arg(V)\ =\ {\pi+\theta\over 2}
\end{equation}
(see figure). Thus it turns anti-clockwise from the positive real axis for
$\theta=-\pi$ to the imaginary axis for $\theta=0$ to the negative real axis
for $\theta=\pi$.
\begin{figure}[h]
\begin{center}
\includegraphics[width=6truecm]{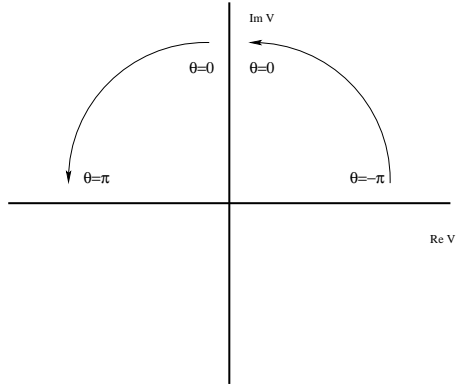}
\caption{Integration path for $V$ as a function of $\theta=\arg(b)$}
\label{appcont2}
\end{center}
\end{figure}

For general $\theta$ the instanton $\vinst_\theta$ is an extremum of
$\CS_\theta[V]$.  For negative coupling ($\theta=\pm\, \pi$), the
instanton is known. It is the solution found and studied in
\cite{DavidWiese1998}, $\vinst_{\pm\pi}=\vinst$; it is real and
negative; it lies on the steepest descent integration path given by
\Eq{argVtheta}.  Let us start from the case $\theta=\pi$, i.e.\ $b$
lies above the discontinuity along the negative real axis, and look at
what happens when $\theta\to 0$.  From the solution for the instanton
at $\theta=\pi$, $\vinst(\rr)$, the instanton for general $\theta <
\pi$ , $\vinst_\theta(\rr)$, is obtained by analytic continuation from
real $\rr$ to complex $\rr$.  Indeed, we know from
\cite{DavidWiese1998} that for a general $V(\rr)$, under a scale
transformation
\begin{equation}
\label{sacling4r}
V(\rr)\ \to\ V_\lambda(\rr)\ =\ \lambda^{{2D\over 2-D}}V(\lambda\rr)
\end{equation}
the two  terms in the effective action $\CS[V]$ scale respectively as
\begin{equation}
\label{scalincEand F}
\CE[V_\lambda]\ =\ \lambda^{{2D\over 2-D}}\,\CE[V]
\qquad;\qquad
\int_\rr V_\lambda^2\ =\ \lambda^{{2\epsilon\over 2-D}}\int_\rr V^2
\ .
\end{equation}

If we assume that the instanton $\vinst(\rr)$, obtained in
\cite{DavidWiese1998} for real $\rr$, can be continued analytically to
complex $\rr$'s,  then it is enough to take instead of a real scaling
factor $\lambda$ a complex phase factor
\begin{equation}
\label{phaselambda}
\lambda\ =\ \rme^{\rmi\omega}
\end{equation}
and to choose as phase
\begin{equation}
\label{omegatheta}
\omega^+_\theta\ =\ (\pi-\theta){2-D\over 2(D-\epsilon)}
\end{equation}
to know that
\begin{equation}
\label{appcont3} \left(\rme^{\rmi\omega}\right)^{{2D\over
2-D}}\,\vinst\left(\rme^{\rmi\omega}\rr\right)
\end{equation}
with $\omega =\omega_{\theta }^{+}$ an extremum of $\CS_\theta[V]$.
Therefore the instanton for $\theta<\pi$ is
\begin{equation}
\label{lf51}
V^\mathrm{inst}_{\theta}(\rr)\ =\ \rme^{\rmi(\pi-\theta) {D\over
(D-\epsilon)}} 
\,V^\mathrm{inst}\left(\rme^{\rmi (\pi-\theta){2-D\over
2(D-\epsilon)}}\rr\right) 
\ .
\end{equation}
It is clear that this instanton $\vinst_\theta$ is now a complex field
configuration, since it involves both a``global Wick rotation" in
$\rr$ space and the multiplication by a global phase.
In particular for $\theta=0$ (real positive coupling constant) the instanton is
\begin{equation}
\label{lf52}
\vinst_{\theta=0}(\rr)\ =\ \rme^{\rmi \pi {D\over (D-\epsilon)}}
\,\vinst\left(\rme^{\rmi \pi{2-D\over 2(D-\epsilon)}}\rr\right)
\ \ .
\end{equation}
The same argument applies for $\theta\in \left]-\pi,0\right]$. If we start from
the same real instanton at $\theta=-\pi$ and deform it to $\theta=0$
we obtain another instanton, which is the complex conjugate
configuration $\overline{V}^\mathrm{inst}_{\theta=0}$ of the instanton
obtained by starting from $\theta=\pi$.

\begin{figure}[h]
\begin{center}
\includegraphics[width=8.truecm]{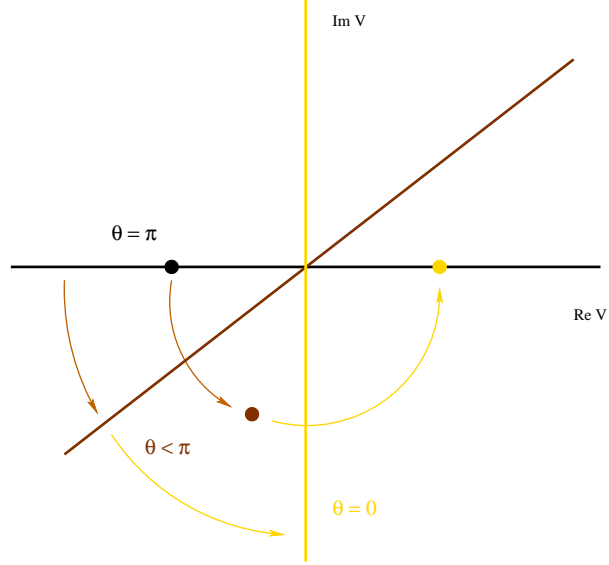} \caption{The
integration path for $\widehat{V}$ and the instanton
$\widehat{V}^{\mathrm{inst}}$ for $\theta=\pi$, $0<\theta<\pi$ and
$\theta=0$ (we have set $\epsilon=0$).} \label{appcont10}
\end{center}
\end{figure}

How is $\vinst_{\theta}$ located with respect to the steepest descent
integration path over $V(\rr)$?  Rather than considering the
functional integral over $V(\rr)$ for real $\rr$'s, it is more
convenient to rotate the space coordinate $\rr$ in the complex
plane. This is equivalent to deforming the time contour in the complex
plane when dealing with time correlation functions in finite
temperature QM and FT. Consider as bulk-space coordinates $\hat
\rr$ defined as
\begin{equation}
\label{appcont4}
\hat\rr\ =\ \rme^{\rmi \omega^+_\theta}\,{\rr}
\end{equation}
and make the change of variables in the functional integral for $V$
\begin{equation}
\label{appcont5}
V(\rr)\ \to\ \widehat{V}(\hat{\rr})=V(\rr)\quad,\qquad \hat{\rr}\ \text{real}
\ .
\end{equation}
The functional measure becomes $\widehat\CD[\widehat{V}]$, the measure
for $\widehat{V}$, and from \eq{normDVb} it is normalized so that
\begin{equation}
\label{appcont6}
\int\widehat\CD[\widehat{V}]\,\exp\left({\rme^{-\rmi\theta}\over 2g}
\,\rme^{-\rmi
d\omega^+_\theta}\int\rmd^d\hat{\rr}\,{\widehat{V}(\hat{\rr})}^2\right)=1
\ .
\end{equation}
The steepest descent integration path for $\widehat{V}$ is therefore
the line with argument $\widehat{\Omega}^+_\theta$
\begin{equation}
\label{argvhat} \arg(\widehat{V})\ =\ \widehat{\Omega}^+_\theta\ =\
{\pi+\theta+d\,\omega^+_\theta\over 2}\ =\ \pi\,+\,{\pi-\theta\over
2}\,{D\over D-\epsilon}
\end{equation}
(remember that $\epsilon=2D-d(2-D)/2$). In the new variable
$\widehat{V}$ the instanton differs from the original real instanton
$\vinst$ by a pure phase (see Eq.~(\ref{lf51}))
\begin{equation}
\label{appcont7}
\widehat{V}_\theta^{\mathrm{inst}}(\hat r)\ =\
\rme^{\rmi(\pi-\theta){D\over D-\epsilon}}\vinst(\hat r) \ .
\end{equation}
Since $\vinst$ is real and negative, its argument is $\pi$ and
therefore $\widehat{V}_\theta^{\mathrm{inst}}(\hat r)$ has a fixed
argument (independent of $\hat r$)
$\widehat{\Omega}^{\mathrm{inst}}_\theta$ given by
\begin{equation}
\label{appcont8}
\arg\left(\widehat{V}_\theta^{\mathrm{inst}}(\hat r)\right)
\ =\ 
\widehat{\Omega}^{\mathrm{inst}}_\theta
\ =\ 
\pi\,+\,{(\pi-\theta)}\,{D\over D-\epsilon}\ .
\end{equation}
For $\theta<\pi$,  $\widehat{\Omega}^{\mathrm{inst}}_\theta$ is larger
than $\widehat{\Omega}^+_\theta$ 
\begin{equation}
\label{appcont9}
\widehat{\Omega}^{\mathrm{inst}}_\theta\ >\
\widehat{\Omega}^+_\theta\quad\text{ if}\quad
\theta\,<\,\pi\quad\text{and}\quad \epsilon\,<\,D \ . 
\end{equation}
This means that the instanton lies below the integration path for
$\widehat{V}$, see figure \ref{appcont10}.%
When $\theta\to\pi$ the integration path becomes the real axis (with
the standard orientation from $-\infty$ to $=\infty$), while the
complex instanton $\widehat{V}^{\mathrm{inst}}$ becomes the real (and
negative) instanton $\vinst$.

\begin{figure}[h]
\begin{center}
\includegraphics[width=10.truecm]{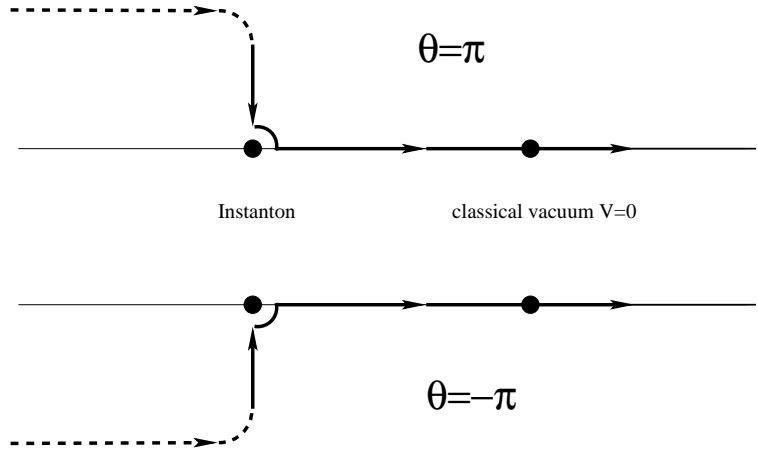}
\end{center}
\caption{Steepest descent integration paths for the unstable mode for
$\theta=\pm\pi$.}    
\label{fig2}
\end{figure}

With this result the steepest descent integration prescription for the
unstable mode around the instanton at $\theta=\pi$ is fixed. We boldly
denote by $V$ this mode.  The integration path from $0$ to $-\infty$
has to start from ${V}=0$ (the real vacuum, minimum of the action
$\CS[V]$), go on the real negative axis up to the instanton $\vinst<0$
which is a local extremum of $\CS[V]$ with action
$\CS^{\mathrm{inst}}=\CS[\vinst]>\CS[V=0]=0$, then ``turn right'' (see
figure \ref{fig2}) in the upper half complex plane in order for
the action to continue to increase, while leaving the instanton below,
then go to $-\infty$.
The first part of the contour (from $\vinst$ to $\infty$) contributes
only to the real part of the partition function $Z$ for negative
coupling (and is dominated by the classical vacuum $V=0$). The second
part of the contour (from $-\infty$ to $\vinst$)  contributes to
the imaginary part of $Z$; in fact the dominant contribution to the
imaginary part comes from half the Gaussian integral in the imaginary
direction at the instanton 
\begin{equation}
\label{appcont12}
\int_{\vinst+\rmi\infty}^{\vinst} \rmd V 
\end{equation}
and gives a factor
\begin{equation}
\label{appcont13}
-\,\rmi\,\,{1\over 2}\left|\det\left(\CS''[\vinst]\right)\right|^{-1/2}
\end{equation}
when compared to the  full contribution of the Gaussian integral
(the factor $-\rmi$ comes from the integration path, the factor $1/2$
from the fact that we integrate the unstable mode from $\rmi\infty$ to
$0$, not from $\rmi\infty$ to $-\rmi\infty$).
 
The same argument shows that for $\theta\to -\pi$ the instanton is
\emph{above} the real axis.  Therefore for the unstable mode the
steepest-descent integration path has to stay below the real axis, with factor
\begin{equation*}
\label{lf62}
\rmi \,{1\over 2}\,\left|\det\left(\CS''[\vinst]\right)\right|^{-1/2}\ .
\end{equation*}
These are the results used in Section \ref{unsteigmode}.

\section{SAW:  $D=1$ SAM versus O($n=0$) field theory}
 \label{sec:o(n)vspolymer}
In this appendix we recall the ``Laplace-De Gennes'' equivalence
between the zero-component O($n=0$) $\phi^4$ field theory and the
(weakly) self-avoiding walk model, which corresponds to the case $D=1$
for the SAM model. The first part of this appendix (sect. 1--3) is
basically textbook material, recalled here to fix the notations and
the normalizations.  We then show that the standard instanton calculus
for the O($n=0$) model gives the same result as our instanton
calculus for the SAM model in the special case $D=1$. This provides an
important check for the consistency of our method.

\subsection{Free Field and Brownian walk:}
\label{subsec:RW} 
The action for the scalar free
field in $d$-dimensional space is  (\emph{note the factor $1/4$}, which is
not the most commonly used 
\footnote{Two choices of normalizations are convenient for polymers:
Here we use $S_{0}[\phi]=\int_{\rr} \frac{1}{4} \left(\nabla_{\rr}\phi
\right)^{2}+ \frac{m^{2}}{2}\phi_{}^{2}$, which corresponds to having
the polymer action
$S_{\mathrm{polymer}}=\int_{\xx}\frac{1}{2}\left(\nabla \rr(\xx)
\right)^{2}$. The other convenient choice is to use
$S_{0}[\phi]=\int_{\rr} \frac{1}{2} \left(\nabla_{\rr}\phi
\right)^{2}+ \frac{m^{2}}{2}\phi^{2}$, which corresponds to
$S_{\mathrm{polymer}}=\int_{\xx}\frac{1}{4}\left(\nabla \rr(\xx)
\right)^{2}$. This is the choice most often taken, see
e.g.\cite{Wiesehabil}. Here we employ the first choice, since we want to
use the most convenient normalization for the polymer action. We also
note that for both choices, $\rme^{-Lm^{2}}$, with $L$ the length of
the polymer is the weight in the Laplace-De Gennes transform.})
\begin{equation}
    S_{0}[\phi]\ =\ \int_{\rr}{1\over 4} (\nabla_{\rr}\phi)^{2} \ +\ 
    {m^{2}\over 2}(\phi)^{2}
    \label{eq:5.1}
\ .
\end{equation}
The 2-points correlation function is
\begin{equation}
G_0(\rr_1,\rr_2;m^2)\ =\ 
    \langle\phi(\rr_{1})\phi(\rr_{2})\rangle_{0}
    \ =\     \langle\rr_{1}|{1\over-\Delta/2+m^{2}}|\rr_{2}\rangle
% \nonumber\\ &\ =\ &
\ =\     \int_{0}^\infty \rmd L\ \rme^{-Lm^{2}}\,
    \langle\rr_{1}|\rme^{L\Delta/2}|\rr_{2} \rangle
     \label{eq:5.2}
\ ,
\end{equation}
and is the Laplace transform with respect to $L$ of the heat kernel
$K(\rr_1, \rr_2;L)=\langle\rr_{1}|\rme^{L\Delta/2}|\rr_{2}\rangle$, which  
admits the random-walk representation
\begin{equation}
\label{rwfrfldeq}
    K(\rr_1, \rr_2; L)\ =\ 
\langle\rr_{1}|\rme^{L\Delta/2}|\rr_{2}\rangle
    \ =\        \int_{\begin{smallmatrix}
   \ \\
   \ \\
{\scriptscriptstyle \rr(0)=\rr_{1}}\\
{\scriptscriptstyle \rr(L)=\rr_{2}}
\end{smallmatrix}
} 
    \!\! \!\!
     \CD[\rr]\,\rme^{-\int_{L}{1\over 2}(\dot\rr)^{2}\rmd s}
\end{equation}
with $\dot r = dr/ds$.  To check the normalization, use the
semi-classical estimate for the small-$L$ limit of the r.h.s.\ of
\eq{rwfrfldeq}, $K\simeq\exp{(-|\rr|^2/2L)}$ and check that $2\partial
K/\partial L=\Delta K$.  In particular at coinciding points
\begin{equation}
    \langle\phi(\rr_{1})^{2}\rangle_{0}\ =\ 
    \langle\rr_{1}|{1\over-\Delta/2+m^{2}}|\rr_{1}\rangle
    \ =\ \int_{0}^\infty \rmd L\ \rme^{-Lm^{2}}\,
    \langle\rr_{1}|\rme^{L\Delta/2}|\rr_{1}\rangle
    \label{eq:5.3}
\end{equation}
and the heat kernel at coinciding points admits the closed random-walk
representation:
\begin{equation}\label{apo4}
 K(\rr_1, \rr_1; L)\ =\ 
\langle\rr_{1}|\rme^{L\Delta/2}|\rr_{1}\rangle
    \ =\        \int_{\begin{smallmatrix}
   \ \\
   \ \\
{\scriptscriptstyle \rr(0)=\rr(L)=\rr_{1}}
\end{smallmatrix}
} 
    \!\!\! \!\!\!\hskip-2.em
     \CD[\rr]\,\rme^{-\int_{L}{1\over 2}(\dot\rr)^{2}\rmd s}
     \ =\ \left.\CZ_{0}(\rr_{1})\right|_{D=1;L}
\ .
\end{equation}
It is the partition function for a closed 1-dimensional membrane
(i.e.\ a closed polymer, or a loop) with length $L$, attached to the
point $\rr_{1}$.  Similarly for the one-loop connected diagram
\begin{eqnarray}
\label{BubbC6}
\frac{1}{2} \langle\phi(\rr_{1})^{2}\phi(\rr_{2})^{2}\rangle_{0}^{\mathrm{conn}} \
& = &\
\frac{1}{2}\left[
\langle\phi(\rr_{1})^{2}\phi(\rr_{2})^{2}\rangle_{0}\,-\,
\langle\phi(\rr_{1})^{2}\rangle_{0}\langle\phi(\rr_{2})^{2}\rangle_{0}\right]
     \ =\ 
\left[\langle\rr_{1}|{1\over-\Delta/2+m^{2}}|\rr_{2}\rangle\right]^{2}
\nonumber\\
&=& 
\int_{0}^\infty \rmd L\ \rme^{-Lm^{2}}\,
\int_{0}^L\,\rmd L_{1}\,
\int \CD[\rr]\,
\delta^d(\rr(0)-\rr_{1})\delta^d(\rr(L_{1})-\rr_{2}) 
\,
\rme^{-\int_{L}{1\over 2}(\dot\rr)^{2}\rmd s}
\nonumber\\
    &=& \int_{0}^\infty \rmd L\ \rme^{-Lm^{2}}\,
     L^{-1}\,\left.\CR_{0}^{(2)}(\rr_{1},\rr_{2})\right|_{D=1,L}
\ .
\end{eqnarray}
This means that the first derivative w.r.t.\ $m^{2}$ of the l.h.s. of
\eq{BubbC6} is the Laplace transform of the 2-point correlation
function $\CR_{0}^{(2)}$ for a free closed loop with length $L$.
Similarly connected correlators of a product of $N$ $\phi^2$ operators
are associated to $N$-point correlation functions for the closed loop.

\subsection{SAW and O($n=0$) field theory:} 
It is well-know that this equivalence extends to the Edwards Model,
defined with the normalizations as in (\ref{ActionSA}).  The
O($n$)-invariant $\phi^{4}$ model is defined by the action
\begin{equation}\label{apo5}
S[\vec\phi]\ =\ \int \rmd^d \rr\, {1\over 
4}\left(\nabla_r\vec\phi\right)^2\, 
+\,{t\over 2}({\vec\phi}^{\ 2})\,+\, {b\over 8}({\vec\phi}^{\ 2})^2
\end{equation}
with $\vec\phi$ a $n$-component real vector field:
\begin{equation}
\vec\phi\ =\ \{\phi^a; a=1,n\}
\label{eq:vecphin}
\end{equation}
$t=m^{2}$ is the squared mass, $b$ is the coupling constant.

The model is equivalent to the Edwards model of polymer with (weak)
2-chain repulsive contact interaction, as defined by the model of
(\ref{ActionSA}) for the $D=1$ case.  The equivalence holds thanks to
the very same Laplace transform between correlation functions as in
the free case ($b=0)$. It is valid to all orders in perturbation
theory, that is as an asymptotic series expansion for small $b$.  The
operator $\frac{1}{2}\vec \phi^{2} (\rr)$ is represented by a
$\delta$-distribution, or more formally\footnote{Note the factor of
$\frac{1}{2}$. Intuitively it is there to compensate for the fact that
the two fields of $\vec \phi^{2}$ can be contracted in two different
ways. This also leads to a relative factor of 4 between the
interactions (\ref{ActionSA}) and (\ref{apo5}).}
\begin{equation}\label{op-ident}
\frac{1}{2} \vec \phi^{2}(\rr) \ \leftrightarrow \ \int \rmd^D \xx\,
\delta (\rr(\xx)-\rr)\ .
\end{equation}
For instance, for the 1-point correlators we have 
\begin{equation}\label{apo6}
\lim_{n\to 0}{2\over n} \left<\frac{1}{2}\vec\phi(\rr_{1})^2\right>
\ =\ \int_{0}^\infty\rmd L\,\rme^{-Lt}\,\CZ(\rr_{1})\big|_{L}
\end{equation}
and for the two-points correlators
\begin{equation}\label{apo7}
\lim_{n\to 0}{2\over n}\left<\frac{1}{2}\vec\phi(\rr_{1})^2\,
\frac{1}{2}\vec\phi(\rr_{2})^2\right> \ =\ \int_{0}^\infty\rmd
L\,\rme^{-Lt}\,L^{-1}\CR^{(2)}(\rr_{1},\rr_{2})\big|_{L}
\end{equation}
etc\dots 

\subsection{Instanton calculus and large-orders for the O($n$) field theory}
We now recall the principle of instanton calculus and large-order
estimates for the scalar O($n$) $\phi^4$ field theory, following the
standard references.  This example is useful, since in the limit of
$n=0$ it describes polymers, i.e.\ $D=1$ membranes.  The field is a
$n$-component real vector field $\vec\phi(r)$,
$\vec\phi=(\phi^{a};a=1,n)$.  The action is the O($n$)-invariant
$\phi^{4}$ action
\begin{equation}\label{apo8}
S[\vec\phi]\ =\ \int \rmd^d r\, {1\over 
2}\left(\nabla_r\vec\phi\right)^2\, 
+\,{t\over 2}({\vec\phi}^{\ 2})\,+\, {b\over 8}({\vec\phi}^{\ 2})^2
\end{equation}
$t=m^{2}$ is the squared mass, $b$ the coupling constant.  We are
interested in observables $O[\vec\phi]$ which are local monomials in
$\phi$ with degree $d_{o}$ in $\phi$, the simplest being 
the energy operator $E$
\begin{equation}\label{apo9}
E[r_{1}]\ =\ (\vec\phi)^{2}(r_{1})\qquad;\qquad \hbox{degree}(E)=d_{E}=2
\ .
\end{equation}
The expectation value for the observable $O$ is given by the standard
formula
\begin{equation}\label{apo10}
\langle O\rangle\ =\ 
\int\CD[\vec\phi]\,O[\vec\phi]\,\rme^{-S[\vec\phi]}
\Big/\int\CD[\vec\phi]\,\rme^{-S[\vec\phi]}
\ .
\end{equation}
We are interested in the large orders of the perturbative series
expansion in $b$.  As we have seen, we have to use the dispersion
relation in the complex-$b$ plane and consider what happens for small
$b$ close to the negative real axis, where $b$ is complex and its
argument is close to $\pm\pi$.  Therefore we rescale the field
\begin{equation}\label{apo11}
\phi\ =\ |b|^{-1/2}\varphi
\qquad;\qquad
\mathrm{with}\qquad \theta\ =\ \mathrm{Arg}(b)
\ .
\end{equation}
This gives
\begin{equation}
\label{rescalesact}
S[\phi]\ =\ {1\over |b|}\,S_{\theta}[\varphi]
\qquad;\qquad
S_{\theta}[\varphi]\ =\ \int \rmd^d r\, {1\over 
2}\left(\nabla_r\varphi\right)^2\, 
+\,{t\over 2}({\varphi}^{\ 2})\,+\, {\rme^{\rmi\theta}\over 8}({\varphi}^{\ 2})^2
\end{equation}
so that
\begin{equation}\label{apo13}
\langle O[\phi]\rangle
\ =\ 
|b|^{-{d_{O}\over 2}}\,\langle O[\varphi]\rangle_{\theta,|b|}
\ =\ 
|b|^{-{d_{O}\over 2}}\,
{\int\CD[\varphi]\,O[\varphi]\,\rme^{-S_{\theta}[\varphi]/|b|}
\over
\int\CD[\varphi]\,\rme^{-S_{\theta}[\varphi]/|b|}}
\ .
\end{equation}
For small positive $b$ the functional integral is dominated by the 
constant classical saddle point
\begin{equation}\label{apo14}
\vec\varphi_{0}(\rr)\ =\ \vec\varphi_{0}\ =\ 0
\end{equation}
constant and absolute minima of $S_{\theta}$ for $\theta=0$.  The
functional integral is in fact well defined as long as
$-\pi<\theta<+\pi$.  Now along the cut at $b<0$, that is for
$\theta=\pm\pi$, another real extremum of the action $S_{\theta}$
becomes important, the instanton
\begin{equation}\label{apo15}
\vec\varphi_{i}\ =\ \vec\varphi_{i}(r;r_{0},\vec u_{0})\ =\ 
\varphi_{i}(r-r_{0})\vec u_{0}
\ .
\end{equation}
The instanton is characterized by its position $r_{0}$ in space, and
its orientation $\vec u_{0}$ in the internal $n$-dimensional space
($\vec u_{0}$ being a unit vector in $\mathbb{R}^n$).
$\varphi_{i}(r)$ is the real finite-action solution of the equation
\begin{equation}\label{apo16}
-\Delta_{r}\varphi_{i}\,+\,t\,\varphi_{i}\,-\,{1\over 
2}\,(\varphi_{i})^{3}
\ =\ 0
\ ,
\end{equation}
which is rotationally invariant around the origin (i.e.\ depends only
on $|r|$) and is non-zero except for $|r|\to\infty$ (or equivalently
$\ge 0$, this is enough to define it uniquely).

The contribution of the instanton in the functional integral is at one
loop proportional to
\begin{equation}\label{totototo}
\rme^{-{1\over |b|}S_{\theta}(\varphi_{i})}
\,
\left[\mathrm{Det'}\left[S_{\theta}''(\vec\varphi_{i})\right]\right]^{-1/2}
\ .
\end{equation}
The measure for the collective coordinate $\rr_{0}$ (the position of
the instanton) is easily obtained, since the metric is
\begin{equation}\label{apo18}
h_{ab}\ =\ 
{1 \over 2\pi |b|}\, \int\rmd^d\rr\,
{1\over d}\,
(\partial_{a}\vec\varphi_{i}\partial_{b}\vec\varphi_{i})
\ =\ 
{1 \over 2\pi |b| d}\, \|\vec\nabla\varphi_{i}\|_{\scriptscriptstyle 2}^2
\,\delta_{ab}
\ .
\end{equation}
Hence the measure is
\begin{equation}\label{apo19}
\rmd\mu(\rr_{0})\ =\ \rmd^d\rr_{0}\,\left[{1\over 2\pi |b|
d}\|\vec\nabla\varphi_{i}\|_{\scriptscriptstyle 2}^2\right]^{d\over 2}
\ .
\end{equation}
The measure for the internal coordinate $\vec u_{0}$ (the orientation)
is
\begin{equation}\label{apo20}
\rmd\mu(\vec u_{0})\ =\ \rmd \vec u_{0}\, \left[{1\over 2\pi |b|}
\int\rmd^d \rr\,{\vec\varphi_{i}}^{\,2}\right]^{{n-1\over 2}} \ =\
\left[{1\over 2\pi |b|}\|\vec\varphi_{i}\|^2_{\scriptscriptstyle
2}\right]^{{n-1\over 2}}
\end{equation}
with $\rmd \vec u_{0}$ the standard measure on the unit sphere ${\cal
S}_{n-1}$ in ${\mathbb R}^n$.  For the O($n$) invariant observables
which are of interest to us, the integration over ${\cal S}_{n-1}$ can
be performed explicitly, giving the factor $\Omega_{n}$ (the volume of
the unit sphere in $\mathbb{R}^n$)
\begin{equation}\label{apo21}
\Omega_{n}
\ =\ 
\int_{{\cal S}_{n-1}}\hskip -1.em\rmd \mu(\vec u_{0})
\ =\ 
\mathrm{Vol}({\cal S}_{n-1})
\ =\ 
{2\, \pi^{{n\over 2}}\over\Gamma(n/2)}
\ .
\end{equation}
Note that this volume factor $\Omega_{n}$ vanishes as $n$ when $n\to
0$.  However, since O($n$) invariant observables such as
$\vec\varphi^2/n$ behave in the background of the instanton
$\vec\varphi_{i}$ as $\varphi_{i}^{\,2}/n$ and are therefore of order
$1/n$, the factors $n$ and $n^{-1}$ compensate to give a finite $n\to
0$ limit.

Now the Hessian is a $n\times n$ matrix in internal space,
\begin{equation}\label{apo22}
S''_{\!ab}(\rr,\rr')\ =\ {\partial
S_{\theta}(\varphi)\over\partial\varphi_{a}(\rr)\partial\varphi_{b}(\rr')}
\ =\ \delta_{ab}(-\Delta+t)\,
-
\,\left(\delta_{ab}\,\vec\varphi^2/2+\varphi^a \varphi^b\right)
\ .
\end{equation}
Hence for the instanton background $\vec\varphi_{i}=\varphi_{i}\,\vec
u_{0}$ the Hessian can be written as the product of the longitudinal
operator
\begin{equation}\label{apo23}
S''_{\!l}\ =\ -\Delta+t
-
{3\over 2}\varphi_{i}^2
\end{equation}
times  $n-1$ transverse operators
\begin{equation}\label{apo24}
S''_{\!\perp}\ =\ -\Delta+t
-
{1\over 2}\varphi_{i}^2\ .
\end{equation}
$S''=S''_{\!l}\otimes\left(S''_{\!\perp}\right)^{n-1}$ .  Note that
$S''_{\!l}$ has $d$ zero modes ${\psi^{\scriptscriptstyle
0}_{l}}_{\mu}=\partial_{\mu}\varphi_{i}$ and that $S''_{\!\perp}$ has
one zero mode ${\psi^{\scriptscriptstyle 0}_{\perp}}=\varphi_{i}$, so
$S''$ has $d+n-1$ zero modes.  Thus
\begin{equation}\label{apo25}
{\det}'(S'')\ =\ {\det}'(S''_{\!l})\,{\det}'(S''_{\!\perp})^{n-1}
\end{equation}
and in the $n=0$ limit 
\begin{equation}\label{apo26}
{\det}'(S'')\big|_{n=0}\ =\
{{\det}'(S''_{\!l})\over{\det}'(S''_{\!\perp})}\ .
\end{equation}
$S''_{\!l}$ has one (and only one) eigenvector $\psi^{-}_{l}$ with
negative eigenvalue $\lambda_{l}^{-}< -2t$. Therefore $\det
S''_{\! l}<0$
\footnote{This is true for $d<4$, for $d=4$ there is no instanton
solution for $t>0$, for $t=0$ there is an instanton with an additional
zero mode ${\psi^{\scriptscriptstyle 0}_{l,\,
s}}=(\rr\nabla_{\rr}+1)\varphi_{i}$ corresponding to the scale
invariance of the massless theory under scale transformation
$\varphi(\rr)\to\lambda\varphi(\lambda\rr)$.  The instanton at $d=4$
is obtained from the instanton for $d<4$ by taking the limit $d\to 4$,
$t\propto 4-d$.}  .

\begin{figure}[h]
\label{contour-1}
\begin{center}
\includegraphics[width=15.cm]{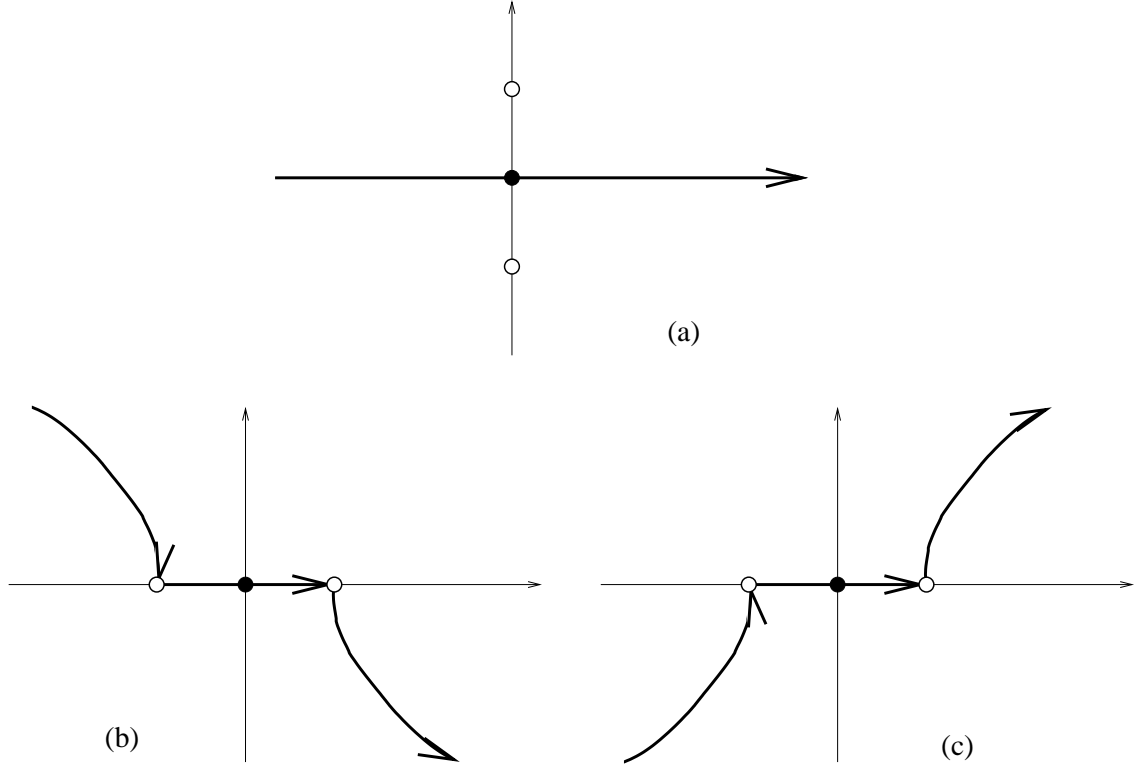} 
\caption{
Steepest descent integration path for the global $\varphi$ variable for $\theta=0$ (a), $\theta=\pi$ (b) and $\theta=-\pi$ (c). The black dot represents the classical vacua $\varphi=0$ and the white dots the 2 instanton saddle-points $\varphi=\pm\varphi_i$.
}  
\end{center}
\end{figure}

To understand which signe must be chosen for the square-root of the
negative determinant $[{\det(S'')}]^{-1/2}$ ( $+\rmi$ or $-\rmi$ ?)
we have to consider the steepest-descent integration path for the
global $\varphi$ variable. It has to go from $\arg \varphi =
\pi\pm\theta/4$ to $\mp\theta/4$, hence for $\theta=+\pi$ it is as
depicted on Fig.~\ref{contour-1}.

Hence the instanton contributes to the imaginary part of an observable
by a coefficient
\begin{equation}\label{apo27}
\mp{\rmi\over 2} \left|{\det}'(S''_{\!l})\right|^{-1/2}
\qquad{\mathrm{if}}\qquad \theta=\pm\pi
\ .
\end{equation}
The rest goes into the real part together with the contribution of the
classical vacuum $\varphi_{0}$.

Putting things together, and using Eq.~\eq{OfromInst} we obtain for
the imaginary part at $b<0$ of the e.v. of the O($n$) invariant
observable $\cal O$ and for $\mathrm{arg}(b)=\theta=\pm\pi$
\begin{eqnarray}
\mathrm{Im}\langle{O}\rangle
&\ =\ &
\mp\,{1\over 2}
\,
|b|^{-d_{\scriptscriptstyle{O}}/2}\,
\rme^{-{{1\over |b|}\big(S[\varphi_{i}]-S[\varphi_{0}]\big)}}
\,
\left|{{\det}'(S''_{\!l}[\varphi_{i}])\,
\big({\det}'(S''_{\!\perp}[\varphi_{i})]\big)^{n-1}
\over\det (S''[\varphi_{0}])^n}\right|^{-{1\over 2}}
\ \times
\nonumber\\
&&\hskip 2em
\,
\left[ {1\over 2\pi |b| d}\|\vec\nabla \varphi_{i}\|_{\scriptscriptstyle{2}}^2\right]^{d\over 2}
\,
\left[{1\over 2\pi |b|}
\|\varphi_{i}\|_{\scriptscriptstyle{2}}^2\right]^{{n-1\over 2}}
\,
\Omega_{n}
\,
\int \rmd^d\rr_{\scriptscriptstyle{0}}\,
\left( 
O[\varphi_{i}[\rr_{0}]]-O[\varphi_{0}]
\right)\nonumber \\
&&
\end{eqnarray}
while of course
\begin{equation}\label{apo28}
\mathrm{Re} (\langle{O}\rangle\big) \ =\ O(\varphi_{0}) \ .
\end{equation}
In particular for $n=0$ and $O$ a product of energy operators $E$ defined as
\begin{eqnarray}
O_1[\rr_1]\ &= &\ \lim_{n\to 0}\, {1\over n}\,\big(\vec\phi(\rr_1)\big)^2 \\
O_2[\rr_1,\rr_2]\ &= &\ \lim_{n\to 0} \, {1\over n}\,
\big(\vec\phi(\rr_1)\big)^2 \big(\vec\phi(\rr_2)\big)^2
\end{eqnarray}
and omitting the $i$ subscript for ``instanton'', we obtain
\begin{eqnarray}
\mathrm{Im}\langle O_1\rangle & = & \mp\,\half \,\ |b|^{-{d+1\over 2}}
\, \rme^{-{1\over |b|}S_i} \, \left| {{\det}'S''_{\! l}\over {\det}'
S''_{\!\perp}} \right|^{-{\half}} \, (2\pi)^{{1-d\over 2}} \,
\left[{\|\nabla\varphi_i\|_{\scriptscriptstyle{2}}^2 \over
d}\right]^{{d\over 2}} \, \|\varphi_i\|_{\scriptscriptstyle{2}}
\\
\mathrm{Im}\langle O_1[\rr_1,\rr_2]\rangle & = & \mp\,\half \,\
|b|^{-{d+3\over 2}} \, \rme^{-{1\over |b|}S} \, \left| {{\det}'S''_{\!
l}\over {\det}' S''_{\!\perp}} \right|^{-{\half}} \! (2\pi)^{{1-d\over
2}} \, \left[{\|\nabla\varphi\|_{\scriptscriptstyle{2}}^2 \over
d}\right]^{{d\over 2}} \!\! \|\varphi\|_{\scriptscriptstyle{2}}^{-1}
\,\varphi^2\!\star\varphi^2(\rr_1-\rr_2)
\nonumber \\
&&\label{apo29}
\end{eqnarray}
$\star$ denotes the usual convolution product $f\star g(\rr)=\int
\rmd\rr' f(\rr')g(\rr'+\rr)$.  Note also a few useful results
\begin{eqnarray}
\label{lf106} &&\varphi_0\ =\ 0\ \rightarrow \ S[\varphi_0]\ =\ 0
\\
&& S[\varphi_i]\ =\ {1\over 8}\,\int \rmd^d\rr\, \varphi_i^4
\\
&&\int \rmd^d\rr\, \big(\nabla\varphi_i\big)^2\ =\ 
-\,t\int \rmd^d\rr\,\varphi_i^2\,+\,{1\over 2} \int \rmd^d\rr\,\varphi_i^4\ .
\end{eqnarray}

\subsection{Instanton calculus for the SAW model of polymers}
Since the Edwards model for SAW is the inverse Laplace transform w.r.t. $t=m^2$ of the O($n$) model, instanton calculus must take into account this transformation and is (slightly) modified, as is explained here.

The action for the SAW model is
\begin{equation}\label{lf108}
S[\phi]=\int {1\over 4}(\nabla\phi)^2+{t\over 2}\phi^2-{|b|\over 8}(\phi^2)^2
\ .
\end{equation}
We have by inverse Laplace transform, for a closed polymer of length $L$
\begin{eqnarray}
\label{lf109} \CZ(\rr;L)\ &=& \ \int_{-\rmi\infty}^{+\rmi\infty}{\rmd
t\over 2\rmi\pi}\,\rme^{Lt}\, \langle O_1(\rr;t)\rangle
\\
\CR(\rr_1,\rr_2;L)\ &=& \ L \int_{-\rmi\infty}^{+\rmi\infty}{\rmd
t\over 2\rmi\pi}\,\rme^{Lt}\, \langle O_2(\rr_1,\rr_2;t)\rangle \ .
\end{eqnarray}
So we consider the effective action
\begin{equation}
\label{lf110} \CS[\phi,t]=\int \left[{1\over 4}(\nabla\phi)^2+{t\over
2}\phi^2-{|b|\over 8}(\phi^2)^2\right]\ -\ t\,L \ .
\end{equation}
To factorize $|b|$ and $L$ we must rescale both $\phi$, $t$ and $\rr$
with
\begin{equation}
\label{rescpolyb}
\phi(\rr)=|b|^{{-d\over 2(d-2)}}\,L^{{-1\over d-2}}\,\sqrt{2}\,\varphi(\rr')
\qquad,\qquad
\rr=|b|^{{1\over d-2}}\,L^{{1\over d-2}}\,\rr'
\qquad,\qquad
t=|b|^{{-2\over d-2}}\,L^{{-2\over d-2}}\tau
\end{equation}
The action becomes
\begin{equation}
\label{lf111} \CS[\phi,t]\ =\ 2\left[|b|L^\epsilon\right]^{-{2\over
d-2}}{\cal S}[\varphi,\tau] \qquad;\qquad\epsilon={4-d\over 2}\ ,
\end{equation}
and the effective action is now
\begin{equation}
\label{lf112} {\cal S}[\varphi,\tau]\ = \ S[\varphi]-{\tau\over 2}
\qquad;\qquad S[\varphi]\ = \ \int \left( {1\over 4}
(\nabla\vec\varphi)^2+{\tau\over 2}\vec\varphi^2- {1\over 4}
\big(\vec\varphi^2\big)^2 \right) \ .
\end{equation}
The effective coupling constant is
\begin{equation}
\label{lf113}
b_{\mathrm{eff}}={1\over 2}\,\Big[|b|L^\epsilon\Big]^{{2\over d-2}}
\end{equation}
instead of $|b|$.

The instanton is given by the saddle point equations 
\begin{equation}
\label{lf115}
\vec\varphi\,=\,\varphi\,\vec u_0
\qquad;\qquad
-{\Delta\over 2}\varphi+\tau\varphi-\varphi^3  =  0
\qquad;\qquad
\int \varphi^2  =  1 
\end{equation}
and the Hessian is
\begin{equation}
{\cal S}''\ =\ \left[
\begin{array}{cc }
      S''& \vec\varphi   \\
  \vec\varphi^t   &  0
\end{array}
\right]
\label{lf116}\ .
\end{equation}
It can still be separated into its transverse part, which is $n-1$
times the transverse operator $S''_{\!\perp}$
\begin{equation}
\label{lf117}
S''_{\!\perp}=-{\Delta\over 2}+\tau-\varphi^2
\end{equation}times the longitudinal part
\begin{equation}
{\cal S}''_{\!l}\ =\ 
\left[
\begin{array}{cc }
      S''_{\! l}& \varphi   \\
     \varphi^t   &  0
\end{array}
\right]
\qquad;\qquad
S''_{\! l}=-{\Delta\over 2}+\tau-3\varphi^2
\label{lf118}
\ ,
\end{equation}
which has $d$ translational zero modes, namely the
$\Psi_\mu=\left[\begin{array}{ c}
      \psi_\mu\\
     0 
\end{array}
\right]$ since $\varphi \cdot\psi_\mu = \int\varphi\partial_\mu\varphi
= \half\int\partial_\mu\varphi^2=0$.  It is then easy to show,
denoting by $P_0$ the projector onto the kernel of $S''_{\! l}$
generated by its zero modes, and defining the ``inverse'' of $S''_{\!
l}$ as
\begin{equation}
\label{lf119}
\left[{1\over S''_{\! l}}\right]' \ =\ \big(S''_{\! l}+P_0\big)^{-1}\,-\,P_0
\end{equation}
that\footnote{ This is an application of the general formula for the
determinant of bloc square matrices $ \det\left[\begin{array}{cc }
 A     &   B \\
 C     &   D
\end{array}
\right]
= \det\left[A\right]\cdot\det\left[D-CA^{-1}B\right]
$
}
\begin{eqnarray}
\label{lf120}
{\det}'{\cal S}''_{\!l}&=&
\det
\begin{pmatrix}
      S''_l+P_0 & \varphi   \\
      \varphi^t &  0 
\end{pmatrix}
\ =\ 
\det\big(S''_l+P_0\big)\ \det\big(-\varphi^t(S''_l+P_0)^{-1}\varphi\big)
\nonumber\\
&=&
-\, {\det}'{S}''_{\!l}\,\left(\varphi\cdot\left[{1\over S''_{\! l}}\right]' 
\cdot\varphi\right)
\ .
\end{eqnarray}
But it turns out that
\begin{equation}
\label{lf121}
\left[{1\over S''_{\! l}}\right]' \cdot\varphi\ =\ 
-\,{1\over 2\tau}\,(\rr\cdot\nabla+1)\varphi
\ .
\end{equation}
Indeed, one can check that $\rr\nabla\varphi$ as well as $\varphi$ are
orthogonal to $\psi_\mu$ and an explicit calculation shows that
\begin{equation}
\label{lf123}
\left[-{\Delta\over 2}+\tau-3\varphi^2\right]
(\rr\cdot\nabla+1)\varphi
\ =\ -2\tau\,\varphi
\ .
\end{equation}
It follows that the additional factor in the determinant (which comes
from the integration over $\tau$ at the saddle point) is
\begin{equation}
\label{lf124} -\, \left(\varphi\cdot\left[{1\over S''_{\! l}}\right]'
\cdot\varphi\right) \ =\ {1\over
2\tau}\,\int\varphi(\rr\nabla+1)\varphi
\ =\ 
-\,{d-2\over 4\tau}\,\int\varphi^2
\ =\ -\,{d-2\over 4\tau}
\end{equation}
$\tau$ is $>0$ if $d<4$, so this factor is negative for $2<d<4$, but
the integration path over $\tau$ is also imaginary (it goes from
$-\rmi\infty$ to $+\rmi\infty$).  Thus the integration over $\tau$
gives the factor
\begin{equation}
\label{lf125}
(2\pi)^{-1}\, |b|^{{-2\over d-2}}\,L^{{-2\over d-2}}
\end{equation}
(coming from the measure ${\rmd\tau\over 2\rmi\pi}$) times
\begin{equation}
\label{lf126}
\left| {2\pi\over {b_{\mathrm{eff}} {d-2\over 4\tau} }}\right|^{1/2}
\end{equation}
coming from the Gaussian integration. This gives
\begin{equation}
\label{lf127}
\left[\pi{d-2\over\tau}\right]^{-1/2} |b|^{{-1\over d-2}} L^{{-d\over 2(d-2)}}
\ .
\end{equation}
For the observable $\phi^2/n$ the integration over the zero modes gives
\begin{equation}
\label{lf128}
2 |b|^{{-d\over d-2}} L^{{-2\over d-2}}\,{\Omega_n\over n}\, \int \varphi^2
\ =\ 
2 |b|^{{-d\over d-2}} L^{{-2\over d-2}}
\end{equation}
times the measure and determinant factors\begin{equation}
\label{lf129}
\left[{\|\varphi\|^2\over 2\pi b_{\mathrm{eff}}}\right]^{-1/2}
\,
\left[{\|\nabla\varphi\|^2\over 2\pi b_{\mathrm{eff}}d}\right]^{d/2}
\,
\left|{{\det}'S''_{l}\over {\det}'S''_\perp}\right|^{-1/2}
\ =\ 
\left[\pi |b|^{{2\over d-2}} L^{{4-d\over d-2}}\right]^{{1-d\over 2}}
\,
\left[{\|\nabla\varphi\|^2\over d}\right]^{d/2}
\,
\left|{{\det}'S''_{l}\over {\det}'S''_\perp}\right|^{-1/2}
\ .
\end{equation}
Putting things together we get
\begin{equation}
\label{ImZpolyfinal}
\mathrm{Im}\CZ(\rr;L)\ =\ \mp\,{1\over 2}
\,L^{{-d\over 2}}\,\big[|b|L^\epsilon\big]^{{-2d\over d-2}}
\rme^{-{(2S-\tau)\big[|b|L^\epsilon\big]^{{-2\over d-2}} }} \, 
\left| {\det}' S''_{\!l} \over {\det}' S''_{\!\perp} \right|^{-\half}\,
\left[{\|\nabla\varphi\|_{\scriptscriptstyle{2}}^{2}\over \pi d}\right]^{{d\over 2}}
\left({d-2\over 4 \tau}\right)^{-\half}
\ .
\end{equation}
Remember that $|b|L^\epsilon$ is dimensionless.
We show below that with our normalizations and the equation for the instanton.
\begin{equation}
\label{lf130}
\tau=(4-d)S\qquad;\qquad \|\vec\nabla\varphi\|^2=2dS
\ .
\end{equation}
This simplifies slightly \eq{ImZpolyfinal}
\begin{equation}
\label{lf131}
\mathrm{Im}\CZ(\rr;L)\ =\ \mp\,{1\over 2}
\,L^{{-d\over 2}}\,\big[|b|L^\epsilon\big]^{{-2d\over d-2}}
\rme^{-{(d-2)S\,\big[|b|L^\epsilon\big]^{{-2\over d-2}} }} \, 
\left| {\det}' S''_{\!l} \over {\det}' S''_{\!\perp} \right|^{-\half}\,
\left[{2S\over \pi }\right]^{{d\over 2}}
\left[{4(4-d)S\over d-2}\right]^{\half}
\ .
\end{equation}

\subsection{$d\to 4$ limit and scale invariance} 
The O($n$) model at
$d=4$ ($\epsilon=0$) becomes scale invariant and the instanton has an
additional zero mode associated with dilations. For the SAW model
nothing special occurs when $\epsilon\to 0$ (as far as global scale
transformations are concerned). Here we show in detail how the
dilation zero mode is absorbed in the transformation O($n$) $\to$ SAW,
and the form of the instanton and of the large-order results at
$\epsilon=0$.

We first derive a few exact results.  If $\varphi$
is the instanton, solution of $(-{\Delta\over
2}+\tau)\varphi-\varphi^3=0$ and the Hessian is
$S''_{\!l}=-{\Delta\over 2}+\tau-{3}\varphi^2$
then\begin{eqnarray}\label{apo31} \left[{1\over S''_{\!l}}\right]'
\varphi& = & -{1\over
2\tau}(\rr\nabla+1)\varphi \\ 
\left[{1\over S''_{\!l}}\right]' \varphi^3 & = & -\half\varphi
\end{eqnarray}
\begin{eqnarray}\label{apo32}
\left(\varphi^3\right\vert \left. (\rr\nabla+1)\varphi\right) 
& = &
\int \varphi^3(\rr\nabla+1)\varphi
= \int \left (1+{\scriptstyle{1\over 4}}\rr\nabla\right)\varphi^4 
= {\displaystyle{4-d\over 4}}\int\varphi^4
\\
 & = & -2\tau \left(\varphi^3 \right.\left[{1\over S''_{\!l}}\right]'
\left.  \varphi\right) = -2\tau \left(\varphi \right. \left[{1\over
S''_{\!l}}\right]' \left.\varphi^3 \right)
\\
& = & \tau \left(\varphi \vert\varphi \right) = \tau \int \varphi^2 \
.
\ .
\end{eqnarray}
The instanton action is
\begin{equation}
\label{lf143} S = \int {1\over 4} (\nabla\varphi)^2+{\tau\over
2}\varphi^2-{{1\over 4}}\varphi^4 = \int \half \varphi (-{\Delta\over
2}+\tau)\varphi -{{1\over 4}}\varphi^4 = {{1\over 4}}\int \varphi^4 
\ .
\end{equation}
Hence
\begin{equation}
\label{lf144}
\int \varphi^2 = {4-d\over \tau} S
\qquad ; \qquad
\int (\nabla\varphi)^2 =  2\,d \,S
\qquad ; \qquad
\int \varphi^4 = 4 S
\ .
\end{equation}
Now remember that $\tau$ is fixed by the normalization
$\int\varphi^2=1$.  In the limit $d\to 4$, we must take the limit
$\tau\to 0$ to get the finite-action instanton.  The general solution
of the equation $-\Delta\varphi=2\varphi^3$ at $d=4$ is
\begin{equation}
\label{lf145}
\varphi(\rr)_{d=4} = {2\rr_0 \over \rr^2+\rr_0^2}
\qquad;\qquad\hbox{with corresponding action}\qquad
S_{d=4} = {2\over 3} \pi^2\ .
\end{equation} 
$\rr_0$ is the instanton size. 
The size is arbitrary for the massless $d=4$ theory by scale invariance.
For $\rr\gg\rr_0$, $\varphi$ satisfies the linearized equation 
$-\varphi''-{\scriptstyle{d-1\over\rr}}\varphi'+2\tau\varphi=0$
and is a Bessel function
\begin{equation}
\label{lf146}
\varphi(\rr) \propto \rr^{1-{d/2}} \mathrm{K}_{d/2-1}(\sqrt{2\tau}\rr)
\end{equation}
and is such that
\begin{equation}
\label{lf147}
\varphi(\rr)\simeq C\, \rr^{2-d}
\qquad{\mathrm{for}}\quad 
\rr_0\ll \rr \ll 1/\sqrt{\tau}
\qquad;\qquad
\varphi(\rr) \propto \rme^{-\rr \sqrt{2\tau}}
\qquad{\mathrm{for}} \quad 
 \rr \gg 1/\sqrt{\tau}
\ .
\end{equation}
To match this behavior with the large-$\rr$ behavior of the instanton
at $d=4$, the constant $C$ must behave as
\begin{equation}
\label{lf148} C(d) = 2\rr_0\,(1+{\cal O}(\varepsilon))
\qquad\hbox{with}\qquad \varepsilon=4-d \ .
\end{equation}
For fixed $\rr_0$, we must let $\tau\to 0$, but at which rate?  The
evaluate this, use the equation (\ref{lf144})  which tells us that
$\tau\int\varphi^2\simeq\varepsilon S_{d=4}$ when $\varepsilon\to 0$.
When evaluating $\int\varphi^2$ in this limit, it is easy to see that
it is the contribution of the domain $\rr_0\ll \rr \ll 1/\sqrt{\tau}$
which dominates the integral, so that
\begin{eqnarray}\label{apo33}
\int\varphi^2 & = & \Omega_d\int_0^\infty\rmd\rr\, \rr^{d-1}
\varphi(\rr)^2 \simeq 2\,\pi^2 \int_{\rr_0} ^{1/\sqrt{\tau}} \rmd\rr\,
\rr^{d-1}\, \left(C\, \rr^{2-d}\right)^2 \simeq 2\,\pi^2\,
(2\rr_0)^2\,\ln(1/\rr_0\sqrt{\tau})
\ .\nonumber \\
\end{eqnarray}
Therefore $\tau$ goes indeed to zero as $d\to 4$ according to
\begin{equation}
\label{lf149}
\varepsilon=4-d\simeq 6\, \tau\rr_0^2\,\ln\left(1/\tau\rr_0^2\right)
\ .
\end{equation}

Now if we consider the polymer, we have to keep its length $L=1$
fixed, hence $\int\varphi^2=1$. Then the instanton size $\rr_0$ has to
vanish together with $\tau$ as $\varepsilon\to 0$.
\begin{equation}
1={\varepsilon\over\tau}{S}\ \implies\
\tau\simeq{\varepsilon}{2\,\pi^2\over 3} \ \implies\
\rr_0\simeq {1\over 2\pi\sqrt{\ln(1/\varepsilon)}}
\label{lf150}
\ .
\end{equation}
Finally, we are interested in the smallest \emph{positive} eigenvalue
$\lambda^+$ of the Hessian $S''_{\!l}$ and the corresponding
eigenvector $\psi^+$.  As $\varepsilon\to 0$ we expect that
$\lambda^+\to 0$ and $\psi^+\to (\rr\nabla+1)\varphi$ the zero-mode
for scale transformations.  In this limit $\lambda^+$ can be estimated
as follows
\begin{eqnarray}
(\varphi\vert\left[{1\over S''_{\!l}}\right]' \varphi)& \simeq &
(\varphi\vert\psi^+){1\over\lambda^+} (\psi^+\vert\varphi) {1\over
(\psi^+\vert\psi^+)} \ .
\end{eqnarray}
But the l.h.s.  is equal to \begin{equation}
\label{lf151}
-{2-d\over 4\tau}(\varphi\vert\varphi)
\simeq {1\over 2 \,\tau} (\varphi\vert\varphi)
\ =\ {1\over 2 \,\tau}
\ .
\end{equation}
Using the asymptotics  obtained for $\varphi$ in the $\varepsilon\to 0$ limit
\begin{eqnarray}\label{apo34}
\varphi(\rr)\simeq 2\,\rr_0\,\rr^{-2} \ \implies\ \psi^+(\rr)\simeq
-2\,\rr_0\,\rr^{-2}& &\qquad\hbox{for} \quad \rr_0\ll \rr \ll
1/\sqrt{\tau}
\end{eqnarray}
we obtain
\begin{eqnarray}\label{apo35}
(\psi^+ \vert \varphi) \simeq -(\varphi \vert \varphi ) =-1
\qquad 
(\psi^+ \vert \psi^+) \simeq (\varphi \vert \varphi ) =1
& & \qquad\hbox{for}\quad \varepsilon\to 0
\ .
\end{eqnarray}
Hence the smallest positive eigenvalue of $\CS''_{\!l}$ vanishes as
$\varepsilon$ when $d\to 4$, as expected
\begin{equation}
\label{lf152}
\lambda^+ \simeq 2\,\tau\,\simeq\,{4\pi^2\over 3}\,\varepsilon
\qquad\hbox{for}\quad \varepsilon\to 0\ .
\end{equation}
Thus in the limit $d\to 4$ the Hessian $S''_{\! l}$ gets an additional
zero mode, so that the zero-mode subtracted determinant $\det '$ is
discontinuous at $d=4$ ($\lim_{d\to 4}{\det}'\left[S''_{\!
l}\right]\neq{\det}'\left[\lim_{d\to 4}S''_{\! l}\right]$),
but we can write in the semiclassical estimates
\begin{equation}
\label{lf153}
{\det}'\left[S''_{\! l}\right]\quad \simeq_{d\to 4}\quad \lambda^+\cdot
{\det}'\left[S''_{\! l} \big\vert_{d=4}\right] 
\ .
\end{equation} 
The singular factor $\tau^{\half}$ in eq.~\eq{ImZpolyfinal}, which
comes from the integration over $t$, is canceled by the $\lambda^+$ in
 ${\det}'S''_{\! l}$, as expected, since we cannot have IR
divergences in the semiclassical estimate at $d=4$.  We get the
IR-finite result
\begin{eqnarray}
\label{apo36}
\mathrm{Im}\, \CZ (\rr;L) & = & \mp\,\half\, L^{-2}\,\vert b\vert^{-4}\,
\rme^{-{4\,\pi^2 \over 3\,\vert b\vert}}
\left|{{\det}' S''_{\! l} \over {\det}' S''_{\!\perp} }\right|^{-\half}
{16\,\pi^2\over 9}
\ ,
\end{eqnarray}
where the IR singular terms coming from the dilation zero mode have
disappeared.  The UV divergences are contained in the two determinants
${\det}'[ S'']$.

\subsection{Comparison SAM versus O($n$) field theory for the
coefficient of the instanton}

We are now ready to check that the determinant factor for the
instanton obtained by our method (non-local SAM model) is equal to the
coefficient \ref{apo36}
obtained by instanton calculus in the O($n=0$)
local field theory.

We have already checked in \cite{DavidWiese1998} that for $D=1$ the instanton
corresponds to the instanton for the $\phi^4$ field theory. 

For $D=1$ the free energy density $\CE[V]$ is nothing but the ground
state energy $E_0$ of a particle with unit mass in the potential $V$,
i.e.\  the lowest eigenvalue $E_0$ of the Hamiltonian operator
\begin{equation}
\label{lf154}
H\ =\ -\ {\Delta_\rr\over 2}\,+\,V(\rr)
\end{equation}
acting on functions over $\mathbb{R}^d$.  We denote $\psi_0$ the
corresponding ground-state wave function.
\begin{equation}
\label{lf155}
\CE[V]\ =\ E_0 \qquad;\qquad H\,\psi_0\ =\ E_0\,\psi_0
\qquad;\qquad \parallel\psi_0\parallel^2\ =\ \int_\rr \psi_0^2\ =\ 1
\ .
\end{equation}
The saddle-point equation is (using first order perturbation theory)
\begin{equation}
\label{lf156}
V(\rr)\ =\ -\, {\delta \CE[V]\over\delta V(\rr)}\ =\ -\,\langle
\psi_0|{\delta H\over \delta V(\rr)}|\psi_0\rangle\ =\
-\,|\psi_0(\rr)|^2 
\ .
\end{equation}
So it can be written as the non-linear Schr\"odinger equation $+$ constraint
\begin{equation}
\label{lf157}
-\half\,\Delta_\rr \psi_0\,-\,E_0\,\psi_0\, -\,\psi_0^3\ =\
0\qquad;\qquad E_0\ \hbox{such that}\ \int_\rr \psi_0^2\ =\ 1 
\ .
\end{equation}
This is equivalent to the saddle-point equation for the polymer instanton
\begin{equation}
\label{lf158}
-\,\Delta\varphi\,+\,\tau\varphi\,-\,\half\,\varphi^3\,=\,0\qquad;\qquad
\tau\ \hbox{such that}\ \int\varphi^2\,=\,1
\end{equation}
by the identification
\begin{equation}
\label{lf159}
\psi_0(\rr)=\varphi(\rr)
\qquad
E_0=-\tau
\ .
\end{equation}
In particular for $L=2$ this gives $\psi_0=\varphi/2$, $\rr=\rr'$ and
$E_0=\tau/2$.

Using second order perturbation theory we have
\begin{equation}
\label{2ndDE0}
{\delta^2 E_0\over\delta V(\rr_1)\delta V(\rr_2)}\ =\ 
2\,\psi_0(\rr_1)\langle\rr_1 |\left[{1\over E_0-H}\right]'
|\rr_2\rangle\psi_0(\rr_2)
\ ,
\end{equation}
where as in a previous section the ``inverse prime'' of an Hermitian
operator means the inverse of this operator restricted to the subspace
orthogonal to its kernel
\begin{equation}
\label{lf160}
\left[{1\over E_0-H}\right]'\ =\ {1\over E_0-H+P_0}-P_0
\qquad;\qquad P_0\ =\ |\psi_0\rangle\langle\psi_0|
\ .
\end{equation}
If we denote by $\phi_0$ the operator which multiplies any function
$\psi$ by $\psi_0$
\begin{equation} \label{lf161} \psi_0\ :\
\psi\,\to\,\psi_0\psi\ ,
\end{equation} 
we rewrite \eq{2ndDE0} as\begin{equation} \label{2ndDE0bis} {\delta^2
E_0\over\delta V(\rr_1)\delta V(\rr_2)}\ =\ \langle\rr_1
|2\,\psi_0\left[{1\over E_0-H}\right]'\psi_0 |\rr_2\rangle \ .
\end{equation}
The second derivative of the effective action $\Gamma$ is thus
\begin{equation}
\label{lf162}
\Gamma''\ =\ 1\ +\ 2\,\psi_0\left[{1\over E_0-H}\right]'\psi_0
\ .
\end{equation}
If there where no problems with the zero modes, we could write
\begin{equation}
\label{lf163}
\det \left[1\ +\ 2\,\psi_0\left[{1\over E_0-H}\right]\psi_0\right]
\ =\ {\det(H-E_0-2\psi_0^2)\over\det(H-E_0)}\ =\ 
{\det(-\Delta/2-E_0-3\psi_0^2)\over\det(-\Delta/2-E_0-\psi_0^2)}~~
\end{equation}
quite similar to the ratio of determinants
\begin{equation}
\label{ }
{\det\left(\CS''_{\! l}\right)\over\det\left(\CS''_{\!\perp}\right)}
\end{equation}
but the zero modes require some care.  Let
\begin{eqnarray}\label{lf164}
A &=& H-E_0=-\Delta/2-E_0-\psi_0^2\\
B &=& H-E_0-2\psi_0^2=-\Delta/2-E_0-3\psi_0^2
\end{eqnarray}
$\psi_0$ is the zero mode of $A$ and since $\partial_\mu A=B$, 
$V_\mu=\partial_\mu\psi_0$ are the $d$ zero modes of $B$, while
$W_\mu=\psi_0\partial_\mu\psi_0$ are the $d$ zero modes of $\Gamma''$
(as can be seen by using $W_\mu=-\partial_\mu V/2$, or by direct
calculation).

We use the following simple result. Let $E$ be a hermitian operator.
If $E$ has zero modes, let $n=dim(Ker(E))$ and $\Phi_i$ a basis of
$Ker(E)$ and $K_0=\sum_i |\Phi_i\rangle\langle\Phi_i|$ the projector
on $Ker(E)$.  Let $F$ be another hermitian operator such that its
restriction to $Ker(E)$, $F'=K_0 F K_0$ is invertible.  Then
\begin{equation}
\label{lf165}
\det[E+\epsilon F]=\epsilon^n {\det}'(E) \det(F')
\qquad\hbox{with of course}\qquad
\det(F')=\det\big[\langle\Phi_i|F|\Phi_j\rangle\big]
\ .
\end{equation}  
Now we consider
\begin{equation}
\label{lf166}
\Gamma''_{\!\epsilon}=\Gamma'' +\epsilon 1\ ,
\end{equation}
and obviously
\begin{equation}
\label{lf167}
\det\big(\Gamma''_{\!\epsilon}\big)=\epsilon^d{\det}'\big(\Gamma''\big)\ .
\end{equation}
Rewrite
\begin{equation}
\label{lf168}
\Gamma''_{\!\epsilon}=1+\epsilon-2\psi_0{1-P_0\over H-E_0+\alpha P_0}\psi_0
\end{equation}
(this does not depend on the real number $\alpha$). Then, since we now
deal with invertible operators, we have
\begin{eqnarray}
\det\big(\Gamma''_{\!\epsilon}\big) & = &
 \det\left[(1+\epsilon)-2\psi_0^2{1-P_0\over H-E_0+\alpha
 P_0}\right]\nonumber \\ & = &
 \det\left[\big((1+\epsilon)(H-E_0+\alpha
P_0)-2\psi_0^2(1-P_0)\big){1\over H-E_0+\alpha P_0}\right] \nonumber \\ 
& =& {\det\left[
(1+\epsilon)(H-E_0)-2\psi_0^2+((1+\epsilon)\alpha+2\psi_0^2)P_0\right]
\over \det\left[H-E_0+\alpha\right]} \ .
\end{eqnarray}
Obviously
\begin{equation}
\label{lf169}
\det\left[H-E_0+\alpha\right]=\alpha\,{\det}'\left[H-E_0\right]
\ .
\end{equation}
Now consider the (non-hermitian) operator in the numerator
\begin{equation}
\label{lf170}
B_\epsilon=(1+\epsilon)(H-E_0)-2\psi_0^2+((1+\epsilon)\alpha+2\psi_0^2)P_0
\ ,
\end{equation}
which is not very different from  the hermitian operator 
\begin{equation}
\label{lf171}
C_\epsilon=(1+\epsilon)(H-E_0)-2\psi_0^2
\ .
\end{equation}
If $\psi$ is a vector orthogonal to $Ker(H-E_0)$, i.e.\
$\langle\psi|\psi_0\rangle=0$ (or $P_0\psi=0$) we have
\begin{equation}
\label{lf172}
B_\epsilon\psi=C_\epsilon\psi\ .
\end{equation}
So the only difference between $B_\epsilon$ and $C_\epsilon$ is when
applied to $\psi_0$
\begin{equation}
\label{lf173}
B_\epsilon\psi_0=(1+\epsilon)\alpha\psi_0
\qquad;\qquad
C_\epsilon\psi_0=-2\psi_0^3
\ .
\end{equation}
In a basis of the eigenvectors $\psi_i$ of $H-E_0$, $B_\epsilon$ and
$C_\epsilon$ have respectively the form
\begin{equation}
\label{B&Cform}
B_\epsilon=\begin{pmatrix}
      (1+\epsilon)\alpha&    b_j \\
       0 &  d_{ij}
\end{pmatrix}
\qquad;\qquad
C_\epsilon=
\begin{pmatrix}
     a &  b_j  \\
     b_i &  d_{ij}
\end{pmatrix}
\end{equation}
with
\begin{equation}
\label{lf174}
a'=(1+\epsilon)\alpha
\qquad
a=-2\int\psi_0^4
\qquad
b_i=-2\int\psi_0^3\psi_i
\qquad
d_{ij}=(1+\epsilon)....
\ .
\end{equation}
Thus
\begin{equation}
\label{lf175}
\det\big(B_\epsilon\big)=(1+\epsilon)\alpha\det(d_{ij})
\end{equation}
while
\begin{equation}
\label{lf176}
\det\big(C_\epsilon\big)=\big(a-b\cdot d^{-1}\cdot b^t\big)\det(d_{ij})
\ .
\end{equation}
Now $B=C_{\epsilon=0}$ has $d$ zero modes, the
$V_\mu=\partial_\mu\psi_0$, hence
\begin{equation}
\label{lf177}
\det\big(C_\epsilon\big)=\epsilon^d{\det}'(B)\det\left[{\langle V_\mu
| H-E_0 | V_\nu \rangle \over \|V_\mu\|^2}\right] \ .
\end{equation}
We have
\begin{equation}
\label{lf178} \|V_\mu\|^2=\int\big(\partial_\mu\psi_0\big)^2 = {1\over
d}\,\int |\vec\nabla \psi_0 |^2\ ,
\end{equation}
and since $(H-E_0 ) V_\mu = 2 \psi_0^2 V_\mu$
\begin{equation}
\label{lf179} \langle V_\mu | H-E_0 | V_\nu \rangle = 2\int \psi_0^2
\partial_\mu \psi_0 \partial_\nu \psi_0 = {2\over d} \int \psi_0^2
\big|\vec\nabla \psi_0 \big|^2 \,\delta_{\mu\nu} \ ,
\end{equation}
since $\psi_0$ is invariant by rotation. Hence
\begin{equation}
\label{lf180} \det\left[{\langle V_\mu | H-E_0 | V_\nu \rangle \over
\|V_\mu\|^2}\right] \ =\ \left[2\,{\int \psi_0 ^2 \,|\vec\nabla \psi_0
|^2 \over \int |\vec\nabla \psi_0 |^2 } \right]^d \ .
\end{equation}
It remains to calculate the coefficient $a-b\cdot d^{-1}\cdot
b^t$. For this we use the fact that
\begin{equation}
\label{lf181} {1\over a-b\cdot d^{-1}\cdot b^t}\ =\
\langle\psi_0|{1\over C_\epsilon}|\psi_0\rangle \ ,
\end{equation}	
which follows from \eq{B&Cform}.
Now a simple calculation shows that
\begin{equation}
\label{lf182} B\big(\vec\rr\cdot\vec\nabla\psi_0+\psi_0\big)\ =\
(-\Delta/2-E_0-3\psi_0^2\big)\big(\vec\rr\cdot\vec\nabla\psi_0+\psi_0\big)\
=\ 2E_0\psi_0
\end{equation}
and since $\psi_0$ is orthogonal to the kernel of $B$ we can write
\begin{equation}
\label{lf183} \lim_{\epsilon \to  0}{1\over C_\epsilon}\psi_0\ =\ {1\over
B}\psi_0\ =\ {1\over 2E_0}
\big(\vec\rr\cdot\vec\nabla\psi_0+\psi_0\big) \ .
\end{equation}
Therefore, integrating by part and using the fact that
$\|\psi_0\|^2=\int\psi_0^2=1$ we obtain
\begin{equation}
\label{lf184}
\lim_{\epsilon=0}{1\over a-b\cdot d^{-1}\cdot b^t}\ =\ 
{1\over 2E_0}\langle\psi_0|\vec\rr\cdot\vec\nabla\psi_0+\psi_0\rangle
\ =\ {1\over 2E_0}\int\psi_0\big(\vec\rr\cdot\vec\nabla\psi_0+\psi_0\big)
\ =\ {2-d\over 4 E_0}
\ .
\end{equation}
Hence finally
\begin{equation}
\label{lf185}
{\det}'\big[\Gamma''\big]\ =\ {2-d\over 4 E_0}
\,\left[2\,
{\int \psi_0 ^2 \,|\vec\nabla \psi_0 |^2 \over \int |\vec\nabla \psi_0 |^2 }
\right]^d
{{\det}'\big[H-E_0-2\psi_0^2\big]\over{\det}'\big[H-E_0\big]}
\ .
\end{equation}
Putting this result into \eq{ImZMem} we get (using the fact that
$V=-\psi_0^2$)
\begin{equation}
\label{ImZpoly}
     \mathrm{Im} Z(b)\ = \ \mp\half\int\rmd^d\rr_0\left[
{\CV\big\|\nabla\psi_0\big\|^2\over\pi d} 
\right]^{d/2}\,  \left[{2-d\over 4 E_0}\right]^{-1/2}\,\rme^{-\CV S}\,
\left|{{\det}'B\over{\det}'A}\right|^{-1/2}
\end{equation}
to be compared to \eq{apo36}.

For this remember that we are dealing with rescaled fields and
couplings (with tildes). So we go back to the original variables by
rescaling
\begin{equation}
\label{backrescD}
\rr\to\big[|b|L^D\big]^{-{2-D\over 2(D-\epsilon)}}\,\rr
\qquad;\qquad
\CV\to|b|^{-{D\over D-\epsilon}}L^{-{D\epsilon\over D-\epsilon}}
\ .
\end{equation}
It gives for $D=1$
\begin{equation}
\label{backresc1}
\rr\to\big[|b|L\big]^{-{1\over d-2}}\,\rr
\qquad;\qquad
\CV\to|b|^{-{2\over d-2}}L^{-{4-d\over d-2}}
\ .
\end{equation}
Since $Z(b)=\int\rmd^d\rr\CZ(b)$ we obtain
\begin{equation}
\label{ImCZpoly}
\mathrm{Im} \CZ(b)\ = \ \mp\half
|b|^{-{2d\over d-2}} L^{-{d(6-d)\over2(d-2)}}\,
\left[
{\big\|\nabla\psi_0\big\|^2\over\pi d} 
\right]^{d/2}\,  \left[{2-d\over 4 E_0}\right]^{-1/2}\,
\rme^{- |b|^{-{2\over d-2}}L^{-{4-d\over d-2}}\,\Gamma}\,
\left|{{\det}'B\over{\det}'A}\right|^{-1/2}
\ .
\end{equation}
This is the same as \eq{apo36} since
\begin{equation}
\label{lf186}
E_0=-\tau\quad ,\quad B =S''_l\quad,\quad A=S''_\perp\quad,\quad \psi_0=\varphi
\quad,\quad \Gamma=E_0+\half\int\psi_0^4= (d-2)S
\end{equation}

\section{Usefull formulas for derivatives of traces and determinants}
\label{a:matrixform} To compute the Hessian matrix \eq{d2Evar} of the
variational energy $\CE_{\mathrm{var}}$ given by \eq{EvarVMr0} we
need to compute the matrix derivatives
\begin{equation}
\label{matrix2der}
{\partial^2\over\partial\JM\partial\JM}\tr\left[\JM_s^{D/2}\right]
\quad\text{and}\quad {\partial^2\over\partial\JM\partial\JM}
\left(\det\left[M_{\mathrm{var}}^{{D-2\over 2}}\JU+\JM_s^{{D-2\over
2}}\right]\right)^{-1/2} \quad\text{with}\quad\JM_s={1\over
2}(\JM+\JM^t)
\end{equation}
the symmetrized of the matrix $\JM$, at the special value
$\JM=M_{\mathrm{var}}\JU$.  To compute these derivatives it is useful
to define the matrix $\JQ_{(ij)}=e_i\otimes e_j$ as the matrix which
on the line $i$ and row $j$ is 1 and is 0
elsewhere\footnote{i.e. $\JQ_{(ij)}
=\bigl\{Q^{kl}_{(ij)}=\delta_{ik}\delta_{jl}\bigr\}$}, so that for any
matrix $\JA=\{A_{ij}\}$
\begin{equation}
	\frac{\p\JA}{\p A_{ij}} =\JQ_{(ij)}
	\quad,\quad
	\frac{\p\JA_s}{\p A_{ij}} ={1\over 2}\bigl(\JQ_{(ij)}+\JQ_{(ji)}\bigr)
\ .
\end{equation}
Using this we can compute the first derivatives
\begin{equation}
\label{ } {\partial\over \partial
A_{ij}}\tr\bigl[\JA^\alpha\bigr]=\alpha\,
\tr\bigl[\JQ_{(ij)}\JA^{\alpha-1}\bigr] \quad,\quad {\partial\over
\partial A_{ij}}\det\bigl[\JA\bigr]=\det\bigl[\JA\bigr]\,
\tr\bigl[\JQ_{(ij)}\JA^{-1}\bigr]
\end{equation}
and using 
the important formula 
\begin{equation}
	\JQ_{(ij)} \JQ_{(kl)} = \JQ_{(il)} \, \delta_{jk} \ .
\end{equation}
the second derivatives  for the trace
\begin{equation}
\label{ }
\frac{\p}{\p A_{ij}}\frac{\p}{\p A_{kl}} \tr\left[ \JA_s^\alpha
\right]\lts_{\JA=A\JU} =
	{ \alpha(\alpha-1)\over 2}\, A^{\alpha-2} \left( \delta_{ik}\delta_{jl}
	 + \delta_{jk}\delta_{il}
	 \right) 
\end{equation}
and for the determinant ($n$ being the dimension of the matrix, so
that $\det[A\JU]=A^n$)
\begin{equation}
\label{ }
\frac{\p}{\p A_{ij}}\frac{\p}{\p A_{kl}}
	\left[ \det (A^\A\JU+\JA_s^\A) \right]^{-\half}\lts_{\JA=A \JU}  =
	\frac{(2A^\A)^{-\frac n2}}{16A^2}
	\left( {\A^2}
	\delta_{ij}\delta_{kl}
	+{\A(2-\A)} (\delta_{il}
	\delta_{jk}+\delta_{ik}\delta_{jl}) \right)
\end{equation}

\if false
For $(n\times n)$-matrices $\JA=\{A_{ij}\}$, which are not necessarily
symmetric, i.e.\ $A_{ij}\not=A_{ji}$, we have for the traces
\begin{eqnarray}
\frac{\p}{\p \JA_{ij}} \tr\left( \JA^l \right)\lts_{\JA=A\JU} &=&
	 l A^{l-1} \delta_{ij}\\
\frac{\p}{\p \JA_{ij}} \tr\left( \JA_s^l \right)\lts_{\JA=A\JU} &=&
	 l A^{l-1} \delta_{ij}\\
\frac{\p}{\p \JA_{ij}}\frac{\p}{\p \JA_{kl}} \tr\left( \JA^l
\right)\lts_{\JA=A\JU} &=&
	 l(l-1) A^{l-2} \delta_{jk}\delta_{li}\\
\frac{\p}{\p \JA_{ij}}\frac{\p}{\p \JA_{kl}} \tr\left( \JA_s^l
\right)\lts_{\JA=A\JU} &=&
	 l(l-1) A^{l-2} \half\left( \delta_{ik}\delta_{jl}
	 + \delta_{jk}\delta_{il}
	 \right) 
\end{eqnarray}
and for the determinants
\begin{eqnarray}
\frac{\p}{\p \JA_{ij}}
\left[ \det \JA \right]^{-\half}\lts_{\JA=A \JU} &=&
	-\half \delta_{ij} A^{-\frac n 2 -1} \\
\frac{\p}{\p \JA_{ij}}
\left[ \det \JA_s \right]^{-\half}\lts_{\JA=A \JU} &=&
	-\half \delta_{ij} A^{-\frac n 2 -1} \\
\frac{\p}{\p \JA_{ij}}\frac{\p}{\p \JA_{kl}}
	\left[ \det \JA \right]^{-\half}\lts_{\JA=A \JU} &=&
	\left( \frac14 \delta_{ij}\delta_{kl} +\half \delta_{il}
	\delta_{jk}\right) A^{-\frac n2 -2}\\
\frac{\p}{\p \JA_{ij}}\frac{\p}{\p \JA_{kl}}
	\left[ \det \JA_s \right]^{-\half}\lts_{\JA=A \JU} &=&
	\frac14 \left(  \delta_{ij}\delta_{kl}
	+\delta_{ik}\delta_{jl}+ \delta_{il}
	\delta_{jk}\right) A^{-\frac n2 -2}\\
\frac{\p}{\p \JA_{ij}}\frac{\p}{\p \JA_{kl}}
	\left[ \det \JA_s^\A \right]^{-\half}\lts_{\JA=A \JU} & =&
	\left( \frac{\A^2} 4
	\delta_{ij}\delta_{kl}
	+\frac{\A}4 (\delta_{il}
	\delta_{jk}+\delta_{ik}\delta_{jl}) \right) A^{-\frac n2\A -2}\\
\frac{\p}{\p \JA_{ij}}\frac{\p}{\p \JA_{kl}}
	\left[ \det (A^\A\JU+\JA_s^\A) \right]^{-\half}\lts_{\JA=A \JU} & =&
	\frac{(2A^\A)^{-\frac n2}}{16A^2}
	\left( {\A^2}
	\delta_{ij}\delta_{kl}
	+{\A(2-\A)} (\delta_{il}
	\delta_{jk}+\delta_{ik}\delta_{jl}) \right)\qquad
\ .
\end{eqnarray}
\fi

\section{Instanton condenstates}\label{condensates} Here we show that
if $V(\rr)$ is the instanton potential, and $\mathfrak{S}$ its action,
we have the exact identities
\begin{equation}
\label{Vcondensate} {\langle V(\rr)\rangle}_V\,=\,-\int_\rr V(\rr)^2\
=\ -2\left(1-{\epsilon\over D}\right)^{-1}\mathfrak{S}
\end{equation}
\begin{equation}
\label{Econdensate} {\langle (\nabla\rr)^2\rangle}_V\,=\,-{d\over
2}\int_\rr V(\rr)^2\ =\ -{d}\left(1-{\epsilon\over
D}\right)^{-1}\mathfrak{S}
\ ,
\end{equation}
where the e.v.~$\langle\ \rangle_V$ refers to the auxiliary model of a
free (non-self-interacting) manifold trapped in the potential
$V(\rr)$, with action
\begin{equation}
\label{auxaction} S_V[\rr]\ =\ \int_\xx {1\over
2}\,(\nabla\rr)^2\,+\,V(\rr)
\end{equation}

The first equality in \eq{Vcondensate} follows from the instanton
equation of motion $\left<\rho\right>+V=0$ and from
\begin{equation}
\label{E-4} {\langle V(\rr(\xx_0))\rangle}_V=\int_\rr
V(\rr)\,{\langle\delta(\rr-\rr(\xx_0))\rangle}_V=\int_\rr
V(\rr)\,{\langle\rho(\rr)\rangle}_V\ , 
\end{equation}
while the second equality comes from a simple result of
\cite{DavidWiese1998}, rederived in Appendix \ref{varbound}, see
Eqs.~\eq{Slambdfa} and \eq{E2F}.

The first equality in \eq{Econdensate} follows from the equations of
motion for the auxiliary model with action \eq{auxaction} and the
instanton equation. If we made the change of variable
\begin{equation}
\label{E-5}
\rr(\xx)\,\to\ \lambda\,\rr(\xx) 
\end{equation}
in the functional integral we obtain (up to contact terms proportional
to $\delta^D(0)$ which vanishe in dimensional regularization, and
which correspond to the normal-product definition of the composite
operator $(\nabla\rr)^2={:\!(\nabla\rr)^2\!:}_0$)
\begin{equation}
\label{E-6} {\langle (\nabla\rr)^2\rangle}_V\,+\,{\langle
\rr\!\cdot\!\nabla_{\!\rr} V(\rr)\rangle}_V\ =\ 0 
\ .
\end{equation}
Now we can rewrite this second term as
\begin{equation}
\label{E-7} {\langle\rr\!\cdot\!\nabla_{\!\rr} V(\rr)\rangle}_V\ =\
\int_\rr \rr\!\cdot\!\nabla_{\!\rr} V(\rr)\
{\langle\delta(\rr-\rr(\xx_0))\rangle}_V
\end{equation}
and using the instanton equation and integrating by part we rewrite it
as
\begin{equation}
\label{E-8} {\langle\rr\!\cdot\!\nabla_{\!\rr} V(\rr)\rangle}_V\ =\
-\int_\rr \rr\!\cdot\!\nabla_{\!\rr} V(\rr)\times V(\rr)\ =\
-\,{1\over 2}\int_\rr\rr\!\cdot\!\nabla_{\!\rr} \left(V(\rr)^2\right)\
=\ {d\over 2}\int_\rr V(\rr)^2
\ .
\end{equation}
Q.E.D.  

Then we use \eq{Vcondensate} to obtain the second equality in
\eq{Econdensate}.

\section{A variational bound for the smallest (negative)
eigenvalue}
\label{varbound} 
In this section we derive a bound for the
(negative) smallest eigenvalue $\lambda_-$ of the Hessian $\CS"$ which
is associated to the unstable mode.  The basic idea is as follows: The
instability is visible by studing a rescaling of $\rr$ and
correspondingly $\xx$, $V$ and $\cal E$. The unstable mode has a
non-vanishing overlap with this dilaton, which leads to a variational
bound.

First of all, we recall the rescaling
\begin{eqnarray}\label{varbound1}
\rr &\quad \longrightarrow\quad& r_{\lambda}= \lambda \rr \\
\xx &\quad \longrightarrow\quad& x_{\lambda}= \lambda^{\frac{2}{2-D}} \xx\\
V (\rr) &\quad \longrightarrow\quad& V_{\lambda} (\rr) =
\lambda^{\frac{2D}{2-D}} V (\lambda \rr)
\ .
\end{eqnarray}
Under this rescaling the two terms of the effective action scale as
\begin{eqnarray}\label{varbound2}
{\cal E}[V] &\quad \longrightarrow\quad& {\cal E} [V_{\lambda }] = 
\lambda^{\frac{2D}{2-D}} {\cal E}[V]\\
{\cal F}[V]
&\quad \longrightarrow\quad &
{\cal F}[V_{\lambda}] = \lambda^{\frac{2\epsilon}{2-D}}  {\cal F}[V]
\ .
\end{eqnarray}
Now we consider the full effective action 
\begin{equation}\label{Slambdfa}
{\cal S} [V_{\lambda}] = {\cal E}[V_{\lambda}] + {\cal F}[V_{\lambda}]
\ .
\end{equation}
The saddle-point equations for the instanton inforce 
\begin{equation}\label{spinstres}
0 = \lambda \frac{\rmd}{\rmd \lambda} {\cal S}[V_{\lambda}
]\big|_{V=V^{\mathrm{inst}}} = \frac{2D}{2-D} {\cal
E}[V^{\mathrm{inst}}] + \frac{2\epsilon}{2-D} {\cal
F}[V^{\mathrm{inst}}] \ , 
\end{equation}
which implies 
\begin{equation}
\label{E2F}
{\cal E}[V^{\mathrm{inst}}]\ =\ -{\epsilon\over D}\,{\cal F}[V^{\mathrm{inst}}]
\end{equation}
The dilaton-mode is 
\begin{equation}\label{dilation}
\left(\lambda \frac{\rmd}{\rmd \lambda} \right)^{2} {\cal
S}[V_{\lambda}^{\mathrm{inst}}] 
= \left(\frac{2D}{2-D} \right)^{2} {\cal E}[V^{\mathrm{inst}}] +
\left(\frac{2 \epsilon}{2-D} \right)^{2} {\cal F}[V^{\mathrm{inst}}]
= \frac{4 \epsilon (\epsilon -D)}{(2-D)^{2}} {\cal F}[V^{\mathrm{inst}}]
\ .
\end{equation}
Note that it does not matter, due to (\ref{spinstres}), of how one
exactly defines the dilaton: one could use $ \lambda^{2} \rmd^2/\rmd
\lambda^{2}$ instead. 

On the other hand 
\begin{equation}\label{;lskfj}
\left(\lambda \frac{\rmd}{\rmd \lambda} \right)^{2} {\cal
S}[V_{\lambda}^{\mathrm{inst}}]  = \left(\psi \cdot 
\CS"
 \cdot \psi  \right)\Big|_{V=V^{\mathrm{inst}}}
\end{equation}
with 
\begin{equation}\label{psi}
\psi (r) = \lambda \frac{\rmd}{\rmd \lambda} V_{\lambda}^{\mathrm{inst}} (r)
\ .
\end{equation}
Expanding in eigenmodes, 
\begin{equation}\label{}
\psi \cdot 
 \CS" 
 \cdot \psi
\Big|_{V=V^{\mathrm{inst}}} = \sum_{i} \left(\psi \cdot e_{i} \right)
\lambda_{i}\left( e_{i}\cdot \psi \right) \ge \lambda_{\mathrm{min}}
\left(\psi \cdot\psi \right)
\ .
\end{equation}
Therefore we have the exact bound
\begin{equation}\label{bound}
\lambda_{\mathrm{min}} \le \frac{\left( \psi \cdot 
\CS"  \cdot \psi \right)}{ \left(\psi \cdot\psi \right)} =
\frac{\left(\lambda \frac{\rmd}{\rmd \lambda} \right)^{2} {\cal
S}[V_{\lambda}^{\mathrm{inst}}] }{ \left(
\frac{\rmd}{\rmd\lambda}V^{\mathrm{inst}}_{\lambda} (r)\cdot 
\frac{\rmd}{\rmd\lambda}V^{\mathrm{inst}}_{\lambda} (r) \right)}
\ .
\end{equation}
Using 
\begin{equation}\label{usethis}
\lambda \frac{\rmd}{\rmd\lambda} V_{\lambda} = \frac{2-D}{2D}
V + r\nabla V (r) 
\end{equation}
one obtains the still exact bound
\begin{equation}\label{resultbound1}
\lambda_{\mathrm{min}} \le - \frac{\epsilon (D-\epsilon )}{2 D^{2}}
\frac{\int_{r}V^{2} (r)}{ \int_{r} \left[V (r)+\frac{2-D}{2D} r\nabla
V (r) \right]^{2}} \ .
\end{equation}
This bound can of course not be calculated exactly, if we do not know
exactly the instanton potential $V$. However, we can use the
variational approximation for $V^{\mathrm {inst}}$ to calculate the
r.h.s.\ of (\ref{resultbound1}) approximately.  Using for $V (r) $ the
Gaussian $V (r)= \exp (-r^{2}/2)$ (all normalizations and the width
cancel at the end), one obtains
\begin{equation}\label{resultbound2}
\lambda_{\mathrm{min}} \le \lambda_{\mathrm{min}}^{\mathrm{var}} = 
\frac{- \, 2 \epsilon (D- \epsilon )}{(2-D) (2D-\epsilon)
+\epsilon^{2}} \ .
\end{equation}

\section{Normalization w.r.t. the variational mass $m_{\mathrm{var}}$
in the variational and post-variational calculations }
\label{appendixnormvar} 
In this appendix we discuss the rescaling used
in the variational and large-$d$ calculations of Sect.~\ref{s:varcal},
where all quantities are expressed in units of the variational mass
scale $m$. This rescaling is in fact quite simple and natural, but it
might become confusing in some calculations, so we present it here
carefully and thoroughly.  

\subsection{The rescaling for $\xx$, $\rr$ and $g$} 
The variational
mass $m$ satisfies the equation \eq{lf74}, which amounts to
\begin{equation}
\label{eq4mvar}
m^{D-\epsilon}=2 c_0(4\pi c_0)^{d/2}
\ ,
\end{equation}
where $c_0=c_0(D)$ is the tadpole 
\begin{equation}
\label{eq4mvar2}
c_0=(4\pi)^{-D/2}\Gamma((2-D)/2)=\includegraphics[width=1.cm]{tadpole}
\ .
\end{equation}
As in \cite{DavidWiese1998} we perform the rescalings
$\xx\to\underline{\xx}$ in $D$-space and $\rr\to\underline{\rr}$ in
$d$-space, with
\begin{equation}
\label{xprkrescal}
\xx=m^{-1}\underline{\xx}\quad,\quad\pp=m\underline{\pp}\qquad;
\qquad\rr=m^{(D-2)/2)}\underline{\rr}\quad,\quad\kk=m^{(2-D)/2}\underline{\kk}
\end{equation}
in order to set the variational mass to unity $m\to \underline{m}=1$.
In the new units the instanton potential $V$ is and its Fourier
transform are rescaled as $V\to \underline{V}$ with
\begin{equation}
\label{Vrescal}
V(\rr)=m^{D}\underline{V}(\underline{\rr})
\ .
\end{equation}
In addition we also redefine the measure over $\rr$ in $d$-space (and
the corresponding measure over $\kk$ in reciprocal space) as
\begin{equation}
\label{resmeasr} \int \rmd^d\rr\ \to\
\int_{\underline{\rr}}\qquad\text{with}\qquad\int_{\underline{\rr}}
=m^{\epsilon-D}\int \rmd^d\underline{\rr}=m^{D}\int\rmd^d\rr
\end{equation}
\begin{equation}
\label{resmeask} \int {\rmd^d\kk\over (2\pi)^d}\ \to\
\int_{\underline{\kk}}\qquad\text{with}\qquad\int_{\underline{\kk}}
=m^{D-\epsilon}\int {\rmd^d\underline{\kk}\over
(2\pi)^d}=m^{-D}\int{\rmd^d\kk\over(2\pi)^d} \ .
\end{equation}
With this new measure the definition of the Fourier transform
$\widehat{\underline{V}}$ of $\underline{V}$ in $d$-space is changed
into
\begin{equation}
\label{FTrescal'} {\widehat{\underline{V}}}(\underline{\kk})=
\int_{\underline{\rr}}\rme^{-\rmi\underline{\kk}\underline{\rr}}
{\underline{V}}(\underline{\rr})\quad;\qquad {\underline{
V}}(\underline{\rr})=
\int_{\underline{\kk}}\rme^{\rmi\underline{\kk}\underline{\rr}}
\widehat{\underline{V}}(\underline{\kk})
\end{equation}
and using (\ref{Vrescal}) the rescaling for the Fourier transform
$\widehat{V}$ of the potential $V$ is
$\widehat{V}\to\widehat{\underline{V}}$ with
\begin{equation}
\label{Vhatrescal}
\widehat{V}(\kk)=\widehat{\underline{V}}(\underline{\kk})
\ .
\end{equation}
Finally since the functional integration measure $\CD[V]$ over $V$ is
normalised by (\ref{normDVb}) which involves the measure over $\rr$
and the effective coupling constant $g$, (this is equivalent to state
that the metric $G(\delta V,\delta V)=(-\rme^{-\rmi\theta}/4\pi
g)\int_\rr \delta V(\rr)^2$ over the space of $V$ configurations
depends on $g$ and the measure over $\rr$), the rescaling of the
$\rr$-integration measure (\ref{resmeasr}) amounts to a rescaling of
the effective coupling constant $g\to\underline{g}$ with
\begin{equation}
\label{gresc}
g=m^D\underline{g}
\end{equation}
or equivalently of the original coupling constant $b\to\underline{b}$
with
\begin{equation}
\label{bresc}
b=m^{D-\epsilon}\underline{b}
\ .
\end{equation}

\subsection{Consequences} 
\subsubsection{Normalization for integrals
and distributions} With these normalizations all powers of $m$
disappear in the variational and post-variational calculations, but we
have to be careful when we perform Gaussian integrals.  Indeed
the following Gaussian integral gives
\begin{equation}
\label{gaussresc}
\int_{\underline{\kk}}\rme^{-\underline{\kk}^2 c_0}=2c_0
\ ,
\end{equation}
where $c_0$ is given by (\ref{eq4mvar}).  Indeed, we have using
(\ref{resmeask}) and the equation (\ref{eq4mvar}) for the variational
mass
\begin{equation}
\label{grespr} \int_{\underline{\kk}}\rme^{-\underline{\kk}^2
c_0}=m^{D-\epsilon}(4\pi c_0)^{-d/2}=m^{-D}(4\pi G_m)^{-d/2}=m^{-D}
2m^2G_m=2c_0 \ .
\end{equation}
One has also to take into account the fact that the Dirac distribution
in $\rr$ space is now
\begin{equation}
\label{diracres}
\underline{\delta}(\underline{\rr})=m^{D-\epsilon}\delta^d(\underline{\rr})
\qquad\text{such
that}\qquad\int_{\underline{\rr}}\underline{\delta}(\underline{\rr})=1
\ .
\end{equation}

\subsubsection{Action and Hessians} 
Once this is done, all the results
for the instanton and the large orders still hold without any factor
$m$, in particular (\ref{ImZfrakMem})-(\ref{ImCZMem}).  The effective
action $\CS$ for the potential $V$ is rescaled into
$\underline{\CS}[\underline{V}]$ given simply by
\begin{equation}
\label{Seffres}
\underline{\CS}[\underline{V}]=\CE[\underline{V}]+{1\over
2}\int_{\underline{\rr}} \underline{V}^2 \quad\text{, and is such
that}\qquad \CS[V]=m^D\underline{\CS}[\underline{V}] \ ,
\end{equation}
as well as its functional derivatives
$\CS'[V]=m^D\underline{\CS}'[\underline{V}]$,
$\CS''[V]=m^D\underline{\CS}''[\underline{V}]$, etc.
The instanton equation \eq{lf36} is still
\begin{equation}
\label{insteqresc}
\widehat{\underline{V}}(\underline{\kk})
+\left<\rme^{\rmi\underline{\kk}\underline{\rr}(\underline{\oo})}
\right>_{\underline{V}}=0
\end{equation}
and the Hessian is still given by (\ref{S2O})- (\ref{Odef}), i.e.\ (in
reciprocal space)
\begin{equation}
\label{CS"resc} {\underline{\CS}''}=\underline{\JU}-\underline{\JO}
\quad\text{with}\quad
{\underline{\JU}}_{\underline{\xx}_1,\underline{\xx}_2}
=\underline{\delta}(\underline{\xx}_1-\underline{\xx}_2) \ ,\quad
\underline{\JO}=-\CE''\quad\text{i.e.}\quad
\widehat{\underline{\mathbb{O}}}_{\underline{\kk}_1,\underline{\kk}_2}
[\underline{V}]=\int_{\underline{\xx}}\left<\rme^{\rmi\underline{\kk}_1
\underline{\rr}(\underline{\oo})}\rme^{\rmi\underline{\kk}_2\underline{\rr}
(\underline{\xx})}\right>_{\underline{V}}^{\mathrm{conn}}
\end{equation}
The logarithm of the Hessian is now
\begin{equation}
\label{Lfresc} \mathfrak{L}\ =\
\underline{\mathfrak{L}}=\log{\det}'[{\underline{\CS}''}]=
\tr\log\left[\underline{\JU}-\underline{\JQ}\right]=-\sum_{k=1}^\infty
{1\over k}\,\tr\left[\underline{\JQ}^k\right]
\end{equation}
with in particular
\begin{equation}
\label{trkresc} \tr\left[\underline{\JQ}\right]=
\int_{\underline{\kk}}\underline{\widehat{\JQ}}_{\underline{\kk},
-\underline{\kk}} \quad,\qquad
\tr\left[\underline{\JQ}^2\right]=\int_{\underline{\kk}_1}\int_{\underline{\kk}_2}\underline{\widehat{\JQ}}_{\underline{\kk}_1,-\underline{\kk}_2}\underline{\widehat{\JQ}}_{\underline{\kk}_2,-\underline{\kk}_1}
\quad,\qquad\text{etc.}  \ .
\end{equation}
Finally the zero-mode measure factor $\mathfrak{W}$ is rescaled as expected
\begin{equation}
\label{Wfresc} \mathfrak{W}=m^d\,\underline{\mathfrak{W}} \quad,\qquad
\underline{\mathfrak{W}}=\left[{1\over 2\pi
d}\int_{\rr}\left(\nabla_{\underline{\rr}}\underline{V}\right)^2\right]^{d/2}
\ .
\end{equation}

\subsubsection{Instanton} 
In particular, this gives the variational
instanton potential (obtained by replacing the e.v. in the instanton
potential ${\langle\quad\rangle}_{\underline{V}}$ by e.v. in the
quadratic potential ${\langle\quad\rangle}_{\underline{m}=1}$ in
(\ref{insteqresc}))
\begin{equation}
\label{varinstresc} \underline{{\widehat{V}}}
\!_{\mathrm{var}}^{\mathrm{\
inst}}(\underline{\kk})=-\rme^{-\underline{\kk}^2 c_0/2}
\quad\text{and by Fourier transform}\quad
{{\underline{V}}}_{\mathrm{var}}^{\mathrm{inst}}(\underline{\rr})
=-2c_0\,2^{d/2}\,\rme^{-\underline{\rr}^2/(2c_0)}\ .
\end{equation}
The instanton potential expanded in normal products w.r.t.\ the unit
variational mass (i.e. $:\quad:=:\quad:_{\underline{m}=1}$) reads
\begin{eqnarray}\label{npVexpres}
\underline{V}(\underline{\rr})&=&2c_0\sum_{n=0}^\infty{1\over 2^n
n!}\,\left({-1\over
2c_0}\right)^n\mu_n\,:\!\left(\underline{\rr}^2\right)^n\!:\\
 \mu_n&=&{1\over d(d+1)\cdots (d+n-1)}\int_{\underline{\kk}}
\left(-\underline{\kk}^2\right)^n\widehat{\underline{V}}(\underline{\kk})
\rme^{-{\underline{\kk}^2\over 2} c_0} \ .
\end{eqnarray}

\subsubsection{Renormalized quantities and counterterms} 
Finally let
us see how the UV counterterms and the renormalized action transform
under this rescaling.  In Sect.~\ref{sss:renSeff} the one-loop
counterterm $\Delta_1\CS$ for the effective action $\CS$ was found to
be given by \eq{DSV}
\begin{equation*}
\label{DSV2}
 \Delta_1\CS[V]\ =\ -\,{\mathtt{C_1}\over\epsilon}{1\over
 2}\left<(\nabla\rr)^2\right>_V-{\mathtt{C_2}\over\epsilon}
 {1\over 4} \int_\rr {V(\rr)}^2
\end{equation*}
and the renormalised effective action $\CS_{\mathrm{ren}}[V]$ was
\begin{equation*}
\label{Seffren2}
\CS_{\mathrm{ren}}[V]\ =\ \CS[V]
- \gren^{{D-\epsilon\over D}}(\mu L)^{-\epsilon} \Delta_1\CS[V]\ .
\end{equation*}
If we now perform the rescalings it is easy to see that
\begin{equation}
\label{resopDS} \left<(\nabla_\xx\rr)^2\right>_V
=m^D\left<(\nabla_{\underline{\xx}}\underline{\rr})^2\right>_{\underline{V}}
\quad\text{and}\quad \int_\rr V(\rr)^2=m^D\int_{\underline{\rr}}
\underline{V}(\underline{\rr})^2 \ .
\end{equation}
We define the rescaled renormalized couplings
$\underline{b}_{\mathrm{r}}$ and $\underline{g}_{\mathrm{r}}$ as for
the bare couplings \eq{gresc}-\eq{bresc}
\begin{equation}
\label{gbrresc} \gren=m^D\underline{g}_{\mathrm{r}}\quad
\text{and}\quad\bren=m^{D-\epsilon}\underline{b}_{\mathrm{r}} \ .
\end{equation}
Then the rescaled renormalized effective action
$\underline{\CS}_{\mathrm{ren}}[\underline{V}]$
defined by
\begin{equation}
\label{Srenresc}
\CS_{\mathrm{ren}}[V]=m^D\underline{\CS}_{\mathrm{ren}}[\underline{V}]
\end{equation}
is given by
\begin{equation}
\label{Seffren3} \underline{\CS}_{\mathrm{ren}}[\underline{V}]\ =\
\underline{\CS}[\underline{V}]
- \underline{g}_{\mathrm{r}}^{{D-\epsilon\over D}}(\mu L)^{-\epsilon}
\Delta_1\underline{\CS}[\underline{V}]
\end{equation}
with the rescaled one-loop counterterm
$\Delta_1\underline{\CS}[\underline{V}]$ given by
\begin{equation}
\label{D1Sres}
\Delta_1\CS[V]=m^\epsilon\Delta_1\underline{\CS}[\underline{V}]
\end{equation}
and using \eq{resopDS} we write
$\Delta_1\underline{\CS}[\underline{V}]$ as
\begin{equation}
\label{DSV3}
 \Delta_1\underline{\CS}[V]\ =\
 -\,{\mathtt{\underline{C}_1}\over\epsilon}{1\over
 2}\left<(\nabla_{\underline{\xx}}\underline{\rr})^2\right>_{\underline{V}}-{\mathtt{\underline{C}_2}\over\epsilon}
 {1\over 4} {\underline{V}(\underline{\rr})}^2
\end{equation}
with the rescaled counterterms
\begin{equation}
\label{Ctt12resc}
\mathtt{\underline{C}_1}=m^{D-\epsilon}\mathtt{C_1}
\quad\text{and}\quad
\mathtt{\underline{C}_2}=m^{D-\epsilon}\mathtt{C_2}
\ .
\end{equation}
Now we use the explicit perturbative results \eq{C1tt}-\eq{C2tt} for
the counterterms $\mathtt{C_1}$ and $\mathtt{C_2}$ and
Eq.~\eq{eq4mvar} for the variational mass $m$ and obtain for the
counterterms $\mathtt{\underline{C}_1}$ and $\mathtt{\underline{C}_2}$
\begin{equation}
\label{Cttresc2}
\mathtt{\underline{C}_1}
=-{\CS_D\over 2D}\left[{c_0\over d_0}\right]^{1+{d\over 2}}
\quad,\quad \mathtt{\underline{C}_2}={2\,\CS_D\over
(2-D)^2}{\Gamma[D/(2-D)]^2\over\Gamma[2D/(2-D)]}\left[{c_0\over
d_0}\right]^{1+{d\over 2}} \ ,
\end{equation}
where we remind that
\begin{equation}
\label{SDcsurd}
\CS_D=2\,\pi^{D/2}/\Gamma[D/2]\quad,\qquad
{c_0/d_0}=-2^{2-D}\,\Gamma[(2-D)/2]/\Gamma[(D-2)/2]
\end{equation}
and in the limit $d\to\infty$, $\epsilon$ fixed,
$\mathtt{\underline{C}_1}$ is of order $\CO(1)$ since
\begin{equation}
\label{C1rescas2} \mathtt{\underline{C}_1}=-\pi\,2^{3-\epsilon}\,
\rme^{-(4-\epsilon){\gamma}_{\mathrm{\scriptscriptstyle{E}}}}\,[1+\CO(1/d)]
\quad,\qquad
{\gamma}_{\mathrm{\scriptscriptstyle{E}}}\quad\text{Euler's constant}
\ .
\end{equation}
while $\mathtt{\underline{C}_2}$ is exponentially small.

\subsection{Final results} 
With these notations, the final results for
the large orders have the same form, with the unrescaled quantities
replaced by the rescales ones.

In the bulk of the paper, when we use these normalisations, we rely on
(\ref{gaussresc}) and (\ref{npVexpres}) and omit the underlinings
$\underline{\ \star\ }$ for all the quantities and the fields such as
$\xx$, $\rr$, $V$, $g$, $\CS$ etc.


\begin{thebibliography}{99}
%
\bibitem{NelPel87}
D.~R.~Nelson and L.~Peliti,
J. de Physique {\bf 48} (1987) 1085.
%
\bibitem{Jerusalem87}
%For an introduction, references and review see:
{\em Statistical Mechanics of Membranes and Surfaces},
Proceedings of the Fifth Jerusalem Winter School for Theoretical Physics
(1987), D. R. Nelson, T. Piran and S. Weinberg Eds., World Scientific,
Singapore (1989).
%
\bibitem{Jerusalem87-II}
{\em Statistical Mechanics of Membranes and Surfaces - 2nd Edition},
D. R. Nelson, T. Piran and S. Weinberg Eds., World Scientific,
Singapore (2004).
%
\bibitem{Wiesehabil}
K. J. Wiese, {\it Polymerized membranes, a review}, in ``Phase
Transitions and Critical phenomena'', vol. 19, 
C.~Domb and J.~Lebowitz, eds., Academic Press, London, 1999.
%
\bibitem{KanKarNel86}
Y. Kantor, M. Kardar and D.~R.~Nelson,
Phys. Rev. Lett. {\bf 57} (1986) 791.
%
\bibitem{KanKarNel87}
Y. Kantor, M. Kardar and D.~R.~Nelson,
Phys. Rev. {\bf A 35} (1987) 3056.
%
\bibitem{AroLub88}
J.~A. Aronovitz and T.~C.~Lubensky,
Phys. Rev. Lett. {\bf 60} (1988) 2634.
%
\bibitem{KarNel87}
M.~Kardar and D.~Nelson,
Phys. Rev. Lett. {\bf 58} (1987) 2774.
%
\bibitem{KarNel88}
M.~Kardar and D.~Nelson,
Phys. Rev. {\bf A 58} (1988) 966.
%
\bibitem{Edw65}
S.~F.~Edwards, 
Proc. Phys. Soc. Lond. {\bf 85} (1965) 613.
%
\bibitem{CloJan90}
For a general review see:\\ 
J. des Cloizeaux and G. Jannink,
{\em Polymers in solution, their modeling and structure}, 
Clarendon Press, Oxford, (1990).
%
\bibitem{DDG3}
F. David, B. Duplantier and E. Guitter, 
Phys. Rev. Lett. {\bf 72} (1994) 311.
%
\bibitem{GBU}
F. David, B. Duplantier and E. Guitter, 
{\em Renormalization Theory for the Self-Avoiding Polymerized membranes},
Saclay Preprint T/97001, cond-mat/9702136.
%
\bibitem{DavidWiese1998}
F. David and K. J. Wiese, %{\it Large Orders for Self-Avoiding Membranes}, 
Nucl. Phys. {\bf B} 535 (1998) 555-595.
%
\bibitem{Dyson52}
F. J. Dyson, Phys. Rev. 85 (1952) 631.
%
\bibitem{Lam68}
C. S. Lam, Nuovo Cimento, 55A (1968) 258.
%
\bibitem{BenderWu69}
C. M. Bender and T. T. Wu, Phys. Rev. 184 (1969) 1231.
%
\bibitem{Lip76}
L.~N. Lipatov,
JETP Lett. {\bf 24} (1976) 157;
Sov. Phys. JETP {\bf 44} (1976) 1055;
JETP Lett. {\bf 25} (1977) 104;
Sov. Phys. JETP {\bf 45} (1977) 216.
%
\bibitem{Zin82}
For a general review on instanton calculus see:\\
J.~Zinn-Justin,
{\em The principles of instanton calculus}, in
{Recent advances in field theory and statistical mechanics},
XXXIX Les Houches Summer School 1982,
J.-B. Zuber and R. Stora Eds.,
North Holland (1984) Amsterdam;\\
J. Zinn-Justin,
{\em Quantum Field Theory and Critical Phenomena}, 
Clarendon Press (1993), Oxford.
%
\bibitem{Gen72}
P.G. de Gennes, 
Phys. Lett. {\bf 38 A} (1972) 339.
%
\bibitem{Clo81} 
J.~des~Cloizeaux, 
J. de Physique {\bf 42} (1981) 635.
%
\bibitem{DavWie96}
F. David and K.~J. Wiese, 
{Phys. Rev. Lett.} {\bf 76} (1996) 4564.
%
\bibitem{WieDav97}
K.~J. Wiese and F. David, 
{Nucl. Phys.} {\bf B 487} (1997) 529.
%
\bibitem{Dup87} 
B.~Duplantier, 
Phys. Rev. Lett. {\bf 58} (1987) 2733;
and in \cite{Jerusalem87}.

%%%%%%%%%%%%%%%%%%%%%%%%%%%%%%%%%%%%%%%%%%%
% Fin de la biblio en principe
%%%%%%%%%%%%%%%%%%%%%%%%%%%%%%%%%%%%%%%%%%%
%%\bibitem{PelLei85}
%%L.~Peliti and S. Leibler,
%%Phys. Rev. Lett. {\bf 54} (1985) 1690.
%%%
%%\bibitem{PacKarNel88}
%%M.~Paczuski, M. Kardar and D.~R.~Nelson,
%%Phys. Rev. Lett. {\bf 60} (1988) 2638.
%%%
%%\bibitem{DavGui88}
%%F.~David and E.~Guitter,
%%Europhys. Lett. {\bf 5} (1988) 709.
%%%%
%%%\bibitem{Clo81} 
%%%J.~des~Cloizeaux, 
%%%J. de Physique {\bf 42} (1981) 635.
%%%%
%%\bibitem{Hwa90}
%%T. Hwa, 
%%{ Phys. Rev.} {\bf A 41} (1990) 1751%-1756.
%%%%
%%%\bibitem{Gen72}
%%%P.G. de Gennes, 
%%%Phys. Lett. {\bf 38 A} (1972) 339.
%%%%
%%\bibitem{BenMah86}
%%M. Benhamou and G. Mahoux, 
%%J. de Physique {\bf 47} (1986) 559.\\
%%see also:\\
%%B. Duplantier,
%%J. de Physique {\bf 47} (1986) 569.
%%%

%%\bibitem{Dup87} 
%%B.~Duplantier, 
%%Phys. Rev. Lett. {\bf 58} (1987) 2733;
%%and in \cite{Jerusalem87}.
%%%
%%%\bibitem{DavWie96}
%%%F. David and K.~J. Wiese, 
%%%{Phys. Rev. Lett.} {\bf 76} (1996) 4564.
%%%%
%%%\bibitem{WieDav97}
%%%K.~J. Wiese and F. David, 
%%%{Nucl. Phys.} {\bf B 487} (1997) 529.
%%%
%%%\bibitem{Dyson52}
%%%F. J. Dyson, Phys. Rev. 85 (1952) 631.
%%%\bibitem{Lam68}
%%%C. S. Lam, Nuovo Cimento, 55A (1968) 258.
%%%\bibitem{BenderWu69}
%%%C. M. Bender and T. T. Wu, Phys. Rev. 184 (1969) 1231.
%%%\bibitem{Lip76}
%%%L.~N. Lipatov,
%%%JETP Lett. {\bf 24} (1976) 157;
%%%Sov. Phys. JETP {\bf 44} (1976) 1055;
%%%JETP Lett. {\bf 25} (1977) 104;
%%%Sov. Phys. JETP {\bf 45} (1977) 216.
%%%%
%%%\bibitem{Zin82}
%%%For a general review see:\\
%%%J.~Zinn-Justin,
%%%{\em The principles of instanton calculus}, in
%%%{Recent advances in field theory and statistical mechanics},
%%%XXXIX Les Houches Summer School 1982,
%%%J.-B. Zuber and R. Stora Eds.,
%%%North Holland (1984) Amsterdam;\\
%%%J. Zinn-Justin,
%%%{\em Quantum Field Theory and Critical Phenomena}, 
%%%Clarendon Press (1989), Oxford.
%%%%

%%


\end{thebibliography}
\end{document}